\documentclass[11pt]{article}

\usepackage[utf8]{inputenc}

\usepackage{wrapfig}
\usepackage{paralist}

\usepackage{xspace}

\usepackage{xifthen}
\usepackage{tabto}
\usepackage{color}
\usepackage{cancel}
\usepackage[toc,page]{appendix}
\usepackage[colorlinks=true, allcolors=blue]{hyperref}
\usepackage{comment}

\usepackage[labelfont=bf]{caption}

\usepackage{authblk}

\usepackage{graphicx}
\usepackage{tikz}
\usepackage{pgfplots}
    \definecolor{palette-0}{RGB}{27,158,119}
    \definecolor{palette-1}{RGB}{217,95,2}
    \definecolor{palette-2}{RGB}{117,112,179}
    \definecolor{palette-3}{RGB}{231,41,138}
    \definecolor{palette-4}{RGB}{102,166,30}
    \definecolor{palette-5}{RGB}{230,171,2}
    \definecolor{palette-6}{RGB}{166,118,29}
    \definecolor{palette-7}{RGB}{102,102,102}

\usepackage{amsthm}
\usepackage{amsmath}
\usepackage{amsfonts}
\usepackage{amssymb}
\usepackage{mathtools}

\usepackage{algorithmicx}
\usepackage{algorithm} 
\usepackage{algpseudocode}

\usepackage{cleveref}
    \crefname{figure}{Figure}{Figures}
    \Crefname{figure}{Figure}{Figures}
    \crefname{section}{Section}{Sections}
    \Crefname{section}{Section}{Sections}
    \crefname{appendix}{Appendix}{Appendices}
    \Crefname{appendix}{Appendix}{Appendices}
    \crefname{definition}{Definition}{Definitions}
    \Crefname{definition}{Definition}{Definitions}
    \crefname{conjecture}{Conjecture}{Conjectures}
    \Crefname{conjecture}{Conjecture}{Conjectures}
    \crefname{lemma}{Lemma}{Lemmas}
    \Crefname{lemma}{Lemma}{Lemmas}
    \crefname{theorem}{Theorem}{Theorems}
    \Crefname{theorem}{Theorem}{Theorems}
    \crefname{corollary}{Corollary}{Corollaries}
    \Crefname{corollary}{Corollary}{Corollaries}
    \crefname{equation}{Equation}{Equations}
    \Crefname{equation}{Equation}{Equations}
    \crefname{algorithm}{Algorithm}{Algorithms}
    \Crefname{algorithm}{Algorithm}{Algorithms}

\newcommand{\pb}{probabilistic broadcast}
\newcommand{\Pb}{Probabilistic broadcast}
\newcommand{\PB}{Probabilistic Broadcast}
\newcommand{\pbab}{ProbabilisticBroadcast}
\newcommand{\pbin}{pb}

\newcommand{\pbal}{\textsf{Murmur}}

\newcommand{\pcb}{probabilistic consistent broadcast}
\newcommand{\Pcb}{Probabilistic consistent broadcast}
\newcommand{\PCB}{Probabilistic Consistent Broadcast}
\newcommand{\pcbab}{ProbabilisticConsistentBroadcast}
\newcommand{\pcbin}{pcb}
\newcommand{\Pcbin}{Pcb}
\newcommand{\pcbal}{\textsf{Sieve}}

\newcommand{\cob}{consistency-only broadcast}
\newcommand{\Cob}{Consistency-only broadcast}
\newcommand{\cobab}{ConsistencyOnlyBroadcast}
\newcommand{\cobin}{cob}
\newcommand{\Cobin}{Cob}
\newcommand{\cobal}{\textsf{Simplified Sieve}}

\newcommand{\prb}{probabilistic reliable broadcast}
\newcommand{\Prb}{Probabilistic reliable broadcast}
\newcommand{\PRB}{Probabilistic Reliable Broadcast}
\newcommand{\prbab}{ProbabilisticReliableBroadcast}
\newcommand{\prbin}{prb}

\newcommand{\prbal}{\textsf{Contagion}}

\newcommand{\contagion}{Threshold Contagion}

\newcommand{\rp}[1]{{\left(#1\right)}}
\newcommand{\qp}[1]{{\left[#1\right]}}
\newcommand{\cp}[1]{{\left\{#1\right\}}}
\newcommand{\ap}[1]{{\left\langle#1\right\rangle}}

\newcommand{\abs}[1]{{\left|#1\right|}}
\newcommand{\floor}[1]{{\lfloor #1 \rfloor}}
\newcommand{\ceil}[1]{{\lceil #1 \rceil}}

\newcommand{\bin}[3]{
    \ifthenelse
    {\equal{#3}{}}
    {\text{Bin}\qp{#1, #2}}
    {\text{Bin}\qp{#1, #2}\rp{#3}}
}

\newcommand{\pois}[2]{
    \ifthenelse
    {\equal{#2}{}}
    {\text{Poisson}\qp{#1}}
    {\text{Poisson}\qp{#1}\rp{#2}}
}

\newcommand{\bern}[2]{
    \ifthenelse
    {\equal{#2}{}}
    {\text{Bern}\qp{#1}}
    {\text{Bern}\qp{#1}\rp{#2}}
}

\newcommand{\prob}[1]{{\mathcal{P}\qp{#1}}}

\newcommand{\true}{\text{\tt True}}
\newcommand{\false}{\text{\tt False}}

\newcommand{\powerset}[2]{
    \ifthenelse
    {\equal{#2}{}}
    {\mathbb{P}\rp{#1}}
    {\mathbb{P}^{#2}\rp{#1}}
}

\DeclareMathOperator*{\argmax}{arg\,max}

\theoremstyle{definition}
    \newtheorem{definition}{Definition}
    \newtheorem{notation}{Notation}

\theoremstyle{plain}
    \newtheorem{lemma}{Lemma}
    \newtheorem{theorem}{Theorem}
    \newtheorem{corollary}{Corollary}

\newcommand{\event}[3]{
    \ifthenelse
    {\equal{#3}{}}
    {\ap{#1.\textrm{#2}}}
    {\ap{#1.\textrm{#2} \mid #3}}
}

\algnewcommand\Instance[2]{\State #1, \textbf{instance} #2}
\algnewcommand\InstanceSystem[3]{\State #1, \textbf{instance} #2, \textbf{system} #3}
\algnewcommand\Trigger[3]{\State \textbf{trigger} $\event{#1}{#2}{#3}$}

\algsetblockdefx[Implements]{Implements}{EndImplements}{}{}{\textbf{Implements:}}{}
\algsetblockdefx[Uses]{Uses}{EndUses}{}{}{\textbf{Uses:}}{}
\algsetblockdefx[Parameters]{Parameters}{EndParameters}{}{}{\textbf{Parameters:}}{}

\algsetblockdefx[Upon]{Upon}{EndUpon}{}{}[3]{\textbf{upon event} $\event{#1}{#2}{#3}$ \textbf{do}}{}
\algsetblockdefx[UponExists]{UponExists}{EndUponExists}{}{}[2]{\textbf{upon exists} $#1$ \textbf{such that} $#2$ \textbf{do}}{}
\algsetblockdefx[UponCondition]{UponCondition}{EndUponCondition}{}{}[1]{\textbf{upon} $#1$ \textbf{do}}

\algsetblockdefx[Procedure]{Procedure}{EndProcedure}{}{}[2]{\textbf{procedure} \emph{#1}($#2$) \textbf{is}}{}

\algdef{SE}[DoUntil]{DoUntil}{EndDoUntil}{\textbf{do}}[1]{\textbf{until} $#1$}
\algsetblockdefx[IfExists]{IfExists}{EndIfExists}{}{}[2]{\textbf{if exists} $#1$ \textbf{such that} $#2$ \textbf{then}}{\textbf{end if}}
\algsetblockdefx[While]{While}{EndWhile}{}{}[1]{\textbf{while} $#1$ \textbf{do}}{\textbf{end while}}
\algsetblockdefx[NTimes]{NTimes}{EndNTimes}{}{}[1]{\textbf{for} $#1$ \textbf{times do}}{\textbf{end for}}
\algsetblockdefx[For]{For}{EndFor}{}{}[3]{\textbf{for} $#1 \in #2..#3$ \textbf{do}}{\textbf{end for}}
\algsetblockdefx[ForAll]{ForAll}{EndForAll}{}{}[2]{\textbf{for all} $#1 \in #2$ \textbf{do}}{\textbf{end for}}

\newcommand{\ignore}[1]{}

\newcommand{\lhs}{\hspace{2em}&&\hspace{-2em}}

\newenvironment{tolerant}[1]{%
  \par\tolerance=#1\relax
}{%
  \par
}

\begin{document}
\pagenumbering{gobble}

\title{\textbf{Scalable Byzantine Reliable Broadcast}
}

\author[1]{Rachid Guerraoui}
\author[2]{Petr Kuznetsov}
\author[1+]{Matteo Monti}
\author[1]{Matej Pavlovic}
\author[1]{Dragos-Adrian Seredinschi}
\author[1]{Yann Vonlanthen}

\affil[1]{École polytechnique fédérale de Lausanne}
\affil[2]{LTCI, T\'el\'ecom ParisTech, Universit\'e Paris-Saclay}
\affil[+]{Corresponding author (\texttt{matteo.monti@epfl.ch})}

\date{}

\maketitle
\newpage

\begin{abstract}
Byzantine reliable broadcast is a powerful primitive that allows a set of processes to agree on a message from a designated sender, even if some processes (including the sender) are Byzantine.
Existing broadcast protocols for this setting scale poorly, as they typically build on \emph{quorum systems} with strong  intersection guarantees, which results in linear per-process communication and computation complexity.

We generalize the Byzantine reliable broadcast abstraction to the \textit{probabilistic} setting, allowing each of its properties to be violated with a fixed, arbitrarily small probability.
We leverage these relaxed guarantees in a protocol where we replace quorums with stochastic \emph{samples}.
Compared to quorums, samples are significantly smaller in size, leading to a more scalable design.
We obtain the first Byzantine reliable broadcast protocol with \textit{logarithmic} per-process communication and computation complexity.

We conduct a complete and thorough analysis of our protocol, deriving bounds on the probability of each of its properties being compromised.
During our analysis, we introduce a novel general technique we call \textit{Adversary Decorators}.
This technique allows us to make claims about the optimal strategy of the Byzantine adversary without having to make any additional assumptions.
We also introduce \contagion, a model of message propagation through a system with Byzantine processes.
To the best of our knowledge, this is the first formal analysis of a probabilistic broadcast protocol in the Byzantine fault model.   
 We show numerically that practically negligible failure probabilities can be achieved with realistic security parameters.

\end{abstract}

\hspace{1cm}
\begin{center}
\end{center}  

\clearpage
\pagenumbering{arabic}


\section{Introduction}

Broadcast is a popular abstraction in the distributed systems toolbox, allowing a process to transmit messages to a set of processes.
The literature defines many flavors of broadcast, with different safety and liveness guarantees~\cite{distributedprogramming,garay2011adaptively,HT93,ma97secure,ped02handling}.
In this paper we focus on Byzantine reliable broadcast, as introduced by Bracha \cite{bra87asynchronous}.
This abstraction is a central building block in practical Byzantine fault-tolerant (BFT) systems~\cite{cach02sintra,du18beat,at2-19-podc}.
We tackle the problem of its scalability, namely reducing the complexity of Byzantine reliable broadcast, and seeking good performance despite a large number of participating processes.

In Byzantine reliable broadcast, a designated sender broadcasts a single message.
Intuitively, the broadcast abstraction ensures that no two correct processes deliver different messages (\emph{consistency}), either all correct processes deliver a message or none does (\emph{totality}), and that, if the sender is correct, all correct processes eventually deliver the broadcast message (\emph{validity}).
This must hold despite a certain fraction of Byzantine processes, potentially including the sender.
We denote by $N$ the number of processes in the system, and $f$  the fraction of processes that are Byzantine.
Existing algorithms for Byzantine reliable broadcast scale poorly
as they typically have $O(N)$ per-process communication complexity~\cite{br85acb,ma97secure,MR97srm,toueg-secure}.
The root cause for the poor scalability of these algorithms is their use of quorums~\cite{ma97bqs,quorum-systems}, i.e., sets of processes that are large enough to always intersect in at least one correct process.
The size of a quorum grows linearly with the size of the system \cite{distributedprogramming}.

To overcome the scalability limitation of quorum-based broadcast, Malkhi \emph{et al.}~\cite{probabilisticquorums} generalized quorums to the probabilistic setting.
In this setting, two random quorums intersect with a fixed, arbitrarily high probability, allowing the size of each quorum to be reduced to $O(\sqrt{N})$.
We are not aware of any Byzantine reliable broadcast algorithm building on probabilistic quorums; nevertheless, such an algorithm could have a per-process communication complexity reduced from $O(N)$ to $O(\sqrt{N})$.
The active$_t$ protocol of Malkhi \emph{et al.}~\cite{ma97secure} uses a form of samples for an optimistic path, but relies on synchrony and has a linear worst-case complexity (that is arguably very likely to occur with only moderate amounts of faulty processes). 

\paragraph{Samples} In this paper, we present a probabilistic gossip-based Byzantine reliable broadcast algorithm having $O(\log{N})$
per-process communication and computation complexity, at the expense of $O(\log{N}/\log\log{N})$ latency. 
Essentially, we propose \emph{samples} as a replacement for quorums.
Like a probabilistic quorum,
a sample is a randomly selected set of processes.
Unlike quorums, samples do not need to intersect. 
Samples can be significantly smaller than quorums, as each sample must be large enough only to be \emph{representative} of the system with high probability.

A process can use its sample to gather information about the global state of the system.
An old Italian saying provides an intuitive understanding of this shift of paradigm: \emph{``To know if the sea is salty, one needs not drink all of it!"}
Intuitively, we leverage the law of large numbers, trading performance for a fixed, arbitrarily small probability of non-representativeness.
To get an intuition of the difference between quorums and samples, consider the emulation of a shared memory in message passing~\cite{atti1995abd}. One writes in a quorum and reads from a quorum to fetch the last value written. Our algorithms are  rather in the vein of "write all, read any". Here we would "write" using a gossip primitive and "sample" the system to seek the last value.

Throughout this paper, we extensively use samples to estimate the number of processes satisfying a set of \emph{yes-or-no} predicates, e.g., the number of processes that are ready to deliver a message $m$. 
Consider the case where a correct process $\pi$ queries $K$ randomly selected processes (a sample) for a predicate $P$.
Assume a fraction $p$ of correct processes from the whole system satisfy predicate $P$. Let $x$ be the fraction of positive responses (out of $K$) that $\pi$ collects.
By the Chernoff bound, the probability of $\abs{x - p} \geq f + \epsilon$ is smaller or equal to $\exp(-\lambda(\epsilon) K)$, where $\lambda$ quickly increases with $\epsilon$.
For sufficient $K$, the probability of $x$ differing from $p$ by more than $f + \epsilon$ can be made exponentially small.

Our algorithms use a \emph{sampling oracle} that returns the identity of a process from the system picked with uniform probability.
In a permissioned system (i.e., one where the set of participating processes is known) sampling reduces to picking with uniform probability an element from the set of processes. In a permissionless system subject to Byzantine failures and slow churn, a (nearly) uniform sampling mechanism is still achievable using gossip~\cite{brahms}.


\paragraph{Scalable Byzantine Reliable Broadcast}
Our probabilistic algorithm, \prbal,  
allows each property of Byzantine reliable broadcast to be violated with an arbitrarily small probability $\epsilon$.
We show that $\epsilon$ scales sub-quadratically with $N$, and decays exponentially in the size of the samples.
As a result, for a fixed value of $\epsilon$, the per-node communication complexity of \prbal\ is logarithmic.


We build \prbal{} incrementally, relying on two sub-protocols, as we describe next.

First, \textbf{\pbal} is a \pb\ algorithm that uses simple message dissemination to establish \emph{validity} and \emph{totality}.
In this algorithm, each correct process relays the sender's message to a randomly picked \emph{gossip sample} of other processes.
For the sample size $\Omega(\log{N})$, the resulting gossip network is a connected graph with $O(\log{N}/\log\log{N})$ diameter, with high probability~\cite{erdos-renyi,random-graph-diameter}. 
In case of a Byzantine sender, however, \pbal{} does not guarantee consistency.

Second, \textbf{\pcbal} is a \pcb\ algorithm that guarantees \emph{consistency}, i.e., no two correct processes deliver different messages.
To do so, each correct process uses a randomly selected \emph{echo sample}.
Intuitively, if enough processes from any echo sample confirm a message $m$, then with high probability no correct processes in the system delivers a different message $m'$.
{\pcbal}, however, does not ensure totality.
If a Byzantine sender broadcasts multiple conflicting messages, a correct process might be unable to gather sufficient confirmations for either of them from its echo sample, and consequently would not deliver any message, even if some correct process delivers a message.

Finally, \textbf{\prbal} is a \prb\ algorithm that guarantees validity, consistency, and totality. The sender uses \pcbal\ to disseminate a consistent message to a subset of the correct processes.
In order to achieve totality, \prbal\ mimics the spreading of a contagious disease in a population.
A process samples the system and if it observes enough other "infected" processes in its sample, it becomes infected itself.
If a critical fraction of processes is initially infected by having received a message from the underlying {\pcbal} layer,  the message spreads to all correct processes with high probability.
If a process observes enough other infected processes, it delivers.
As in the original deterministic implementation by Bracha~\cite{bra87asynchronous}, the crucial point here is that "enough" for becoming infected is less than  "enough" for delivering.
This way, with high probability, either all correct processes deliver a message or none does---\prbal\ satisfies totality.
The other two important properties (validity and consistency) are inherited from the underlying ({\pbal} and {\pcbal})  layers. 

\paragraph{Probability Analysis and Applications}
A major technical contribution of this work is  a complete, formal analysis of the properties of our three algorithms.
To the best of our knowledge, this is the first analysis of a probabilistic broadcast algorithm in the Byzantine fault model, and this turned out to be very challenging.    
Intuitively, providing a bound on the probability of a property being violated reduces to studying a joint distribution between the inherent randomness of the system and the behavior of the Byzantine adversary.
Since the behavior of the adversary is arbitrary, the marginal distribution of the Byzantine's behavior is unknown.

We develop two novel strategies to bound the probability of a property being violated, which we use in the analysis of \pcbal\ and \prbal\ respectively.

    \textbf{(1)} When evaluating the consistency of \pcbal, we show that a bound holds for every possibly optimal adversarial strategy. Essentially, we identify a subset of adversarial strategies that we prove to include the optimal one, i.e., the one that has the highest probability of compromising the consistency of \pcbal. We then prove that every possibly optimal adversarial strategy has a probability of compromising the consistency of \pcbal\ smaller than some $\epsilon$.

    \textbf{(2)} When evaluating the totality of \prbal, we show that the adversarial strategy does not affect the outcome of the execution. Here, we show that any adversarial strategy reduces to a well-defined sequence of choices. We then prove that, due to the limited knowledge of the Byzantine adversary, every choice is equivalent to a random one.


    Our analysis shows that, for a practical choice of parameters, the probability of violating the properties of our algorithm can be brought down to $10^{-16}$ for systems with thousands of processes.

In the rest of this paper, we state our system model and assumptions (\cref{section:model}), and then present our \pbal,
\pcbal, and
\prbal\ algorithms (\Cref{section:pb,section:pcb,section:prb}).
While describing our algorithms, we give high-level ideas about their analyses and refer the interested reader to the corresponding appendices containing all details including pseudocode and formal proofs.
We discuss related work in \Cref{section:related}. 




\section{Model and Assumptions}
\label{section:model}


We assume an asynchronous message-passing system where the set $\Pi$ of $N=|\Pi|$ processes partaking in an algorithm is fixed.
Any two processes can communicate via a reliable authenticated point-to-point link.

We assume that each correct process has access to a local, unbiased, independent source of randomness. We assume that every correct process has direct access to an oracle $\Omega$ that, provided with an integer $n \leq N$, yields the identities of $n$ distinct processes, chosen uniformly at random from $\Pi$. 
Implementing $\Omega$ is beyond the scope of this paper, but it is straightforward in practice.
In a system where the set of participating processes is known, sampling reduces to picking with uniform probability an element from the set of processes.
In a system without a global membership view that may even be subject to slow churn, a (nearly) uniform sampling mechanism is available in literature due to Bortnikov \emph{et al.}~\cite{brahms}.

At most a fraction $f$ of the processes are Byzantine, i.e., subject to arbitrary failures~\cite{la82byzantine}. Byzantine processes may collude and coordinate their actions. 
Unless stated otherwise, we denote by $\Pi_C \subseteq \Pi$ the set of correct processes and by $C = \abs{\Pi_C} = \rp{1 - f} N$ the number of correct processes.
We assume a static Byzantine adversary controlling the faulty processes, i.e., the set of processes controlled by the adversary is fixed at the beginning and does not change throughout the execution of the protocols.

We make standard cryptographic assumptions regarding the power of the adversary, namely that it cannot subvert cryptographic primitives, e.g., forge a signature.
We also assume that Byzantine processes are not aware of (1)~the output of the local source of randomness of any correct process; and (2)~which correct processes are communicating with each other.
The latter assumption is important to prevent the adversary from poisoning the view of the system of a targeted correct process without having to bias the local randomness source of any correct process.
Even against ISP-grade adversaries, we can implement this assumption in practice by means such as onion routing~\cite{tor} or private messaging~\cite{vuvuzela}.

\section{\PB{} with \pbal}
\label{section:pb}

In this section, we introduce the \emph{\pb} abstraction and its implementation, \pbal. 
Briefly, \pb{} ensures validity and totality.
We use this abstraction in \pcbal\ (\Cref{section:pcb}) to initially distribute the message from a sender to all correct processes.




The \pb{} interface assumes a specific sender process $\sigma$. An instance $\pbin$ of \pb\  exports two events.
First, process $\sigma$ can request through $\event{\pbin}{Broadcast}{m}$ to broadcast a message $m$.
Second, the indication event $\event{\pbin}{Deliver}{m}$ is an upcall for delivering message $m$ broadcast by $\sigma$. For any $\epsilon \in [0, 1]$, we say that \pb\ is $\epsilon$-secure if:
\begin{itemize}
    \item \textbf{No duplication}: No correct process delivers more than one message.
    \item \textbf{Integrity}: If a correct process delivers a message $m$, and $\sigma$ is correct, then $m$ was previously broadcast by $\sigma$.
    \item $\epsilon$-\textbf{Validity}: If $\sigma$ is correct, and $\sigma$ broadcasts a message $m$, then $\sigma$ eventually delivers $m$ with probability at least $(1 - \epsilon)$.
    \item $\epsilon$-\textbf{Totality}: If a correct process delivers a message, then every correct process eventually delivers a message with probability at least $(1 - \epsilon)$.
\end{itemize}

\subsection{Gossip-based Algorithm}

%
\pbal\ (presented in detail in Appendix~\ref{appendix:pbal}, \cref{algorithm:pbal}) distributes a single message across the system by means of gossip: upon reception, a correct process relays the message to a set of randomly selected neighbors. The algorithm depends on one parameter: \emph{expected gossip sample size} $G$.

Upon initialization, every correct process uses the sampling oracle $\Omega$ to select (on average) $G$ other processes to gossip with. Gossip links are reciprocated, making the gossip graph undirected.

To broadcast a message $m$, the designated sender $\sigma$
signs $m$ and sends it to all its neighbors.
Upon receiving a correctly signed message $m$ from $\sigma$ 
for the first time, 
each correct process delivers $m$ 
and forwards $m$ to every process in its neighborhood.

\subsection{Analysis Using Erd{\"o}s-Rényi Graphs}
\label{subsection:pbanalysis-erg}

%
The detailed analysis, provided in Appendix~\ref{appendix:pbal}, Section \ref{subsection:pb-nd-int-val} and \ref{subsection:pbtotality}, formally proves the correctness of \pbal\ by deriving a bound on $\epsilon$ as a function of the algorithm and system parameters.
Here we give a very high-level sketch of our probabilistic analysis of \pbal.

\paragraph{No duplication, integrity and $\epsilon$-validity (\cref{subsection:pb-nd-int-val})}
\pbal\ satisfies these properties:
\begin{itemize}
    
    \item
\textbf{No duplication}: A correct process maintains a $delivered$ variable that it checks and updates when delivering a message, preventing it from delivering more than one message.
    \item
\textbf{Integrity}: Before broadcasting a message, the sender signs that message with its private key. Before delivering a message $m$, a correct process verifies $m$'s signature. This prevents any correct process from delivering a message that was not previously broadcast by the sender.
    \item
$\epsilon$-\textbf{Validity}: Upon broadcasting a message, the sender also immediately delivers it. Since this happens \emph{deterministically}, \pbal\ satisfies $0$-validity, independently from the parameter $G$.
\end{itemize}

\paragraph{$\epsilon$-Totality (\cref{subsection:pbtotality})}
\pbal\ satisfies $\epsilon$-totality with $\epsilon$ upper-bounded by a function that decays exponentially with $G$, and polynomially increases with $f$. We prove that the network of connections established among the correct processes is an undirected Erdős–Rényi graph~\cite{erdos-renyi}. Totality is satisfied if such graph is connected. 

Erdős–Rényi graphs are well known in literature \cite{erdosrenyi} to display a connectivity phase transition: when the expected number of connections each node has exceeds the logarithm of the number of nodes, the probability of the graph being connected steeply increases from $0$ to $1$ (in the limit of infinitely large systems, this increase becomes a step function). We use this result to compute the probability of the sub-graph of correct processes being connected and, consequently, of \pbal\ satisfying totality (\cref{theorem:pbtotality}).
\section{\PCB{} with \pcbal}
\label{section:pcb}

In this section, we first introduce the \pcb{} abstraction, which allows (a subset of) the correct processes to agree on a single message from a (potentially Byzantine) designated sender. 
We then discuss \pcbal, an implementation of this abstraction.
We use \pcb\ in the implementation of \prbal\ (see \cref{section:prb}) as a way to consistently disseminate messages. \pcbal\ itself builds on top of \pb\ (see \cref{section:pb}).

\Pcb{} does not guarantee totality, but it does guarantee consistency: despite a Byzantine sender, no two correct processes deliver different messages. 
If the sender is Byzantine, however, it may happen with a non-negligible probability that only a proper subset of the correct processes deliver the message.




%
For any $\epsilon \in [0, 1]$, we say that \pcb\ is $\epsilon$-secure if it satisfies the properties of \textbf{No duplication} and \textbf{Integrity} as defined above, and:
\begin{itemize}
\item $\epsilon$-\textbf{Total validity}: If $\sigma$ is correct, and $\sigma$ broadcasts a message $m$, every correct process eventually delivers $m$ with probability at least $(1 - \epsilon)$.
\item $\epsilon$-\textbf{Consistency}: Every correct process that delivers a message delivers the same message with probability at least $(1 - \epsilon)$.
\end{itemize}

\subsection{Sample-Based Algorithm}

\pcbal\ (presented in detail in Appendix~\ref{appendix:pcbal}, \cref{algorithm:pcbal}) uses {\tt Echo} messages to consistently distribute a single message to (a subset of) the correct processes: before delivering a message, a correct process samples the system to estimate how many other processes received the same message. The algorithm depends on two parameters: the \emph{echo sample size} $E$ and the \emph{delivery threshold} $\hat E$.

Upon initialization, every correct process uses the sampling oracle $\Omega$ to select an \emph{echo sample} $\mathcal{E}$ of size $E$, and sends an {\tt EchoSubscribe} message to every process in $\mathcal{E}$.
Upon broadcasting, the sender uses the underlying \pb\ (e.g., \pbal) to initially distribute a message to every correct process.
This step does not ensure consistency, so processes may see conflicting messages if the sender $\sigma$ is Byzantine.
Upon receiving a message $m$ from \pb, a correct process $\pi$ sends an $(\texttt{Echo}, m)$ message to every process that sent an {\tt EchoSubscribe} message to $\pi$.
(Note that, due to the no duplication property of {\pb}, this can happen only once per process.) 
Upon collecting $\hat E$ $(\texttt{Echo}, m)$ messages from its echo sample $\mathcal{E}$, $\pi$ delivers $m$.
Notably, if $\pi$ delivers $m$, then with high probability every other correct process either also delivers $m$, or does not deliver anything at all, but never delivers $m' \neq m$.



\subsection{Analysis Using Adversary Decorators}
\label{subsection:pcbanalysis}

Here we present a high-level outline of the analysis of \pcbal; for a full formal treatment, see \cref{appendix:pcbal}, where we prove the correctness of \pcbal\ by deriving a bound on $\epsilon$.

\paragraph{No duplication and integrity (\cref{subsection:pcb-nd-int})}
\pcbal\ deterministically satisfies these properties the same way as \pbal\ does.

\paragraph{$\epsilon$-Total Validity (\cref{subsection:pcb-total-validity})}
Since we assume a correct sender $\sigma$ (by the premise of total validity), a bound on the probability $\epsilon$ of violating total validity can easily be derived from the probability of the underlying \pb\ failing and from the probability of some process' random echo sample having more than $E - \hat E$ Byzantine processes.


\paragraph{$\epsilon$-Consistency (Appendices \ref{subsection:pcb-preliminary-lemmas}-\ref{subsection:pcb-consistency})}
While the intuition why \pcbal\ satisfies consistency is rather simple, proving it formally is the most technically involved part of this paper. We now provide the intuition and present the techniques we use to prove it, while deferring the full body of the formal proof to the appendix.

In order for \pcbal\ to violate consistency, two correct processes must deliver two different messages (which can only happen if the sender $\sigma$ is malicious).
This, in turn, means that two correct processes $\pi$ and $\pi'$ must observe two different messages $m$ and $m'$ sufficiently represented in their respective echo samples.
I.e., $\pi$ receives $(\texttt{Echo}, m)$ at least $\hat E$ times and $\pi'$ receives $(\texttt{Echo}, m')$ at least $\hat E$ times.

Note that a correct process only sends $(\texttt{Echo}, m)$ for a single message $m$ received from the underlying \pb\ layer.
The intuition of \pcbal\ is the same as in quorum-based algorithms.
With quorums, if enough correct processes issue $(\texttt{Echo}, m)$ to make at least one correct process deliver $m$, the remaining processes (regardless of the behavior of the Byzantine ones) are not sufficient to make any other correct process deliver $m'$.
For \pcbal, this holds with high probability as long as $\hat E$ is sufficiently high and the fraction $f$ of Byzantine processes is limited.

To prove these intuitions, we first describe \cobal\ (\cref{subsection:cobal}), a strawman variant of \pcbal\ that is easier to analyze. We prove that \cobal\ guarantees consistency with strictly lower probability than \pcbal\ does (\cref{subsection:simplifiedadversarialpower}, \cref{lemma:cobalweakerthanpcbal}).
Thus, an upper bound on the probability of \cobal\ failing is also an upper bound on the probability of \pcbal\ failing.

Next, we analyze \cobal\ using a novel technique that involves modeling the adversary as an algorithm that interacts with the system through a well-defined interface (\cref{subsection:adversarialexecution}).
We start from the set of all possible adversarial algorithms and gradually reduce this set, while proving that the reduced set still includes an \emph{optimal} adversary (\cref{subsection:twophaseadversaries}).
(An adversary is optimal if it maximizes the probability $\epsilon$ of violating consistency.)
Intuitively, we prove that certain actions of the adversary always lead to strictly lowering $\epsilon$, and thus need not be considered.
For example, an adversary can only decrease its chance of compromising consistency when omitting {\tt Echo} messages.

To this end, we introduce the concept of \emph{decorators}.
A decorator is an algorithm that lies between an adversary and a system.
It emulates a system and exposes the corresponding interface to the decorated adversary.
At the same time, the decorator also exposes the interface of an adversary to interact with a system.
The purpose of a decorator is to alter the interaction between the adversary and the system. For any decorated adversary, we prove that the decorator does not decrease the probability $\epsilon$ of the adversary compromising the system.
Thus, a decorator effectively transforms an adversary into a stronger one.
Each decorator maps a set of adversaries into one of its proper subsets that is easier to analyze (\cref{appendix:decorators}).

Through a series of decorators, we obtain a tractable set of adversaries that provably contains an optimal one.
Then we derive the bound on $\epsilon$ under these adversaries (\cref{theorem:cobconsistency}).
\section{\PRB{} with \prbal}
\label{section:prb}

Our main algorithm, \prbal, implements the \prb\ abstraction.
This abstraction is strictly stronger than \pcb, as it additionally guarantees $\epsilon$-totality.
Despite a Byzantine sender, either none or every correct process delivers the broadcast message.




For any $\epsilon \in [0, 1]$, we say that \prb\ is $\epsilon$-secure if it satisfies the properties of \textbf{No duplication}, \textbf{Integrity}, $\epsilon$-\textbf{Validity}, $\epsilon$-\textbf{Consistency} and $\epsilon$-\textbf{Totality}, as already defined in previous sections.

\subsection{Feedback-Based Algorithm}
\label{subsection:prb-alg-preview}

Our algorithm implementing \prb\ is called \prbal\ and we present it in detail in Appendix \ref{appendix:prbal} (\cref{algorithm:prbal}).
It uses a feedback mechanism to securely distribute a single message to every correct process.
The main challenge of \prbal\ is to ensure totality; we prove that the other properties are easily inherited from the underlying layer with high probability.

The basic idea of \prbal\ roughly corresponds to the last stage of Bracha's broadcast algorithm \cite{bra87asynchronous}.
During the execution of \prbal\ for message $m$, processes first become \emph{ready} for $m$.
A correct process $\pi$ can become ready for $m$ in two ways: 
\begin{compactenum}
\item $\pi$ receives $m$ from the underlying consistent broadcast layer.
\item $\pi$ observes a certain fraction of other processes being ready for $m$.
\end{compactenum}
A correct process delivers $m$ only after it observes enough other processes being ready for $m$.

Unlike Bracha, we use samples (as opposed to quorums) to assess whether enough nodes are ready for $m$ (and consequently our results are all  probabilistic in nature).
Upon initialization, every correct process selects a ready sample $\mathcal{R}$ of size $R$ and a delivery sample $\mathcal{D}$ of size $D$.
Our algorithm depends on four parameters: the \emph{ready and delivery sample sizes} $R$ and $D$, and the \emph{ready and delivery thresholds} $\hat R$ and $\hat D$.

The delivery sample $\mathcal{D}$ is the sample used to assess whether $m$ can be delivered.
A correct process $\pi$ delivers $m$ if at least $\hat D$ out of the $D$ processes in $\pi$'s delivery sample are ready for $m$.

The purpose of the ready sample $\mathcal{R}$ is to create a feedback loop, a crucial part of the \prbal\ algorithm.
When a correct process $\pi$ observes at least $\hat R$ out of the $R$ other processes in $\pi$'s ready sample to be ready for $m$, $\pi$ itself becomes ready for $m$.
A direct consequence of such a feedback loop is the existence of a critical fraction of processes that, when ready for $m$, cause all the other correct processes become ready for $m$ with high probability.

We require that $\hat R/R < \hat D/D$, i.e., the fraction of ready processes $\pi$ needs to observe in order to become ready itself is smaller than the fraction of ready processed required for $\pi$ to deliver $m$.
Totality is then implied by the following intuitive argument. 
If a correct process $\pi$ delivers $m$, it must have observed a fraction of at least $\hat D/D$ other processes being ready for $m$.
As this fraction is higher than the critical fraction required for all correct processes to become ready for $m$, all correct processes will eventually become ready for $m$.
Consequently, all correct processes will eventually deliver $m$.
On the other hand, if too few processes are initially ready for $m$, such that the critical fraction is not reached, with high probability no correct process will observe the (even higher) fraction $\hat D/D$ of ready processes in its sample.
Consequently, no correct process delivers $m$.

To broadcast a message $m$, the sender $\sigma$ initially uses \pcb\ (\Cref{section:pcb}) to disseminate $m$ consistently to (a subset of) the correct processes.
All correct processes that receive $m$ through \pcb\ become ready for $m$.
If their number is sufficiently high, according to the mechanism described above, all correct processes deliver $m$ with high probability.
If only a few correct processes deliver $m$ from \pcb, with high probability no correct process delivers $m$.

\subsection{\contagion\ Game}

Before presenting the analysis of \prbal, we  overview  the \emph{\contagion} game, an important tool in our analysis.
In this game, we simulate the spreading of a contagious disease (without a cure) among members of a population, the same way the ``readiness'' for a message spreads among correct processes that execute our \prbal\ algorithm.

\contagion\ is played on the nodes of a directed multigraph, where each node represents a member of a population (whose state is either \emph{infected} or \emph{healthy}), and each edge represents a \emph{can-infect} relation.
An edge $(a, b)$ means that $a$ can infect $b$.
We also call $a$ the \emph{predecessor} of $b$.
In our \prbal\ algorithm, this corresponds to $a$ being in the ready sample of $b$.
Analogously to \prbal, a node becomes infected when enough of its predecessors are infected.

\contagion\ is played by one player in one or more \emph{rounds}.
At the beginning of each round, the player infects a subset of the healthy nodes.
In the rest of the round, the infection (analogous to the readiness for a message) propagates as follows.
A healthy node that reaches a certain threshold ($\hat R$) of infected predecessors becomes infected as well (potentially contributing to the infection of more nodes).
The round finishes when no healthy node has $\hat R$ or more infected predecessors, or when all nodes are infected.

\begin{figure}
\centering
\includegraphics[scale=0.80]{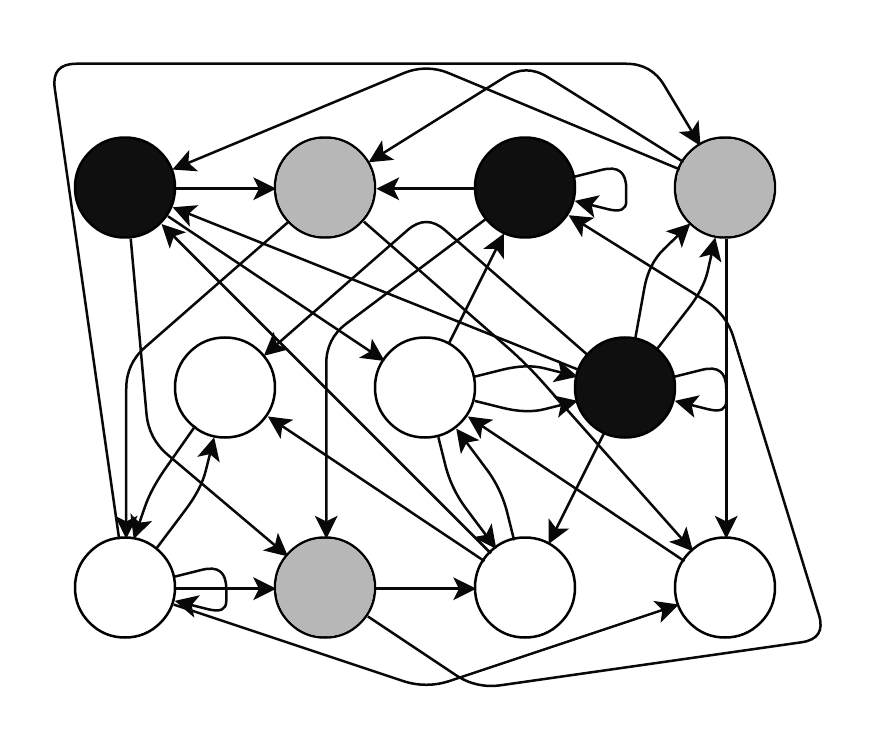}
\caption{A possible instance of a Threshold Contagion game. Black nodes represent currently infected nodes, grey nodes will get infected in the next step as at least $\hat R = 2$ of their predecessors are infected.}
\label{figure:thresholdcontagion}
\end{figure}

In the analogy with our \prbal\ algorithm, infection by a player at the start of each round corresponds to a process receiving a message from the underlying \pcb\ layer.
Infection through other nodes is analogous to observing $\hat R$ ready processes in the ready sample.

We analyze the \contagion\ game, and compute the probability distribution underlying the number of nodes that are infected at the end of a each round, depending on the number of healthy nodes infected by the player.
Applying this analysis to the \prbal\ algorithm (the adversary being the player), we obtain the probability distribution of the number of processes ready for a message, which, in turn, allows us to compute a bound on the probability of violating the properties of \prbal.
We provide all details on the \contagion\ game itself in \cref{appendix:contagion}.

\subsection{Analysis Using \contagion}
\label{subsection:prbanalysis}

Here we present an outline of the analysis of \prbal; for a full formal treatment, see \cref{appendix:prbal}.

\paragraph{No duplication and integrity (\cref{subsection:prb-nd-int})}
\prbal\ deterministically satisfies these properties the same way as our previous algorithms do.

\paragraph{$\epsilon$-Validity (\cref{subsection:prb-validity})}
Assuming a correct sender $\sigma$ (by the premise of validity), we derive a bound on the probability $\epsilon$ of violating validity from the probability of the underlying \pcb\ failing and from the probability of $\sigma$'s random delivery sample containig more than $D-\hat D$ Byzantine processes.



\paragraph{$\epsilon$-Consistency (\cref{subsection:prb-consistency})}
When computing the upper bound on the probability $\epsilon$ of compromising consistency, we assume that if the consistency of the underlying \pcb\ is compromised, then the consistency of \prb\ is compromised as well.
The rest of the analysis assumes that \prb\ is consistent.

In such case, every correct process receives at most one message $m^*$ from the underlying \pcb.
Simply by acting correctly, Byzantine processes can cause any correct process to eventually deliver $m^*$.
Consistency is compromised if the adversary can also cause at least one correct process to deliver a message $m \neq m^*$, given that no correct process becomes ready for $m$ by receiving it through the underlying \pcb.

We start by noting that, since a correct process $\pi$ can be ready for an arbitrary number of messages, the set of processes that are eventually ready for $m$ is not affected by which processes are eventually ready for a message $m^*$.
If enough processes in $\pi$'s delivery sample are eventually ready both for $m$ and $m^*$, then $\pi$ can deliver either $m$ or $m^*$.
In this case, the adversary (who controls the network scheduling, see \cref{section:model}) decides which message $\pi$ delivers.

The probability of $m$ being delivered by any correct process is maximized when every Byzantine process behaves as if it was ready for $m$ (\cref{subsection:prb-consistency}, \cref{lemma:forcedisoptimal}).
Note that a Byzantine process being ready for $m$ behaves identically to a correct process that receives $m$ through \pcb.
We model the adversarial system using a single-round game of \contagion\ where both correct and Byzantine processes are represented as nodes in the multigraph and all nodes representing Byzantine processes are initially infected (\cref{subsection:thresholdcontagionprb}, \cref{lemma:forcedequalsthresholdcontagion}).

Given the distribution of the number of correct processes that are ready for $m$ at the end \contagion, we compute the probability that at least one correct process will deliver $m \neq m^*$.
This probability, combined with the probability that the consistency of \pcb\ is violated, yields the probability $\epsilon$ of violating the consistency of \prbal.

\paragraph{$\epsilon$-Totality} Again, to compute an upper bound on the probability of our algorithm compromising totality, we assume that compromising the consistency of \pcb\ also compromises the totality of \prb.
Assuming that \pcb\ satisfies consistency, at most one message $m^*$ is received by any correct process through the underlying \pcb.
We loosen the bound on the probability of compromising totality (and simplify analysis) by considering totality to be compromised if any message $m \neq m^*$ is delivered by any correct process.
This allows us to focus on message $m^*$.
We further loosen the bound by assuming that the Byzantine adversary can arbitrarily cause any correct process to become ready for $m^*$.
Whenever this happens, zero or more additional correct processes will also become ready for $m^*$ as a result of the feedback loop described in \cref{subsection:prb-alg-preview}.
To compromise totality, there must exists at least one correct process that delivers $m^*$ and at least one correct process does not.

We prove (\cref{subsubsection:cstepthresholdcontagion}, \cref{lemma:blackjack}) that the optimal adversarial strategy to compromise totality is to repeat the following.
(1) Make a correct node ready for $m^*$. (2) Wait until the ``readiness'' propagates to zero or more correct nodes. (3) Have specific Byzantine processes behave as correct processes ready for $m^*$, if this leads to some (but not all) correct processes delivering $m^*$.
Totality is satisfied if, after every step of the adversary, either the feedback loop makes all correct processes deliver $m^*$ (relying only on correct processes' ready samples), or no correct process delivers $m^*$ (even with the ``support'' of Byzantine processes) (\cref{theorem:prbtotality}).
Otherwise, totality is violated.



We study this behavior with a multi-round game of \contagion, where only correct processes are represented as nodes in the multigraph and, at the beginning of each round, the player (i.e., the adversary) infects one uninfected node.
From the probability distribution of the number of infected nodes after each round, we derive the probability of compromising totality by message $m^*$.
This probability equals to the probability that there is at least one round after which the number of infected nodes allows some but not all the processes to deliver $m^*$.



\section{Security and Complexity Evaluation}
\label{section:eval}

\begin{figure}
    \centering
    \includegraphics[width=12cm]{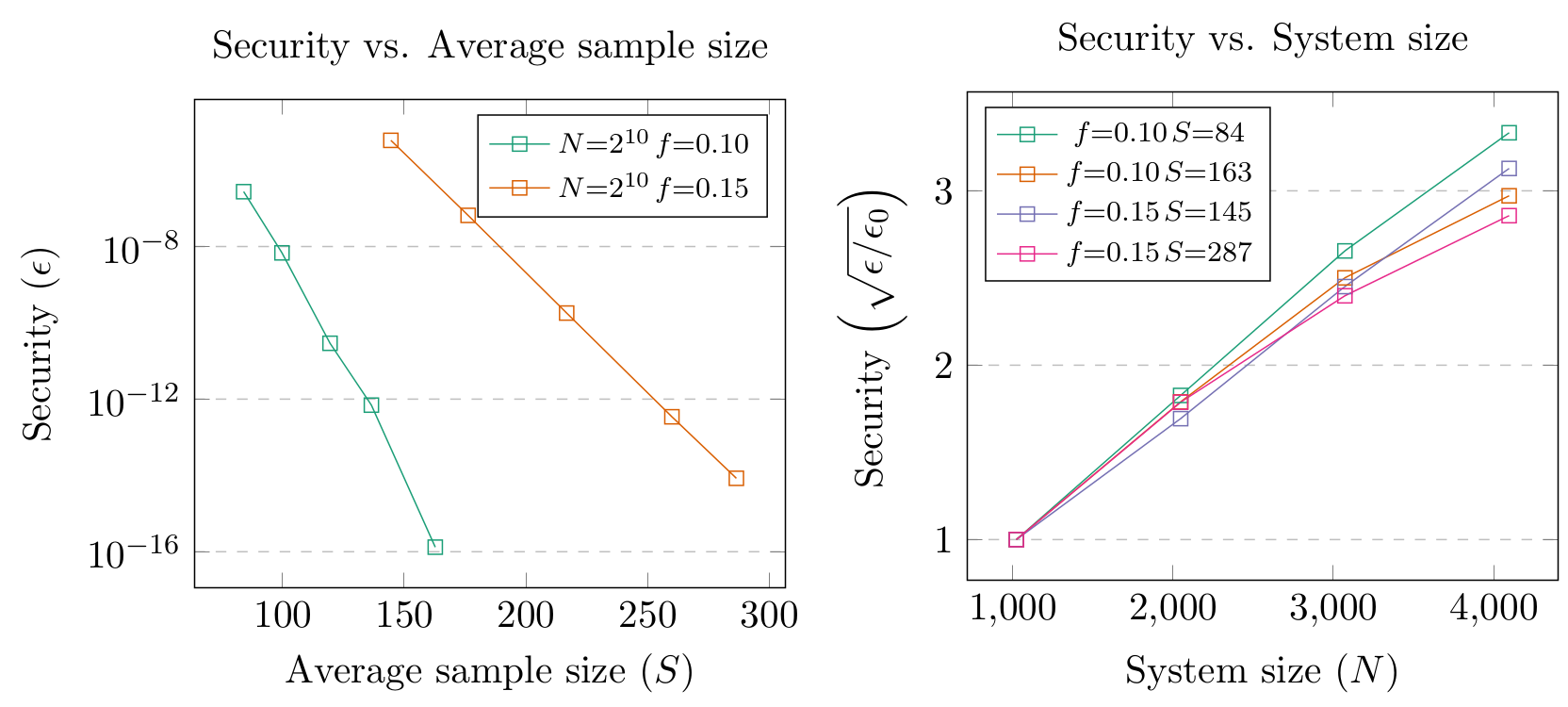}
    \caption{\textbf{Left} -- $\epsilon$-security of \prbal, as a function of the average sample size $S = \ap{G, E, R, D}$. We use a system size of $1024$ processes and fractions of tolerated Byzantine processes $f = 0.1$ and $f = 0.15$.
    \textbf{Right} -- Square root of the normalized $\epsilon$-security of \prbal, as a function of the system size $N$, for various fractions of Byzantine processes ($f$) and average sample sizes ($S$). We normalize the values in each series by the first element of that series. All lines appearing to grow sub-linearly with a square-rooted y-axis demonstrates that the normalized $\epsilon$ security grows sub-quadratically.
    }
    \label{fig:security-eval}
\end{figure}

In~\cref{section:pb,section:pcb,section:prb}, we introduced three algorithms, \pbal, \pcbal\ and \prbal, and outlined their analysis (deferring the formal details to the appendices).

The modular design of our algorithm allows us to study its components independently. 
We employ numerical techniques to maximize the $\epsilon$-security of \prbal, under the constraint that the sum of all the sample sizes of a process is constant ($G + E + R + D = \textit{const}$). Since a process communicates with all the processes in its samples, this corresponds to a fixed communication complexity.

For a given system size $N$ and fraction of Byzantine processes $f$, we relate this per-process communication complexity to the $\epsilon$-security of \prbal. As \Cref{fig:security-eval} (left) shows, the probability $\epsilon$ of compromising the security of \prbal\ decays exponentially in the average sample size $S$.

We also study how the $\epsilon$-security of \prbal\ changes as a function of the system size $N$, for a fixed set of parameters ($G, E, R, D$).
\Cref{fig:security-eval} (right) shows that the $\epsilon$-security is bounded by a quadratic function in $N$.
Thus, for a fixed security $\epsilon$, the average sample size (and consequently, the communication complexity of our algorithm) grows logarithmically with the system size $N$.

Given that a process $\pi$ only exchanges a constant number of messages with each member of $\pi$'s samples, and the sample size is logarithmic in system size, each node needs to exchange $O(\log{N})$ messages.
Thus, for $N$ nodes in the system, the overall message complexity is $O(N \log{N})$.
The latency in terms of message delays between broadcasting and delivery of a message is $O(\log{N} / \log\log{N})$. Specifically, the latency converges to $O(\log{N} / \log\log{N})$ message delays for gossip-based dissemination with \pbal\ (we prove this in \cref{subsection:pbtotality}, \cref{theorem:pblatency}), and $2$ message delays in total for {\tt Echo} (\pcbal) and {\tt Ready} (\prbal) messages.


\section{Related Work}
\label{section:related}

At its base, our broadcast algorithm relies on gossip.
There is a great body of literature studying various aspects of gossip, proposing flavors of gossip protocols for different environments and analyzing their complexities \cite{Alistarh2010,Berenbrink2008,Berenbrink2010MemRand,Berenbrink2010Compl,Avin2011,Elssser2015OnTI,Fernandess2007,Giakkoupis2016,Sourav2018SlowLF,Georgiou2011,Georgiou2008,Georgiou2013,Voulgaris2003,Zhang2008,Ghaffari2016,Haeupler2012}.
However, to the best of our knowledge, we propose the first highly scalable gossip-based reliable broadcast protocol resilient to Byzantine faults with a thorough probabilistic analysis.

The communication pattern in the implementation of both our \pcbal\ and \prbal\ algorithms can be traced back to the Asynchronous Byzantine Agreement (ABA) primitive of Bracha and Toueg~\cite{br85acb} and the subsequent line of work~\cite{bra87asynchronous,cach02sintra,ma97secure,reit94}.
Indeed, our echo-based mechanism in \pcbal\ resembles algorithms from classic quorum-based systems for Byzantine consistent broadcast~\cite{toueg-secure,re94ramp}.
The ready-based mechanism in \prbal\ is inspired by a two-phase protocol appearing in several practical (quorum-based) systems~\cite{cach02sintra,du18beat,highthroughputmulticast}.
Compared to classic work on this topic, the key feature of \prbal\ and \pcbal\ is that they replace the building block of quorum systems with stochastic samples, thus enabling better scalability for the price of abandoning deterministic guarantees.

There is significant prior work on using epidemic algorithms to implement
scalable \emph{reliable} broadcast~\cite{bimodal,lpbcast,jelasity-gossip,marzullo-gossip}.
Under benign failures or constant churn, these algorithms
ensure, with high probability, that every broadcast
message reaches all or none, and that all messages from correct
senders are delivered.
Our goal is to additionally provide \emph{consistency} for broadcast messages, and tolerate \emph{Byzantine} environments~\cite{br85acb,MR97srm,toueg-secure}.
To the best of our knowledge, we are the first to apply the epidemic
sample-based methodology in this context.
Our main algorithm \prbal\ scales well to dynamic systems of  thousands
of nodes, some of which may be Byzantine.
This makes it a suitable choice for \emph{permissionless} settings that are gaining popularity with the advent of blockchains~\cite{nakamotobitcoin}.

Distributed clustering techniques seek to group the processes of a system into clusters, sometimes called shards or quorums, of size $O(log N)$~\cite{awerbuch2009towards,gue13highly,king2011load,king06scalableleader,scheideler2005spread}.
This line of work has various goals (e.g., leader election, ``almost everywhere'' agreement, building an overlay network) and they also aim for scalable solutions.
The overarching principle in clustering techniques is similar to our use of samples: build each cluster in a provably random manner so that the adversary cannot dominate any single cluster.
Samples in our solution are private and individual on a per-process basis, in contrast to clusters which are typically public and global for the whole system.

The idea of \emph{communication locality} 
appears in the context of secure multi-party computation (MPC) protocols~\cite{elet13localit,cha15hgm,garay2017price}.
This property captures the intuition that, in order to obtain scalable distributed protocols and permit a large number of participants, it is desirable to limit the number of participants each process must communicate with.
All of our three algorithms have this communication locality property, since each process coordinates only with logarithmically-sized samples.
In contrast to secure MPC protocols, our algorithms have different goals, system model, or assumptions (e.g., we do not assume a client-server model~\cite{garay2017price}, nor do we seek to address privacy issues).
Our algorithms can be used as building blocks towards helping tackle scalability in MPC protocols, and we consider this an interesting avenue for future work.






\newpage
\bibliographystyle{plain}
\bibliography{bibliography}

\begin{thebibliography}{10}

\bibitem{erdosrenyi}
Daron Acemoglu and Asu Ozdaglar.
\newblock 6.207/14.15: Networks - lecture 4: Erdős–rényi graphs and phase
  transitions.
\newblock \url{https://economics.mit.edu/files/4622}, 2009.

\bibitem{Alistarh2010}
Dan Alistarh, Seth Gilbert, Rachid Guerraoui, and Morteza Zadimoghaddam.
\newblock How efficient can gossip be? (on the cost of resilient information
  exchange).
\newblock In {\em Proceedings of the 37th International Colloquium Conference
  on Automata, Languages and Programming: Part II}, ICALP'10, pages 115--126,
  Berlin, Heidelberg, 2010. Springer-Verlag.

\bibitem{atti1995abd}
Hagit Attiya, Amotz Bar-Noy, and Danny Dolev.
\newblock Sharing memory robustly in message-passing systems.
\newblock {\em JACM}, 42(1), 1995.

\bibitem{Avin2011}
Chen Avin, Michael Borokhovich, Keren Censor-Hillel, and Zvi Lotker.
\newblock Order optimal information spreading using algebraic gossip.
\newblock In {\em Proceedings of the 30th Annual ACM SIGACT-SIGOPS Symposium on
  Principles of Distributed Computing}, PODC '11, pages 363--372, New York, NY,
  USA, 2011. ACM.

\bibitem{awerbuch2009towards}
Baruch Awerbuch and Christian Scheideler.
\newblock {Towards a scalable and robust DHT}.
\newblock {\em Theory of Computing Systems}, 45(2):234--260, 2009.

\bibitem{Berenbrink2008}
Petra Berenbrink, Robert Elsaesser, and Tom Friedetzky.
\newblock Efficient randomised broadcasting in random regular networks with
  applications in peer-to-peer systems.
\newblock In {\em Proceedings of the Twenty-seventh ACM Symposium on Principles
  of Distributed Computing}, PODC '08, pages 155--164, New York, NY, USA, 2008.
  ACM.

\bibitem{Berenbrink2010Compl}
Petra Berenbrink, Robert Els\"{a}sser, and Thomas Sauerwald.
\newblock Communication complexity of quasirandom rumor spreading.
\newblock In {\em Proceedings of the 18th Annual European Conference on
  Algorithms: Part I}, ESA'10, pages 134--145, Berlin, Heidelberg, 2010.
  Springer-Verlag.

\bibitem{Berenbrink2010MemRand}
Petra Berenbrink, Robert Elsässer, and Thomas Sauerwald.
\newblock Randomised broadcasting: Memory vs. randomness.
\newblock {\em Theoretical Computer Science}, 520:306--319, 04 2010.

\bibitem{bimodal}
Kenneth~P. Birman, Mark Hayden, Oznur Ozkasap, Zhen Xiao, Mihai Budiu, and
  Yaron Minsky.
\newblock Bimodal multicast.
\newblock {\em ACM Trans. Comput. Syst.}, 17(2):41--88, May 1999.

\bibitem{brahms}
Edward Bortnikov, Maxim Gurevich, Idit Keidar, Gabriel Kliot, and Alexander
  Shraer.
\newblock Brahms: Byzantine resilient random membership sampling.
\newblock {\em Computer Networks}, 53(13):2340 -- 2359, 2009.
\newblock Gossiping in Distributed Systems.

\bibitem{elet13localit}
Elette Boyle, Shafi Goldwasser, and Stefano Tessaro.
\newblock Communication locality in secure multi-party computation.
\newblock In {\em Theory of Cryptography}, 2013.

\bibitem{bra87asynchronous}
Gabriel Bracha.
\newblock Asynchronous {B}yzantine agreement protocols.
\newblock {\em Information and Computation}, 75(2):130--143, 1987.

\bibitem{br85acb}
Gabriel Bracha and Sam Toueg.
\newblock {Asynchronous Consensus and Broadcast Protocols}.
\newblock {\em JACM}, 32(4), 1985.

\bibitem{distributedprogramming}
Christian Cachin, Rachid Guerraoui, and Lu{\'i}s Rodrigues.
\newblock {\em Introduction to Reliable and Secure Distributed Programming}.
\newblock Springer Publishing Company, Incorporated, 2nd edition, 2011.

\bibitem{cach02sintra}
Christian Cachin and Jonathan~A. Poritz.
\newblock Secure intrusion-tolerant replication on the internet.
\newblock In {\em DSN}, 2002.

\bibitem{cha15hgm}
Nishanth Chandran, Wutichai Chongchitmate, Juan~A. Garay, Shafi Goldwasser,
  Rafail Ostrovsky, and Vassilis Zikas.
\newblock The hidden graph model: Communication locality and optimal resiliency
  with adaptive faults.
\newblock In {\em ITCS '15}, 2015.

\bibitem{random-graph-diameter}
Fan Chung and Linyuan Lu.
\newblock The diameter of sparse random graphs.
\newblock {\em Advances in Applied Mathematics}, 26:257–279, 2001.

\bibitem{tor}
Roger Dingledine, Nick Mathewson, and Paul Syverson.
\newblock Tor: The second-generation onion router.
\newblock In {\em Proceedings of the 13th Conference on USENIX Security
  Symposium - Volume 13}, SSYM'04, pages 21--21, Berkeley, CA, USA, 2004.
  USENIX Association.

\bibitem{du18beat}
Sisi Duan, Michael~K. Reiter, and Haibin Zhang.
\newblock {BEAT: Asynchronous BFT Made Practical}.
\newblock In {\em CCS}, 2018.

\bibitem{Elssser2015OnTI}
Robert Els{\"a}sser and Dominik Kaaser.
\newblock On the influence of graph density on randomized gossiping.
\newblock {\em 2015 IEEE International Parallel and Distributed Processing
  Symposium}, pages 521--531, 2015.

\bibitem{erdos-renyi}
Paul Erd{\"o}s and Alfr{\'e}d R{\'e}nyi.
\newblock On random graphs.
\newblock {\em Publicationes Mathematicae}, 6:290–297, 1959.

\bibitem{lpbcast}
P.~Th. Eugster, R.~Guerraoui, S.~B. Handurukande, P.~Kouznetsov, and A.-M.
  Kermarrec.
\newblock Lightweight probabilistic broadcast.
\newblock {\em ACM Trans. Comput. Syst.}, 21(4):341--374, November 2003.

\bibitem{Fernandess2007}
Yaacov Fernandess, Antonio Fern\'{a}ndez, and Maxime Monod.
\newblock A generic theoretical framework for modeling gossip-based algorithms.
\newblock {\em SIGOPS Oper. Syst. Rev.}, 41(5):19--27, October 2007.

\bibitem{garay2017price}
Juan Garay, Yuval Ishai, Rafail Ostrovsky, and Vassilis Zikas.
\newblock The price of low communication in secure multi-party computation.
\newblock In {\em Annual International Cryptology Conference}, pages 420--446.
  Springer, 2017.

\bibitem{garay2011adaptively}
Juan~A Garay, Jonathan Katz, Ranjit Kumaresan, and Hong-Sheng Zhou.
\newblock {Adaptively Secure Broadcast, Revisited}.
\newblock In {\em PODC}, pages 179--186. Citeseer, 2011.

\bibitem{Georgiou2008}
Chryssis Georgiou, Seth Gilbert, Rachid Guerraoui, and Dariusz~R. Kowalski.
\newblock On the complexity of asynchronous gossip.
\newblock In {\em Proceedings of the Twenty-seventh ACM Symposium on Principles
  of Distributed Computing}, PODC '08, pages 135--144, New York, NY, USA, 2008.
  ACM.

\bibitem{Georgiou2013}
Chryssis Georgiou, Seth Gilbert, Rachid Guerraoui, and Dariusz~R. Kowalski.
\newblock Asynchronous gossip.
\newblock {\em J. ACM}, 60(2):11:1--11:42, May 2013.

\bibitem{Georgiou2011}
Chryssis Georgiou, Seth Gilbert, and Dariusz~R. Kowalski.
\newblock Meeting the deadline: on the complexity of fault-tolerant continuous
  gossip.
\newblock {\em Distributed Computing}, 24(5):223--244, Dec 2011.

\bibitem{Ghaffari2016}
Mohsen Ghaffari and Merav Parter.
\newblock A polylogarithmic gossip algorithm for plurality consensus.
\newblock In {\em Proceedings of the 2016 ACM Symposium on Principles of
  Distributed Computing}, PODC '16, pages 117--126, New York, NY, USA, 2016.
  ACM.

\bibitem{Giakkoupis2016}
George Giakkoupis, Yasamin Nazari, and Philipp Woelfel.
\newblock How asynchrony affects rumor spreading time.
\newblock In {\em Proceedings of the 2016 ACM Symposium on Principles of
  Distributed Computing}, PODC '16, pages 185--194, New York, NY, USA, 2016.
  ACM.

\bibitem{gue13highly}
Rachid Guerraoui, Florian Huc, and Anne-Marie Kermarrec.
\newblock {Highly dynamic distributed computing with byzantine failures}.
\newblock In {\em PODC}, 2013.

\bibitem{at2-19-podc}
Rachid Guerraoui, Petr Kuznetsov, Matteo Monti, Matej Pavlovic, and Dragos
  Seredinschi.
\newblock {The Consensus Number of a Cryptocurrency}.
\newblock In {\em PODC}, 2019.
\newblock (to appear).

\bibitem{HT93}
Vassos Hadzilacos and Sam Toueg.
\newblock Fault-tolerant broadcasts and related problems.
\newblock In Sape~J. Mullender, editor, {\em Distributed Systems}, chapter~5,
  pages 97--145. Addison-Wesley, 1993.

\bibitem{Haeupler2012}
Bernhard Haeupler, Gopal Pandurangan, David Peleg, Rajmohan Rajaraman, and
  Zhifeng Sun.
\newblock Discovery through gossip.
\newblock In {\em Proceedings of the Twenty-fourth Annual ACM Symposium on
  Parallelism in Algorithms and Architectures}, SPAA '12, pages 140--149, New
  York, NY, USA, 2012. ACM.

\bibitem{jelasity-gossip}
M\'{a}rk Jelasity, Alberto Montresor, and Ozalp Babaoglu.
\newblock T-man: Gossip-based fast overlay topology construction.
\newblock {\em Comput. Netw.}, 53(13):2321--2339, August 2009.

\bibitem{king2011load}
Valerie King, Steven Lonargan, Jared Saia, and Amitabh Trehan.
\newblock {Load Balanced Scalable Byzantine Agreement through Quorum Building,
  with Full Information}.
\newblock In {\em International Conference on Distributed Computing and
  Networking}, pages 203--214. Springer, 2011.

\bibitem{king06scalableleader}
Valerie King, Jared Saia, Vishal Sanwalani, and Erik Vee.
\newblock Scalable leader election.
\newblock In {\em SODA}, 2006.

\bibitem{la82byzantine}
Leslie Lamport, Robert Shostak, and Marshall Pease.
\newblock The byzantine generals problem.
\newblock {\em TOPLAS}, 4(3), 1982.

\bibitem{marzullo-gossip}
Meng-Jang Lin, Keith Marzullo, and Stefano Masini.
\newblock Gossip versus deterministically constrained flooding on small
  networks.
\newblock In {\em Proceedings of the 14th International Conference on
  Distributed Computing}, DISC '00, pages 253--267, London, UK, UK, 2000.
  Springer-Verlag.

\bibitem{ma97secure}
Dahlia Malkhi, Michael Merritt, and Ohad Rodeh.
\newblock Secure {R}eliable {M}ulticast {P}rotocols in a {WAN}.
\newblock In {\em ICDCS}, 1997.

\bibitem{ma97bqs}
Dahlia Malkhi and Michael Reiter.
\newblock Byzantine quorum systems.
\newblock In {\em Proceedings of the twenty-ninth annual ACM symposium on
  Theory of computing}, pages 569--578. ACM, 1997.

\bibitem{highthroughputmulticast}
Dahlia Malkhi and Michael~K. Reiter.
\newblock A high-throughput secure reliable multicast protocol.
\newblock In {\em CSFW}, 1996.

\bibitem{MR97srm}
Dahlia Malkhi and Michael~K. Reiter.
\newblock A high-throughput secure reliable multicast protocol.
\newblock {\em Journal of Computer Security}, 5(2):113--128, 1997.

\bibitem{probabilisticquorums}
Dahlia Malkhi, Michael~K Reiter, Avishai Wool, and Rebecca~N Wright.
\newblock Probabilistic quorum systems.
\newblock {\em Inf. Comput.}, 170(2):184--206, November 2001.

\bibitem{nakamotobitcoin}
Satoshi Nakamoto.
\newblock Bitcoin: A peer-to-peer electronic cash system, 2008.

\bibitem{ped02handling}
Fernando Pedone and Andr{\'e} Schiper.
\newblock Handling message semantics with generic broadcast protocols.
\newblock {\em Distributed Computing}, 15(2):97--107, 2002.

\bibitem{re94ramp}
Michael~K. Reiter.
\newblock {Secure Agreement Protocols: Reliable and Atomic Group Multicast in
  Rampart}.
\newblock In {\em CCS}, 1994.

\bibitem{reit94}
Michael~K. Reiter and Kenneth~P. Birman.
\newblock {How to securely replicate services}.
\newblock {\em ACM Transactions on Programming Languages and Systems (TOPLAS)},
  16(3), 1994.

\bibitem{scheideler2005spread}
Christian Scheideler.
\newblock {How to Spread Adversarial Nodes? Rotate!}
\newblock In {\em STOC}, pages 704--713. ACM, 2005.

\bibitem{Sourav2018SlowLF}
Suman Sourav, Peter Robinson, and Seth Gilbert.
\newblock Slow links, fast links, and the cost of gossip.
\newblock {\em 2018 IEEE 38th International Conference on Distributed Computing
  Systems (ICDCS)}, pages 786--796, 2018.

\bibitem{toueg-secure}
Sam Toueg.
\newblock Randomized byzantine agreements.
\newblock In {\em Proceedings of the Third Annual ACM Symposium on Principles
  of Distributed Computing}, PODC '84, pages 163--178, New York, NY, USA, 1984.
  ACM.

\bibitem{vuvuzela}
Jelle van~den Hooff, David Lazar, Matei Zaharia, and Nickolai Zeldovich.
\newblock Vuvuzela: Scalable private messaging resistant to traffic analysis.
\newblock In {\em Proceedings of the 25th Symposium on Operating Systems
  Principles}, SOSP '15, pages 137--152, New York, NY, USA, 2015. ACM.

\bibitem{Voulgaris2003}
Spyros Voulgaris, M\'{a}rk Jelasity, and Maarten van Steen.
\newblock A robust and scalable peer-to-peer gossiping protocol.
\newblock In {\em Proceedings of the Second International Conference on Agents
  and Peer-to-Peer Computing}, AP2PC'03, pages 47--58, Berlin, Heidelberg,
  2004. Springer-Verlag.

\bibitem{quorum-systems}
Marko Vukolic.
\newblock The origin of quorum systems.
\newblock {\em Bulletin of the {EATCS}}, 101:125--147, 2010.

\bibitem{Zhang2008}
B.~{Zhang}, K.~{Han}, B.~{Ravindran}, and E.~D. {Jensen}.
\newblock Rtqg: Real-time quorum-based gossip protocol for unreliable networks.
\newblock In {\em 2008 Third International Conference on Availability,
  Reliability and Security}, pages 564--571, March 2008.

\end{thebibliography}

\newpage

\tableofcontents
\clearpage

\appendix

\section{\pbal}
\label{appendix:pbal}

In this appendix, we present in greater detail the \textbf{\pb} abstraction and discuss its properties. We then present \pbal, an algorithm that implements \pb, and evaluate its \textbf{security} and \textbf{complexity} as a function of its \textbf{parameters}. 

The \pb\ abstraction serves the purpose of reliably broadcasting a single message from a designated correct sender to all correct processes (\textbf{validity}, \textbf{totality}). \\

We use \pb\ in the implementation of \pcbal\ (see \Cref{section:pcb}) to initially distribute the message from the designated sender to all correct processes.

\subsection{Definition}

The \textbf{\pb} interface (instance $\pbin$, sender $\sigma$) exports the following \textbf{events}:
\begin{itemize}
    \item \textbf{Request}: $\event{\pbin}{Broadcast}{m}$: Broadcasts a message $m$ to all processes. This is only used by $\sigma$.
    \item \textbf{Indication} $\event{\pbin}{Deliver}{m}$: Delivers a message $m$ broadcast by process $\sigma$.
\end{itemize}

For any $\epsilon \in [0, 1]$, we say that \pb\ is $\epsilon$-secure if:
\begin{enumerate}
    \item \textbf{No duplication}: No correct process delivers more than one message.
    \item \textbf{Integrity}: If a correct process delivers a message $m$, and $\sigma$ is correct, then $m$ was previously broadcast by $\sigma$.
    \item $\epsilon$-\textbf{Validity}: If $\sigma$ is correct, and $\sigma$ broadcasts a message $m$, then $\sigma$ eventually delivers $m$ with probability at least $(1 - \epsilon)$.
    \item $\epsilon$-\textbf{Totality}: If a correct process delivers a message, then every correct process eventually delivers a message with probability at least $(1 - \epsilon)$.
\end{enumerate}

\subsection{Algorithm}
\label{subsection:pbalgorithm}

\begin{algorithm}
\begin{algorithmic}[1]
\Implements
    \Instance{\pbab}{\pbin}
\EndImplements

\Uses
    \Instance{AuthenticatedPointToPointLinks}{al}
\EndUses

\Parameters
    \State $G$: expected gossip sample size
\EndParameters

\Upon{pb}{Init}{}
    \State $\mathcal{G} = \Omega(\pois{G}{})$; \label{line:pbinitializesample}
    \ForAll{\pi}{\mathcal{G}}
        \Trigger{al}{Send}{\pi, [\text{\tt GossipSubscribe}]}; \label{line:pbsubscribe}
    \EndForAll
    \State $delivered = \bot$;
\EndUpon

\Upon{al}{Deliver}{\pi, [\text{\tt GossipSubscribe}]} \label{line:pbreceivesubscribe}
    \If{$delivered \neq \bot$}
        \State $(message, signature) = delivered$;
        \Trigger{al}{Send}{\pi, [\text{\tt Gossip}, message, signature]}; \label{line:pbcatchup}
    \EndIf
    \State $\mathcal{G} \leftarrow \mathcal{G} \cup \{\pi\}$; \label{line:pbupdatesample}
\EndUpon

\Procedure{dispatch}{message, signature}
    \If{$delivered = \bot$} \label{line:pbcheckdelivered}
        \State $delivered \leftarrow (message, signature)$; \label{line:pbsetdelivered}
        \ForAll{\pi}{\mathcal{G}}
            \Trigger{al}{Send}{\pi, [\text{\tt Gossip}, message, signature]}; \label{line:pbforward}
        \EndForAll
        \Trigger{\pbin}{Deliver}{message} \label{line:pbdeliver}
    \EndIf
\EndProcedure

\Upon{pb}{Broadcast}{message} \Comment{only process $\sigma$}
    \State $dispatch(message, sign(message))$; \label{line:pbbroadcast}
\EndUpon

\algstore{murmur}
\end{algorithmic}
\caption{\pbal}
\label{algorithm:pbal}
\end{algorithm}

\begin{algorithm}
\begin{algorithmic}[1]
\algrestore{murmur}

\Upon{al}{Deliver}{\pi, [\text{\tt Gossip}, message, signature]}
    \If{$verify(\sigma, message, signature)$} \label{line:pbchecksignature}
        \State $dispatch(message, signature)$;
    \EndIf
\EndUpon

\end{algorithmic}
\end{algorithm}

\pbal\ (\Cref{algorithm:pbal}) distributes a single message across the system by means of \textbf{gossip}: upon reception, a correct process relays the message to a set of randomly selected neighbors. The algorithm depends on one integer parameter, $G$ (\emph{expected gossip sample size}), whose value we discuss in \cref{subsection:pbtotality}.

\paragraph{Initialization} Upon initialization, (\cref{line:pbinitializesample}) every correct process randomly samples a value $\bar G$ from a \emph{Poisson} distribution with expected value $G$, and uses the sampling oracle $\Omega$ to select $\bar G$ distinct processes that it will use to initialize its \textbf{gossip sample} $\mathcal{G}$. 

\paragraph{Link reciprocation} Once its gossip sample is initialized, a correct process sends a {\tt GossipSubscribe} message to all the processes in $\mathcal{G}$ (\cref{line:pbsubscribe}). Upon receiving a {\tt GossipSubscribe} message from a process $\pi$ (\cref{line:pbreceivesubscribe}), a correct process adds $\pi$ to its own gossip sample (\cref{line:pbupdatesample}), and sends back the gossiped message if it has already received it (\cref{line:pbcatchup}).

\paragraph{Gossip} When broadcasting the message (\cref{line:pbbroadcast}), a correct designated sender $\sigma$ signs the message and sends it to every process in its gossip sample $\mathcal{G}$ (\cref{line:pbforward}). Upon receiving a correctly signed message from $\sigma$ (\cref{line:pbchecksignature}) for the first time (this is enforced by updating the value of $delivered$, \cref{line:pbcheckdelivered}), a correct process delivers it (\cref{line:pbdeliver}) and forwards it to every process in its gossip sample (\cref{line:pbforward}).

\subsection{No duplication, integrity and validity}
\label{subsection:pb-nd-int-val}
We start by verifying that \pbal\ satistifes \textbf{no duplication}, \textbf{integrity} and $0$-\textbf{validity}, independently of $G$.

\begin{theorem}
\label{theorem:pbnoduplication}
\pbal\ satisfies no duplication.
\begin{proof}
Procedure $dispatch$ explicitly checks (\cref{line:pbcheckdelivered}) if the variable $delivered$ is equal to $\bot$ before delivering any message. Before a message is delivered (\cref{line:pbdeliver}), $delivered$ is updated to a value different from $\bot$ (\cref{line:pbsetdelivered}). Therefore a correct process only delivers one message.
\end{proof}
\end{theorem}

\begin{theorem}
\label{theorem:pbintegrity}
\pbal\ satistifes integrity.
\begin{proof}
Upon receiving a {\tt Gossip} message, a correct process checks its signature against the public key of the designated sender $\sigma$ (\cref{line:pbchecksignature}). Moreover, if $\sigma$ is correct, it only signs $message$ when broadcasting (\cref{line:pbbroadcast}). Since we assume that cryptographic signatures cannot be forged, this implies that the message was previously broadcast by $\sigma$.
\end{proof}
\end{theorem}

\begin{theorem}
\label{theorem:pbvalidity}
\pbal\ satisfies 0-validity.
\begin{proof}
Upon broadcasting a message $m$, a correct sender calls the procedure $dispatch(m, sign(m))$ (\cref{line:pbbroadcast}). Since $delivered$ is initialized to $\bot$, this immediately results in the delivery of $m$ (\cref{line:pbdeliver}).

Since the validity property is satisfied deterministically, \pbal\ satisfies  $\epsilon$-validity for $\epsilon = 0$.
\end{proof}
\end{theorem}

\subsection{Totality}
\label{subsection:pbtotality}

We now compute, given the parameter $G$, the $\epsilon$-\textbf{totality} of \pbal. To this end, we first prove some preliminary lemmas.

\begin{lemma}
\label{lemma:gossipreciprocation}
Let $\rho$ and $\pi$ be two correct processes, let $\rho$ be in $\pi$'s gossip sample. Then $\pi$ is eventually in $\rho$'s gossip sample.
\begin{proof}
A gossip sample is updated only upon initialization (\cref{line:pbinitializesample}) or when a {\tt GossipSubscribe} message is received (\cref{line:pbupdatesample}). 

If $\pi$ selected $\rho$ upon initialization, then it also sent it a {\tt GossipSubscribe} message (\cref{line:pbsubscribe}). Since Byzantine network scheduling can only finitely delay the messages between correct processes, $\rho$ eventually receives $\pi$'s message (\cref{line:pbreceivesubscribe}) and adds $\pi$ to its gossip sample. 

If $\pi$ received a {\tt GossipSubscribe} message from $\rho$, then (\cref{line:pbsubscribe}) $\rho$ selected $\pi$ upon initialization, which means that $\pi$ is already in $\rho$'s gossip sample.
\end{proof}
\end{lemma}

\begin{definition}[Correct gossip network]
Let $\pi$, $\rho$ be two correct processes, let $\pi \leftrightarrow \rho$ denote the condition \emph{$\rho$ is eventually in $\pi$'s gossip sample}. \cref{lemma:gossipreciprocation} proves that
\begin{equation*}
    \rp{\pi \leftrightarrow \rho} \Leftrightarrow \rp{\rho \leftrightarrow \pi}
\end{equation*}

We define \textbf{correct gossip network} to be the undirected graph
\begin{equation}
    \mathbb{G} = \rp{\Pi_C, \cp{\rp{\pi, \rho} \in \Pi^2_C \mid \pi \leftrightarrow \rho}}
\end{equation}
\end{definition}

\begin{lemma}
\label{lemma:gossipconnectedness}
If the correct gossip network is connected, then \pbal\ satisfies totality.
\begin{proof}
We start by noting that a correct process eventually delivers a message (\cref{line:pbdeliver}) if and only if it eventually sets $delivered$ to a value different from $\bot$ (\cref{line:pbsetdelivered}).

Let $\pi$ be a correct process for which eventually $delivered \neq \bot$. Upon setting $delivered \leftarrow (m \neq \bot)$, $\pi$ sends $m$ to all the processes in its gossip sample (\cref{line:pbforward}). Moreover, upon receiving a {\tt GossipSubscribe} message \emph{after} setting $delivered \leftarrow m$, $\pi$ replies with $m$ (\cref{line:pbcatchup}).

Therefore, every correct process that is eventually in $\pi$'s gossip sample eventually satisfies $delivered \neq \bot$. If $\mathbb{G}$ is connected, then a path exists in $\mathbb{G}$ between $\pi$ and every other correct process, and they all eventually satisfy $delivered \neq \bot$, i.e., they deliver a message.
\end{proof}
\end{lemma}

From \cref{lemma:gossipconnectedness} it follows that \pbal\ satisfies $\epsilon$-totality if the probability of $\mathbb{G}$ being disconnected is at most $\epsilon$.

\begin{notation}[Binomial distribution]
We use $\bin{N}{p}{}$ to denote the \textbf{binomial distribution} with $N$ trials and $p$ probability of success.
\end{notation}

\begin{notation}[Poisson distribution]
We use $\pois{\lambda}{}$ to denote the \textbf{Poisson distribution} with expected value $\lambda$.
\end{notation}

\begin{notation}[Probability]
Let $E$, $F$ be events. We use $\prob{E}$ to denote the probability of E. We use $\prob{E \mid F}$ to denote the probability of $E$, conditioned on the occurrence of $F$.

Let $X$, $Y$, $Z$ be random variables. For example, we use the following expressions interchangeably:
\begin{equation*}
    \prob{\bar X} \longleftrightarrow \prob{X = \bar X}
\end{equation*}

Note how $X$ is a random variable, while $\bar X$ is an element in the codomain of $X$. Stand-ins can be combined. For example, we use the following expressions interchangeably:
\begin{eqnarray*}
\prob{\bar X, \bar Y} &\longleftrightarrow& \prob{X = \bar X, Y = \bar Y} \\
\prob{\bar X \mid \bar Y} &\longleftrightarrow& \prob{X = \bar X \mid Y = \bar Y} \\
\prob{\bar X, \bar Y \mid \bar Z} &\longleftrightarrow& \prob{X = \bar X, Y = \bar Y \mid Z = \bar Z}
\end{eqnarray*}

Stand-ins are only used to express exact values. Whenever non-trivial expressions are needed, we use their explicit form. Explicit notation and stand-ins can be combined. For example, we use the following expressions interchangeably:
\begin{eqnarray*}
\prob{\bar X \mid Y < K} &\longleftrightarrow& \prob{X = \bar X \mid Y < K} \\
\prob{\bar X \mid X < K} &\longleftrightarrow& \prob{X = \bar X \mid X < K}
\end{eqnarray*}

\end{notation}

\begin{lemma}
\label{lemma:gossipconnectionprobability}
In the limit $N \rightarrow \infty$, $\mathbb{G}$ is a $G\rp{C, p}$ Erdős–Rényi graph, with
\begin{equation*}
    p = 1 - \rp{1 - \frac{G}{N}}^2
\end{equation*}
\begin{proof}
It is a known result that, for large samples and small probabilities, a binomial distribution converges to a Poisson distribution:
\begin{eqnarray*}
    \lhs
    \lim_{\substack{N \rightarrow \infty \\ Np = \text{const}}} \qp
    {
        \bin{N}{p}{n} = \binom{N}{n} p^n \rp{1 - p}^{N - n}
    }
    \\ 
    &=&
    \qp
    {
        \frac{\rp{Np}^n}{n!}e^{-Np} = \pois{Np}{n}
    }
\end{eqnarray*}
therefore, in the limit $N \rightarrow \infty$,
\begin{equation}
    \label{equation:poissontobinomial}
    \pois{G}{n} \simeq \bin{N}{\frac{G}{N}}{n}
\end{equation}

As we discussed in \cref{subsection:pbalgorithm}, a gossip sample $\mathcal{G}$ is initialized upon initialization (\cref{line:pbinitializesample}) by first sampling a value $\bar G$ from a $\pois{G}{}$ distribution, then selecting $\bar G$ distinct processes from $\Pi$ with uniform probability. 

Let $\pi \in \Pi_C, \rho \in \Pi$, let $\mathcal{G}^{in}_\pi$ be $\pi$'s initial gossip sample, let $q = G/N$. By the law of total probability, and using \cref{equation:poissontobinomial}, we have for large $N$
\begin{eqnarray*}
\prob{\rho \in \mathcal{G}^{in}_\pi} &=& \sum_{\bar G = 0}^N \rp{\prob{\rho \in \mathcal{G}^{in}_\pi \mid \bar G} \prob{\bar G}} \\
&=& \sum_{\bar G = 0}^N \rp{\frac{\bar G}{N} \pois{G}{\bar G}} \simeq \sum_{\bar G = 0}^N \rp{\frac{\bar G}{N} \bin{N}{q}{\bar G}} \\
&=& \sum_{\bar G = 0}^N \rp{\frac{\bar G}{N} \binom{N}{\bar G} q^{\bar G}\rp{1 - q}^{N - \bar G}} \\
&=& \sum_{\bar G = 0}^N \rp{\frac{\bar G}{N} \frac{N!}{\bar G! \rp{N - \bar G}!} q^{\bar G} \rp{1 - q}^{N - \bar G}} \\
&=& \sum_{\bar G = 1}^N \rp{\frac{\rp{N - 1}!}{\rp{\bar G - 1}!\rp{N - \bar G}!} q q^{\bar G - 1}\rp{1 - q}^{N - \bar G}} \\
&=& q \sum_{\bar G' = 0}^{N - 1} \rp{\frac{\rp{N - 1}!}{\bar G'! \rp{N - 1 - \bar G'}!} q^{\bar G'} \rp{1 - q}^{N - 1 - \bar G'}} \\
&=& q \sum_{\bar G' = 0}^{N - 1} \bin{N - 1}{q}{\bar G'} = q
\end{eqnarray*}

Let $\rho_1, \ldots, \rho_R$ be distinct processes, with $R \leq N$. Similar calculations yield
\begin{equation}
    \label{equation:independentgossipsampling}
    \prob{\rho_1 \in \mathcal{G}^{in}_\pi, \ldots, \rho_R \in \mathcal{G}^{in}_\pi} = q^R
\end{equation}

\cref{equation:independentgossipsampling} proves that every process $\rho \in \Pi$ has an independent probability $q$ of being in $\mathcal{G}^{in}_\pi$. Since for any two $\pi, \xi \in \Pi_C$ we have
\begin{equation*}
    \rp{\pi \leftrightarrow \xi} \Leftrightarrow \rp{\pi \in \mathcal{G}^{in}_\xi \vee \xi \in \mathcal{G}^{in}_\pi}
\end{equation*}
we can derive the probability $p$ of any two correct processes being connected:
\begin{equation}
    \label{equation:gossipconnectionprobability}
    p = 1 - \rp{1 - q}^2 = 1 - \rp{1 - \frac{G}{N}}^2
\end{equation}

Therefore, following \cref{equation:independentgossipsampling,equation:gossipconnectionprobability}, $\mathbb{G} = G(C, p)$ is an Erdős – Rényi graph with $H$ nodes and $p$ probability of connection between any two nodes.
\end{proof}
\end{lemma}

\cref{lemma:gossipconnectionprobability} allows us to bound the $\epsilon$-totality of \pbal, given $G$.

\begin{theorem}
\label{theorem:pbtotality}
\pbal\ satisfies $\epsilon_t$-totality, with $\epsilon_t$ bound by
\begin{equation}
    \label{equation:pbtotalitysecurity}
    \epsilon_t \leq \sum_{k = 1}^{C/2} \rp{\binom{C}{k}\rp{1 - p}^{k\rp{C - k}}}
\end{equation}
\begin{proof}
It follows immediately from \cref{lemma:gossipconnectionprobability} and a known result \cite{erdosrenyi} on the connectivity of Erdős–Rényi graphs.
\end{proof}
\end{theorem}

We prove an additional result on the latency of \pbal.

\begin{theorem}
\label{theorem:pblatency}
\begin{tolerant}{300}
The latency of \pbal\ is asymptotically sub-logarithmic. More formally, the diameter $D(C, G)$ of the correct gossip network limits to
\end{tolerant}
\begin{equation*}
    \lim_{C \rightarrow \infty} D(C, G) = \frac{\log(C)}{\log\rp{2 - 2f} + \log(G)}
\end{equation*}
\begin{proof}
It is a known result \cite{random-graph-diameter} that the diameter of an Erdős–Rényi graph $G(C, p)$ converges, for $Cp \rightarrow \infty$, to $\log(C) / \log(Cp)$. 

Noting that
\begin{equation*}
    \lim_{C \rightarrow \infty} 1 - \rp{1 - \frac{G}{N}}^2 = \frac{2G}{N}
\end{equation*}
we get
\begin{eqnarray*}
\lim_{C \rightarrow \infty} D(C, G) &=& \frac{\log(C)}{\log(C) + \log(p)} \\
&=& \frac{\log(C)}{\log(C) + \log(2) + \log(G) - \log(N)} \\
&=& \frac{\log(C)}{\log\rp{\frac{2C}{N}} + \log(G)} \\
&=& \frac{\log(C)}{\log(2 - 2f) + \log(G)}
\end{eqnarray*}
which proves the lemma. For a fixed security $\epsilon$, we showed in \cref{theorem:pbtotality} that $G$ must scale logarithmically with the size of the system. As a result, for a fixed security $\epsilon$, the latency scales as $O(\log(N) / \log(\log(N)))$
\end{proof}
\end{theorem}

\clearpage
\section{\pcbal}
\label{appendix:pcbal}

In this appendix, we present in greater detail the \textbf{\pcb} abstraction and discuss its properties. We then present \pcbal, an algorithm that implements \pcb, and evaluate its \textbf{security} and \textbf{complexity} as a function of its \textbf{parameters}.

The \pcb\ abstraction allows a subset of the correct processes to agree on a single message from a potentially Byzantine designated sender. \Pcb\ is a distinct from \pb. \Pb\ guarantees (\textbf{totality}) that if any correct process delivers a message, every correct process delivers a message. \Pcb, instead, guarantees (\textbf{consistency}) that, even if the sender is Byzantine, no two correct processes deliver different messages. However, if the sender is Byzantine, it may happen with a non-negligible probability that only an intermediate fraction of the correct processes deliver the message. 

We use \pcb\ in the implementation of \prbal\ (see \cref{section:prb}) as a way to consistently broadcast messages.

\subsection{Definition}
\label{subsection:pcbdefinition}

The \textbf{\pcb} interface (instance $\pcbin$, sender $\sigma$) exposes the following two \textbf{events}:
\begin{itemize}
\item \textbf{Request}: $\event{\pcbin}{Broadcast}{m}$: Broadcasts a message $m$ to all processes. This is only used by $\sigma$.
\item \textbf{Indication}: $\event{\pcbin}{Deliver}{m}$: Delivers a message $m$ broadcast by process $\sigma$.
\end{itemize}

For any $\epsilon \in [0, 1]$, we say that \pcb\ is $\epsilon$-secure if:
\begin{enumerate}
\item \textbf{No duplication}: No correct process delivers more than one message.
\item \textbf{Integrity}: If a correct process delivers a message $m$, and $\sigma$ is correct, then $m$ was previously broadcast by $\sigma$.
\item $\epsilon$-\textbf{Total validity}: If $\sigma$ is correct, and $\sigma$ broadcasts a message $m$, every correct process eventually delivers $m$ with probability at least $(1 - \epsilon)$.
\item $\epsilon$-\textbf{Consistency}: Every correct process that delivers a message delivers the same message with probability at least $(1 - \epsilon)$.
\end{enumerate}

\subsection{Algorithm}
\label{subsection:pcbalgorithm}

\begin{algorithm}
\begin{algorithmic}[1]
\Procedure{sample}{message, size}
    \State $\psi = \emptyset$;
    \NTimes{size}
        \State $\psi \leftarrow \psi \cup \Omega(1)$; \label{line:sampleselection}
    \EndNTimes
    \ForAll{\pi}{\psi}
        \Trigger{al}{Send}{\pi, [message]}; \label{line:samplesubscribe}
    \EndForAll
    \State \Return $\psi$;
\EndProcedure
\end{algorithmic}

\caption{Procedure $sample$}
\label{algorithm:sample}
\end{algorithm}

\begin{algorithm}
\begin{algorithmic}[1]
\Implements
    \Instance{\pcbab}{\pcbin}
\EndImplements

\Uses
    \Instance{AuthenticatedPointToPointLinks}{al}
    \Instance{\pbab}{\pbin}
\EndUses

\Parameters
    \State $E$: echo sample size
    \State $\hat E$: delivery threshold
\EndParameters

\Upon{\pcbin}{Init}{} \label{line:pcbinitialization}
    \State $echo = \bot$; \tabto*{2.5cm} $delivered = \false$; \tabto*{6cm} $\tilde{\mathcal{E}} = \emptyset$;
    \State
     \State $\mathcal{E} = sample(\text{\tt EchoSubscribe}, E)$; 
    \State $replies = \{\bot\}^E$; 
\EndUpon

\Upon{al}{Deliver}{\pi, [\text{\tt EchoSubscribe}]}
    \If{$echo \neq \bot$}
        \State $(message, signature) = echo$;
        \Trigger{al}{Send}{\pi, [{\tt Echo}, message, signature]}; \label{line:pcbechocatchup}
    \EndIf
    \State $\tilde{\mathcal{E}} \leftarrow \tilde{\mathcal{E}} \cup \{\pi\}$; \label{line:pcbechosubscribe}
\EndUpon

\Upon{\pcbin}{Broadcast}{message} \Comment{only process $\sigma$}
    \Trigger{\pbin}{Broadcast}{[\text{\tt Send}, message, sign(message)]}; \label{line:pcbbroadcast}
\EndUpon

\Upon{\pbin}{Deliver}{[\text{\tt Send}, message, signature]} \label{line:pcbpbdelivery}
    \If{$verify(\sigma, message, signature)$} \label{line:pcbchecksendsignature}
        \State $echo \leftarrow (message, signature)$; \label{line:pcbsetecho}
        \ForAll{\rho}{\tilde{\mathcal{E}}}
            \Trigger{al}{Send}{\rho, [\text{\tt Echo}, message, signature]}; \label{line:pcbsendecho}
        \EndForAll
    \EndIf
\EndUpon

\algstore{sieve}
\end{algorithmic}
\caption{\pcbal}
\label{algorithm:pcbal}
\end{algorithm}

\begin{algorithm}
\begin{algorithmic}[1]
\algrestore{sieve}

\Upon{al}{Deliver}{\pi, [\text{\tt Echo}, message, signature]} \label{line:pcbreceiveecho}
    \If{$\pi \in \mathcal{E}\; \textbf{and}\; replies[\pi] = \bot \;\textbf{and}\; verify(\sigma, message, signature)$} \label{line:pcbcheckechosignature}
        \State $replies[\pi] \leftarrow (message, signature)$;
    \EndIf
\EndUpon

\UponCondition{\abs{\cp{\rho \in \mathcal{E} \mid replies[\rho] = echo}} \geq \hat E\;\textbf{and}\; delivered = \false} \label{line:pcbthreshold}
    \State $delivered \leftarrow \true$;
    \Trigger{\pcbin}{Deliver}{message};  \label{line:pcbdeliver}
\EndUponCondition

\end{algorithmic}
\end{algorithm}

\Cref{algorithm:sample} implements a $sample$ procedure that we use both in the implementation of \pcbal\ and \prbal. Procedure $sample(message, size)$ uses $\Omega$ to pick $size$ processes with replacement, and sends them $message$.

\Cref{algorithm:pcbal} implements \pcbal. \pcbal\ consistently distributes a single message across the system as follows:
\begin{itemize}
    \item Initially, \pb\ distributes potentially conflicting copies of the message to every correct process.
    \item Upon receiving a message $m$ from \pb, a correct process issues an {\tt Echo} message for $m$.
    \item Upon receiving enough {\tt Echo} messages for the message $m$ it {\tt Echo}ed, a correct process delivers $m$.
\end{itemize}

A correct process collects {\tt Echo} messages from a randomly selected \emph{echo sample} of size $E$, and delivers the message it {\tt Echo}ed upon receiving $\hat E$ {\tt Echo}es for it. We discuss the values of the two parameters of \pcbal\ in \cref{subsection:pcbanalysis}.

\paragraph{Sampling} Upon initialization (\cref{line:pcbinitialization}), a correct process randomly selects an \textbf{echo sample} $\mathcal{E}$ of size $E$. Samples are selected with replacement by repeatedly calling $\Omega$ (\cref{algorithm:sample}, \cref{line:sampleselection}). A correct process sends an {\tt EchoSubscribe} message to all the processes in its echo sample (\cref{algorithm:sample}, \cref{line:samplesubscribe}).

\paragraph{Publish-subscribe} Unlike in the deterministic version of {\tt Authenticated Echo Broadcast}, where a correct process broadcasts its {\tt Echo} messages to the whole system, here each process only listens for messages coming from its echo sample (\cref{line:pcbcheckechosignature}). 

A correct process maintains an \textbf{echo subscription set} $\tilde{\mathcal{E}}$. Upon receiving an {\tt EchoSubscribe} message from a process $\pi$, a correct process adds $\pi$ to $\tilde{\mathcal{E}}$ (\cref{line:pcbechosubscribe}). If a correct process receives an {\tt EchoSubscribe} message \emph{after} publishing its {\tt Echo} message, it also sends back the previously published message (\cref{line:pcbechocatchup}).

A correct process  only sends its {\tt Echo} messages (\cref{line:pcbsendecho}) to its echo subscription set.

\paragraph{Echo} The designated sender $\sigma$ initially broadcasts its message using \pb\ (\cref{line:pcbbroadcast}). Upon \pbin.Delivery of a message $m$ (correctly signed by $\sigma$) (\cref{line:pcbpbdelivery}), a correct process sends an {\tt Echo} message for $m$ to all the nodes in its echo subscription set (\cref{line:pcbsendecho}).

\paragraph{Delivery} 

A correct process $\pi$ that {\tt Echo}ed a message $m$ delivers $m$ (\cref{line:pcbdeliver}) upon collecting at least $\hat E$ {\tt Echo} messages for $m$ (\cref{line:pcbthreshold}) from the processes in its echo sample.

\subsection{No duplication and integrity}
\label{subsection:pcb-nd-int}

We start by verifying that \pcbal\ satisfies both \textbf{no duplication} and \textbf{integrity}.

\begin{theorem}
\label{theorem:pcbnoduplication}
\pcbal\ satisfies no duplication.
\begin{proof}
A message is delivered (\cref{line:pcbdeliver}) only if the variable $delivered$ is equal to $\false$ (\cref{line:pcbthreshold}). Before any message is delivered, $delivered$ is set to $\true$. Therefore no more than one message is ever delivered.
\end{proof}
\end{theorem}

\begin{theorem}
\label{theorem:pcbintegrity}
\pcbal\ satisfies integrity.
\begin{proof}
Upon receiving an {\tt Echo} message, a correct process checks its signature against the public of the designated sender $\sigma$ (\cref{line:pcbcheckechosignature}), and the $(message, signature)$ pair is added to the $replies$ variable only if this check succeeds. Moreover, a message is delivered only if it is represented at least $\hat E > 0$ times in $replies$ (\cref{line:pcbthreshold}).

If $\sigma$ is correct, it only signs $message$ when broadcasting (\cref{line:pcbbroadcast}). Since we assume that cryptographic signatures cannot be forged, this implies that the message was previously broadcast by $\sigma$.
\end{proof}
\end{theorem}

\subsection{Total validity}
\label{subsection:pcb-total-validity}

We now compute, given $E$ and $\hat E$, the $\epsilon$-\textbf{total validity} of \pcbal. To this end, we prove some preliminary lemmas.

\begin{lemma}
\label{lemma:pcbvaliditygivennotpbtotality}
In an execution of \pcbal, if \pbin\ does not satisfy totality, then \pcbin\ does not satisfy total validity.
\begin{proof}
A correct process delivers a message (\cref{line:pcbdeliver}) only if the $echo$ variable is different from $\bot$. Moreover, the $echo$ variable is set to a value different from $\bot$  (\cref{line:pcbsetecho}) only upon \pbin.Delivery of a message (\cref{line:pcbpbdelivery}).

Let $m$ be the message broadcast by the correct sender $\sigma$. If \pbin\ does not satisfy totality, then at least one correct process never sets $echo$ to $m$. Therefore, at least one correct process does not deliver the $m$, and the total validity of \pcbin\ is comrpomised.
\end{proof}
\end{lemma}

\begin{lemma}
\label{lemma:pcbvalidityoverwhelmed}
In an execution of \pcbal, if \pbin\ satisfies totality and no correct process has more than $E - \hat E$ Byzantine processes in its echo sample, then \pcbin\ satisfies total validity.
\begin{proof}
Let $m$ be the message broadcast by the correct sender $\sigma$. Since \pbin\ satisfies totality (it always satisfies validity), every correct process eventually issues an {\tt Echo}($m$) message (i.e., an {\tt Echo} message for $m$) (\cref{line:pcbsendecho}).

Let $\pi$ be a correct process that has no more than $E - \hat E$ Byzantine processes in its echo sample. Obviously, $\pi$ has at least $\hat E$ correct processes in its echo sample. Therefore, $\pi$ eventually receives at least $\hat E$ {\tt Echo}($m$) messages (\cref{line:pcbreceiveecho}), and delivers $m$ (\cref{line:pcbthreshold}).
\end{proof}
\end{lemma}

\cref{lemma:pcbvaliditygivennotpbtotality,lemma:pcbvalidityoverwhelmed} allow us to bound the $\epsilon$-total validity of \pcbal, given $E$ and $\hat E$.

\begin{theorem}
\label{theorem:pcbtotalvalidity}
\pcbal\ satisfies $\epsilon_v$-total validity, with
\begin{equation}
\begin{split}
    \epsilon_v &\leq \epsilon_t^{\pbin} + \rp{1 - \epsilon_t^{\pbin}} \rp{1 - \rp{1 - \epsilon_o}^C}\\
    \epsilon_o &= \sum_{\bar F = E - \hat E + 1}^E \bin{E}{f}{\bar F}
\end{split}
\end{equation}
if the underlying abstraction of \pb\ satisfies $\epsilon_t^{\pbin}$-totality.
\begin{proof}
Following from \cref{lemma:pcbvaliditygivennotpbtotality,lemma:pcbvalidityoverwhelmed}, the total validity of \pcbin\ can be compromised only if the totality of \pbin\ is compromised as well, or if at least one correct process has more than $E - \hat E$ Byzantine processes in its echo sample.

Since procedure $sample$ independently picks $E$ processes with replacement, each element of a correct process' echo sample has an independent probability $f$ of being Byzantine, i.e., the number of Byzantine processes in a correct echo sample is binomially distributed.

Therefore, a correct process has a probability $\epsilon_o$ of having more than $E - \hat E$ Byzantine processes in its echo sample. Since every correct process picks its echo sample independently, the probability of at least one correct process having more than $E - \hat E$ Byzantine processes in its echo sample is $1 - \rp{1 - \epsilon_o}^C$. 
\end{proof}
\end{theorem}

\subsection{Preliminary lemmas}
\label{subsection:pcb-preliminary-lemmas}

In order to compute an upper bound for the probability of the consistency of \pcbal\ being compromised, we will make use of some preliminary lemmas. The statements of these lemmas are independent from the context of \pcbal. For the sake of readability, we therefore gather them in this section, and use them throughout the rest of this appendix.

\begin{lemma}
\label{lemma:binomialmerge}
Let $A, B \in \mathbb{N}$, let $x, y \in \mathbb{N}$ such that $x + y \leq B$. Let $X$, $Y$ be random variables defined by
\begin{eqnarray*}
\prob{\bar X} &=& \bin{A}{\frac{x}{B}}{\bar X} \\
\prob{\bar Y \mid \bar X} &=& \bin{A - \bar X}{\frac{y}{B - x}}{\bar Y}
\end{eqnarray*}

We have
\begin{equation*}
    \prob{X + Y = K} = \bin{A}{\frac{x + y}{B}}{K}
\end{equation*}

\begin{proof}
Since $X$ is binomially distributed, it can be expressed as a sum of independent Bernoulli random variables:
\begin{eqnarray*}
X &=& X_1 + \ldots + X_A \\
X_i &\sim& \text{Bern}\qp{\frac{x}{B}}
\end{eqnarray*}

Given the value of $\bar X$, $Y$ is also binomially distributed with probability $y / (B - x)$ and $E - \bar X$ trials. We can therefore express $Y$ as the sum of $E$ Bernoulli variables $Y_1, \ldots, Y_E$:
\begin{eqnarray*}
Y &=& Y_1 + \ldots + Y_E \\
\prob{Y_i = 1 \mid \bar X_i} &=&
\begin{cases}
    0 &\text{iff}\; \bar X_i = 1 \\
    \frac{y}{B - x} &\text{otherwise}
\end{cases}
\end{eqnarray*}

We indeed note how, out of $Y_1, \ldots, Y_E$:
\begin{itemize}
\item Only $E - \bar X$ variables have a non-null probability of being equal to $1$.
\item Those variables that have a non-null probability of being equal to $1$ have a probability $y / (B - x)$ of being equal to $1$.
\end{itemize}

We therefore have
\begin{equation*}
    X + Y = \rp{X_1 + Y_1} + \ldots + \rp{X_A + Y_A}
\end{equation*}
and from the law of total probability we have
\begin{eqnarray*}
    \prob{X_i + Y_i = 1} &=& \prob{X_i = 1} + \prob{Y_i = 1 \mid X_i = 0}\prob{X_i = 0} \\
    &=& \frac{x}{B} + \rp{1 - \frac{x}{B}}\rp{\frac{y}{B - x}} \\
    &=& \frac{x}{B} + \rp{\frac{\rp{B - x}}{B} \frac{y}{\rp{B - x}}} \\
    &=& \frac{x + y}{B}
\end{eqnarray*}
therefore
\begin{equation*}
    \rp{X_i + Y_i} \sim \text{Bern}\qp{\frac{x + y}{B}}
\end{equation*}
which proves the lemma.
\end{proof}
\end{lemma}

\begin{lemma}
\label{lemma:narrowbinomialsample}
Let $A, B \in \mathbb{N}$ such that $A \geq B$, let $p \in [0, 1]$. Let $X_1, \ldots, X_B$ be random variables defined by
\begin{equation*}
    X_i \sim \bin{A - i}{p}{}
\end{equation*}

We have that
\begin{equation*}
    \prob{X_i \geq B - i}
\end{equation*}
is an increasing function of $i$.

\begin{proof}

We prove the lemma by induction by showing that, for any $i < B$,
\begin{equation*}
    \prob{X_i \geq B - i} \leq \prob{X_{i + 1} \geq B - \rp{i + 1}}
\end{equation*}

In order to obtain the above, we expand
\begin{eqnarray*}
\lhs \prob{X_i \geq B - i} - \prob{X_{i + 1} \geq B - i - 1} \\
&=& \sum_{n = B - i}^{A - i} \rp{\frac{\rp{A - i}!}{\rp{A - i - n}! n!} p^n \rp{1 - p}^{A - i - n}} \\
&& - \sum_{n = B - i - 1}^{A - i - 1} \rp{\frac{\rp{A - i - 1}!}{\rp{A - i - 1 - n}! n!} p^n \rp{1 - p}^{A - i - 1 - n}} \\
&=& \rp{\star_1}
\end{eqnarray*}

By shifting the index in the second sum we get
\begin{eqnarray*}
\rp{\star_1} &=& \sum_{n = B - i}^{A - i} \rp{\frac{\rp{A - i}!}{\rp{A - i - n}! n!} p^n \rp{1 - p}^{A - i - n}} \\
&& - \sum_{n = B - i}^{A - i} \left ( \frac{\rp{A - i - 1}!}{\rp{A - i - 1 - \rp{n - 1}}! \rp{n - 1}!} \right . \\
&& \phantom{- \sum_{n = B - i}^{A - i} \bigg (}  p^{\rp{n - 1}} \rp{1 - p}^{A - i - 1 - \rp{n - 1}} \bigg ) \\
&=& \sum_{n = B - i}^{A - i} \left( \frac{\rp{A - i}!}{\rp{A - i - n}! n!} p^n \rp{1 - p}^{A - i - n} \right. \\ 
&& \phantom{\sum_{n = B - i}^{A - i} \big(} \left. - \frac{\rp{A - i}! n}{\rp{A - i}\rp{A - i - n}! n!} \frac{p^n}{p}\rp{1 - p}^{A - i - n} \right) \\
&=& \sum_{n = B - i}^{A - i} \rp{\rp{\frac{\rp{A - i}!}{\rp{A - i - n}! n!} p^n \rp{1 - p}^{A - i - n}} \rp{1 - \frac{n}{\rp{A - i}p}}} \\
&=& \rp{\star_2}
\end{eqnarray*}
and by letting $N = A - i$, $M = B - i$ we get
\begin{equation*}
    \rp{\star_2} = \sum_{n = M}^N \rp{\bin{N}{p}{n} \rp{1 - \frac{n}{Np}}} = \rp{\star_3}
\end{equation*}

Noticing that $(1 - n / Np)$ is positive for $n < Np$, we have
\begin{eqnarray*}
    \rp{\star_3} &\leq& \sum_{n = 0}^N \rp{\bin{N}{p}{n}} - \frac{1}{Np}\sum_{n = 0}^N \rp{n \bin{N}{p}{n}} \\
    &=& 1 - \frac{Np}{Np} = 0
\end{eqnarray*}
which proves the lemma.
\end{proof}
\end{lemma}

\begin{notation}[Ranges]
Let $a, b \in \mathbb{N}$, with $b \geq a$. We use $a..b$ to denote the range of integers $\cp{a, \ldots, b}$.
\end{notation}

\begin{lemma}
\label{lemma:shiftweight}
Let $f, g: 0..K \rightarrow \mathbb{R}$, with $f$ increasing, $g$ positive and
\begin{equation*}
    \sum_{x = 0}^K g(x) = 1
\end{equation*}
we have
\begin{equation*}
    \sum_{x = 0}^K \rp{f(x)g(x)} \geq \sum_{x = 0}^{K - 1}\rp{\frac{f(x)g(x)}{1 - g(K)}}
\end{equation*}

\begin{proof}
We have
\begin{eqnarray*}
\lhs \sum_{x = 0}^K \rp{f(x)g(x)} - \sum_{x = 0}^{K - 1} \rp{\frac{f(x)g(x)}{1 - g(K)}} \\ 
&=& f(K)g(K) + \sum_{x = 0}^{K - 1} \rp{f(x)g(x)}\rp{1 - \frac{1}{1 - g(K)}} \\
&=& f(K)g(K) - \sum_{x = 0}^{K - 1} \rp{f(x)g(x)} \rp{\frac{1 - \rp{1 - g(K)}}{1 - g(K)}} \\
&=& g(K) \rp{f(K) - \frac{1}{1 - g(K)} \sum_{x = 0}^{K - 1} \rp{f(x)g(x)}} \\
&=& \frac{g(K)}{1 - g(K)} \rp{f(K) - f(K)g(K) - \sum_{x = 0}^{K - 1}\rp{f(x)g(x)}} \\
&=& \frac{g(K)}{1 - g(K)} \rp{f(K) \sum_{x = 0}^K\rp{f(x)g(x)}} \\
&=& \rp{\star_1}
\end{eqnarray*}
and noting that $g(K) \geq 0$, $1 - g(K) \geq 0$, and $f$ is increasing, we have
\begin{equation*}
\rp{\star_1} \geq \frac{g(K)}{1 - g(K)} \rp{f(K) - f(K) \sum_{x = 0}^K g(x)} = \rp{\star_2}
\end{equation*}
and since $\sum g(x) = 1$ we get
\begin{equation*}
    \frac{g(K)}{1 - g(K)}\rp{f(k) - f(K)} = 0
\end{equation*}
\end{proof}
\end{lemma}

\begin{corollary}
\label{corollary:shiftweight}
Let $f, g: 0..K \rightarrow \mathbb{R}$, with $f$ increasing, $g$ positive and
\begin{equation*}
    \sum_{x = 0}^K g(x) = 1
\end{equation*}
for any $l \in 0..(K - 1)$, we have
\begin{equation*}
    \sum_{x = 0}^K \rp{f(x)g(x)} \geq \frac{\sum_{x = 0}^{K - l}f(x)g(x)}{\sum_{x = 0}^{K - l} g(x)}
\end{equation*}

\begin{proof}
It follows immediately from applying \cref{lemma:shiftweight} $l$ times.
\end{proof}
\end{corollary}

\begin{lemma}
\label{lemma:cumulativebound}
Let $f: -1..C \rightarrow \mathbb{R}$, let $g, h: -1..C \rightarrow [0, 1]$, with:
\begin{itemize}
    \item $f$ decreasing.
    \item $g, h$ increasing.
    \item $g(x) \leq h(x)$ for all $x$.
    \item $g(-1) = h(-1) = 0$.
    \item $g(C) = h(C) = 1$.
\end{itemize}

We have
\begin{equation*}
    \sum_{x = 0}^C \rp{f(x) \rp{g(x) - g(x - 1)}} \leq \sum_{x = 0}^C \rp{f(x) \rp{h(x) - h(x - 1)}}
\end{equation*}

\begin{proof}
We have
\begin{eqnarray*}
\lhs \sum_{x = 0}^C f(x)\rp{g(x) - g(x - 1)} - \sum_{x = 0}^C f(x) \rp{h(x) - h(x - 1)} \\
&=& \sum_{x = 0}^C f(x) \rp{\rp{g(x) - h(x)} - \rp{g(x - 1) - h(x - 1)}} \\
&=& \sum_{x = 0}^C f(x) \rp{g(x) - h(x)} - \sum_{x = 0}^C f(x) \rp{g(x - 1) - h(x - 1)} \\
&=& \rp{\star_1}
\end{eqnarray*}

By shifting the index in then second sum we get
\begin{eqnarray*}
\rp{\star_1} &=& \sum_{x = 0}^C f(x) \rp{g(x) - h(x)} - \sum_{x = -1}^{C - 1} f(x + 1) \rp{g(x) - h(x)} \\
&=& \sum_{x = 0}^{C - 1} \rp{f(x) - f(x + 1)} \rp{g(x) - h(x)} \\
&& + f(C)\rp{g(C) - h(C)} - f(-1)\rp{g(-1) - h(-1)} \\
&=& \rp{\star_2}
\end{eqnarray*}
and by noting that:
\begin{itemize}
    \item Since $f$ is decreasing, $f(x) - f(x + 1) \geq 0$.
    \item By hypothesis, $g(x) - h(x) \leq 0$.
    \item By hypothesis, $g(C) - h(C) = 1 - 1 = 0$.
    \item By hypothesis, $g(-1) - h(-1) = 0 - 0 = 0$.
\end{itemize}

Consequently, all the terms of the sum in $\rp{\star_2}$ are negative, and the two terms out of the sum are null. Therefore, $\rp{\star_2} \leq 0$.
\end{proof}
\end{lemma}

\begin{lemma}
\label{lemma:chernoffconvexity}
Let $N \in \mathbb{N}$, let $K < N$, let $h$, $p_1, \ldots, p_T \in [0, 1]$ such that
\begin{equation*}
    h = \sum_i p_i \leq \frac{K - \sqrt{K}}{N}
\end{equation*}
let $X_1, \ldots, X_T$ be independent random variables defined by
\begin{equation*}
    \prob{\bar X_i} = \bin{N}{p_i}{\bar X_i}
\end{equation*}

We have
\begin{equation*}
\prob{\bigvee_i \rp{X_i > K}} \leq \rp{\frac{eNh}{K}}^K e^{-Nh}
\end{equation*}

\begin{proof}
Let $p \in [0, 1]$, let $X \sim \bin{N}{p}{}$. From the multiplicative form of the Chernoff bound we have
\begin{eqnarray*}
    \prob{X > K} &=& \prob{X > (1 + \delta) \mu} < \rp{\frac{e^\delta}{(1 + \delta)^{(1 + \delta)}}}^\mu \\
    \delta &=& \rp{\frac{K}{N p} - 1} \\
    \mu &=& N p
\end{eqnarray*}

From the above follows
\begin{eqnarray*}
    \prob{X > K} &<& \rp{\frac{\exp \rp{\frac{K}{Np} - 1}}{\rp{\frac{K}{Np}}^{\frac{K}{Np}}}}^\mu \\
    &=& \exp \rp{Np \rp{\frac{K}{Np} - 1 - \frac{K}{Np} \log \rp{\frac{K}{Np}}}} \\
    &=& \exp \rp{K - Np - K \log \rp{\frac{K}{Np}}} \\
    &=& \underbrace{\exp \rp{K - K \log K + K \log N}}_{\rp{\star_a}} \underbrace{\exp \rp{K \log p - Np}}_{\rp{\star_b}}
\end{eqnarray*}

We now study the domain where $\rp{\star_b}$ is convex:
\begin{eqnarray*}
    \frac{\partial^2}{\partial^2 p} \exp \rp{K \log p - N p} &=&
    \underbrace{p^{K - 2}e^{-Np}}_{\geq 0}\rp{K^2 -K\rp{2Np + 1} + N^2 p^2} \\
\end{eqnarray*}

Therefore we require
\begin{eqnarray*}
    N^2p^2 - \rp{2KN}p + \rp{K^2-K} &\geq& 0
\end{eqnarray*}

Which reduces to
\begin{eqnarray*}
    p &\leq& \frac{2KN - \sqrt{4 K^2 N^2 - 4 \rp{N^2 K^2 - N^2 K}}}{2N^2} \\
    &=& \frac{K - \sqrt{K}}{N}
\end{eqnarray*}

From Boole's inequality we have
\begin{equation*}
    \prob{\bigvee_i \rp{X_i > K}} \leq \sum_i \prob{X_i > K} = \rp{\star_1}
\end{equation*}
which we can expand into
\begin{equation*}
    \rp{\star_1} = \underbrace{\exp \rp{K - K \log K + K \log N}}_{\rp{\star_a}} \sum_i \underbrace{\exp \rp{K \log p_i - N p_i}}_{\rp{\star_b}} = \rp{\star_2}
\end{equation*}
as we established, $\rp{\star_b}$ is convex on the range $\qp{0, \sum_i p_i}$. Consequently,
\begin{eqnarray*}
    \rp{\star_2} &\leq& \exp \rp{K - K \log K + K \log N}\exp \rp{K \log h - Nh} \\
    &=& \rp{\frac{eNh}{K}}^K e^{-Nh}
\end{eqnarray*}
which proves the lemma.
\end{proof}
\end{lemma}

\subsection{\cobal}
\label{subsection:cobal}

In this section, we introduce \cobal, a modified version of \pcbal.

\cobal\ is a strawman both from a performance and a safety point of view. Indeed, on the one hand \cobal\ has $O(N^2)$ per-process communication complexity, which makes it unfit for any real-world, scalable deployment. On the other, we prove that it is strictly easier for any Byzantine adversary to compromise the consistency of \cobal\ than that of \pcbal.

Unlike \pcbal, however, \cobal\ allows for an analytic probabilistic analysis. A critical goal of this appendix is to compute a bound $\epsilon_c$ on the probability of compromising the consistency of \cobal. Since the consistency of \cobal\ is weaker than that of \pcbal, $\epsilon_c$ is also a bound on the probability of compromising the consistency of \pcbal.

\subsubsection{\Cob}
\label{subsubsection:cobinterface}

\cobal\ implements \cob, a minimal version of the \pcb\ abstraction, designed to only provide $\epsilon$-consistency. In particular, we drop the no duplication property, i.e., we allow a correct process to deliver more than one message.\\

The \textbf{\cob} interface (instance $\cobin$, sender $\sigma$) exposes the following two \textbf{events}:
\begin{itemize}
    \item \textbf{Request}: $\event{\cobin}{Broadcast}{m}$: Broadcasts a message $m$ to all processes. This is only used by $\sigma$.
    \item \textbf{Indication} $\event{\cobin}{Deliver}{m}$: Delivers a message $m$ broadcast by process $\sigma$.
\end{itemize}

For any $\epsilon \in [0, 1]$, we say that \cob\ is $\epsilon$-secure if:
\begin{enumerate}
    \item $\epsilon$-\textbf{Consistency}: With probability at least $(1 - \epsilon)$, at most one message $m$ exists, such that $m$ is delivered by any correct process.
\end{enumerate}

We note how the above definition of $\epsilon$-consistency is equivalent to the one we provided in \cref{subsection:pcbdefinition}, but adapted for a context where no duplication is not guaranteed. In \cob, consistency is compromised even if a single correct process delivers two or more different messages.

\subsubsection{Byzantine oracle}

In order to implement \cobal, we make an additional assumption about the system:
\begin{itemize}
    \item (\textbf{Byzantine oracle}) \label{assumption:byzantineoracle} Every correct process has direct access to an oracle $\Psi$ that, provided with a process $\pi$, returns $\true$ if $\pi$ is Byzantine, and $\false$ if $\pi$ is correct.
\end{itemize}

This assumption is obviously unsatisfiable in any realistic distributed system. Indeed, a system subject to Byzantine failures where every correct process can tell correct processes from Byzantine failures is hardly a Byzantine system. It is therefore critical to underline that Assumption \ref{assumption:byzantineoracle} is \textbf{not} a requirement for the implementation of \pcbal. Indeed, no correct process invokes $\Psi$ throughout any execution of \cref{algorithm:pcbal}. Assumption \ref{assumption:byzantineoracle} is purely a theoretical artifice to aid in our proof of correctness.

\subsubsection{Algorithm}
\label{subsubsection:cobalgorithm}

Before introducing the design principles behind \cobal, we prove a simple preliminary result.

\begin{lemma}
No execution of \pb\ results in more than $C$ different messages being delivered.
\begin{proof}
Following from \cref{theorem:pbnoduplication}, \pb\ satisfies no duplication, i.e., no correct process delivers more than one message. As we discussed in \cref{section:model}, the system is composed of $C$ correct processes.
\end{proof}
\end{lemma}

Since the set of messages that are \pbin.Delivered by at least one correct process has no more than $C$ elements, and noting that a correct process \pcbin.Delivers a message $m$ only if it \pbin.Delivered $m$, it is not restrictive to introduce the following definition.

\begin{definition}[Message]
A \textbf{message} is an element of the set
\begin{equation*}
    \mathcal{M} = 1..C
\end{equation*}
\end{definition}

\begin{algorithm}
\begin{algorithmic}[1]
    \Procedure{correct}{{}} \label{line:mimiccorrect}
        \DoUntil
            \State $\rho = \Omega(1)$ \label{line:mimiccorrectpick}
        \EndDoUntil{\Psi(\rho) = \false} \label{line:mimiccorrectcheck}
        \State \Return $\rho$;
    \EndProcedure
    \Procedure{mimic}{reference} \label{line:mimicmimic}
        \State $\psi = \emptyset$;
        \ForAll{\rho}{reference}
            \If{$\Psi(\rho) = \true$} 
                \State $\psi \leftarrow \psi \cup \cp{\rho}$; \label{line:mimicmimicbyzantine}
            \Else
                \State $\psi \leftarrow \psi \cup correct()$; \label{line:mimicmimiccorrect}
            \EndIf
        \EndForAll
        \State \Return $\psi$;
    \EndProcedure
\end{algorithmic}
\caption{Procedure $mimic$}
\label{algorithm:mimic}
\end{algorithm}

\begin{algorithm}
\begin{algorithmic}[1]
\Implements
    \Instance{\cobab}{\cobin}
\EndImplements

\Uses
    \Instance{AuthenticatedPointToPointLinks}{al}
    \Instance{\pbab}{\pbin}
\EndUses

\Parameters
    \State $E$: echo sample size
    \State $\hat E$: delivery threshold
\EndParameters

\Upon{\cobin}{Init}{} \label{line:cobinit}
    \State $delivered = \cp{\false}^C$; $reveal = \cp{\false}^C$;
    \State $revealed = \cp{\false}^C$; 
    \State $replies = \cp{\bot}^{C \times E}$; \Comment{$C \times E$ table filled with $\bot$.}
    \State $\mathcal{E} = \cp{\emptyset}^{C}$; 
    \State $\mathcal{E}[1] \leftarrow sample({\tt EchoSubscribe}, E)$; \label{line:cobpickfirstsample}
    \State
    \For{j}{2}{C}
        \State $\mathcal{E}[j] \leftarrow mimic(\mathcal{E}[1])$; \label{line:cobpickothersamples}
    \EndFor
\EndUpon

\Upon{\cobin}{Broadcast}{message} \Comment{only process $\sigma$} \label{line:cobbroadcast}
    \Trigger{\pbin}{Broadcast}{[\text{\tt Send}, message]};
\EndUpon

\Upon{\pbin}{Deliver}{[\text{\tt Send}, message]} \label{line:cobpbdeliver}
    \ForAll{\rho}{\Pi}
        \ForAll{sample}{\mathcal{M}}
            \Trigger{al}{Send}{\rho, [\text{\tt Echo}, sample, message]}; \label{line:cobsendecho}
        \EndForAll
    \EndForAll
\EndUpon

\Upon{al}{Deliver}{\rho, [\text{\tt Echo}, sample, message]} \label{line:cobreceiveecho}
    \If{$\rho \in \mathcal{E}[sample] \;\text{and}\; replies[sample][\rho] = \bot$}
        \State $replies[sample][\rho] \leftarrow message$; \label{line:cobsetreply}
        \State $revealed[sample] \leftarrow \false$; \label{line:cobsetrevealed}
    \EndIf
\EndUpon
\algstore{cobal}
\end{algorithmic}
\caption{\cobal}
\label{algorithm:cobal}
\end{algorithm}

\begin{algorithm}
\begin{algorithmic}[1]
\algrestore{cobal}
\UponExists{message}{|\{\rho \in \mathcal{E}[message] \mid replies[message][\rho] = \allowbreak message\}| \geq \hat E \textbf{ and } delivered[message] = \false} \label{line:cobdeliverythreshold}
    \State $delivered[message] \leftarrow \true$;
    \State $reveal[message] \leftarrow \true$; \label{line:cobrevealbecausedeliver}
    \Trigger{\cobin}{Deliver}{message}; \label{line:cobdelivery}
\EndUponExists

\UponExists{message}{reveal[message] = \true \textbf{ and } \allowbreak revealed[message] = \false} 
\label{line:cobreveal}
    \State $revealed[message] \leftarrow \true$;
    \State $sample = \cp{\rho \in \mathcal{E}[message] \mid replies[message][\rho] \neq \bot}$;
    \ForAll{\pi}{\Pi}
        \Trigger{al}{Send}{\pi, [\text{\tt Reveal}, message, sample]}; \label{line:cobrevealsend}
    \EndForAll
\EndUponExists

\Upon{al}{Deliver}{\rho, [{\tt Reveal}, message, sample]}
    \If{$\Psi(\rho) = \false$}
        \State $reveal[message] \leftarrow \true$; \label{line:cobrevealfeedback}
    \EndIf
\EndUpon
\end{algorithmic}
\end{algorithm}

\cref{algorithm:cobal} implements \cobal. \cobal\ bears multiple differences to \pcbal:
\begin{itemize}
    \item A correct process can deliver more than one message. No correct process, however, delivers the same message more than once.
    \item In order to \cobin.Deliver a message, a correct process does not need to \pbin.Deliver any message.
    \item A correct process maintains $C$ \textbf{echo samples} $\mathcal{E}[1..C]$. The {\tt Echo} messages collected from the processes in the $i$-th echo sample $\mathcal{E}[i]$ determine whether or not message $i \in \mathcal{M}$ is delivered.
    \item {\tt Echo} messages have two fields: $sample$ and $message$. Intuitively, an {\tt Echo}$(s, m)$ message (i.e., an {\tt Echo} message with fields $s$ and $m$) represents the following statement: ``within the context of message $s$, consider my {\tt Echo} to be for message $m$".

    Upon \pbin.Delivering a message $m$, a correct process sends $C$ {\tt Echo} messages to each other process, one {\tt Echo}$(s, m)$ message for every $s \in \mathcal{M}$. In other words, the correct behavior is to echo $m$ across all contexts $s \in \mathcal{M}$. A Byzantine process, however, can in principle send to the same process a set of {\tt Echo} messages echoing different messages in different contexts (e.g, {\tt Echo}$(s, m)$ and {\tt Echo}$(s', m' \neq m)$).
    \item When a correct process $\pi$ collects at least $\hat E$ {\tt Echo}$(m, m)$ messages from the processes in $\mathcal{E}[m]$, $\pi$ delivers $m$.
\end{itemize}

\paragraph{Mimic} \cref{algorithm:mimic} presents two utility procedures for manipulating samples with respect to their Byzantine component:
\begin{itemize}
    \item ($correct$, \cref{line:mimiccorrect}) Procedure $correct$ returns a correct process, picked with uniform probability. It does so by invoking $\Omega$ to select a process $\rho$ with uniform probability (\cref{line:mimiccorrectpick}), then using $\Psi$ to repick if $\rho$ is Byzantine (\cref{line:mimiccorrectcheck}).
    \item ($mimic$, \cref{line:mimicmimic}) Provided with a sample $reference$, procedure $mimic$ returns a sample $\psi$ that shares with $reference$ all Byzantine processes. It does so by looping over each process $\rho$ in $reference$. If $\rho$ is Byzantine (\cref{line:mimicmimicbyzantine}), $\rho$ is added to $\psi$. If $\rho$ is correct (\cref{line:mimicmimiccorrect}), procedure $correct()$ is used to add a random correct process to $\psi$.
\end{itemize}

\paragraph{Samples} Upon initialization (\cref{line:cobinit}), a correct process initializes $C$ echo samples $\mathcal{E}[1..C]$ that share the same set of Byzantine processes. It does so by using procedure $sample(\ldots)$ to randomly pick $\mathcal{E}[1]$ (\cref{line:cobpickfirstsample}), then using $mimic(\mathcal{E}[1])$ to pick samples $2$ to $C$ (\cref{line:cobpickothersamples}).

We underline how $\mathcal{E}[1]$ is selected using the $sample$ procedure we defined in \cref{algorithm:sample}. As a result, upon initialization, a correct process sends an {\tt EchoSubscribe} message to each process in $\mathcal{E}[1]$. However, a correct process does not handle the al.Delivery of an {\tt EchoSubscribe} message. This is done on purpose. The only goal of those {\tt EchoSubscribe} messages is to let the Byzantine adversary know which Byzantine processes are in $\mathcal{E}[1]$ (and, consequently, in every other sample).

\paragraph{Broadcast} Upon \cobin.Broadcasting a message $message$ (\cref{line:cobbroadcast}), the correct designated sender uses \pbin.Broadcast to distribute $message$.

\paragraph{Echo} When a correct process \pbin.Delivers a message $message$ (\cref{line:cobpbdeliver}), it sends to each process $\rho$ an {\tt Echo}($sample, message$) message, for every $sample$ in $\mathcal{M}$ (\cref{line:cobsendecho}). In other words, the correct behavior of a correct process that \pbin.Delivered $message$ is to echo $message$ across all samples.

We note how \cobal\ does not make use of echo subscription sets. A correct process sends its {\tt Echo} messages to every process in the system. The goal of \cobal, indeed, is not performance, but probabilistic tractability.

\paragraph{Delivery} A correct process maintains a table $replies$ to keep track of the {\tt Echo} messages received by each node in its echo samples. Upon receiving an {\tt Echo}($sample$, $message$) message from a process $\rho$ for the first time (\cref{line:cobreceiveecho}), if $\rho$ is in $\mathcal{E}[sample]$, a correct process sets $replies[sample][\rho]$ to $message$ (\cref{line:cobsetreply}). 

Upon receiving at least $\hat E$ {\tt Echo}($message$, $message$) messages from the processes in $\mathcal{E}[message]$ (\cref{line:cobdeliverythreshold}) (this is checked using the $replies$ table), a correct process \cobin.Delivers $message$ (\cref{line:cobdelivery}).

\paragraph{Reveal} A correct process maintains a $reveal$ array to keep track of  which echo samples it should \emph{reveal}. When, for some $message$, $reveal[message] = \true$ (\cref{line:cobreveal}), a correct process sends to every process a {\tt Reveal} message, containing the set of processes in $\mathcal{E}[message]$ that issued an {\tt Echo}($message$, $message'$) message for some $message' \in \mathcal{M}$ (\cref{line:cobrevealsend}). In other words, whenever $reveal[message] = \true$, a correct process reveals the set of processes in its echo sample for $message$ that issued a an {\tt Echo} message for that sample.

If, after revealing its sample for $message$, a correct process receives additional {\tt Echo} messages from the processes in $\mathcal{E}[message]$, the reveal procedure is performed again. This is enforced by setting a $revealed$ flag back to $\false$ (\cref{line:cobsetrevealed}) every time a new {\tt Echo} message is received.

A correct process sets $reveal[message]$ to $\true$ under two circumstances: when it \cobin.Delivers $message$ (\cref{line:cobrevealbecausedeliver}) and when it receives a {\tt Reveal} message for $message$ from a correct process (\cref{line:cobrevealfeedback}). As a result, whenever any correct process delivers $message$, every correct process reveals its sample for $message$, regardless of whether or not it delivered $message$.

Like {\tt EchoSubscribe}, the {\tt Reveal} message serves the only purpose to provide information to the Byzantine adversary.

\subsection{Adversarial execution}
\label{subsection:adversarialexecution}

In this section, we define the model underlying an adversarial execution of \pcbal\ and \cobal, and identify the set of Byzantine adversaries for each algorithm. Here, a Byzantine adversary is an agent that acts upon a system with the goal to compromise its consistency. Throughout the rest of this appendix, we use the term \textbf{\pcbin\ adversary} to denote a Byzantine adversary for \pcbal, and the term \textbf{\cobin\ adversary} (or just \textbf{adversary}) to denote a Byzantine adversary for \cobal.

The main goal of this section is to formalize the information available both to the \pcbin\ and the \cobin\ adversary, and the set of actions that they can perform on the system throughout an adversarial execution of either algorithm. \\

Throughout the rest of this appendix, we bound the probability of compromising the consistency of \pcbal\ by assuming that, if the totality of \pbin\ is compromised, then the consistency of \pcbin\ is compromised as well. In what follows, therefore, we assume that \pbin\ satisfies totality.

\subsubsection{Model (\pcbal)}
\label{subsubsection:modelpcbadversary}

Let $\pi$ be any correct process. We make the following assumptions about an adversarial execution of \pcbal:
\begin{itemize}
    \item As we established in \cref{section:model}, the \pcbin\ adversary does not know which correct processes are in $\pi$' echo sample. The \pcbin\ adversary knows, however, which Byzantine processes are in $\pi$'s echo sample.
    \item At any time, the \pcbin\ adversary knows if $\pi$ delivered a message. If $\pi$ delivered a message, then the \pcbin\ adversary knows which message did $\pi$ deliver.
    \item The \pcbin\ adversary can cause $\pi$ to \pbin.Deliver any message. As we established with \cref{theorem:pbnoduplication}, $\pi$ will, however, \pbin.Deliver only one message throughout an execution of \pcbal.
\end{itemize}

Throughout an adversarial execution of \pcbal, an adversary performs a sequence of minimal operations on the system. Each operation consists of either of the following:
\begin{itemize}
    \item Selecting a correct process that did not \pbin.Deliver any message, and causing it to \pbin.Deliver a message.
    \item Selecting a Byzantine process and causing it to send an {\tt Echo} message to a correct process.
\end{itemize}
As a result of each operation, zero or more correct processes deliver a message. The \pcbin\ adversary is successful if, at the end of the adversarial execution, at least two different messages are delivered by at least one correct process.

\subsubsection{Model (\cobal)}
\label{subsubsection:modelcobadversary}

Let $\pi$ be any correct process. We make the following assumptions about an adversarial execution of \cobal:
\begin{itemize}
    \item As we established in \cref{section:model}, the \cobin\ adversary does not know which correct processes are in $\pi$'s echo samples. The \cobin\ adversary knows, however, which Byzantine processes are in $\pi$'s echo samples.
    \item At any time, the \cobin\ adversary knows if $\pi$ delivered a message. If $\pi$ delivered a message, then the \cobin\ adversary knows which message did $\pi$ deliver. Moreover, if $\pi$ delivered a message $m$, then at any time the \cobin\ adversary also knows the processes in $\pi$'s echo sample for $m$ that sent an {\tt Echo}($m$, $m'$) message to $\pi$, for some message $m'$.
    \item The \cobin\ adversary can cause $\pi$ to \pbin.Deliver any message. As we established with \cref{theorem:pbnoduplication}, $\pi$ will, however, \pbin.Deliver only one message throughout an execution of \cobal.
\end{itemize}

Throughout an adversarial execution of \cobal, an adversary performs a sequence of minimal operations on the system. Each operation consists of either of the following:
\begin{itemize}
    \item Selecting a correct process that did not \pbin.Deliver any message, and causing it to \pbin.Deliver a message.
    \item Selecting a Byzantine process and causing it to send an {\tt Echo} message to a correct process.
\end{itemize}
As a result of each operation, zero or more correct processes deliver a message. The \cobin\ adversary is successful if, at the end of the adversarial execution, at least two different messages are delivered by at least one correct process.

\subsubsection{Network scheduling}
\label{subsubsection:pcbcobnetworkscheduling}

In this section, we discuss the behavior of the adversary in relation to network scheduling. As we discussed in \cref{section:model}, the system is asynchronous, i.e., every message is eventually delivered but can be delayed by an arbitrary, finite amount of time.

\paragraph{{\tt Gossip} messages}

As we stated in \cref{subsection:adversarialexecution}, throughout this appendix we assume that the \pbin\ instance used by \pcbal\ and \cobal\ satisfies totality. While this means that the adversary cannot prevent any correct process from eventually \pbin.Delivering a message, the adversary can indeed arbitrarily choose which correct process \pbin.Delivers which message. 

This can be achieved by delaying the delivery of the {\tt Gossip} messages issued by correct processes. Noting that a correct process will accept a {\tt Gossip} message from any source, the adversary can then cause any of the processes it controls to quickly send a {\tt Gossip} message with arbitrary content to any correct process, effectively causing it to \pbin.Deliver an arbitrary message.

\paragraph{{\tt Echo} messages}

As we stated in \cref{subsubsection:modelpcbadversary,subsubsection:modelcobadversary}, the two minimal operations a (\pcbin\ or \cobin) adversary can perform essentially reduce to causing a Byzantine process to either send a {\tt Gossip} or an {\tt Echo} message to a correct process. We can see that those operations are indeed minimal: a correct process atomically al.Delivers a message (i.e., a message is the minimal amount of information that can be meaningfully transferred on the network), and a correct process will ignore any message that is not a {\tt Gossip} or an {\tt Echo} message.

Upon \pbin.Delivering a message, a correct process will issue zero or more {\tt Echo} messages. As we discussed in \cref{section:model}, the adversary can arbitrarily delay those messages, but they will eventually be delivered. As a result, the outcome of an adversarial execution is solely determined by the sequence of operations performed by the adversary, and is not affected by network scheduling.

While the adversary could delay the delivery of {\tt Echo} messages issued by correct processes, the only effect this would have is to prevent the adversary from knowing the effect of an operation on the system before performing the next one. An optimal adversary, therefore, performs an operation, then waits until all the {\tt Echo} messages issued by correct processes are delivered before performing the next operation.

\subsubsection{Interfaces}
\label{subsubsection:interfaces}

In \cref{subsubsection:modelpcbadversary,subsubsection:modelcobadversary}, we defined the model underlying an adversarial execution of \pcbal\ and \cobal\ respectively. In \cref{subsubsection:pcbcobnetworkscheduling}, we discussed the behavior of the Byzantine adversary in relation to network scheduling. Throughout the rest of this appendix, we concretely model a (\pcbin\ or \cobin) adversary as an \emph{algorithm} that interacts with a \emph{system}.

As we discussed, a (\pcbin\ or \cobin) adversary works in steps: at every step, the adversary either performs one operation on the system, or queries the system for information about its state. In this section, we model this interaction by defining four interfaces, respectively implemented by the (\pcbin\ or \cobin) adversary and the (\pcbin\ or \cobin) system. \\

Both the \textbf{\pcbin\ adversary} and the \textbf{\cobin\ adversary} interfaces (instance $adv$) expose the following \textbf{procedures}:
\begin{tolerant}{300}
\begin{itemize}
    \item $Init()$: It is called once, at the beginning of the adversarial execution, before any operation is performed on the system. Here the (\pcbin\ or \cobin) adversary setups its internal state.
    \item $Step()$: It is called repeatedly, until the adversarial execution is completed. Here the (\pcbin\ or \cobin) adversary performs one operation on the system. The execution fails (e.g., an exception is raised) if a call to $adv.Step()$ does not result in one, and only one, call to $sys.Deliver(\ldots)$, $sys.Echo(\ldots)$ or $sys.End()$ (as we define them below).
\end{itemize}
\end{tolerant}

The \textbf{\pcbin\ system} interface (instance $sys$) exposes the following \textbf{procedures}:
\begin{itemize}
    \item $Byzantine(process \in \Pi_C)$: Returns a list of all the Byzantine processes in $process$' echo sample. The \pcbin\ adversary can invoke this procedure an unlimited number of times both from the $Init()$ and the $Step()$ procedure.
    \item $State()$: Returns a list of pairs $(\pi \in \Pi_C, m \in \mathcal{M})$, representing which correct process currently delivered which message. The \pcbin\ adversary can invoke this procedure an unlimited number of times from the $Step()$ procedure.
    \item $Deliver(process \in \Pi_C, message \in \mathcal{M})$: Causes $process$ to \pbin.Deliver $message$. The execution fails if $Deliver$ is provided with the same $process$ argument more than once: a correct process does not \pbin.Deliver more than one message. The procedure does not return any value.
    \item $Echo(process \in \Pi_C, source \in \Pi \setminus \Pi_C, message \in \mathcal{M})$: Causes $source$ to send an {\tt Echo}($message$) message to $process$. The execution fails if $Echo$ is provided with the same $process$ and $source$ arguments more than once: a correct process does not consider more than one {\tt Echo} message from the same source. The procedure does not return any value.
    \item $End()$: Causes the execution to gracefully terminate. The execution fails if $End()$ is called before $Deliver(\ldots)$ is invoked exactly $C$ times: under the assumption that \pbin\ satisfies totality, every correct process eventually \pbin.Delivers a message. The procedure does not return any value.
\end{itemize}

The \textbf{\cobin\ system} interface (instance $sys$) exposes the following \textbf{procedures}:
\begin{itemize}
    \item $Byzantine(process \in \Pi_C)$: Returns a list of all the Byzantine processes in the first echo sample of $process$. The \cobin\ adversary can invoke this procedure an unlimited number of times both from the $Init()$ and the $Step()$ procedure.
    \item $State()$: Returns a list of pairs $(\pi \in \Pi_C, m \in \mathcal{M})$, representing which correct process currently delivered which message. The \cobin\ adversary can invoke this procedure an unlimited number of times from the $Step()$ procedure.
    \item $Sample(process \in \Pi_C, message \in \mathcal{M})$: Returns the processes that are in the echo sample for message $message$ of process $process$ and that sent an {\tt Echo}($message$, \allowbreak $message'$) to $process$, for some message $message'$. The \cobin\ adversary can invoke this procedure an unlimited number of times from the $Step()$ procedure. The execution fails if no correct process has \cobin.Delivered $message$: a correct process does not \emph{reveal} its echo sample for $message$ before $message$ is delivered by at least one correct process.
    \item $Deliver(process \in \Pi_c, message \in \mathcal{M})$: Causes $process$ to \pbin.Deliver $message$. The execution fails if $Deliver$ is provided with the same $process$ argument more than once: a correct process does not \pbin.Deliver more than one message. The procedure does not return any value.
    \item $Echo(process \in \Pi_C, sample \in \mathcal{M}, source \in \Pi \setminus \Pi_C, message \in \mathcal{M})$: Causes $source$ to send an {\tt Echo}($sample$, $message$) message to $process$. The execution fails if, throughout an execution, $Echo$ is provided with the same $process$, $sample$ and $source$ arguments more than once: a correct process does not consider more than one {\tt Echo} message for the same sample from the same source. The procedure does not return any value.
    \item $End()$: Causes the execution to gracefully terminate. The execution fails if $End()$ is called before $Deliver(\ldots)$ is invoked exactly $C$ times: under the assumption that \pbin\ satisfies totality, every correct process eventually \pbin.Delivers a message. The procedure does not return any value.
\end{itemize}

\subsection{Simplified adversarial power}
\label{subsection:simplifiedadversarialpower}

In this section, we prove that an optimal \cob\ adversary is more powerful than an optimal \pcb\ adversary. This result is intuitive: a correct proces in \cobal\ can deliver more than one message, and in general more information is available to the \cobin\ adversary than to the \pcbin\ adversary.

\subsubsection{Preliminary definitions}

Before proving that an optimal \cobin\ adversary is more powerful than an optimal \pcbin\ adversary, we provide some definitions on \pcbin\ and \cobin\ systems and adversaries.

\begin{definition}[\Pcbin\ system]
A \textbf{\pcbin\ system} $\sigma$ is an element of the set
\begin{eqnarray*}
    \mathcal{S}_{\pcbin} &=& \mathcal{E}_{\pcbin}^C \\
    \mathcal{E}_{\pcbin} &=& \Pi^E
\end{eqnarray*}

Intuitively, a system $\sigma \in \mathcal{S}_{\pcbin}$ is defined by the echo sample of each of its $C$ correct processes. The echo sample of a correct process is a vector of $E$ processes. 

Let $\sigma \in \mathcal{S}_{pcb}$, we use $\sigma[\pi \in \Pi_C][i \in 1..E]$ to denote the $i$-th process in $\pi$'s echo sample.
\end{definition}

\begin{definition}[\Cobin\ system]
A \textbf{\cobin\ system} $\sigma$ is an element of the set
\begin{eqnarray*}
    \mathcal{S}_{\cobin} &=& \mathcal{E}_{\cobin}^C \\
    \mathcal{E}_{\cobin} &=& \cp{\rp{e_1, \ldots, e_C \in \Pi^E} \mid \mathcal{M}\rp{e_i, e_j} \;\;\forall i, j \in 1..C} \\
    \mathcal{M}(e, e') &:& \forall k, \rp{e_k \in \Pi \setminus \Pi_C} \implies \rp{e'_k = e_k}
\end{eqnarray*}

Intuitively, a system $\sigma \in \mathcal{S}_{\cobin}$ is defined by the echo samples of each of its $C$ correct processes. Each correct process has $C$ echo samples $e_1, \ldots, e_C$ (one per message), each represented by a vector of $E$ processes. Any two echo samples $e_i$, $e_j$ of a given process satisfy $\mathcal{M}(e_i, e_j)$, i.e., they share the same set of Byzantine processes.

We also use just $\mathcal{S}$ to denote the set of \cobin\ systems $\mathcal{S}_{\cobin}$. Let $\sigma \in \mathcal{S}_{\cobin}$, we use $\sigma[\pi \in \Pi_C][m \in \mathcal{M}][i \in 1..E]$ to denote the $i$-th process in $\pi$'s echo sample for $m$.
\end{definition}

\begin{definition}[Adversary]
A \textbf{\pcbin\ adversary} (\textbf{\cobin\ adversary}) is a terminating algorithm that exposes the \pcbin\ adversary (\cobin\ adversary) interface and does not cause the adversarial execution to fail (see \cref{subsubsection:interfaces}) when coupled with any system $\sigma \in \mathcal{S}_{\pcbin}$ ($\sigma \in \mathcal{S}_{\cobin}$). 

Let $\alpha$, $\alpha'$ be two \pcbin\ (\cobin) adversaries such that, for every $\sigma \in \mathcal{S}_{\pcbin}$ ($\sigma \in \mathcal{S}_{\cobin}$), the execution of $\alpha$ coupled with $\sigma$ is identical to the execution of $\alpha'$ coupled with $\sigma$. We consider $\alpha$ and $\alpha'$ to be functionally the same adversary.

We use $\mathcal{A}_{\pcbin}$ to denote the set of \pcbin\ adversaries. We use $\mathcal{A}_{\cobin}$ (or just $\mathcal{A}$) to denote the set of \cobin\ adversaries.
\end{definition}

\begin{definition}[Adversarial power]
Let $\alpha$ be a \pcbin\ (\cobin) adversary. The \textbf{adversarial power} of $\alpha$ is the probability of $\alpha$ compromising the consistency of a \pcbin\ (\cobin) system, picked with uniform probability from $\mathcal{S}_{\pcbin}$ ($\mathcal{S}_{\cobin}$).
\end{definition}

\begin{definition}[Optimal adversary]
\label{definition:optimaladversary}
Let $\alpha$ be a \pcbin\ (\cobin) adversary. We say that $\alpha$ is an \textbf{optimal adversary} if its adversarial power is greater or equal to that of any other \pcbin\ (\cobin) adversary.
\end{definition}

We note that \cref{definition:optimaladversary} is well defined: indeed, both $\mathcal{A}_{\pcbin}$ and $\mathcal{A}_{\cobin}$ are finite sets, and therefore admit a maximum for the adversarial power.

\begin{definition}[Optimal set of adversaries]
Let $\mathcal{A}'$ be a set of \pcbin\ (\cobin) adversaries. We say that $\mathcal{A}'$ is an \textbf{optimal set of adversaries} if $\mathcal{A}'$ includes an optimal \pcbin\ (\cobin) adversary.
\end{definition}

\begin{definition}[\Pcbin\ invocation/response pair]
The pair $(i, r)$ is a \textbf{\pcbin\ invocation/response pair} if
\begin{eqnarray*}
    i = ({\tt Byzantine}, \pi \in \Pi_C) && r = (\xi_1, \ldots, \xi_k \in \Pi \setminus \Pi_C) \\
    i = ({\tt State}) && r = ((\pi_1, m_1), \ldots, \\
                    && \phantom{r = (} (\pi_k \in \Pi_C, m_k \in \mathcal{M})) \\
    i = ({\tt Deliver}, \pi \in \Pi_C, m \in \mathcal{M}) && r = \bot \\
    i = ({\tt Echo}, \pi \in \Pi_C, \xi \in \Pi \setminus \Pi_C, m \in \mathcal{M}) && r = \bot
\end{eqnarray*}
\end{definition}

\begin{definition}[\Cobin\ invocation/response pair]
The pair $(i, r)$ is a \textbf{\cobin\ invocation/response pair} if
\begin{eqnarray*}
    i = ({\tt Byzantine}, \pi \in \Pi_C) && r = (\xi_1, \ldots, \xi_k \in \Pi \setminus \Pi_C) \\
    i = ({\tt Sample}, \pi \in \Pi_C, m \in \mathcal{M}) && r = (\rho_1, \ldots, \rho_k \in \Pi) \\
    i = ({\tt State}) && r = ((\pi_1, m_1), \ldots, \\ 
                    && \phantom{r = (} (\pi_k \in \Pi_C, m_k \in \mathcal{M})) \\
    i = ({\tt Deliver}, \pi \in \Pi_C, m \in \mathcal{M}) && r = \bot \\
    i = ({\tt Echo}, \pi \in \Pi_C, s \in \mathcal{M}, \xi \in \Pi \setminus \Pi_C, m \in \mathcal{M}) && r = \bot
\end{eqnarray*}
\end{definition}

\begin{definition}[Trace]
A \textbf{\pcbin\ trace} (\textbf{\cobin\ trace}) is a finite sequence of \pcbin\ (\cobin) invocation/response pairs. Let $\alpha$ be a (\pcbin\ or \cobin) adversary, let $\sigma$ be a (\pcbin\ or \cobin, correspondingly) system. We use $\tau(\alpha, \sigma)$ to denote the trace produced by $\alpha$ coupled with $\sigma$. We use $\mathcal{T}$ to denote the set of traces.
\end{definition}

\begin{notation}[Power set]
Let $X$ be a set. We use $\powerset{X}{}$ to denote the \textbf{power set} of $X$. We use $\powerset{X}{K} = \cp{x \in \powerset{X}{} \mid \abs{x} = K}$ to denote the elements in $\powerset{X}{}$ that have $K$ elements. We use $\powerset{X}{K+} = \cp{x \in \powerset{X}{} \mid \abs{x} \geq K}$ to denote the elements in $\powerset{X}{}$ that have at least $K$ elements.
\end{notation}

\subsubsection{Consistency of \cobal}

We can now prove that the $\epsilon$-consistency of \cobal\ is strictly weaker than that of \pcbal.

\begin{algorithm}
\begin{algorithmic}[1]
\Implements
    \Instance{\Cobin Adversary + \Pcbin System}{cadv}
\EndImplements

\Uses
    \InstanceSystem{\Pcbin Adversary}{padv}{cadv}
    \Instance{\Cobin System}{sys}
\EndUses

\Procedure{cadv.Init}{{}}
    \State $deliveries = \cp{\bot}^C$;
    \State $padv.Init()$;
\EndProcedure

\Procedure{cadv.Step}{{}}
    \State $padv.Step()$;
\EndProcedure

\Procedure{cadv.Byzantine}{process}
    \State \Return $sys.Byzantine(process)$;
\EndProcedure

\Procedure{cadv.State}{{}}
    \State $state = \emptyset$;
    \State
    \ForAll{(\pi, m)}{sys.State()}
        \If{$deliveries[\pi] = m$}
            \State $state \leftarrow state \cup \cp{(\pi, m)}$;
        \EndIf
    \EndForAll
    \State
    \State \Return $state$;
\EndProcedure

\Procedure{cadv.Deliver}{process, message}
    \State $deliveries[process] \leftarrow message$;
    \State $sys.Deliver(process, message)$;
\EndProcedure

\Procedure{cadv.Echo}{process, source, message}
    \State $sys.Echo(process, message, source, message)$;
\EndProcedure

\Procedure{cadv.End}{{}}
    \State $sys.End()$;
\EndProcedure
\end{algorithmic}
\caption{{\tt \Cobin\ decorator}}
\label{algorithm:cobdecorator}
\end{algorithm}

\begin{lemma}
\label{lemma:cobalweakerthanpcbal}
An optimal \cobin\ adversary is more powerful than an optimal \pcbin\ adversary.
\begin{proof}
Let $\alpha^*$ be an optimal \pcbin\ adversary. In order to prove that an optimal \cobin\ adversary is more powerful than $\alpha^*$, we just need to find a \cobin\ adversary $\alpha^+$ that is more powerful than $\alpha^*$. We achieve this using a \textbf{\pcbin-to-\cobin\ decorator}, i.e., an algorithm that acts as an interface between a \pcbin\ adversary and \cobin\ system. A \pcbin\ adversary coupled with a \pcbin-to-\cobin\ decorator effectively implements a \cobin\ adversary. Here we show that a \pcbin-to-\cobin\ decorator $\Delta_{cob}$ exists such that, for every $\alpha \in \mathcal{A}_{pcb}$, the \cobin\ adversary $\alpha' = \Delta_{cob}(a)$ is more powerful than $\alpha$. If this is true, the lemma is proved: indeed, $\alpha^+ = \Delta_{cob}(\alpha^*)$ is more powerful than $\alpha^*$.

\paragraph{Decorator}

\cref{algorithm:cobdecorator} implements {\tt \Cobin\ decorator}, a \pcbin-to-\cobin\ decorator. Provided with a \pcbin\ adversary $padv$, {\tt \Cobin\ decorator} acts as an interface between $padv$ and a \cobin\ system $sys$, effectively implementing a \cobin\ adversary $cadv$. {\tt \Cobin\ decorator} exposes both the \cobin\ adversary and the \pcbin\ system interfaces: the underlying \pcbin\ adversary $padv$ uses $cadv$ as its system.

{\tt \Cobin\ decorator} works as follows:
\begin{itemize}
    \item Procedure $cadv.Init()$ initializes a $deliveries$ array that is used to keep track of the message \pbin.Delivered by each correct process, and a $gap$ set that it uses to keep track of the messages \cobin.Delivered by each correct process in $sys$.
    \item Procedure $cadv.Step()$ simply forwards the call to $padv.Step()$.
    \item Procedure $cadv.State()$ returns a list of pairs $(\pi \in \Pi_C, m \in \mathcal{M})$ such that $\pi$ both \pbin.Delivered and delivered $m$ in $sys$. This is achieved by querying $sys.State()$, then looping over each element $(\pi, m)$ of the response and checking if $deliveries[\pi] = m$.
    \item Procedure $cadv.Byzantine(process)$ simply forwards the call to \\ $sys.Byzantine(process)$.
    \item Procedure $cadv.Deliver(process, message)$ sets $deliveries[process]$ to $message$ (to signify that $process$ \pbin.Delivered $message$). It then forwards the call to $sys.Deliver(process, message)$, causing $process$ to \pbin.Deliver $message$. 
    \item Procedure $cadv.Echo(process, source, message)$ forwards the call to \\ $sys.Echo(process, message, source, message)$, causing $source$ to send an {\tt Echo}($message$, $message$) message to $process$.
    \item Procedure $cadv.End()$ simply forwards the call to $sys.End()$.
\end{itemize}

\begin{figure}
\centering
\includegraphics[scale=0.75]{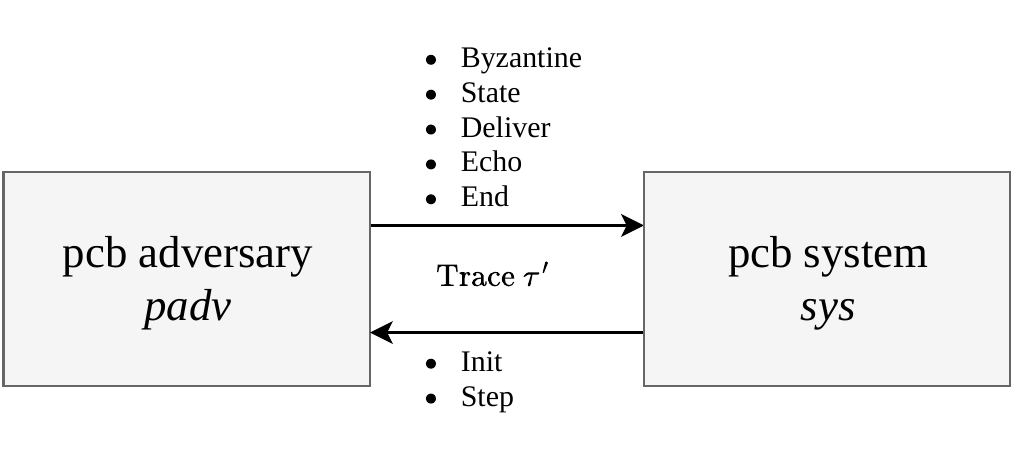}
\caption{An execution without decorator.}
\label{figure:sieve:nodecorator}
\end{figure}

\begin{figure}
\centering
\includegraphics[scale=0.75]{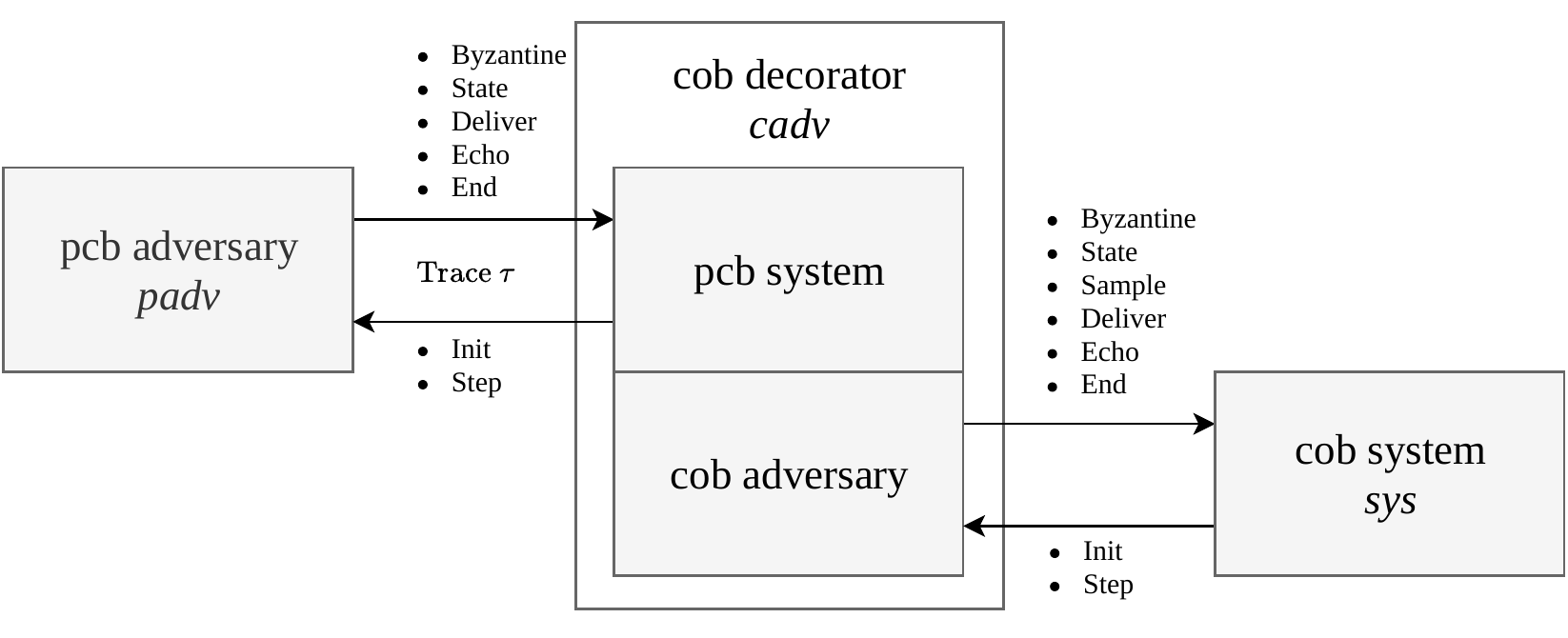}
\caption{A decorator exposing both its system interface to the pcb adversary and its adversary interface to the cob system.}
\label{figure:sieve:decorator}
\end{figure}

Let $\Delta_{\cobin}: \mathcal{A}_{\pcbin} \rightarrow \mathcal{A}_{\cobin}$ denote the function that {\tt \Cobin \ decorator} implements, mapping \pcbin\ adversaries into \cobin\ adversaries. We want to prove that, for every $\alpha \in \mathcal{A}_{\pcbin}$, the adversarial power of $\alpha' = \Delta_{\cobin}(\alpha)$ is greater than that of $\alpha$.

\paragraph{System translation}

Let $\alpha$ be a \pcbin\ adversary. We start by noting that, since $\alpha$ is correct, $\alpha$ always causes every correct process to \pbin.Deliver a message. We can therefore define a function 
\begin{equation*}
    \mu: \mathcal{A}_{\pcbin} \times \mathcal{S}_{\pcbin} \times \Pi_C \rightarrow \mathcal{M}
\end{equation*} 
such that $\mu(\alpha, \sigma, \pi) = m$ if and only if $\alpha$ eventually causes $\pi$ to \pbin.Deliver $m$, when $\alpha$ is coupled with $\sigma$.

We then define a \textbf{system translation} function $\Psi[\alpha]: \mathcal{S}_{\pcbin} \rightarrow \powerset{\mathcal{S}_{\cobin}}{}$ that maps a \pcbin\ system into a set of \cobin\ systems:
\begin{equation*}
    \rp{\sigma' \in \Psi[\alpha](\sigma)} \Longleftrightarrow \rp{\forall \pi \in \Pi_C,\,\sigma[\pi] = \sigma'[\pi][\mu(\alpha, \sigma, \pi)]}
\end{equation*}

Let $\sigma$ be a \pcbin\ system, let $\sigma'$ be a \cobin\ system, let $\pi$ be any correct process, let $m$ be the message that $\alpha$ eventually causes $\pi$ to \pbin.Deliver, when $\alpha$ is coupled with $\sigma$. Intuitively, $\sigma'$ is in $\Psi[\alpha](\sigma)$ if $\pi$'s echo sample for $m$ in $\sigma'$ is identical to $\pi$'s echo sample in $\sigma$.

\paragraph{Roadmap}

Let $\alpha \in \mathcal{A}_{\pcbin}$, $\alpha' = \Delta_{\cobin}(\alpha)$. Let $\sigma \in \mathcal{S}_{\pcbin}$ such that $\alpha$ compromises the consistency of $\sigma$. In order to prove that $\alpha'$ is more powerful than $\alpha$, we prove that:
\begin{itemize}
    \item For every $\sigma' \in \Psi[\alpha](\sigma)$, $\alpha'$ compromises the consistency of $\sigma'$.
    \item The probability of $\Psi[\alpha](\sigma)$ is equal to the probability of $\sigma$.
    \item For every $\hat \sigma \in \mathcal{S}_{\pcbin}$ such that $\hat \sigma \neq \sigma$, the sets $\Psi[\alpha](\sigma)$ and $\Psi[\alpha](\hat \sigma)$ are disjoint.
\end{itemize}

Indeed, if all of the above are true, then the probability of $\alpha'$ compromising the consistency of a random \cobin\ system $\sigma'$ is greater or equal to the probability of $\alpha$ compromising the consistency of a random system $\sigma$, and the lemma is proved.

\begin{figure}
\centering
\includegraphics[scale=0.45]{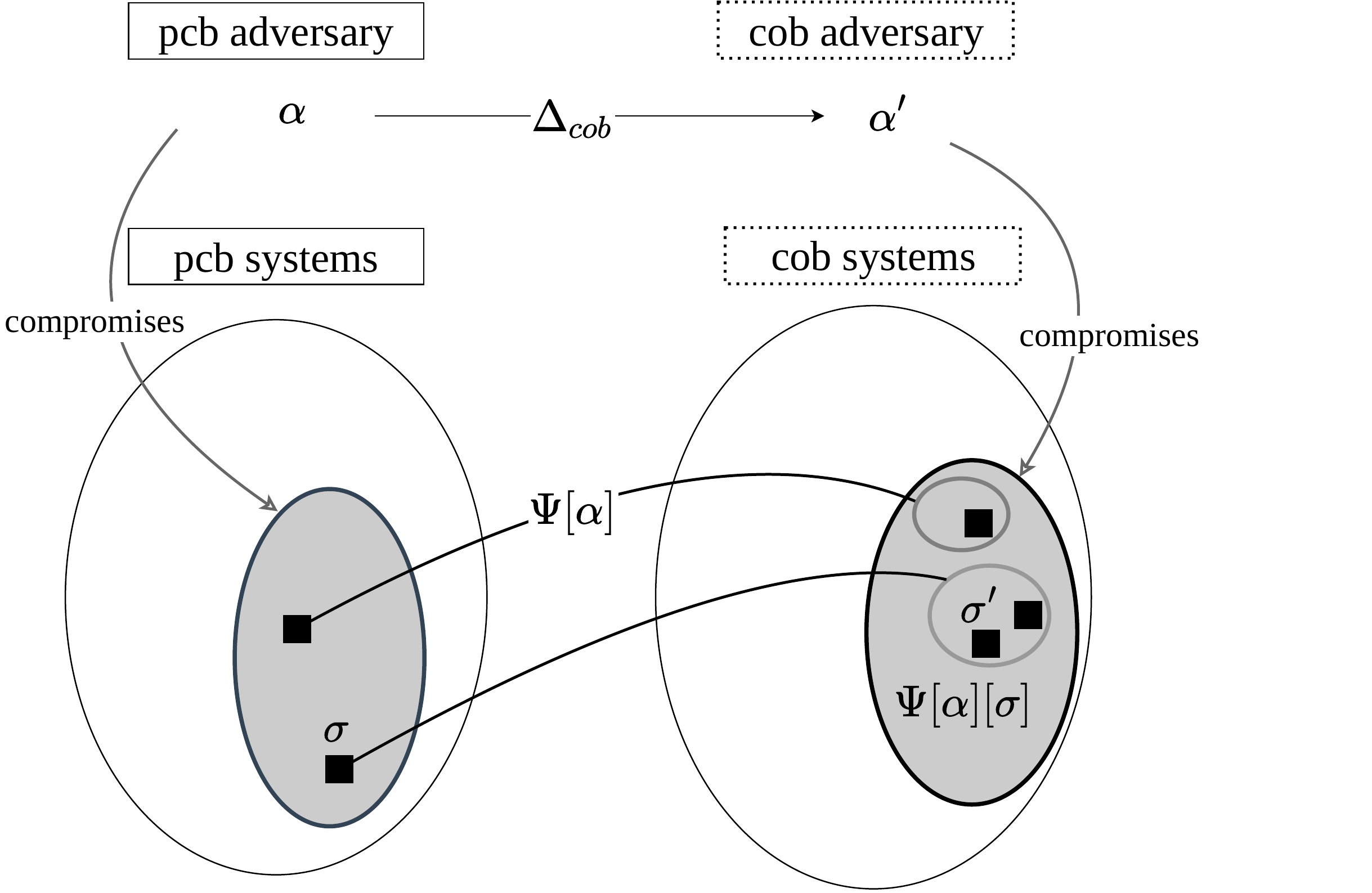}
\caption{An illustration of the steps needed to prove that the adversarial power of $\alpha$ is greater than that of $\alpha'$.}
\label{figure:sieve:roadmap}
\end{figure}

\paragraph{Trace}

We start by noting that, if we couple {\tt \Cobin\ decorator} with $\sigma'$, we effectively obtain a \pcbin\ system interface $\delta$ with which $\alpha$ directly exchanges invocations and responses. Here we show that the trace $\tau(\alpha, \sigma)$ is identical to the trace $\tau(\alpha, \delta)$. Intuitively, this means that $\alpha$ has no way of \emph{distinguishing} whether it has been coupled directly with $\sigma$, or it has been coupled with $\sigma'$, with {\tt \Cobin\ decorator} acting as an interface. We prove this by induction.

Let us assume
\begin{eqnarray*}
    \tau(\alpha, \sigma) &=& ((i_1, r_1), \ldots) \\
    \tau(\alpha, \delta) &=& ((i'_1, r'_1), \ldots) \\
    i_j = i'_j, r_j = r'_j && \forall j \leq n
\end{eqnarray*}
with $n \geq 0$ (here $n = 0$ means that this is $\alpha$'s first invocation). We start by noting that, since $a$ is a deterministic algorithm, we immediately have
\begin{equation*}
    i_{n + 1} = i'_{n + 1}
\end{equation*}
and we need to prove that $r_{n + 1} = r'_{n + 1}$. \\

Let us assume that $i_{n + 1} = ({\tt Byzantine}, \pi)$. By hypothesis, at least one of the echo samples of $\pi$ in $\sigma'$ is identical to the echo sample of $\pi$ in $\sigma$. Moreover, all $\pi$'s echo samples in $\sigma'$ share the same set of Byzantine processes. Therefore, the first of $\pi$'s echo samples in $\sigma'$ contains the same Byzantine processes as $\pi$'s echo sample in $\sigma$. Finally, the decorator simply forwards the call to $cadv.Byzantine(\pi)$ to $sys.Byzantine(\pi)$. Consequently, $r_{n + 1} = r'_{n + 1}$. \\

Before considering the case $i_{n + 1} = ({\tt State})$, we prove some auxiliary results. Let $\pi$ be a correct process, let $\rho$ be a process, let $\xi$ be a Byzantine process, let $m$ be a message. For every $j \leq n + 1$, as we established, we have $i_j = i'_j$. Therefore, after the $(n + 1)$-th invocation, the following hold true:
\begin{itemize}
    \item $\pi$ \pbin.Delivered $m$ in $\sigma'$ if and only if $\pi$ \pbin.Delivered $m$ in $\sigma$. Indeed, $cadv.Deliver(\pi, m)$ was invoked if and only if $sys.Deliver(\pi, m)$ was invoked as well.
    \item $\pi$ \pbin.Delivered $m$ in $\sigma'$ if and only if $deliveries[\pi] = m$. Indeed, $cadv.Deliver(\pi, m)$ was invoked if and only if $deliveries[\pi]$ was set to $m$.
    \item $\xi$ sent an {\tt Echo}($m$) to $\pi$ in $\sigma'$ if and only if $\xi$ sent an {\tt Echo}($m$, $m$) message to $\pi$ in $\sigma$. Indeed, $cadv.Echo(\pi, \xi, m)$ was invoked if and only if $sys.Echo(\pi, m, \xi, m)$ was invoked as well.
    \item If $\pi$ \pbin.Delivered $m$ in $\sigma$, then $\pi$'s echo sample for $m$ in $\sigma'$ is identical to $\pi$'s echo sample in $\sigma$. This follows from the definition of $\Psi$ (we recall that $\sigma' \in \Psi[\alpha](\sigma)$).
    \item If $\pi$ delivered $m$ in $\sigma$, it also delivered $m$ in $\sigma'$. Indeed, since $\pi$ \pbin.Delivered $m$ in $\sigma$, $\pi$'s echo sample for $m$ in $\sigma'$ is identical to $\pi$'s echo sample in $\sigma$. Moreover, if $\pi$ received an {\tt Echo}($m$) message from $\rho$ in $\sigma$, then it also received an {\tt Echo}($m$, $m$) message from $\rho$ in $\sigma'$.
    \item If $\pi$ both \pbin.Delivered and delivered $m$ in $\sigma'$, it also delivered $m$ in $\sigma$. Indeed, since $\pi$ \pbin.Delivered $m$ in $\sigma'$, then it also \pbin.Delivered $m$ in $\sigma$, and $\pi$'s echo sample in $\sigma$ is identical to $\pi$'s echo sample for $m$ in $\sigma'$. Moreover, if $\pi$ received an {\tt Echo}($m$) message from $\rho$ in $\sigma'$, then it also received an {\tt Echo}($m$, $m$) message from $\rho$ in $\sigma$.
\end{itemize}

Let us assume $i_{n + 1} = ({\tt State})$. We start by noting that $cadv.State()$ returns all the pairs $(\pi', m')$ in $sys.State()$ that satisfy $deliveries[\pi'] = m'$. If $(\pi, m) \in r_{n + 1}$, then $\pi$ both \pbin.Delivered and delivered $m$ both in $\sigma$. Therefore, $\pi$ both \pbin.Delivered and delivered $m$ in $\sigma'$, and $deliveries[\pi] = m$. Consequently, $(\pi, m) \in r'_{n + 1}$. If $(\pi, m) \in r'_{n + 1}$, then $(\pi, m)$ was returned from $sys.State()$, and $deliveries[\pi] = m$. Therefore, $\pi$ both \pbin.Delivered and delivered $m$ in $\sigma'$. Consequently, $\pi$ delivered $m$ in $\sigma$, and $(\pi, m) \in r_{n + 1}$.

Noting that procedures $Deliver(\ldots)$ and $Echo(\ldots)$ never return a value, we trivially have that if $i_{n + 1} = ({\tt Deliver}, \pi, m)$ or $i_{n + 1} = ({\tt Echo}, \pi, s, \xi, m)$ then $r_{n + 1} = \bot = r'_{n + 1}$. By induction, we have $\tau(\alpha, \sigma) = \tau(\alpha, \delta)$.

\paragraph{Consistency of $\sigma'$}

We proved that $\tau(\alpha, \sigma) = \tau(\alpha, \delta)$. Moreover, we proved that if a correct process $\pi$ eventually \pcbin.Delivers a message $m$ in $\sigma$, then $\pi$ also \cobin.Delivers $m$ in $\sigma'$.

Since $\alpha$ compromises the consistency of $\sigma$, two correct processes $\pi$, $\pi'$ and two distinct messages $m$, $m' \neq m$ exist such that, in $\sigma$, $\pi$ \pcbin.Delivered $m$ and $\pi'$ \pcbin.Delivered $m'$. Therefore, in $\sigma'$, $\pi$ \cobin.Delivered $m$ and $\pi'$ \cobin.Delivered $m'$. Therefore $\alpha'$ compromises the consistency of $\sigma'$.

\paragraph{Translation probabilities}

We now prove that, for every $\sigma \in \mathcal{S}_{\pcbin}$, the probability of $\Psi[\alpha](\sigma)$ is equal to the probability of $\sigma$.

The probability of $\sigma$ is
\begin{equation*}
    \prob{\sigma} = \prob{\sigma[\pi_1][1] = \pi_{1, 1}, \ldots , \sigma[\pi_C][E] = \pi_{C, E}} = N^{-EC}
\end{equation*}
and the probability of $\Psi[\alpha](\sigma)$ is
\begin{eqnarray*}
    \prob{\Psi[\alpha](\sigma)} &=& \\
    \prob{\sigma[\pi_1][\mu(\alpha, \sigma, \pi_1)][1] = \pi_{1, 1}, \ldots , \sigma[\pi_C][\mu(\alpha, \sigma, \pi_C)[E] = \pi_{C, E}]} &=& N^{-EC}
\end{eqnarray*}
which proves the result.

\paragraph{Translation disjunction}

We now prove that, for any two $\sigma_a$, $\sigma_b \neq \sigma_a$, we have $\Psi[\alpha](\sigma_a) \cap \Psi[\alpha](\sigma_b) = \emptyset$. We prove this by contradiction. Suppose a system $\sigma'$ exists such that $\sigma' \in \Psi[\alpha](\sigma_a)$ and $\sigma' \in \Psi[\alpha](\sigma_b)$. We want to prove that $\sigma_a = \sigma_b$.

We start by noting that, if $\tau(\alpha, \sigma_a) = \tau(\alpha, \sigma_b)$, then $\sigma_a = \sigma_b$. Indeed, we have
\begin{eqnarray*}
\tau(\alpha, \sigma_a) &=& \tau(\alpha, \sigma_b) \\
\implies \mu(\alpha, \sigma_a, \pi) &=& \mu(\alpha, \sigma_b, \pi) \;\forall \pi \in \Pi_C \\
\implies \sigma_a[\pi] &=& \sigma'[\pi][\mu(\alpha, \sigma_a, \pi)] \\
&=& \sigma'[\pi][\mu(\alpha, \sigma_b, \pi)] \\ 
&=& \sigma_b[\pi] \; \forall \pi \\
\implies \sigma_a &=& \sigma_b
\end{eqnarray*}

We prove that $\tau(\alpha, \sigma_a) = \tau(\alpha, \sigma_b)$ by induction. Let us assume
\begin{eqnarray*}
    \tau(\alpha, \sigma_a) &=& ((i_1, r_1), \ldots) \\
    \tau(\alpha, \sigma_b) &=& ((i'_1, r'_1), \ldots) \\
    i_j = i'_j, r_j = r'_j && \forall j \leq n
\end{eqnarray*}
with $n \geq 0$ (here $n = 0$ means that this is $\alpha$'s first invocation). We start by noting that, since $a$ is a deterministic algorithm, we immediately have
\begin{equation*}
    i_{n + 1} = i'_{n + 1}
\end{equation*}
and we need to prove that $r_{n + 1} = r'_{n + 1}$.

Let us assume that $i_{n + 1} = ({\tt Byzantine}, \pi)$. By hypothesis, among the echo samples of $\pi$ in $\sigma'$, at least one is identical to the echo sample of $\pi$ in $\sigma_a$, and at least one is identical to the echo sample of $\pi$ in $\sigma_b$. Noting that $\pi$'s echo samples share the same set of Byzantine processes, we immediately have that the Byzantine processes in $\sigma_a[\pi]$ are the same as in $\sigma_b[\pi]$, and $r_{n + 1} = r'_{n + 1}$.

Before considering the case $i_{n + 1} = ({\tt State})$, we prove some auxiliary results. Let $\pi$ be a correct process, let $\rho$ be a process, let $\xi$ be a Byzantine process, let $m$ be a message. For every $j \leq n + 1$, as we established, we have $i_j = i'_j$. Therefore, after the $(n + 1)$-th invocation, the following hold true:
\begin{itemize}
    \item $\pi$ \pbin.Delivered $m$ in $\sigma_a$ if and only if $\pi$ \pbin.Delivered $m$ in $\sigma_b$.
    \item $\xi$ sent an {\tt Echo}($m$) message to $\pi$ in $\sigma_a$ if and only if $\xi$ send an {\tt Echo}($m$) message to $\pi$ in $\sigma_b$.
    \item If $\pi$ \pbin.Delivered $m$ (both in $\sigma_a$ and $\sigma_b$), then $\sigma_a[\pi] = \sigma_b[\pi]$. Indeed, 
    \begin{eqnarray*}
        \sigma_a[\pi] &=& \sigma'[\pi][\mu(\alpha, \sigma_a, \pi)] \\
        &=& \sigma'[\pi][\mu(\alpha, \sigma_b, \pi)] \\
        &=& \sigma_b[\pi]
    \end{eqnarray*}
    \item $\pi$ delivered $m$ in $\sigma_a$ if and only if $\pi$ delivered $m$ in $\sigma_b$. Indeed, if $\pi$ delivered $m$ in $\sigma_a$, then it also \pbin.Delivered $m$ in $\sigma_a$ and, consequently, $\sigma_b$. Therefore, $\pi$'s echo sample in $\sigma_a$ is identical to $\pi$'s echo sample in $\sigma_b$. Since $\pi$ received the same {\tt Echo} messages in $\sigma_a$ and $\sigma_b$ then $\pi$ delivered $m$ in $\sigma_b$. The argument can be trivially reversed to prove that, if $\pi$ delivered $m$ in $\sigma_b$, then $\pi$ also delivered $m$ in $\sigma_a$.
\end{itemize}

Let us consider the case $i_{n + 1} = ({\tt State})$. From the above follows $r_{n + 1} = r'_{n + 1}$.

Noting that procedures $Deliver(\ldots)$ and $Echo(\ldots)$ never return a value, we trivially have that if $i_{n + 1} = ({\tt Deliver}, \pi, m)$ or $i_{n + 1} = ({\tt Echo}, \pi, s, \xi, m)$ then $r_{n + 1} = \bot = r'_{n + 1}$. By induction, we have $\tau(\alpha, \sigma_a) = \tau(\alpha, \sigma_b)$. 

Therefore, $\sigma_a = \sigma_b$, which contradicts the hypothesis and thus proves that the sets $\Psi[\alpha](\sigma_a)$ and $\Psi[\alpha](\sigma_b)$ are disjoint.

\end{proof}
\end{lemma}

\subsection{Two-phase adversaries}
\label{subsection:twophaseadversaries}

In \cref{subsection:simplifiedadversarialpower} we proved the important result that it is easier to compromise the consistency of \cobal\ than that of \pcbal. Throughout the rest of this appendix, we compute a bound on the $\epsilon$-security of \cobal.

It is easy to see that the $\epsilon$-security of \cobal\ is equal to the adversarial power of an optimal adversary. Therefore, $\epsilon_c$ is a bound on the $\epsilon$-security of \cobal\ if $\epsilon_c$ bounds the adversarial power of every adversary in an optimal set of adversaries.

In this section, we derive a set $\mathcal{A}_{tp} \subseteq \mathcal{A}$ of \emph{two-phase} adversaries that we prove to be optimal. Unlike $\mathcal{A}$, $\mathcal{A}_{tp}$ is small enough to be probabilistically tractable. In the next sections, we compute a bound on the adversarial power of every $a \in \mathcal{A}_{tp}$.

In a similar way to \cref{lemma:cobalweakerthanpcbal}, the proofs of optimality of most of the sets of adversaries presented in this section make extensive use of decorators, and are in general lengthy and non-trivial. For the sake of readability, in this section we only state our results, and defer each explicit proof to \cref{appendix:decorators}.

\subsubsection{Auto-echo adversary}
\label{subsubsection:autoechoadversary}

As we introduced in \cref{subsubsection:cobalgorithm}, an {\tt Echo} message in \cobal\ has two fields: a sample $s$ and a message $m$. Intuitively, an {\tt Echo}($s$, $m$) message represents the following statement: ``within the context of message $s$, consider my {\tt Echo} to be for message $m$".

Upon \pbin.Delivering a message $m$, a correct process sends to every other process an {\tt Echo}($s$, $m$) for every $s$. In other words, a correct process supports the message it \pbin.Delivers across all samples. A Byzantine process, however, is not constrained to do this.

A correct process \cobin.Delivers a message $m$ upon collecting enough {\tt Echo}($m$, $m$) messages from its echo sample for $m$. It is easy to see, therefore, that the probability of a correct process $\pi$ \cobin.Delivering $m$ increases if all the Byzantine processes send an {\tt Echo}($m$, $m$) message to $\pi$.

\begin{definition}[Auto-echo adversary]
An adversary $a \in \mathcal{A}$ is an \textbf{auto-echo adversary} if, at the beginning of its execution, it causes $\xi$ to send an {\tt Echo}($m$, $m$) message to $\pi$, for every $\pi \in \Pi_c$, $\xi \in \Pi \setminus \Pi_C$ and $m \in \mathcal{M}$. We use $\mathcal{A}_{ae}$ to denote the set of auto-echo adversaries.
\end{definition}

In \cref{subsection:autoechoadversaryproof}, we formally prove this intuition, i.e., we prove that the set of auto-echo adversaries $\mathcal{A}_{ae}$ is optimal.

\subsubsection{Process-sequential adversary}

As we discussed in \cref{subsubsection:cobalgorithm}, a correct process reveals its sample for a message $m$ only after delivering $m$. At the beginning of the execution, the adversary only knows which Byzantine processes are in each correct process' echo samples. In \cref{subsubsection:autoechoadversary}, however, we proved that this does not affect the optimal adversary's strategy: the set of Byzantine processes in a correct process' echo samples don't play any role in an optimal adversarial execution.

Intuitively, therefore, an optimal adversary has effectively no meaningful way to distinguish any two correct processes based on the outcome that their actions will have on the system.

\begin{definition}[Correct process enumeration]
We define a bijection
\begin{equation*}
    \zeta: 1..C \leftrightarrow \Pi_C
\end{equation*}
that uniquely maps an integer identifier $i \in 1..C$ to a correct process.
\end{definition}

\begin{definition}[Process-sequential adversary]
An auto-echo adversary $\alpha \in \mathcal{A}_{ae}$ is a \textbf{process-sequential adversary} if it never causes $\zeta(i)$ to \pbin.Deliver a message before any $\zeta(j < i)$. We use $\mathcal{A}_{ps}$ to denote the set of process-sequential adversaries.
\end{definition}

In \cref{subsection:processsequentialadversaryproof}, we formally prove this intuition, i.e., we prove that the set of process-sequential adversaries $\mathcal{A}_{ps}$ is optimal.

\subsubsection{Sequential adversary}
\label{subsubsection:sequentialadversary}

As we introduced in \cref{subsubsection:cobalgorithm}, in \cobal\ a correct process independently selects $C$ echo samples, one for every message in $\mathcal{M}$. Moreover, every echo sample shares the same set of Byzantine processes. Finally, let $\pi$ be a correct process, let $m$ be a message, no correct process in $\pi$'s echo sample for $m$ is known to the adversary before $\pi$ delivers $m$.

Intuitively, therefore, an adversary has effectively no meaningful way of distinguishing two messages, based on the outcome that their \pbin.Delivery will have on the system.

\begin{definition}[Poisoned process]
Let $\sigma$ be a system, let $\pi$ be a correct process. We say that \textbf{$\pi$ is poisoned in $\sigma$} if and only if at least $\hat E$ processes in $\pi$'s first echo sample in $\sigma$ are Byzantine.
\end{definition}

\begin{definition}[Sequential adversary]
A process-sequential adversary $\alpha \in \mathcal{A}_{ps}$ is a \textbf{sequential adversary} if it never causes a correct process to \pbin.Deliver $m \in \mathcal{M}$ before causing every $l < m \in \mathcal{M}$ to be \pbin.Delivered by at least one correct process. We use $\mathcal{A}_{sq}$ to denote the set of sequential adversaries.
\end{definition}

In \cref{subsection:sequentialadversaryproof}, we formally prove this intuition, i.e., we prove that the set of sequential adversaries $\mathcal{A}_{sq}$ is optimal.

\subsubsection{Non-redundant adversary}
\label{subsubsection:nonredundantadversary}

As we established in \cref{subsubsection:cobinterface}, the consistency of \cob\ is compromised if and only if at least two messages are delivered by at least one correct process.

It is easy to see, therefore, that an adversary that has already caused at least one correct process to deliver a message $m$ gains no advantage from causing more correct processes to \pbin.Deliver $m$. Indeed, doing so would not increase the probability of at least one correct process delivering $m$ (that condition is verified with probability $1$): an optimal adversary should focus its remaining \pbin.Deliveries on achieving the goal to cause at least one other message to be delivered by at least one correct process.

\begin{definition}[Non-redundant adversary]
A sequential adversary $\alpha \in \mathcal{A}_{sq}$ is a \textbf{non-redundant adversary} if, whenever exactly one message $m$ has been delivered, it never causes any additional correct process to \pbin.Deliver $m$. We use $\mathcal{A}_{nr}$ to denote the set of non-redundant adversaries.
\end{definition}

In \cref{subsection:nonredundantadversaryproof}, we formally prove this intuition, i.e., we prove that the set of non-redundant adversaries $\mathcal{A}_{nr}$ is optimal.

\subsubsection{Sample-blind adversary}

In \cref{subsubsection:cobalgorithm}, we discussed how, in \cobal, a correct process reveals its echo sample for a message after at least one correct process delivered that message. Throughout \cref{subsection:twophaseadversaries}, we extensively used {\tt Reveal} messages (through the $Sample(\ldots)$ system interface) to build a sequence of decorators that improved the power of any adversary in their domain.

In this section, we prove the counter-intuitive result that the information contained in a {\tt Reveal} message is actually useless to an optimal adversary. Indeed, the decorators we developed leveraged {\tt Reveal} messages to \emph{correct} the sub-optimal behavior of a generic adversary. However, for every decorator that we developed, we argue that we could develop an adversary in the codomain of that decorator, that never uses the information provided by {\tt Reveal} messages.

An intuitive insight on {\tt Reveal} messages can be provided by the observation that the information they provide is disclosed in the moment it ceases to actually be useful. Indeed, a correct process reveals the content of its echo sample for a message $m$ only after at least one correct process delivered $m$. As we proved in \cref{subsubsection:nonredundantadversary}, causing additional processes to deliver $m$ gives no advantage to the adversary. Moreover, since the correct processes in each echo sample are picked independently from each other, the knowledge of a correct process $\pi$'s echo sample for $m$ does not grant any advantage in causing $\pi$ to deliver $m' \neq m$.

\begin{notation}[Undefined minima and maxima]
Let $X \subset \mathbb{N}$, with $X$ finite, let $S: X \rightarrow \cp{\true, \false}$ be a predicate on $X$. We use
\begin{eqnarray*}
    (\min \: n \in X \mid S(n)) &=& \bot \\
    (\max \: n \in X \mid S(n)) &=& \bot
\end{eqnarray*}
to denote that
\begin{equation*}
    \nexists \: n \in X \mid S(n)
\end{equation*}
\end{notation}

\begin{definition}[Trace compatibility]
Let $\tau$ be a trace, let $\sigma$ be a system. We say that \textbf{$\tau$ is compatible with $\sigma$}, or $\tau \sim \sigma$, if the sequence of invocations in $\tau$, applied in order to $\sigma$, produces the corresponding sequence of responses in $\tau$.
\end{definition}

\begin{notation}[Consistency compromission]
Let $\alpha$ be an adversary, let $\sigma$ be a system, let $\tau$ be a trace. We use $\alpha \searrow \sigma$ to signify that $\alpha$ compromises the consistency of $\sigma$. We use $\tau \searrow \sigma$ to signify that the sequence of invocations in $\tau$ compromises the consistency of $\sigma$.
\end{notation}

\begin{definition}[Sample-blind adversary]
A non-redundant adversary $\alpha \in \mathcal{A}_{nr}$ is a \textbf{sample-blind adversary} if it never invokes $Sample(\ldots)$. We use $\mathcal{A}_{sb}$ to denote the set of sample-blind adversaries.
\end{definition}

In \cref{subsection:sampleblindadversaryproof}, we formally prove this intuition, i.e., prove that the set of sample-blind adversaries $\mathcal{A}_{sb}$ is optimal.

\subsubsection{Byzantine-counting adversary}

In \cref{subsubsection:modelcobadversary} we discussed how an adversary for \cobal\ knows which Byzantine processes are in the first echo sample of any correct process. In \cref{subsubsection:autoechoadversary}, however, we proved that the optimal adversarial behavior with respect to {\tt Echo} messages is always to cause every Byzantine process to send an {\tt Echo}($m$, $m$) message to every correct process, for every message $m \in \mathcal{M}$.

Intuitively, therefore, a correct process gains no advantage from knowing specifically which Byzantine processes are in the first echo sample of any correct process.

\begin{definition}[Byzantine-counting adversary]
A sample-blind adversary $\alpha \in \mathcal{A}_{sb}$ is a \textbf{Byzantine-counting adversary} if, whenever it invokes $Byzantine(\pi \in \Pi_C)$, it invokes $\abs{Byzantine(\pi)}$. In other words, the behavior of a Byzantine-counting adversary does not depend on the specific set of Byzantine processes in the first echo sample of any correct process. We use $\mathcal{A}_{bc}$ to denote the set of Byzantine-counting adversaries.
\end{definition}

In \cref{subsection:byzantinecountingadversaryproof}, we formally prove this intuition, i.e., we prove that the set of Byzantine-counting adversaries $\mathcal{A}_{bc}$ is optimal.

\subsubsection{Single-response adversary}
\label{subsubsection:singleresponseadversary}

As we introduced in \cref{subsubsection:modelcobadversary}, the goal of a \cobin\ adversary is to compromise the consistency of a \cobin\ system by causing two distinct messages to be delivered by at least one correct process each. In order to achieve this, it acts upon the system in steps, causing correct processes to \pbin.Deliver a sequence of messages, until the consistency is compromised.

We distinguish two phases of an adversarial execution.

\begin{definition}[Trace phases]
Let $\alpha$ be an adversary, let $\sigma$ be a system. We call \textbf{first phase of $\tau(\alpha, \sigma)$} the sequence $\tau(\alpha, \sigma)_1, \ldots, \tau(\alpha, \sigma)_n$ with $n$ given by
\begin{eqnarray*}
    n &=& 
    \begin{cases}
        \min_j \mid S(j) &\text{iff}\; \exists j \mid S(j) \\
        \abs{\tau(\alpha, \sigma)} &\text{otherwise}
    \end{cases} \\
    S(j) &=& \rp{\tau(\alpha, \sigma)_j = ({\tt State}, r_h), \; r_h \neq \emptyset}
\end{eqnarray*}

We call $\tau(\alpha, \sigma)_{n + 1}, \ldots, \tau(\alpha, \sigma)_{\abs{\tau(\alpha, \sigma)}}$ the \textbf{second phase of $\tau(\alpha, \sigma)$}. We call $\eta(\alpha, \sigma)$ the first phase of $\tau(\alpha, \sigma)$. We call $\theta(\alpha, \sigma)$ the second phase of $\tau(\alpha, \sigma)$.

The first phase of a trace ends when, for the first time, a call to $State()$ returns a non-empty set. Intuitively, the first phase ends when the adversary becomes aware that at least one correct process delivered a message.
\end{definition}

Let us focus on the second phase of an adversarial execution carried out by a Byzantine-counting adversary. We know that, at the beginning of the second phase, at least one message has been delivered by at least one correct process. If more than one message has been delivered, the adversary already compromised the consistency of the system, and the invocations in the second phase are irrelevant to its success. 

If exactly one message has been delivered, an optimal adversary will issue a sequence of invocations that, given the information available on the system, maximizes the probability of at least one more message being delivered by at least one correct process. Since the adversary is non-redundant, the response provided by any invocation to $State()$ will not change until the consistency is compromised. Intuitively, therefore, the information available to the adversary throughout the second phase does not change until consistency is compromised. Since any invocation issued by the adversary after consistency is compromised is irrelevant to its success, an optimal adversary does not need to invoke $State()$ throughout the second phase of any adversarial execution.

\begin{definition}[Single-response adversary]
A Byzantine-counting adversary $\alpha \in \mathcal{A}_{bc}$ is a \textbf{single-response adversary} if it never invokes $State()$ throughout the second phase of any adversarial execution. We use $\mathcal{A}_{sr}$ to denote the set of single-response adversaries.
\end{definition}

In \cref{subsection:singleresponseadversaryproof}, we formally prove this intuition, i.e., we prove that the set of single-response adversaries $\mathcal{A}_{sr}$ is optimal.

\subsubsection{State-polling adversary}
\label{subsubsection:statemonitoringadversary}

In \cref{subsubsection:singleresponseadversary}, we proved that an optimal adversary does not need to invoke $State()$ in the second phase of an adversarial execution, i.e., after at least one message has been delivered by at least one correct process. 

It is easy to see, however, that, throughout the first phase, the information provided by $State()$ is useful to the adversary. Intuitively, the sooner a single-response adversary becomes aware that at least one correct process delivered a message, the sooner it can focus its strategy to cause the delivery of a second, distinct message.

In this section, we prove this intuition, i.e., we formally prove that the set of \emph{state-polling adversaries} is optimal.

\begin{definition}[State-polling adversary]
A single-response adversary $\alpha \in \mathcal{A}_{sr}$ is a \textbf{state-polling adversary} if it invokes $State()$ before the first invocation of $Deliver(\ldots)$ and after each invocation of $Deliver(\ldots)$, until $State()$ returns a non-empty set. We use $\mathcal{A}_{sp}$ to denote the set of state-polling adversaries.
\end{definition}

\begin{lemma}
The set of state-polling adversaries $\mathcal{A}_{sp}$ is optimal.

\begin{proof}
It follows immediately from the observation that, for any adversary, not invoking $State()$ is equivalent to invoking $State()$ and ignoring its response.
\end{proof}
\end{lemma}

\subsubsection{Two-phase adversary}
\label{subsubsection:twophaseadversary}

In \cref{subsubsection:statemonitoringadversary}, we proved that: throughout the first phase, an optimal adversary invokes $State()$ before the first invocation of $Deliver(\ldots)$ and after each invocation of $Deliver(\ldots)$; throughout the second phase, an optimal adversary never needs to invoke $State()$.

As we discussed, if the first phase is concluded with more than one message being delivered, the adversary already compromised the consistency of the system, and the invocations in the second phase are irrelevant to its success.  

Let us consider the case where, at the beginning of the second phase, exactly one message $m^*$ has been delivered. In \cref{subsubsection:cobalgorithm}, we discussed how a correct process selects the correct component of each echo sample independently. Intuitively, therefore, the knowledge of which processes delivered $m^*$ is useless to the adversary, as it provides no information about the correct component of any echo sample for a message $m \neq m^*$. In other words, an optimal adversary only needs to know \emph{when} the first phase of the execution is concluded, but not \emph{how}.

\begin{definition}[Two-phase adversary]
A state-polling adversary $\alpha \in \mathcal{A}_{sp}$ is a \textbf{two-phase adversary} if, whenever it invokes $State()$, it only invokes $(State() \neq \emptyset)$. In other words, the behavior of a two-phase adversary does not depend on the content of $State()$, but only on whether or not $State()$ is empty.
\end{definition}

\begin{lemma}
\label{lemma:firstphaselength}
Let $\alpha$ be a state-polling adversary, let $\sigma, \sigma'$ be systems such that
\begin{equation*}
    \abs{\eta(\alpha, \sigma)} = \abs{\eta(\alpha, \sigma')}
\end{equation*}
and, for all $\pi \in \Pi_C$, $m \in \mathcal{M}$,
\begin{equation*}
    \abs{\cp{n \in 1..E \mid \sigma[\pi][m][n] \in \Pi_C}} = \abs{\cp{n \in 1..E \mid \sigma'[\pi][m][n] \in \Pi_C}}
\end{equation*}

We have
\begin{equation*}
    \forall n < \abs{\eta(\alpha, \sigma)}, \tau(\alpha, \sigma)_n = \tau(\alpha, \sigma')_n
\end{equation*}

\begin{proof}
The lemma is proved by induction. Let us assume
\begin{eqnarray*}
    \tau(\alpha, \sigma) &=& ((i_1, r_1), \ldots) \\
    \tau(\alpha, \sigma') &=& ((i'_1, r'_1), \ldots) \\
    i_j = i'_j, r_j = r'_j && \forall j \leq n
\end{eqnarray*}

We start by noting that, since $\alpha$ is a deterministic algorithm, we immediately have
\begin{equation*}
    i_{n + 1} = i'_{n + 1}
\end{equation*}
and we need to prove that $r_{n + 1} = r'_{n + 1}$. \\

Let us assume that $i_{n + 1} = ({\tt Byzantine}, \pi)$. By hypothesis, the number of Byzantine processes in $\pi$'s first echo sample is identical in $\sigma$ and $\sigma'$: with a minor abuse of notation we effectively have $r_{n + 1} = r'_{n + 1}$.

Let us assume that $i_{n + 1} = ({\tt State})$. By hypothesis, $n + 1 < \abs{\eta(\alpha, \sigma)} = \abs{\eta(\alpha, \sigma')}$, and we immediately get $r_{n + 1} = r'_{n + 1} = \emptyset$.

Since $\alpha$ is a sample-blind adversary, we have $i_{n + 1} \neq ({\tt Sample}, \pi, m)$.

Noting that procedures $Deliver(\ldots)$ and $Echo(\ldots)$ never return a value, we trivially have that if $i_{n + 1} = ({\tt Deliver}, \pi, m)$ or $i_{n + 1} = ({\tt Echo}, \pi, s, \xi, m)$ then $r_{n + 1} = \bot = r'_{n + 1}$. By induction, we have that, for every $n < \abs{\eta(\alpha, \sigma)}$, $\tau(\alpha, \sigma)_n = \tau(\alpha, \delta)_n$.
\end{proof}
\end{lemma}

\begin{lemma}
\label{lemma:firstphasematch}
Let $\alpha$ be a state-polling adversary, let $\sigma, \sigma'$ be systems such that 
\begin{equation*}
    \eta(\alpha, \sigma) = \eta(\alpha, \sigma')
\end{equation*}
and, for all $\pi \in \Pi_C$, $m \in \mathcal{M}$,
\begin{equation*}
    \abs{\cp{n \in 1..E \mid \sigma[\pi][m][n] \in \Pi_C}} = \abs{\cp{n \in 1..E \mid \sigma'[\pi][m][n] \in \Pi_C}}
\end{equation*}

We have
\begin{equation*}
    \tau(\alpha, \sigma) = \tau(\alpha, \sigma')
\end{equation*}

\begin{proof}
The proof is similar to the proof of \cref{lemma:firstphaselength}, and we omit it for the sake of brevity. The lemma is proved by induction and noting that, since $\alpha$ is a single-response adversary, it never invokes $State()$ throughout the second phase of an adversarial execution.
\end{proof}
\end{lemma}

In \cref{subsection:twophaseadversaryproof}, we formally prove that the set of two-phase adversaries $\mathcal{A}_{tp}$ is optimal. \\

Before moving on to computing a bound on the adversarial power of $\mathcal{A}_{tp}$, we prove two additional lemmas on the behavior of two-phase adversaries.

\begin{lemma}
\label{lemma:twophaseprefix}
Let $\alpha$ be a two-phase adversary. Let $\eta^{(i)}(\alpha, \sigma)$ denote the sequence of invocations in $\eta(\alpha, \sigma)$. Let $\sigma, \sigma'$ be systems such that, for all $\pi \in \Pi_C$, 
\begin{equation*}
    \abs{\sigma.Byzantine(\pi)} = \abs{\sigma'.Byzantine(\pi)}
\end{equation*}

We have
\begin{equation*}
    \forall n \leq \min(\abs{\eta(\alpha, \sigma)}, \abs{\eta(\alpha, \sigma')}), \; \eta^{(i)}(\alpha, \sigma)_n = \eta^{(i)}(\alpha, \sigma')_n
\end{equation*}

\begin{proof}
The proof is similar to the proof of \cref{lemma:firstphaselength}, and we omit it for the sake of brevity. The lemma is proved by induction and noting that, except for the last one, every response to $({\tt State})$ in $\eta(\alpha, \sigma)$, $\eta(\alpha, \sigma')$ is, by definition, $\emptyset$.
\end{proof}
\end{lemma}

\begin{lemma}
\label{lemma:twophasesuffix}
Let $\alpha$ be a two-phase adversary. Let $\sigma, \sigma'$ be systems such that $\abs{\eta(\alpha, \sigma) = \abs{\eta(\alpha, \sigma')}}$. Let $\theta^{(i)}(\alpha, \sigma)$ denote the sequence of invocations in $\theta(\alpha, \sigma)$ and, for all $\pi \in \Pi_C$,

\begin{equation*}
    \abs{\sigma.Byzantine(\pi)} = \abs{\sigma'.Byzantine(\pi)}
\end{equation*}

We have
\begin{equation*}
    \theta^{(i)}(\alpha, \sigma) = \theta^{(i)}(\alpha, \sigma')
\end{equation*}

\begin{proof}
The proof is again similar to the proof of \cref{lemma:firstphaselength}, and we omit it for the sake of brevity. The lemma is proved by induction and noting that:
\begin{itemize}
    \item Since $\alpha$ is two-phase, it only invokes $State() \neq \emptyset$, the content of the $\abs{\eta(\alpha, \sigma)}$-th response does not affect its behavior.
    \item Since $\alpha$ is single-response, it never invokes $State()$ throughout the second phase. 
\end{itemize}
\end{proof}
\end{lemma}

\subsection{Consistency}
\label{subsection:pcb-consistency}

In this section, we finally achieve the main goal of this appendix, i.e., to compute a bound on the $\epsilon$-consistency of \pcbal. In order to achieve this, in \cref{subsection:cobal}, we introduced \cobal, a strawman algorithm designed to be analytically tractable. 

In \cref{subsection:simplifiedadversarialpower}, we proved that the consistency of \cobal\ is weaker than the consistency of \pcbal. More precisely, we proved that an optimal adversary has a greater probability of compromising the consistency of \cobal\ than that of \pcbal. 

In doing so, we reduced the problem of bounding the $\epsilon$-consistency of \pcbal\ to that of bounding the adversarial power of a set of adversaries for \cobal\ that provably includes an optimal adversary.

Throughout \cref{subsection:twophaseadversaries}, we employed a sequence of decorators to iteratively reduce the size of the set that provably includes an optimal adversary. Specifically, we proved that the set $\mathcal{A}_{tp}$ of two-phase adversaries is optimal. Intuitively, we proved that the behavior of an optimal adversary reduces to:
\begin{itemize}
    \item (Echo phase): Causing every Byzantine process to send an {\tt Echo}($m$, $m$) message to every correct process, for every message $m$.
    \item (First phase): In sequence, causing correct processes to deliver a predefined sequence of messages until at least one correct process delivers a message.
    \item (Second phase): In sequence, causing the remaining set of correct process to deliver a predefined sequence of messages, determined only by the number of correct processes that \pbin.Delivered a message throughout the first phase.
\end{itemize}

In particular, the only information that we did not prove to be unnecessary to the Byzantine adversary is:
\begin{itemize}
    \item The number of Byzantine processes in the first echo sample of each correct process $\pi$. This information is available to the adversary from the beginning of the adversarial execution, and does not change throughout the execution. We conjecture this information to still be of no use to the adversary, but we don't rely on this conjecture in proving what follows.
    \item The number of correct processes that \pbin.Deliver a message throughout the first phase of the adversarial execution, i.e., before at least one correct process delivers a message.
\end{itemize}

In this section, we redefine a two-phase adversary as a table of messages. In doing so, we provide a sound structure to a set of adversaries that provably includes an optimal one. We then use this structure to analitically bound the probability of any two-phase adversary compromising the consistency of a random \cobal\ system.

First, we focus on the second phase of an adversarial execution, and study the probability of any two-phase adversary compromising the consistency of \cobal, given the number of correct processes that \pbin.Delivered, throughout the first phase, the message that was delivered by at least one correct process at the end of the first phase.

We then focus on the first phase of an adversarial execution, and study the probability of any two-phase adversary concluding the first phase of an adversarial execution having caused less than $n$ correct processes to \pbin.Deliver $m$, $m$ being the message that at least one correct process delivers at the end of the first phase. 

We finally join the two above results to compute a bound $\epsilon_c$ on the probability of a two-phase adversary compromising the consistency of \cobal. Since at least one two-phase adversary is provably optimal, \cobal\ satisfies $\epsilon_c$-consistency. Since the $\epsilon$-consistency of \pcbal\ is provably bound by the $\epsilon$-consistency of \cobal, \pcbal\ satisfies $\epsilon_c$-consistency.

\subsubsection{Two-phase adversaries}
\label{subsubsection:twophaseadversaries}

In \cref{subsubsection:twophaseadversary}, we proved that the set $\mathcal{A}_{tp}$ is optimal. In this section, we use \cref{lemma:twophaseprefix,lemma:twophasesuffix} to re-define the set of two-phase adversaries as a set of \emph{triangular message tables}.

\begin{definition}[Byzantine population]
A \textbf{Byzantine population} is a vector in the set
\begin{equation*}
    \mathcal{F} = \rp{0..E}^{\Pi_C}
\end{equation*}

Let $\sigma$ be a system. We define the \textbf{Byzantine population} of $\sigma$ by
\begin{equation*}
    \forall \pi,\;F(\sigma)_{\pi} = \abs{\sigma.Byzantine(\pi)}
\end{equation*}
\end{definition}

\begin{definition}[Two-phase adversary]
A \textbf{two-phase adversary} $\alpha \in \mathcal{A}_{tp}$ is a \emph{triangular table} defined by:
\begin{eqnarray*}
\alpha[F]_i \in \mathcal{M} &\;& F \in \mathcal{F}, \;i \in 1..C \\
\alpha[F]^n_i \in \mathcal{M} &\;& F \in \mathcal{F}, \; n \in 0..C, \; i \in 1..(C - n)
\end{eqnarray*}
\end{definition}

Coupled with a system $\sigma$, a two-phase adversary $\alpha$:
\begin{itemize}
    \item (Echo phase) Causes every Byzantine process to send an {\tt Echo}($m$, $m$) message to every correct process in $\sigma$, for every message $m$.
    \item (First phase) Sequentially causes $\zeta(1)$ to \pbin.Deliver $\alpha[F(\sigma)]_1$, $\zeta(2)$ to \pbin.Deliver $\alpha[F(\sigma)]_2$, $\ldots$ in $\sigma$, until, as a result of the $n$-th \pbin.Delivery, at least one correct process delivers a message in $\sigma$. We note that, if $\sigma$ is poisoned, then at least one correct process delivers a mesage in $\sigma$ as a result of the echo phase, and $n = 0$.
    \item (Second phase) Sequentially causes $\zeta(n + 1)$ to \pbin.Deliver $\alpha[F(\sigma)]^n_1$, $\ldots$, $\zeta(C)$ to \pbin.Deliver $\alpha[F(\sigma)]^n_{C - n}$ in $\sigma$. 
\end{itemize}

\subsubsection{Random variables}
\label{subsubsection:randomvariables}

Let $\alpha$ be a two-phase adversary. In the next sections, we compute a bound on the probability of $\alpha$ compromising the consistency of a random, non-poisoned system. To this end, in this section we introduce a set of random variables. 

\begin{notation}[Delivery indicator]
Let $\sigma$ be a system, let $m$, $m_1, \ldots, m_n$ be messages. We use
\begin{equation*}
    \delta_m[m_1, \ldots, m_n](\sigma) \in \cp{\true, \false}
\end{equation*}
to indicate whether or not at least one correct process delivers $m$ in $\sigma$, if $\zeta(1)$ \pbin.Delivers $m_1$, $\ldots$, $\zeta(n)$ \pbin.Delivers $m_n$ in $\sigma$. We additionally define
\begin{equation*}
    \delta[m_1, \ldots, m_n](\sigma) = \bigvee_{m \in \mathcal{M}} \delta[m_1, \ldots, m_n](\sigma)
\end{equation*}
\end{notation}

Let $\sigma$ be a random, non-poisoned system. We define:
\begin{itemize}
    \item \textbf{Byzantine population} $F_{\pi \in \Pi_C}(\sigma)$: represents the number of Byzantine processes in the first echo sample of $\pi$ in $\sigma$.
    \item \textbf{First phase duration} $\eta(\sigma)$: represents the number of correct processes that \pbin.Deliver a message in the first phase, when $\alpha$ is coupled with $\sigma$. More formally,
    \begin{equation*}
        \eta(\sigma) = \min n \mid \rp{\delta[\alpha[F(\sigma)]_1, \ldots, \alpha[F(\sigma)]_n] = \true \vee n = C}
    \end{equation*}
    \item \textbf{First-phase deliveries} $S_{m \in \mathcal{M}}(\sigma)$: represents the number of correct processes that \pbin.Deliver message $m$ throughout the first phase, when $\alpha$ is coupled with $\sigma$. More formally,
    \begin{equation*}
        S_m(\sigma) = \abs{\cp{n \in 1..\eta(\sigma) \mid \alpha[F(\sigma)]_n = m}}
    \end{equation*}
    \item \textbf{Second-phase deliveries} $T_{m \in \mathcal{M}}(\sigma)$: represents the number of correct processes that \pbin.Deliver message $m$ throughout the second phase, when $\alpha$ is coupled with $\sigma$. More formally,
    \begin{equation*}
        T_m(\sigma) = \abs{\cp{n \in 1..(C - \eta(\sigma)) \mid \alpha[F(\sigma)]^{\eta(\sigma)}_n = m}}
    \end{equation*}
    \item \textbf{Deliveries} $N_{m \in \mathcal{M}}(\sigma)$: represents the number of correct processes that \pbin.Deliver message $m$, when $\alpha$ is coupled with $\sigma$. More formally,
    \begin{equation*}
        N_m(\sigma) = S_m(\sigma) + T_m(\sigma)
    \end{equation*}
    \item \textbf{First delivered message} $H(\sigma) \in \mathcal{M} \cup \cp{\bot}$: if, when $\alpha$ is coupled with $\sigma$, at least one correct process delivers a message, $H(\sigma)$ represents the first message to be delivered by at least one correct process in $\sigma$. Otherwise, $H(\sigma) = \bot$. More formally,
    \begin{equation*}
        H(\sigma) =
        \begin{cases}
            \alpha[F(\sigma)]_{\eta(\sigma)} &\text{iff}\; \delta[\alpha[F(\sigma)]_1, \ldots, \alpha[F(\sigma)]_C] = \true \\
            \bot &\text{otherwise}
        \end{cases}
    \end{equation*}
    \item \textbf{Correct echoes} $E^{k \in 0..C}_{m \in \mathcal{M}}[\pi](\sigma) \in 0..E \cup \cp{\bot}$: if $k \leq N_i(\sigma)$, then $E^k_i[\pi](\sigma)$ represents the number of correct processes in $\pi$'s echo sample for $m$ that sent an {\tt Echo}($m$, $m$) message to $\pi$ in $\sigma$, when exactly $k$ correct processes \pbin.Delivered $m$ in $\sigma$. Otherwise, $E^k_m[\pi](\sigma) = \bot$.
    \item \textbf{Delivery} $A^{k \in 0..C}_{m \in \mathcal{M}}[\pi \in \Pi_C](\sigma) \in \cp{\true, \false, \bot}$: if $k \leq N_m(\sigma)$, $A^k_m[\pi]$ represents, when $\alpha$ is coupled with $\sigma$, whether or not $\pi$ delivered $m$ after $k$ correct processes \pbin.Delivered $m$. More formally,
    \begin{equation*}
        A^k_m[\pi](\sigma) =
        \begin{cases}
        E^k_m[\pi] \geq \hat E - F_{\pi} &\text{iff}\; k \leq N_m(\sigma) \\
        \bot &\text{otherwise}
        \end{cases}
    \end{equation*}
    \item \textbf{Global delivery} $A^{k \in 0..C}_{m \in \mathcal{M}}(\sigma)$: if $k \leq N_m(\sigma)$, $A^k_i$ represents, when $\alpha$ is coupled with $\sigma$, whether or not at least one process delivered $m$ after $k$ correct processes \pbin.Delivered $m$. More formally,
    \begin{equation*}
        A^k_m(\sigma) =
        \begin{cases}
            \bigvee_{\pi \in \Pi_C} A^k_m[\pi](\sigma) &\text{iff}\; k \leq N_m(\sigma) \\
            \bot &\text{otherwise}
        \end{cases}
    \end{equation*}
    \item \textbf{First phase plan} $L_m(\sigma)$: represents the number of times $m$ appears in the sequence
    \begin{equation*}
        \alpha[F(\sigma)]_1, \ldots, \alpha[F(\sigma)]_C
    \end{equation*}
    Intuitively, $L_m$ represents the number of correct processes that $\alpha$ would eventually cause to \pbin.Deliver $m$, if no correct process ever delivered any message.
    \item \textbf{Adversarial success} $W$: $W$ represents whether or not the adversary successfully compromises the consistency of the system.
\end{itemize}

We additionally define:
\begin{eqnarray*}
    E_m[\pi](\sigma) &=& E^{N_m(\sigma)}_m[\pi](\sigma) \\
    E^{(s)}_m[\pi](\sigma) &=& E^{S_m(\sigma)}_m[\pi](\sigma) \\
    E^{(t)}_m[\pi](\sigma) &=& E_m[\pi](\sigma) - E^{(s)}_m[\pi](\sigma) \\
    A_m[\pi](\sigma) &=& A^{N_m(\sigma)}_m[\pi](\sigma) \\
    A_m(\sigma) &=& A^{N_m(\sigma)}(\sigma)
\end{eqnarray*}

\subsubsection{Byzantine population, correct echoes, delivery}

In this section, we compute the probability distributions underlying Byzantine population. Given the Byzantine population, we then compute the number of correct echoes and the probability of delivery.

\paragraph{Byzantine population}

As we discussed in \cref{subsubsection:cobalgorithm}, every correct process selects its first echo sample using the $Sample(\ldots)$ procedure, which, in turn, picks each element independently from the set of processes. Therefore, the number of correct processes in the first echo sample of each correct process is independently binomially distributed:
\begin{equation*}
    \prob{\bar F_\pi} = \bin{E}{f}{\bar F_\pi}
\end{equation*}

\paragraph{Correct echoes}

Let $\pi$ be a correct process, let $m$ be a message. If $\pi$ has $\bar F_\pi$ Byzantine processes in its first echo sample, and exactly $k$ correct processes \pbin.Delivered $m$, then each of the $E - \bar F_\pi$ correct process in $\pi$'s echo sample for $m$ has an independent probability $k/C$ of having \pbin.Delivered $m$. 

Consequently, we have
\begin{equation*}
    \prob{\bar E^k_m[\pi] \mid \bar F_\pi} = 
    \begin{cases}
        \bin{E - \bar F_\pi}{\frac{k}{C}}{\bar E^k_m[\pi]} \prob{k \leq N_m \mid \bar F_\pi} &\text{iff}\;\bar E^k_m[\pi] \neq \bot \\
        \prob{k > N_m \mid \bar F_\pi} &\text{otherwise}
    \end{cases}
\end{equation*}

We underline that the above holds true only because the adversary $\alpha$ is non-redundant. Indeed, since $\alpha$ knows the first phase duration $\eta$, it also knows $H$ (this immediately follows from $H = \alpha[F]_\eta$). Therefore, if $\alpha$ was not non-redundant, the value of $E^k_m[\pi]$ would not necessarily be independent from the event $k \leq N_m$. \\

We can see this with an example. With a minor slip of notation, consider an adversary $\alpha$ such that
\begin{eqnarray*}
    \alpha[\cp{0}^{\Pi_C}]_1 &=& 1 \\
    \alpha[\cp{0}^{\Pi_C}]_{i > 1} &\neq& 1\\
    \alpha[\cp{0}^{\Pi_C}]^1_1 &=& 1
\end{eqnarray*}

We can immediately see that $\alpha$ is not non-redundant: if no correct process has any Byzantine process in its echo samples, and at least one correct process delivers $1$ as an immediate result of $\zeta(1)$ \pbin.Delivering $1$, $\alpha$ causes $\zeta(2)$ to \pbin.Deliver $1$ again. If $\eta = 1$, then $l \geq 1$ correct process $\pi^*_1, \ldots, \pi^*_l$ exists such that $\zeta(1)$ appears at least $\hat E$ times in $\pi^*_i$'s echo sample for $1$. Since a correct process $\pi$ has a probability $l/C$ of being among $\pi^*_1, \ldots, \pi^*_l$, if $N_1 \geq 2$ the distribution of $\bar E^k_m[\pi]$ becomes
\begin{eqnarray*}
    \lhs \prob{\bar E^k_m[\pi] \mid F[\pi] = 0, N_1 \geq 2} \\
    &=& \bin{E}{\frac{k}{C}}{\bar E^k_m[\pi]} 
    \rp
    {
        \frac{C - l}{C} + \frac{l}{C} \frac{I(\bar E^k_m[\pi] \geq \hat E)}{\sum_{e = \hat E}^E \bin{E}{\frac{k}{C}}{e}}
    }
\end{eqnarray*}
which is clearly not a binomial. Intuitively, if $\alpha$ was not non-redundant, it could cause the value of $N_m$ to depend on whether or not $m$ was delivered by at least one correct process, which obviously correlates with the value of $E^k_m[\pi]$. \\

Since $\alpha$ is non-redundant, however, and every correct process picks each echo sample independently, the value of $E^k_m[\pi]$ is indeed independent from the event $k \leq N_m$.

\paragraph{Delivery}

Noting that a correct process $\pi$ delivers a message $m$ if it collects at least $\hat E$ {\tt Echo}($m$, $m$) messages from its echo sample for $m$, we can use the distribution underlying the correct echoes to obtain
\begin{equation*}
    \prob{A^k_m[\pi] \mid \bar F_\pi} = \sum_{\bar E^k_m[\pi] = \hat E - \bar F_\pi}^{E - \bar F_\pi} \prob{\bar E^k_m[\pi] \mid \bar F_\pi}
\end{equation*}
and, using the law of total probability, we get
\begin{equation*}
    \prob{A^k_m[\pi]} = \sum_{\bar F_\pi = 0}^E \prob{A^k_m[\pi] \mid \bar F_\pi} \prob{\bar F_\pi}
\end{equation*}

Finally, since the above holds independently for every process $\pi$, we have
\begin{equation*}
    \prob{A^k_m} = 1 - \prod_{\pi \in \Pi_C} \rp{1 - \prob{A^k_m[\pi]}} = 1 - \rp{1 - \prob{A^k_m[\zeta(1)]}}^C
\end{equation*}

\subsubsection{Second phase}

In the previous sections, we computed the probability of any correct process delivering a message $m$, given that $k$ correct processes \pbin.Delivered $m$. In \cref{subsubsection:twophaseadversaries}, we discussed how an optimal adversarial execution unfolds in two phases: the first takes place before any correct process delivers any message; throughout the second, the goal of the adversary is to cause at least one correct process to deliver one additional message. 

In this section, we focus on the second phase. We assume that a message $H$ has already been delivered by at least one correct process. Given the number of correct processes that \pbin.Delivered each message throughout the first phase, we compute (where possible) a bound on the probability of any message different from $H$ being delivered before the end of the adversarial execution, i.e., the probability of the adversary successfully compromising the consistency of the system.

\paragraph{Correct echoes for a non-delivered message}

Let $\pi$ be a correct process that has $\bar F$ Byzantine processes in its first echo sample. Let $m$ be a message such that $\pi$ does not deliver $m$ after $k$ correct processes \pbin.Delivered $m$. Here we use Bayes' theorem to compute the probability distribution underlying the number of correct echoes received by $\pi$ for $m$.

\begin{notation}[Indicator function]
We use $I$ to denote the \textbf{indicator function}. Let $c$ be a predicate, then
\begin{equation*}
    I(c) = 
    \begin{cases}
        1 &\text{iff}\; c\; \text{is true} \\
        0 &\text{otherwise}
    \end{cases}
\end{equation*}
\end{notation}

\begin{eqnarray*}
    \prob{\bar E^k_m[\pi] \mid \cancel{A^k_m[\pi]}, \bar F_\pi} &=& \prob{\bar E^k_m[\pi] \mid E^k_m[\pi] < \hat E - \bar F_\pi, \bar F_\pi} \\
    &=& \frac{\prob{E^k_m[\pi] < \hat E - \bar F_\pi \mid \bar E^k_m[\pi], \bar F_\pi} \prob{\bar E^k_m[\pi] \mid \bar F_\pi}}{\prob{E^k_m[\pi] < \hat E - \bar F_\pi \mid \bar F_\pi}} \\
    &=& \frac{I(\bar E^k_m[\pi] < \hat E - \bar F_\pi) \prob{\bar E^k_m[\pi] \mid \bar F_\pi}}{\sum_{e = 0}^{\hat E - \bar F_\pi - 1} \prob{E^k_m[\pi] = e \mid \bar F_\pi}}
\end{eqnarray*}
where the numerator of the last term includes an indicator function because any condition $A < B$, given $\bar A$ and $\bar B$, is always satisfied deterministically.

\paragraph{Conditions}

Let $\pi$ be a correct process, let $m$ be a message. Throughout the rest of this section, we compute the probability of $\pi$ eventually delivering $m$ under the following conditions:
\begin{itemize}
    \item $\bar F_\pi$ processes in $\pi$'s first echo sample are Byzantine.
    \item $m$ is not the message that is delivered at the end of the first phase, i.e., $H \neq m$.
    \item $\bar S_m$ correct processes \pbin.Deliver $m$ throughout the first phase.
    \item $\bar T_m$ correct processes \pbin.Deliver $m$ throughout the second phase.
\end{itemize}

\paragraph{First phase correct echoes}

Here we compute, under the above conditions, the probability distribution underlying $E^{(s)}_m[\pi]$, i.e., the number of correct echoes that $\pi$ collects for $m$ throughout the first phase.

Since $H \neq m$, $\pi$ does not deliver $m$ throughout the first phase. In other words, $\pi$ does not deliver $m$ after $\bar S_m$ correct processes \pbin.Delivered $m$, and we immediately have
\begin{equation*}
    \prob{\bar E^{(s)}_m[\pi] \mid H \neq m, \bar S_m, \bar F_\pi} = \prob{E^{\bar S_m}_m[\pi] = \bar E^{(s)}_m[\pi] \mid \cancel{A^{\bar S_m}_m[\pi]}, \bar F_\pi}
\end{equation*}

\paragraph{Second phase correct echoes}

Here we compute, under the above conditions and given $\bar E^{(s)}_m[\pi]$, the probability distribution underlying $\bar E^{(t)}_m[\pi]$, i.e., the number of correct echoes that $\pi$ collects for $m$ throughout the second phase.

We start by noting that, out of the $E$ elements in $\pi$'s echo sample for $m$: 
\begin{itemize}
    \item $\bar F_\pi$ are Byzantine.
    \item $\bar E^{(s)}_m[\pi]$ belong to the set of $\bar S_m$ processes that \pbin.Delivered $m$ throughout the first phase.
    \item $E - \bar F_\pi - \bar E^{(s)}_m[\pi]$ belong to the set of $C - \bar S_m$ processes that did not \pbin.Deliver $m$ throughout the first phase.
\end{itemize}

Moreover, out of the $C - \bar S_m$ processes that did not \pbin.Deliver $m$ throughout the first phase, $\bar T_m$ \pbin.Delivered $m$ throughout the second phase. Therefore, each of the processes in $\pi$'s echo sample for $m$ that did not \pbin.Deliver $m$ throughout the first phase has an independent probability $\bar T_m / (C - \bar S_m)$ of \pbin.Delivering $m$ throughout the second phase. 

Consequently, $\bar E^{(t)}_m$ is binomially distributed:
\begin{equation*}
    \prob{\bar E^{(t)}_m \mid \bar E^{(s)}_m, \bar S_m, \bar T_m, \bar F_\pi} = \bin{E - \bar F_\pi - \bar E^{(s)}_m}{\frac{\bar T_m}{C - \bar S_m}}{\bar E^{(t)}_m[\pi]}
\end{equation*}

\paragraph{Delivery probability (given message)}

We can finally compute, under the above conditions, the probability of $\pi$ eventually delivering $m$. 

We start by expanding the definition of $A_m[\pi]$ to get
\begin{eqnarray*}
    \lhs \prob{A_m[\pi] \mid H \neq m, \bar S_m, \bar T_m, \bar F_\pi} \\
    &=& \prob{E_m[\pi] \geq \hat E - \bar F_\pi \mid H \neq m, \bar S_m, \bar T_m, \bar F_\pi} = \rp{\star_1}
\end{eqnarray*}
and then expand the definition of $E_m[\pi]$ to get
\begin{equation*}
    \rp{\star_1} = \prob{E^{(t)}_m[\pi] \geq \hat E - \bar F_\pi - E^{(s)}_m[\pi] \mid H \neq m, \bar S_m, \bar T_m, \bar F_\pi} = \rp{\star_2}
\end{equation*}

Finally, using the law of total probability on each possible value of $E^{(s)}_m[\pi]$, we get
\begin{eqnarray*}
    \rp{\star_2} &=& \sum_{\bar E^{(s)}_m[\pi] = 0}^{E - \bar F_\pi} \left ( \underbrace{\prob{E^{(t)}_m[\pi] \geq \hat E - \bar F_\pi - \bar E^{(s)}_m[\pi] \mid \bar E^{(s)}_m[\pi], \bar S_m, \bar T_m, \bar F_\pi}}_{\rp{\star_a}} \right . \\
    && \phantom{\sum_{\bar E^{(s)}_m[\pi] = 0}^{E - \bar F_\pi}} \left . \cdot \underbrace{\prob{\bar E^{(s)}_m[\pi] \mid H \neq m, \bar S_m, \bar F_\pi}}_{\rp{\star_b}} \right )
\end{eqnarray*}

As we previously established,
\begin{equation*}
    \prob{\bar E^{(t)}_m[\pi] \mid E^{(s)}_m[\pi] = i, \bar S_m, \bar T_m, \bar F_\pi} = \prob{X_i = \bar E^{(t)}_m[\pi]}
\end{equation*}
with
\begin{eqnarray*}
    X_i &\sim& \bin{A - i}{p}{} \\ 
    A &=& E - \bar F_\pi \\ 
    p &=& \frac{\bar T_m}{(C - \bar S_m)}
\end{eqnarray*}

Moreover,
\begin{equation*}
    \rp{\star_a} = \prob{X_i \geq B - i}
\end{equation*}
with
\begin{equation*}
    B = \hat E - \bar F_\pi \leq E - \bar F_\pi = A
\end{equation*}

Therefore, following from \cref{lemma:narrowbinomialsample}, $\rp{\star_a}$ is an increasing function of $\bar E^{(s)}_m[\pi]$. Moreover, as we previously established,
\begin{eqnarray*}
    \prob{\bar E^{(s)}_m[\pi] \mid H \neq m, \bar S_m, \bar F_\pi} &=& \prob{E^{\bar S_m}_m[\pi] = \bar E^{(s)}_m[\pi] \mid \cancel{A^{\bar S_m}_m[\pi]}, \bar F_\pi} \\ 
    &=& \frac{I(\bar E^{\bar S_m}_m[\pi] < \hat E - \bar F_\pi) \prob{\bar E^{\bar S_m}_m[\pi] \mid \bar F_\pi}}{\sum_{e = 0}^{\hat E - \bar F_\pi - 1} \prob{E^{\bar S_m}_m[\pi] = e \mid \bar F_\pi}}
\end{eqnarray*}
and $\rp{\star_2}$ can be restated as
\begin{eqnarray*}
    \rp{\star_2} &=& \frac{\sum_{x = 0}^{K - l} f(x)g(x)}{\sum_{x = 0}^{K - l} g(x)} \\
    K &=& E - \bar F_\pi \\
    l &=& E - \hat E + 1 \\
    f(x) &=& \prob{E^{(t)}_m[\pi] \geq \hat E - \bar F_\pi - x \mid E^{(s)}_m[\pi] = x, \bar S_m, \bar T_m, \bar F_\pi} \\
    g(x) &=& \prob{E^{\bar S_m}_m[\pi] = x \mid \bar F_\pi}
\end{eqnarray*}
with $f(x)$ increasing and $\sum_{x = 0}^K g(x) = 1$. Following from \cref{corollary:shiftweight}, we therefore have
\begin{eqnarray*}
    \lhs \prob{A_m[\pi] \mid H \neq m, \bar S_m, \bar T_m, \bar F_\pi} \\
    &\leq& \sum_{\bar E^{(s)}_m[\pi] = 0}^{E - \bar F_\pi} \prob{\bar E^{(s)}_m[\pi] + E^{(t)}_m[\pi] \geq \hat E - \bar F_\pi \mid \bar E^{(s)}_m[\pi], \bar S_m, \bar T_m, \bar F_\pi} \\
    && \phantom{\sum_{\bar E^{(s)}_m[\pi] = 0}^{E - \bar F_\pi}} \cdot \prob{E^{\bar S_m}_m[\pi] = \bar E^{(s)}_m[\pi] \mid \bar F_\pi} \\ 
    &=& \rp{\star_3}
\end{eqnarray*}
which, as we previously established, can be restated as
\begin{eqnarray*}
    \rp{\star_3} &=& \prob{X + Y \geq H} = \sum_{K = H}^A \prob{X + Y = K} \\
    \prob{\bar X} &=& \bin{A}{\frac{x}{B}}{\bar X} \\
    \prob{\bar Y \mid \bar X} &=& \bin{A - \bar X}{\frac{y}{B - x}}{\bar Y} \\ 
    H &=& \hat E - \bar F_\pi \\
    A &=& E - \bar F_\pi \\
    B &=& C \\
    x &=& \bar S_m \\
    y &=& \bar T_m
\end{eqnarray*}
which, using \cref{lemma:binomialmerge}, yields the bound
\begin{equation}
\label{equation:probdelivergivenprocessmessage}
    \prob{A_m[\pi] \mid H \neq m, \bar S_m, \bar T_m, \bar F_\pi} \leq \sum_{e = \hat E - \bar F_\pi} \bin{E - \bar F_\pi}{\frac{\bar S_m + \bar T_m}{C}}{e}
\end{equation}

\paragraph{Delivery probability (any message)} We now move on to compute the probability that a correct process $\pi$ will eventually deliver any message other than $H$, under the following assumptions:
\begin{itemize}
    \item The first phase of the adversarial execution is concluded.
    \item The number $\bar F_\pi$ of Byzantine processes in the first echo sample of $\pi$ is given.
    \item The number $\bar S_m$, $\bar T_m$ of correct processes that \pbin.Delivered each message $m$ throughout the first and second phase respectively is given.
\end{itemize}

Since every echo sample is picked independently, from \cref{equation:probdelivergivenprocessmessage} follows
\begin{eqnarray*}
    \prob{\bigvee_{m \neq \bar H} A_m[\pi] \mid \bar S_1, \ldots, \bar S_C, \bar T_1, \ldots, \bar T_C, \bar F_\pi} &\leq& \prob{\bigvee_{i \neq m} \rp{X_i \geq K}} \\
    \prob{\bar X_i} &=& \bin{N}{p_i}{\bar X_i} \\
    N &=& E - \bar F_\pi \\
    K &=& \hat E - \bar F_\pi \\
    p_i &=& \frac{\bar S_i + \bar T_i}{C}
\end{eqnarray*}
and noting that
\begin{equation*}
    \sum_{m \neq \bar H} \frac{\bar S_m + \bar T_m}{C} = \sum_{n \neq \bar H} \bar N_m = C - \bar N_{\bar H}
\end{equation*}
we can use \cref{lemma:chernoffconvexity} to obtain the bound
\begin{equation*}
    \prob{\bigvee_{m \neq \bar H} A_m[\pi] \mid \bar S_1, \ldots, \bar S_C, \bar T_1, \ldots, \bar T_C, \bar F_\pi} \leq \phi(\bar N_{\bar H}, \bar F_\pi)
\end{equation*}
with
\begin{equation}
    \phi(\bar N_{\bar H}, \bar F_\pi) = \begin{cases}
         \alpha(\bar N_{\bar H}, \bar F_\pi) \cdot \beta(\bar N_{\bar H}, \bar F_\pi)
         &\text{iff}\; \frac{C - \bar N_{\bar H}}{C} \leq \frac{\rp{\hat E - \bar F_\pi} - \sqrt{\hat E - \bar F_\pi}}{E - \bar F_\pi}\\
        1 &\text{otherwise}
    \end{cases}
\end{equation}
where
\begin{equation*}
    \alpha(\bar N_{\bar H}, \bar F_\pi) = 
        \rp
        {
            \frac
            {
                e (E - \bar F_\pi) \frac{C - \bar N_{\bar H}}{C}
            }
            {
                \hat E - \bar F_\pi
            }
        }^{(\hat E - \bar F_\pi)}
\end{equation*}
\begin{equation*}
    \beta(\bar N_{\bar H}, \bar F_\pi) = 
        \exp\rp{-(E - \bar F_\pi) \frac{C - \bar N_{\bar H}}{C}}
\end{equation*}

At a first glance, the second branch of the bound above could seem unreasonably lax. We underline, however, that for a large enough $(\hat E - \bar F_\pi)$,
\begin{equation*}
    \frac{\rp{\hat E - \bar F_\pi} - \sqrt{\hat E - \bar F_\pi}}{E - \bar F_\pi} \simeq \frac{\hat E - \bar F_\pi}{E - \bar F_\pi}
\end{equation*}
and, since the median of $\bin{N}{p}{}$ is either $\ceil{Np}$ or $\floor{Np}$,
\begin{equation*}
    \sum_{e = \hat E - \bar F_\pi}^{E - \bar F_\pi} \bin{E - \bar F_\pi}{\frac{\hat E - \bar F_\pi}{E - \bar F_\pi}}{e} \simeq \frac{1}{2}
\end{equation*}

Therefore, even in the second branch, the bound introduces a limited multiplicative error. Moreover, as we will see in the numerical analysis, the error introduced by the bound is non-negligible only for extremely unlikely values of $\bar N_{\bar H}$.

\paragraph{Adversarial success probability}

Throughout this section, we computed the probability that a correct process $\pi$ will deliver a message different from the message that was delivered throughout the first phase.

We showed that such probability can be bound by a function that only depends on the number of Byzantine processes in the first echo sample of $\pi$, and the number of correct processes that \pbin.Delivered $H$ throughout the first phase.

We therefore have
\begin{equation*}
    \prob{\bigvee_{m \neq H} A_m[\pi] \mid \bar N_H, \bar F_\pi} \leq \phi(\bar N_H, \bar F_\pi)
\end{equation*}

By the law of total probability we have
\begin{eqnarray*}
    \prob{\bigvee_{m \neq H} A_m[\pi] \mid \bar N_H} &=& \sum_{m \neq H}  \prob{\bigvee_{m \neq H} A_m[\pi] \mid \bar N_H, \bar F_\pi} \prob{\bar F_\pi \mid \bar N_H} \\
    &\leq& \sum_{m \neq \bar H} \phi(\bar N_H, \bar F_\pi) \prob{\bar F_\pi \mid \bar N_H}
\end{eqnarray*}

Since  $H$ was delivered by at least one correct process at the end of the first phase, we know that:
\begin{itemize}
    \item One correct process $\pi^+$ delivered $H$ immediately after $\bar N_H$ correct processes \pbin.Delivered $H$.
    \item Every other correct process did not deliver $H$ before $\bar N_H$ correct processes \pbin.Delivered $H$.
\end{itemize}

We start by computing the probability distribution underlying $\bar F_{\pi^+}$. Using Bayes' theorem we get
\begin{eqnarray*}
    \prob{\bar F_{\pi^+} \mid \bar N_H} &=& \prob{\bar F_{\pi^+} \mid A^{\bar N_H}_H[\pi^+], \cancel{A^{\bar N_H - 1}_H[\pi^+]}} \\
    &=& \frac{\prob{A^{\bar N_H}_H[\pi^+], \cancel{A^{\bar N_H - 1}_H[\pi^+]} \mid \bar F_{\pi^+}}\prob{\bar F_{\pi^+}}}{\prob{A^{\bar N_H}_H[\pi^+], \cancel{A^{\bar N_H - 1}_H[\pi^+]}}}
\end{eqnarray*}
and noting that $A^{\bar N_H - 1}_H[\pi^+] \implies A^{\bar N_H}_H[\pi^+]$, we have
\begin{eqnarray*}
    \prob{A^{\bar N_H}_H[\pi^+], \cancel{A^{\bar N_H - 1}_H[\pi^+]}} &=& \prob{A^{\bar N_H}_H[\pi^+]} - \prob{A^{\bar N_H - 1}_H[\pi^+]} \\ 
    \prob{A^{\bar N_H}_H[\pi^+], \cancel{A^{\bar N_H - 1}_H[\pi^+]} \mid \bar F_{\pi^+}} &=& \prob{A^{\bar N_H}_H[\pi^+] \mid \bar F_{\pi^+}} \\
    && - \prob{A^{\bar N_H - 1}_H[\pi^+] \mid \bar F_{\pi^+}} 
\end{eqnarray*}

Similarly, for $\pi^- \neq \pi^+$, we get
\begin{eqnarray*}
    \prob{\bar F_{\pi^-} \mid \bar N_H} &=& \prob{\bar F_{\pi^-} \mid \cancel{A^{\bar N_H - 1}_H[\pi^-]}} \\
    &=& \frac{\prob{\cancel{A^{\bar N_H - 1}_H[\pi^-]} \mid \bar F_{\pi^-}}\prob{\bar F_{\pi^-}}}{\prob{\cancel{A^{\bar N_H - 1}_H[\pi^-]}}}
\end{eqnarray*}

Since each correct process picks its echo sample independently, we have
\begin{equation}
\label{equation:boundsecondphase}
\begin{split}
    \prob{W \mid \bar N_H} &= 1 - \prod_{\pi \in \Pi_C} \rp{1 - \prob{\bigvee_{m \neq H} A_m[\pi] \mid \bar N_H}} \\
    &\leq 1 - \rp{1 - \phi^+(\bar N_H)}\rp{1 - \phi^-(\bar N_H)}^{C - 1} 
\end{split}
\end{equation}
with
\begin{eqnarray*}
    \phi^+(\bar N_H) &=& \sum_{\bar F_{\pi^+} = 0}^E \phi(\bar N_H, \bar F_{\pi^+}) \prob{\bar F_{\pi^+} \mid \bar N_H} \\
    \phi^-(\bar N_H) &=& \sum_{\bar F_{\pi^-} = 0}^E \phi(\bar N_H, \bar F_{\pi^-}) \prob{\bar F_{\pi^-} \mid \bar N_H}
\end{eqnarray*}

\subsubsection{First phase}

In the previous section, we computed, given the number of correct processes that \pbin.Delivered the first delivered message, the probability of a two-phase adversary successfully compromising the consistency of a system. 

In this section, we compute the probability distribution underlying the number of correct processes that \pbin.Deliver the first delivered message.

\begin{definition}[Deafened adversary]
\label{definition:deafenedadversary}
Let $\alpha$ be a two-phase adversary. We define $\Delta(\alpha)$ the \textbf{deafened version of $\alpha$} if:
\begin{itemize}
    \item $\Delta(\alpha)$ is a process-sequential adversary.
    \item Coupled with a system $\sigma$, $\Delta(\alpha)$ sequentially causes the \pbin.Delivery of $\alpha[F(\sigma)]_1, \ldots, \alpha[F(\sigma)]_C$.
\end{itemize}

Intuitively, the deafened version of a two-phase adversary $\alpha$ is an adversary whose adversarial execution would be identical to $\alpha$'s, if no correct process ever delivered any message. 
\end{definition}

\begin{lemma}
\label{lemma:deafphase}
Let $\alpha$ be a two-phase adversary, let $\sigma$ be a system. We have
\begin{equation*}
    \eta(\alpha, \sigma) = \eta(\Delta(\alpha), \sigma)
\end{equation*}

\begin{proof}
It follows immediately from \cref{definition:deafenedadversary}: $\Delta(\alpha)$ causes the same processes to \pbin.Deliver the same messages as $\alpha$ throughout the first phase.
\end{proof}
\end{lemma}

\begin{definition}[Delivery cost]
Let $\alpha$ be an auto-echo adversary, let $\sigma$ be a non-poisoned system, let $m$ be a message such that, when $\alpha$ is coupled with $\sigma$, at least one correct process delivers $m$. We define the \textbf{delivery cost of $m$} $\lambda(\alpha, \sigma, m)$ as the minimum $\lambda \in 1..C$ such that, when $\alpha$ is coupled with $\sigma$, at least one correct process delivers $m$ after $\lambda$ correct processes \pbin.Delivered $m$.
\end{definition}

\begin{lemma}
\label{lemma:boundNH}
Let $\alpha$ be a two-phase adversary, let $\sigma$ be a non-poisoned system such that, when coupled with $\sigma$, $\alpha$ causes at least one correct process to deliver one message. Let $\bar H$ be the first message delivered by at least one correct process, when $\alpha$ is coupled with $\sigma$.

We have that
\begin{equation*}
    \lambda(\alpha, \sigma, \bar H) \geq \min_{m \in \mathcal{M}} \lambda(\Delta(\alpha), \sigma, m)
\end{equation*}

\begin{proof}
Following from \cref{lemma:deafphase} $\eta(\alpha, \sigma) = \eta(\Delta(\alpha), \sigma)$. Therefore, at least one correct process delivers $\bar H$ after $\lambda(\alpha, \sigma, \bar H)$ processes \pbin.Deliver $\bar H$, when $\Delta(\alpha)$ is coupled with $\sigma$. 
\end{proof}
\end{lemma}

In this section, we bound the cumulative probability $\prob{N_H \leq L}$ for an adversary $\alpha$ by bounding the probability that the deafened adversary $\Delta(\alpha)$ will cause the delivery of at least one message $m$, with a cost smaller or equal to $L$.

Let $m$ be a message. We start by noting that, by definition, $\Delta(\alpha)$ eventually causes $L_m$ correct processes to \pbin.Deliver $m$. Let $\pi$ be a correct process, let $\bar F_\pi$ be the number of Byzantine processes in $\pi$'s first echo sample. \\

We denote with $\Lambda_m[\pi]$ the random variable representing the minimum number of correct processes that \pbin.Deliver $m$, before $\pi$ delivers $m$. If $\pi$ never delivers $m$, we set $\Lambda_m[\pi] = \infty$. \\

Let $L \in 1..C$. Using the tools we developed in the previous section, we immediately get
\begin{equation*}
    \prob{\Lambda_m[\pi] \leq L \mid \bar L_m, \bar F_\pi} = \sum_{e = \hat E - \bar F_\pi}^{E - \bar F_\pi} \bin{E - \bar F_\pi}{\frac{\min\rp{\bar L_m, L}}{C}}{e}
\end{equation*}
and using the independence of echo samples, we get
\begin{eqnarray*}
    \prob{\bigvee_{m \in \mathcal{M}} \Lambda_m[\pi] \leq L \mid \bar L_1, \ldots, \bar L_C, \bar F_\pi} &=& \prob{\bigvee_{i \in \mathcal{M}} (X_i \geq K)} \\
    \prob{\bar X_i} &=& \bin{N}{p_i}{\bar X_i} \\
    N &=& E - \bar F_\pi \\
    K &=& \hat E - \bar F_\pi \\ 
    p_i &=& \frac{\min\rp{\bar L_i, L}}{C}
\end{eqnarray*}
and we can use \cref{lemma:chernoffconvexity} to obtain the bound
\begin{eqnarray*}
    \lhs \prob{\bigvee_{m \in \mathcal{M}} \Lambda_m[\pi] \leq L \mid \bar L_1, \ldots, \bar L_C, \bar F_\pi} \\
    &\leq& 1 - \rp{1 - \psi(L, \bar F_\pi)}^{\floor{\frac{C}{L}}} \rp{1 - \psi(C \bmod L, \bar F_\pi)}
\end{eqnarray*}
with
\begin{equation}
    \psi(M, \bar F_\pi) = \begin{cases}
        \alpha(M, \bar F_\pi) \cdot \beta(M, \bar F_\pi)
         &\text{iff}\; \frac{M}{C} \leq \frac{\rp{\hat E - \bar F_\pi} - \sqrt{\hat E - \bar F_\pi}}{E - \bar F_\pi}\\
        1 &\text{otherwise}
    \end{cases}
\end{equation}
where
\begin{equation*}
    \alpha(M, \bar F_\pi) =
        \rp
        {
            \frac
            {
                e (E - \bar F_\pi) \frac{M}{C}
            }
            {
                \hat E - \bar F_\pi
            }
        }^{(\hat E - \bar F_\pi)}
\end{equation*}
\begin{equation*}
    \beta(M, \bar F_\pi) = 
    \exp\rp{-(E - \bar F_\pi) \frac{M}{C}}
\end{equation*}

Noting that the bound holds for any value of $\bar L_1, \ldots, \bar L_C$, we can use again the law of total probability to obtain
\begin{equation*}
    \prob{\bigvee_{m \in \mathcal{M}} \Lambda_m[\pi] \leq L} \leq \psi(L)
\end{equation*}
with
\begin{equation}
    \psi(L) = \sum_{\bar F_\pi = 0}^E 1 - \rp{1 - \psi(L, \bar F_\pi)}^{\floor{\frac{C}{L}}} \rp{1 - \psi(C \bmod L, \bar F_\pi)} \prob{\bar F_\pi}
\end{equation}
and using the independence of echo samples across correct processes we finally get
\begin{equation}
    \label{equation:boundadaptivefirstphase}
    \prob{\bigvee_{\pi \in \Pi_C, m \in \mathcal{M}} \Lambda_m[\pi] \leq L} \leq 1 - (1 - \psi(L))^C
\end{equation}

We now have all the elements to prove
\begin{theorem}
\label{theorem:cobconsistency}
\pcbal\ satisfies $\epsilon_c$-consistency, with
\begin{eqnarray*}
    \epsilon_c &\leq& \epsilon_p + \sum_{L = 0}^C \tilde{\psi}(L) \tilde{\phi}(L) \\
    \tilde{\psi}(L) &=& 
    \begin{cases}
        \rp{1 - (1 - \psi(L))^C} - \rp{1 - (1 - \psi(L - 1))^C} &\text{iff}\; L \in 1..C \\
        0 &\text{iff}\; L \in \cp{-1, 0} \\
        1 &\text{iff}\; L = C
    \end{cases} \\
    \tilde{\phi}(L) &=& \rp{1 - (1 - \phi^+(L))(1 - \phi^-(L))^{C - 1}} \\ 
    \epsilon_p &=& 1 - \rp{1 - \sum_{\bar F = \hat E}^E \bin{E}{f}{\bar F}}^C
\end{eqnarray*}
\begin{proof}
Following from \cref{lemma:boundNH}, we have
\begin{equation}
    \label{equation:boundfirstphase}
    \prob{N_H \leq L} \leq \prob{\bigvee_{\pi \in \Pi_C, m \in \mathcal{M}} \Lambda_m[\pi] \leq L}
\end{equation}

By the law of total probability, we have
\begin{eqnarray*}
    \prob{W} &=& \sum_{x = 0}^C \rp{f(x) \rp{g(x) - g(x - 1)}} \\ 
    f(x) &=& \prob{W \mid \bar N_H = x} \\
    g(x) &=& \prob{N_H \leq x}
\end{eqnarray*}
and from \cref{lemma:cumulativebound} we get
\begin{eqnarray*}
    \prob{W} &\leq& \sum_{x = 0}^C \rp{f(x) \rp{h(x) - h(x - 1)}} \\ 
    h(x) &=& \prob{\bigvee_{\pi \in \Pi_C, m \in \mathcal{M}} \Lambda_m[\pi] \leq x}
\end{eqnarray*}

The probability of compromising a non-poisoned system is obtained by applying the bounds in \cref{equation:boundfirstphase,equation:boundadaptivefirstphase,equation:boundsecondphase}. 

It is easy to see that $\epsilon_p$ represents the probability of a random system being poisoned: indeed, each correct process has an independent probability
\begin{equation*}
    \sum_{\bar F = \hat E}^E \bin{E}{f}{\bar F}
\end{equation*}
of having more than $\hat E$ Byzantine processes in its first echo sample, i.e., of being poisoned.

Therefore, the bound on $\epsilon_c$ bounds the probability of any two-phase adversary compromising the consistency of a \cobin\ system. Due to \cref{lemma:twophaseadversaries}, the set $\mathcal{A}_{tp}$ of two-phase adversaries is optimal. Therefore, \cobal\ satisfies $\epsilon_c$-consistency.

Due to \cref{lemma:cobalweakerthanpcbal}, the adversarial power of an optimal \pcbin\ adversary is bound by the adversarial power of an optimal \cobin\ adversary, and the theorem is proved.
\end{proof}
\end{theorem}

\clearpage
\section{\prbal}
\label{appendix:prbal}

In this section, we present in greater detail the \textbf{\prb} abstraction and discuss its properties. We then present \prbal, an algorithm that implements \prb, and evaluate its \textbf{security} and \textbf{complexity} as a function of its \textbf{parameters}.

The \prb\ abstraction allows the entire set of correct processes to agree on a single message from a potentially Byzantine designated sender. \Prb\ is a strictly stronger abstraction than \pcb: in the case of a Byzantine sender, while \pcb\ only guarantees that every correct process that delivers a message delivers the same message (\textbf{consistency}), \prb\ also guarantees that either no or every correct process delivers a message (\textbf{totality}).

\subsection{Definition}

The \textbf{\prb} interface (instance $\prbin$, sender $\sigma$) exposes the following two \textbf{events}:
\begin{itemize}
\item \textbf{Request}: $\event{\prbin}{Broadcast}{m}$: Broadcasts a message $m$ to all processes. This is only used by $\sigma$. 
\item \textbf{Indication}: $\event{\prbin}{Deliver}{m}$: Delivers a message $m$ broadcast by process $\sigma$.
\end{itemize}

For any $\epsilon \in [0, 1]$, we say that \prb\ is $\epsilon$-secure if:
\begin{enumerate}
\item \textbf{No duplication}: No correct process delivers more than one message.
\item \textbf{Integrity}: If a correct process delivers a message $m$, and $\sigma$ is correct, then $m$ was previously broadcast by $\sigma$.
\item $\epsilon$-\textbf{Validity}: If $\sigma$ is correct, and $\sigma$ broadcasts a message $m$, then $\sigma$ eventually delivers $m$ with probability at least $(1 - \epsilon)$.
\item $\epsilon$-\textbf{Totality}: If a correct process delivers a message, then every correct process eventually delivers a message with probabiity at least $(1 - \epsilon)$.
\item $\epsilon$-\textbf{Consistency}: Every correct process that delivers a message delivers the same message with probability at least $(1 - \epsilon)$.
\end{enumerate}

\subsection{Algorithm}
\label{subsection:prbalgorithm}

\begin{algorithm}
\begin{algorithmic}[1]
\Implements
    \Instance{\prbab}{\prbin}
\EndImplements

\Uses
    \Instance{AuthenticatedPointToPointLinks}{al}
    \Instance{\pcbab}{\pcbin}
\EndUses

\Parameters
    \State $R$: ready sample size \tabto*{5cm} $\hat R$: contagion threshold
    \State $D$: delivery sample size \tabto*{5cm} $\hat D$: delivery threshold
\EndParameters

\Upon{\prbin}{Init}{}
\label{line:prbinitialization}
\State $ready = \emptyset$; \tabto*{3cm} $delivered = \false$; \tabto*{6.5cm} $\tilde{\mathcal{R}} = \emptyset$;
\State
\State $\mathcal{R} = sample({\tt ReadySubscribe}, R)$;
\State $\mathcal{D} = sample({\tt ReadySubscribe}, D)$;
\State
\State $replies.ready = \cp{\emptyset}^R$; \tabto*{5cm} $replies.delivery = \cp{\emptyset}^D$
\EndUpon

\Upon{al}{Deliver}{\pi, [\text{\tt ReadySubscribe}]}
    \ForAll{(message, signature)}{ready}
        \Trigger{al}{Send}{\pi, [{\tt Ready}, message, signature]}; \label{line:prbreadycatchup}
    \EndForAll
    \State $\tilde{\mathcal{R}} \leftarrow \tilde{\mathcal{R}} \cup \{\pi\}$; \label{line:prbreadysubscribe}
\EndUpon

\Upon{\prbin}{Broadcast}{message} \Comment{only process $\sigma$}
    \Trigger{\pcbin}{Broadcast}{[\text{\tt Send}, message, sign(message)]}; \label{line:prbbroadcast}
\EndUpon

\Upon{\pcbin}{Deliver}{[\text{\tt Send}, message, signature]} \label{line:prbpcbdelivery}
    \If{$verify(\sigma, message, signature)$}
        \State $ready \leftarrow ready \cup \cp{(message, signature)}$;
        \ForAll{\rho}{\tilde{\mathcal{R}}}
        \label{line:prbsendreadyfrompcbloop}
            \Trigger{al}{Send}{\rho, [\text{\tt Ready}, message, signature]}; \label{line:prbsendreadyfrompcb}
        \EndForAll
    \EndIf
\EndUpon

\algstore{prbal}
\end{algorithmic}
\caption{\prbal}
\label{algorithm:prbal}
\end{algorithm}

\begin{algorithm}
\begin{algorithmic}[1]
\algrestore{prbal}

\Upon{al}{Deliver}{\pi, [\text{\tt Ready}, message, signature]} \label{line:prbreceiveready}
    \If{$verify(\sigma, message, signature)$} \label{line:prbchecksignature}
        \State $reply = (message, signature)$;
        \If{$\pi \in \mathcal{R}$} \label{line:prbreadymessagesourcecheck}
            \State $replies.ready[\pi] \leftarrow replies.ready[\pi] \cup \cp{reply}$;
        \EndIf
        \If{$\pi \in \mathcal{D}$} \label{line:prbdeliverymessagesourcecheck}
            \State $replies.delivery[\pi] \leftarrow replies.delivery[\pi] \cup \cp{reply}$
        \EndIf
    \EndIf
\EndUpon

\UponExists{message}{|\{\rho \in \mathcal{R} \mid (message, signature) \in replies.ready[\rho]\}| \geq \hat R} \label{line:prbreadyfromready}
    \State $ready \leftarrow ready \cup \cp{(message, signature)}$;
    \ForAll{\rho}{\tilde{\mathcal{R}}}
    \label{line:prbsendreadyfromreadyloop}
        \Trigger{al}{Send}{\rho, [\text{\tt Ready}, message, signature]};
        \label{line:prbsendreadyfromready}
    \EndForAll
\EndUponExists

\UponExists{message}{|\{\rho \in \mathcal{D} \mid (message, signature) \in replies.delivery[\rho]\}| \geq \hat D\; \textbf{and}\; delivered = \false} \label{line:prbdeliverycondition}
    \State $delivered \leftarrow \true$;
    \Trigger{\prbin}{Deliver}{message}; \label{line:prbdeliver}
\EndUponExists

\end{algorithmic}
\end{algorithm}

\Cref{algorithm:prbal} implements \prbal. Let $\pi$ be a correct process, let $m$ be a message. \prbal\ securely distributes a single message across the system as follows:
\begin{itemize}
    \item Initially, \pcb\ consistently distributes the same message to a subset of the correct processes.
    \item $\pi$ can issue a {\tt Ready} message for more than one message. $\pi$ issues a {\tt Ready} message $m$ when either:
    \begin{itemize}
        \item $\pi$ receives $m$ from \pcb, or
        \item $\pi$ collects enough {\tt Ready} messages for $m$ from its \emph{ready sample}.
    \end{itemize}
    \item $\pi$ delivers $m$ if $m$ is the first message for which $\pi$ collected enough {\tt Ready} messages from its delivery sample.
\end{itemize}

A correct process collects {\tt Ready} messages from two randomly selected samples, the \emph{ready sample} of size $R$, and the \emph{delivery sample} of size $D$. A correct process issues a {\tt Ready} message for $m$ upon collecting $\hat R$ {\tt Ready} messages for $m$ from its ready sample, and it delivers $m$ upon collecting $\hat D$ {\tt Ready} messages for $m$ from its delivery sample. We discuss the values of the four parameters of \prbal\ in \cref{subsection:prbanalysis}.

\paragraph{Sampling} Upon initialization (\cref{line:prbinitialization}), a correct process randomly selects a \textbf{ready sample} $\mathcal{R}$ of size $R$, and a \textbf{delivery sample} $\mathcal{D}$ of size $D$. Samples are selected with replacement by repeatedly calling $\Omega$ (\cref{algorithm:sample}, \cref{line:sampleselection}).

\paragraph{Publish-subscribe} Like \pcbal, \prbal\ uses publish-subscribe to reduce its communication complexity. This is achieved by having each correct process send {\tt Ready} messages only to its ready subscription set (\cref{line:prbsendreadyfrompcbloop,line:prbsendreadyfromreadyloop}), and accept {\tt Ready} messages only from its ready and delivery samples (\cref{line:prbreadymessagesourcecheck,line:prbdeliverymessagesourcecheck}). 

\paragraph{Consistent broadcast} The designated sender $\sigma$ initially broadcasts its message using \pcb\ (\cref{line:prbbroadcast}). When message $m$ is \pcbin.Delivered (correctly signed by $\sigma$) (\cref{line:prbpcbdelivery}), a correct process sends a {\tt Ready} message for $m$ (\cref{line:prbsendreadyfrompcb}) to all the processes in its ready subscription set.

\paragraph{Contagion} Upon collecting $\hat R$ {\tt Ready} messages for a message $m$ (\cref{line:prbreadyfromready}), a correct process sends a {\tt Ready} message for $m$ (\cref{line:prbsendreadyfromready}) to all the nodes in its ready subscription set.

\paragraph{Delivery} Upon collecting $\hat D$ {\tt Ready} messages for a message $m$ for the first time, (\cref{line:prbdeliverycondition}), a correct process delivers $m$ (\cref{line:prbdeliver}).

\subsection{No duplication and integrity}
\label{subsection:prb-nd-int}

We start by verifying that \prbal\ satisfies both \textbf{no duplication} and \textbf{integrity}.

\begin{theorem}
\prbal\ satisfies no duplication.

\begin{proof}
A message is delivered (\cref{line:prbdeliver}) only if the variable $delivered$ is equal to $\false$ (\cref{line:prbdeliverycondition}). Before any message is delivered, $delivered$ is set to $\true$. Therefore no more than one message is ever delivered.
\end{proof}
\end{theorem}

\begin{theorem}
\prbal\ satisfies integrity.

\begin{proof}
Upon receiving a {\tt Ready} message, a correct process checks its signature against the public key of the designated sender $\sigma$ (\cref{line:prbchecksignature}), and the $(message, signature)$ pair is added to the $replies.delivery$ variable only if this check succeeds. Moreover, a message is delivered only if it is represented at least $\hat D$ times in $replies.delivery$ (\cref{line:prbdeliverycondition}).

If $\sigma$ is correct, it only signs $message$ when broadcasting (\cref{line:prbbroadcast}). Since we assume that cryptographic signatures cannot be forged, this implies that the message was previously broadcast by $\sigma$.
\end{proof}
\end{theorem}

\subsection{Validity}
\label{subsection:prb-validity}

We now compute, given $D$ and $\hat D$, the $\epsilon$-\textbf{validity} of \prbal. To this end, we prove one preliminary lemma.

\begin{lemma}
\label{lemma:prbvalidityoverwhelmed}
In an execution of \prbal, if \pcbin\ satisfies total validity and the sender has no more than $D - \hat D$ Byzantine processes in its delivery sample, then \prbin\ satisfies validity.

\begin{proof}
Let $m$ be the message broadcast by the correct sender $\sigma$. Since \pcbin\ satisfies total validity, every correct process eventually issues a {\tt Ready}($m$) message (i.e., a {\tt Ready} message for $m$) (\cref{line:prbsendreadyfrompcb}).

By hypothesis, $\sigma$ has no more than $D - \hat D$ Byzantine processes in its echo sample. Obviously, $\sigma$ has at least $\hat D$ correct processes in its echo sample. Therefore, $\sigma$ eventually receives at least $\hat D$ {\tt Ready}($m$) messages (\cref{line:prbreceiveready}), and delivers $m$ (\cref{line:prbdeliver}). 
\end{proof}
\end{lemma}

\cref{lemma:prbvalidityoverwhelmed} allows us to bound the $\epsilon$-validity of \prbal, given $D$ and $\hat D$.

\begin{theorem}
\label{theorem:prbvalidity}
\prbal\ satisfies $\epsilon_v$-validity, with
\begin{equation}
\begin{split}
    \epsilon_v &\leq \epsilon^{\pcbin}_v + \rp{1 - \epsilon^{\pcbin}_v} \epsilon_o \\
    \epsilon_o &= \sum_{\bar F = D - \hat D + 1}^D \bin{D}{f}{\bar F}
\end{split}
\end{equation}
if the underlying abstraction of \pcbin\ satisfies $\epsilon^{\pcbin}_v$-total validity.

\begin{proof}
We compute a bound on $\epsilon_v$ by assuming that, if the total validity of the underlying \pcbin\ instance is compromised, the validity of \prbin\ is compromised as well. Following from \cref{lemma:prbvalidityoverwhelmed}, the validity of \prbin\ can be compromised only if the total validity of \pcbin\ is compromised as well, or if $\sigma$ has more than $D - \hat D$ Byzantine processes in its delivery sample.

Since procedure $sample$ independently picks $D$ processes with replacement, each element of a correct process' echo sample has an independent probability $f$ of being Byzantine, i.e., the number of Byzantine processes in a correct delivery sample is binomially distributed.

Therefore, $\sigma$ has a probability $\epsilon_o$ of having more than $D - \hat D$ Byzantine processes in its delivery sample.
\end{proof}
\end{theorem}

\subsection{Adversarial execution}

In this section, we define the model underlying an adversarial execution of \prbal. Here, a Byzantine adversary is an agent that acts upon a system with the goal to compromise its consistency and / or totality. The main goal of this section is to formalize the information available to the adversary, and the set of actions that it can perform on the system throughout an adversarial execution.

Throughout the rest of this appendix, we bound the probability of compromising the consistency and totality of \prbal\ by assuming that, if the consistency of the \pcbin\ instance used in \prbal\ is compromised, then both the consistency and the totality of \prbal\ are compromised as well. In what follows, therefore, we assume that \pcbal\ satisfies consistency.

\subsubsection{Model}
\label{subsubsection:prbadversarymodel}

Let $\pi$ be any correct process. We make the following assumptions about an adversarial execution of \prbal:
\begin{itemize}
    \item As we established in \cref{section:model}, the adversary does not know which correct processes are in $\pi$'s ready or delivery samples. The adversary knows, however, which Byzantine processes are in $\pi$'s ready sample, and which Byzantine processes are in $\pi$'s delivery sample.
    \item At any time, the adversary knows the set of messages for which $\pi$ sent a {\tt Ready} message.
    \item At any time, the adversary knows if $\pi$ delivered a message. If $\pi$ delivered a message, then the adversary knows which message did $\pi$ deliver.
    \item The adversary can arbitrarily cause $\pi$ to \pcbin.Deliver a given message $m^*$. Since we assume that the underlying \pcbin\ instance satisfies consistency, the adversary cannot cause two correct processes to \pcbin.Deliver two different messages.
\end{itemize}

Throughout an adversarial execution of \prbal, an adversary performs a sequence of minimal operations on the system. Each operation consists of either of the following:
\begin{itemize}
    \item Selecting a correct process that did not \pcbin.Deliver $m^*$ and causing it to \pcbin.Deliver $m^*$.
    \item Selecting a Byzantine process and causing it to issue a {\tt Ready} message to a correct process.
\end{itemize}

As a result of each operation, zero or more processes send a {\tt Ready} message and/or deliver a message. The adversary is successful if, at the end of the adversarial execution, either the consistency or the totality of the system is compromised.

\subsection{Epidemic processes}

In the next sections, we compute bounds for the $\epsilon$-consistency and $\epsilon$-totality of \prbal. In order to do so, in this section we study the \emph{feedback mechanism} produced by {\tt Ready} messages in an execution of \prbal. 

As we discussed in \cref{subsection:prbalgorithm}, a correct process issues a {\tt Ready} message for a message $m$ after either \pcbin.Delivering $m$ (\cref{line:prbsendreadyfrompcb}) or collecting at least $\hat R$ {\tt Ready}($m$) messages from its ready sample (\cref{line:prbsendreadyfromready}). We formalize this observation in the following definition.

\begin{definition}[Ready, E-ready, R-ready]
Let $\pi$ be a correct process, let $m$ be a message. Throughout an execution of \prbal, $\pi$ is \textbf{E-ready} for $m$ if $\pi$ eventually \pcbin.Delivers $m$; $\pi$ is \textbf{R-ready} for $m$ if $\pi$ eventually receives at least $\hat R$ {\tt Ready}($m$) messages from its ready sample; $\pi$ is \textbf{ready} for $m$ if $\pi$ is either E-ready or R-ready for $m$.

We note how a correct process can simultaneously be E-ready and R-ready for the same message.
\end{definition}

It is easy to observe that the R-ready condition creates a feedback process: as a result of a correct process being R-ready for a message $m$, it issues a {\tt Ready}($m$) message that might cause other correct processes to become R-ready for $m$ as well.

Intuitively, this feedback process is designed to have two stable configurations:
\begin{itemize}
    \item \textbf{Few processes are ready}: the fraction of correct processes that are E-ready for a message $m$ is significantly smaller than $\hat R / R$. As a result, the probability of a correct process being R-ready for $m$ becomes very small, and the set of processes that are ready for $m$ is, with high probability, nearly identical to the set of processes that are E-ready for $m$.
    \item \textbf{All processes are ready}: the fraction of correct processes that are E-ready for a message $m$ is not significantly smaller than $\hat R / R$. As a result, a correct process that is not E-ready for $m$ has a significant probability of becoming R-ready for $m$. If this happens, the probability of a correct process becoming R-ready for $m$ further increases, and eventually every correct proces is ready for $m$.
\end{itemize}

In this section, we show that the R-ready feedback mechanism is isomorphic to an epidemic process as we define it in \cref{appendix:contagion}. In summary, an epidemic process depends on one parameter (contagion threshold $\hat R$) to mimic the spread of a disease in a population:
\begin{itemize}
    \item A population is represented on the nodes of a directed multigraph, allowing multi-edges and loops. Intuitively, an $a \rightarrow b$ edge represents the relation \emph{$a$ can infect $b$}.
    \item Each member of the population (or node) can be in either of two states: healthy or infected. An infected node stays infected: there is no cure for the infection.
    \item A set of nodes is initially infected. The epidemic process evolves in steps. At every step, all the nodes that have at least $\hat R$ infected predecessors become infected as well. The process is completed when either all nodes are infected, or no healthy node has at least $\hat R$ infected predecessors.
\end{itemize}

We refer the reader to \cref{appendix:contagion} for a more formal discussion of epidemic processes. In this section, we prove the critical result that the R-ready feedback mechanism in \prbal\ is isomorphic to an epidemic process.

\begin{definition}[Adversarial execution]
A \textbf{adversarial execution} (or just \textbf{execution}) is the sequence of events produced by an execution of \prbal\ on $N$ processes, a fraction $f$ of which are under the control of the adversary described in Section \ref{subsubsection:prbadversarymodel}. For the sake of brevity, we omit a more formal definition.

Let $x, x'$ be executions. We say that $x$ is equivalent to $x'$ ($x = x'$) if:
\begin{itemize}
    \item The sequences of messages exchanged are identical in $x$ and $x'$.
    \item The values produced by each correct, local source of randomness are identical in $x$ and $x'$.
\end{itemize}
\end{definition}

\begin{definition}[Ready sample matrix]
A \textbf{ready sample matrix} is an element of the set
\begin{equation*}
    \mathcal{J} = \rp{\Pi^R}^{\Pi_C}
\end{equation*}
\end{definition}

\begin{definition}[Ready sample matrix of an execution]
Let $x$ be an execution, let $j$ be a ready sample matrix. \textbf{$j$ is $x$'s sample matrix} if, for every correct process $\pi$, $\pi$'s ready sample in $x$ is $j_\pi$.
\end{definition}

\begin{definition}[Random ready sample matrix]
\label{definition:randomsamplematrix}
A \textbf{random ready sample matrix} is a random variable representing the sample matrix of a random execution.
\end{definition}

\begin{lemma}
\label{lemma:randomsamplematrix}
Random sample matrices are uniformly distributed. More formally, if $j$ is a random sample matrix, then
\begin{equation*}
    \prob{\bar j} = \rp{\frac{1}{N}}^{R C}
\end{equation*}
\begin{proof}
As we discussed in \cref{section:model}, the adversary has no control over the local source of randomness of each correct process. Each correct process independently selects with uniform probability $R$ elements for its ready sample.
\end{proof}
\end{lemma}

\begin{lemma}
\label{lemma:prbscheduling}
Let $j$ be a ready sample matrix. Let $x, x'$ be executions of \prbal\ such that: 
\begin{itemize}
    \item No Byzantine process issues any {\tt Ready} message in $x$ or $x'$.
    \item The ready sample matrix of both $x$ and $x'$ is $j$.
\end{itemize}  

Let $\rho_E, \rho_E'$ denote the set of correct processes that are E-ready for $m$ in $x$, $x'$ respectively. Let $\rho$, $\rho'$ denote the set of correct processes that are ready for $m$ in $x$, $x'$ respectively.

We have
\begin{equation*}
    \rp{\rho_E = \rho_E'} \implies \rp{\rho = \rho'}
\end{equation*}

\begin{proof}

Let us assume $\rho_E = \rho_E'$. Let $\pi$ be a correct process. As we established, $\pi$ is ready for $m$ if $\pi$ is either E-ready or R-ready for $m$. Since $\rho_E = \rho_E'$, we immediately have that $\pi$ is E-ready for $m$ in $x$ if and only if $\pi$ is E-ready for $m$ in $x'$.

By definition, $\pi$ is R-ready for $m$ in $x$ ($x'$) if it eventually receives at least $\hat R$ {\tt Ready}($m$) messages from its ready sample in $x$ ($x'$). By hypothesis, no Byzantine process issues any {\tt Ready} message in $x$ ($x'$). Therefore, $\pi$ is eventually R-ready for $m$ in $x$ ($x'$) if $\pi$ receives at least $\hat R$ {\tt Ready}($m$) messages from the correct processes in its ready sample in $x$ ($x'$). 

As we discussed in \cref{section:model}, we assume that every message is eventually delivered in an unbounded but finite amount of time. Therefore, $\pi$ is eventually R-ready for $m$ in $x$ ($x'$) if at least $\hat R$ correct processes in $\pi$'s sample eventually issue a {\tt Ready}($m$) message in $x$ ($x'$), i.e., if at least $\hat R$ correct processes in $\pi$'s sample are eventually ready for $m$ in $x$ ($x'$).

Since the above condition does not depend on the network scheduling, a correct process $\pi$ is eventually ready for $m$ in $x$ if and only if $\pi$ is also eventually ready for $m$ in $x'$. Therefore, $\rho = \rho'$.
\end{proof}
\end{lemma}

\begin{lemma}
\label{lemma:unassistedequalscontagion}

Let $x$ be an execution of \prbal\ where no Byzantine process ever issues any {\tt Ready} message. Let $j$ be $x$'s ready sample matrix. Let $m$ be a message, let $\rho_E$ denote the set of correct processes that are E-ready for $m$ in $x$. Let $\rho$ denote the set of correct processes that are eventually ready for $m$ in $x$.

Let $s_0 = ((v, e), w_0)$ be a contagion state (as defined in \cref{definition:contagionstate}), with
\begin{eqnarray*}
    v &=& \Pi_C \\
    \rp{\pi, \pi'} \in e &\Longleftrightarrow& \pi \in j_{\pi'} \\
    w_0 &=& \rho_E
\end{eqnarray*}

Let $s_\infty = ((v, e), w_\infty)$ be the contagion state resulting from the epidemic process with input $s_0$. We have
\begin{equation*}
    \rho = w_\infty
\end{equation*}

\begin{proof}
Following from \cref{lemma:prbscheduling}, $\rho$ does not depend on $x$'s network scheduling. Without loss of generality, we can therefore make a synchrony assumption for $x$, and assume that every message delay is unitary.

Let $\rho_t$ denote the set of correct processes that are ready for $m$ in $x$ at time $t$. We have
\begin{equation*}
    \rho_0 = w_0 = \rho_E
\end{equation*}

In $x$, a correct process that is not ready for $m$ at time $t$ becomes ready for $m$ at time $t + 1$ if at least $\hat R$ processes in its ready sample are ready for $m$ at time $t$. Therefore
\begin{equation*}
    \pi \in \rho_{t + 1} \Longleftrightarrow \rp{\pi \in \rho_t \; \vee \; \abs{j_\pi \cap R_t} \geq \hat R}
\end{equation*}

As we discuss in \cref{appendix:contagion}, at step $t + 1$, all the healthy nodes in an epidemic process that have at least $\hat R$ predecessors infected at time $t$ become infected. Therefore
\begin{equation*}
    \pi \in w_{t + 1} \Longleftrightarrow \rp{\pi \in w_t \; \vee \; \abs{j_\pi \cap w_t} \geq \hat R}
\end{equation*}

Therefore, if $\rho_t = w_t$, then $\rho_{t + 1} = w_{t + 1}$, and, by induction, for all $t$, $\rho_t = w_t$. In \cref{appendix:contagion}, we prove that an epidemic process identically converges in a finite number of steps. Consequently, $\rho_\infty = w_\infty$, which proves the lemma.
\end{proof}
\end{lemma}

\begin{lemma}
\label{lemma:forcedequalscontagion}
Let $m$ be a message. Let $x$ be an execution of \prbal\ where every Byzantine process sends a {\tt Ready}($m$) message to every correct process from which it received a {\tt ReadySubscribe} message. Let $j$ be $x$'s ready sample matrix. Let $\rho_E$ denote the set of correct processes that are E-ready for $m$ in $x$. Let $\rho$ denote the set of correct processes that are eventually ready for $m$ in $x$. \\

Let $s_0 = ((v, e), w_0)$ be a contagion state (as defined in \cref{definition:contagionstate}), with
\begin{eqnarray*}
    v &=& \Pi \\
    \rp{\pi, \pi'} \in e &\Longleftrightarrow& \rp{\pi' \in \Pi_C} \wedge \rp{\pi \in j_{\pi'}} \\
    w_0 &=& \rho_E \cup \rp{\Pi \setminus \Pi_C}
\end{eqnarray*}

Let $s_\infty = ((v, e), w_\infty)$ be the contagion state resulting from the epidemic process with input $s_0$. We have
\begin{equation*}
    \rho = w_\infty \setminus \rp{\Pi \setminus \Pi_C}
\end{equation*}

\begin{proof}
It follows immediately from \cref{lemma:unassistedequalscontagion} and the observation that, in $x$, a Byzantine process sends the same {\tt Ready} messages as a correct process that is E-ready for $m$.
\end{proof}
\end{lemma}

\subsection{Threshold contagion}
\label{subsection:thresholdcontagionprb}

As we discussed in the previous section, in \cref{appendix:contagion} we introduce epidemic processes, an abstract model of the feedback mechanism produced by {\tt Ready} messages in an execution of \prbal. 
Given the multigraph on which it occurs, an epidemic process is deterministic. In \cref{appendix:contagion}, we also generalize epidemic processes to the probabilistic setting: we introduce and analyze \contagion, a game where a player infects in rounds arbitrary subsets of a population, causing a sequence of epidemic processes on a random, unknown multigraph.

\contagion\ depends on six parameters: node count $N$, sample size $R$, link probability $l$, round count $K$, infection batch $S$, and contagion threshold $\hat R$. \\ 

In summary, a game of \contagion\ is played as follows:
\begin{itemize}
    \item A random multigraph with $N$ nodes is generated. The number of predecessors of each node follows a $\bin{R}{l}{}$ distribution. Each predecessor of a node is independently picked with uniform probability from the set of nodes.
    
    The topology of the network is not disclosed to the adversary.
    \item For $K$ rounds:
    \begin{itemize}
        \item The player infects an arbitrary set of $S$ healthy nodes.
        \item An epidemic process with contagion threshold $\hat R$ is ran on the resulting contagion state.
    \end{itemize}
\end{itemize}

We refer the reader to \cref{appendix:contagion} for a more formal discussion of \contagion. There we introduce the random variable
\begin{equation*}
    \gamma(N, R, l, K, S, \hat R)
\end{equation*}
representing the number of nodes that are infected at the end of a game of \contagion. We then prove that, by arbitrary choosing which nodes to infect, the adversary has no way to bias $\gamma$. Finally, we analitically compute the probability distribution underlying $\gamma$.

In this section, we prove the critical result that a game of \contagion\ can be used to model two classes of adversarial executions of \prbal.

\begin{lemma}
\label{lemma:unassistedequalsthresholdcontagion}
Let $m^*$ be a message. Let $x$ be an adversarial execution of \prbal\ where:
\begin{itemize}
    \item No Byzantine process issues any {\tt Ready} message.
    \item For $K$ rounds:
    \begin{itemize}
        \item The adversary selects, if possible, $S$ correct process that are not ready for $m^*$, and causes them to \pcbin.Deliver $m^*$.
        \item The adversary waits until every resulting {\tt Ready} message is delivered.
    \end{itemize}
\end{itemize}

Let $\rho$ denote the number of correct processes in $\sigma$ that, at the end of the adversarial execution, are ready for $m$. We have
\begin{equation*}
    \prob{\bar \rho} = \prob{\gamma(C, R, 1 - f, K, S, \hat R) = \bar \rho}
\end{equation*}

\begin{proof}

We start by defining a function $\mathfrak{g}: \mathcal{J} \rightarrow \mathcal{G}$ (see \cref{definition:predecessormatrix} for a definition of $\mathcal{G}$) by
\begin{equation*}
    \mathfrak{g}(j)_{i, k} =
    \begin{cases}
        \zeta^{-1}\rp{j_{\zeta(i), k}} &\text{iff}\; j_{\zeta(i), k} \in \Pi_C \\
        \bot &\text{otherwise}
    \end{cases}
\end{equation*}

We start by noting that, for every $\bar g \in \mathcal{G}$,
\begin{equation*}
    \prob{\bar g} = \prob{\mathfrak{g}^{-1}(g)}
\end{equation*}

Indeed, following from \cref{lemma:randompredecessorvector,lemma:randompredecessormatrix}:
\begin{eqnarray*}
    \prob{\bar g} &=& \prod_{i, k} \prob{\bar g_{i, k}} \\
    \prob{g_{i, k} = \bot} &=& (1 - l) = f \\
    \prob{g_{i, k} = \rp{\bar g_{i, k} \in 1..C}} &=& \frac{1}{C}
\end{eqnarray*}
and following from \cref{definition:randomsamplematrix,lemma:randomsamplematrix} we have
\begin{eqnarray*}
    \prob{\bar j} &=& \prod_{\pi, k} \prob{\bar j_{\pi, k}} \\
    \prob{j_{\pi, k} \in \Pi \setminus \Pi_C} &=& f \\
    \prob{j_{\pi, k} = \rp{\bar \pi' \in \Pi_C}} &=& \frac{1}{C}
\end{eqnarray*}

We now build from $x$ a game of \contagion\ $y$, played on $\mathfrak{g}(j)$. At the beginning of each round, if the adversary causes a correct process $\pi$ to \pcbin.Deliver $m^*$, then $\zeta(\pi)$ is infected. 

We can prove that, if $\pi$ is eventually ready for $m^*$ in $x$, then $\zeta(\pi)$ is eventually infected in $y$. Indeed, following from \cref{lemma:unassistedequalscontagion}, if $\pi$ is ready for $m^*$ at the end of a round in $x$, then $\zeta(\pi)$ is infected at the end the same round in $y$. 

Therefore, the following hold true:
\begin{itemize}
    \item The probability of $\bar j$ is identical to the probability of $\mathfrak{g}(\bar j)$.
    \item The number of correct processes that are eventually ready for $m^*$ in $x$ is identical to the number of nodes that are eventually infected in $y$.
\end{itemize}

\end{proof}
\end{lemma}

\begin{lemma}
\label{lemma:forcedequalsthresholdcontagion}
Let $m$ be a message. Let $x$ be an adversarial execution of \prbal\ where:
\begin{itemize}
    \item No correct process \pcbin.Delivers $m$.
    \item Every Byzantine process sends a {\tt Ready}($m$) message to every correct process from which it received a {\tt ReadySubscribe} message.
\end{itemize}

Let $\rho$ denote the number of correct processes in $\sigma$ that, at the end of the adversarial execution, are ready for $m$. We have
\begin{equation*}
    \prob{\bar \rho} = \prob{\gamma(N, R, 0, 1, N - C, \hat R) = \bar \rho + \rp{N - C}}
\end{equation*}

\begin{proof}
The proof is similar to the proof of \cref{lemma:unassistedequalsthresholdcontagion}, using \cref{lemma:forcedequalscontagion} instead of \cref{lemma:unassistedequalscontagion}.
\end{proof}
\end{lemma}

\subsection{Preliminary lemmas}

In order to compute an upper bound for the probability of the consistency of \prbal\ being compromised, we will make use of some preliminary lemmas. The statements of those lemmas are independent from the context of \prbal. For the sake of readability, we therefore gather them in this section, and use them throughout the rest of this appendix.

\begin{lemma}
\label{lemma:tailgrowingprob}
Let $N, K \in \mathbb{N}$ such that $K \in 0..N$. Let $X$ be a random variable defined by
\begin{equation*}
    \prob{\bar X} = \bin{N}{p}{\bar X}
\end{equation*}

We have that
\begin{equation*}
    \prob{X \geq K}
\end{equation*}
is an increasing function of $p$.

\begin{proof}
We expand
\begin{equation*}
    \prob{X \geq K} = \sum_{\bar X = K}^N \bin{N}{p}{\bar X}
\end{equation*}
and take the derivative
\begin{eqnarray*}
    \lhs \frac{\partial}{\partial p} \sum_{\bar X = K}^N \binom{N}{\bar X} p^{\bar X}(1 - p)^{N - \bar X} \\ &=& \sum_{\bar X = K}^N \binom{N}{\bar X} \rp{p^{\bar X} \rp{1 - p}^{N - \bar X}} \rp{\frac{\bar X}{p} - \frac{N - \bar X}{1 - p}} \\
    &=& \underbrace{\frac{1}{p(1 - p)}}_{\geq 1} \sum_{\bar X = K}^N \bin{N}{p}{\bar X}(\bar X - pN) \\
    &\geq&  \sum_{\bar X = K}^N \bin{N}{p}{\bar X}(\bar X - pN)
\end{eqnarray*}

We now prove that, for every $K \in [0, N]$,
\begin{equation}
\label{equation:tailgrowingprob.induction}
\sum_{\bar X = K}^N \bin{N}{p}{\bar X}(\bar X - pN) \geq 0
\end{equation}

We start by noting that \cref{equation:tailgrowingprob.induction} holds true for every $K > pN$. Indeed, if $K > pN$, then every term of the sum in \cref{equation:tailgrowingprob.induction} is positive. 

\medskip

We prove that \cref{equation:tailgrowingprob.induction} holds true for every $K < pN$ by induction. For $K = 0$ we have
\begin{eqnarray*}
    \lhs \sum_{\bar X = 0}^N \bin{N}{p}{\bar X}(\bar X - pN) \\
    &=& \underbrace{\sum_{\bar X = 0}^N \bar X \bin{N}{p}{\bar X}}_{= pN} - pN \underbrace{\sum_{\bar X = 0}^N \bin{N}{p}{\bar X}}_{= 1} \\
    &=& 0
\end{eqnarray*}

Let us assume that \cref{equation:tailgrowingprob.induction} holds true for some $K < pN$. We have
\begin{eqnarray*}
    \lhs \sum_{\bar X = K + 1}^N \bin{N}{p}{\bar X} (\bar X - pN) \\
    &=& \underbrace{\sum_{\bar X = K}^N \bin{N}{p}{\bar X} (\bar X - pN)}_{\geq 0 \text{ by IH}} - \bin{N}{p}{K}(K - pN) \\
    &\geq& 0
\end{eqnarray*}
which proves that \cref{equation:tailgrowingprob.induction} holds true for $K + 1$ as well. By induction, \cref{equation:tailgrowingprob.induction} holds true for every $K < pN$. This proves that the derivative is positive for all $p \in [0, 1]$ which proves the lemma.
\end{proof}
\end{lemma}

\subsection{Consistency}
\label{subsection:prb-consistency}

In this section, we compute a bound on the $\epsilon$-consistency of \prbal. As we discussed in \cref{subsubsection:prbadversarymodel}, here we bound the probability of compromising the consistency of \prbal\ by assuming that, if the consistency of the \pcbin\ instance used in \prbal\ is compromised, the consistency of \prbal\ is compromised as well.

Let $m^*$ denote the only message that any correct process can \pcbin.Deliver. We start by noting that, simply by having every Byzantine process behave like a correct process, an adversary can cause any correct process to deliver $m^*$: indeed, with $f = 0$, \prbal\ satisfies validity deterministically \footnote{Here we are slightly abusing the result of \cref{theorem:prbvalidity}, as it only guarantees that a correct sender will eventually deliver its message. The result, however, independently holds for any other correct process as well.}.

As we discussed in \cref{subsection:prbalgorithm}, a correct process can issue a {\tt Ready} message for an arbitrary number of messages. In other words, causing a correct process to become E-ready for $m^*$ does not affect its behavior with respect to a message $m \neq m^*$.

Therefore, if an adversary can cause at least one correct process $\pi$ to eventually receive at least $\hat R$ {\tt Ready} messages for a message $m \neq m^*$, it can also compromise the consistency of \prbal. 

Indeed, as we discussed in \cref{section:model}, the adversary has arbitrary control over the network scheduling. Even if $\pi$ would eventually receive enough {\tt Ready}($m^*$) messages to deliver $m^*$, the adversary can slow those messages down, and cause $\pi$ to first receive enough {\tt Ready}($m$) messages to deliver $m$. Every other correct process will eventually deliver $m^*$, thus compromising the consistency of the system.

We formalize the above intuition in the following lemma.

\begin{lemma}
\label{lemma:forcedisoptimal}
Let $m^*$ denote the only message that any correct process can \pcbin.Deliver. An optimal adversary causes every Byzantine process to send a {\tt Ready}($m$) message, for some $m \neq m^*$, to every correct process from which it received a {\tt ReadySubscribe} message.

\begin{proof}
Let $B$ denote the number of Byzantine processes that eventually issue a {\tt Ready}($m$) message. Let $\pi$ be a correct process, let $Q$ denote the number of {\tt Ready}($m$) messages that $\pi$ eventually collects. Since $\pi$ picks each element of its delivery sample independently, $Q$ is binomially distributed:
\begin{equation*}
    \prob{\bar Q} = \bin{D}{\frac{B}{N}}{\bar Q}
\end{equation*}

Following from \cref{lemma:tailgrowingprob},
\begin{equation*}
    \prob{Q \geq \hat D} = \sum_{\bar Q = \hat D}^D \prob{\bar Q}
\end{equation*}
is an increasing function of $B$, and maximized by $B = (N - C)$. Therefore, the probability of $\pi$ eventually receiving enough {\tt Ready}($m$) messages to deliver $m$ is maximized if every Byzantine process issues a {\tt Ready}($m$) message.

As we previously established, the adversary can cause every correct process to also receive at least $\hat D$ {\tt Ready}($m^*$) messages. Since the adversary has control over network scheduling, it can cause $\pi$ to deliver $m$, and every other process to deliver $m^*$, thus compromising the consistency of the system.
\end{proof}
\end{lemma}

\begin{lemma}
\label{lemma:worthlessretrying}
Let $m^*$ denote the only message that any correct process can \pcbin.Deliver, let $m \neq m^*$. If, throughout an optimal adversarial execution, no correct process eventually collects enough {\tt Ready}($m$) messages to deliver $m$, then no correct process eventually collects enough {\tt Ready}($m$) messages to deliver any message $m' \neq m$.

\begin{proof}
Following from \cref{lemma:forcedisoptimal}, the optimal adversary causes every Byzantine process to issue a {\tt Ready}($m$) message. In \cref{lemma:forcedequalscontagion}, we use the fact that this strategy makes the Byzantine processes behave identically to correct processes that are E-ready for $m$ to show that the set of correct processes that are eventually ready for $m$ only depends on the ready sample matrix of the execution.

Since a correct process does not change its ready or delivery samples throughout an execution, the set of processes that will eventually be ready for $m'$ is at most the same as the set of processes that will eventually be ready for $m$. In turn, this means that if no correct process eventually delivers $m$, no correct process eventually delivers $m'$ either.
\end{proof}
\end{lemma}

We can now use \cref{lemma:forcedisoptimal} to compute a bound on the $\epsilon$-consistency of \prbal.

We introduce the random variable $\gamma^+$ by
\begin{equation*}
    \prob{\bar \gamma^+} = \prob{\gamma(N, R, 0, 1, N - C, \hat R) = \bar \gamma^+}
\end{equation*}

Following from \cref{lemma:forcedisoptimal,lemma:forcedequalsthresholdcontagion}, $\gamma^+$ represents the number of processes (Byzantine or correct) that eventually issue a {\tt Ready} message for a message $m \neq m^*$, when an optimal adversary is trying to compromise the consistency of the system.

We can finally compute a bound for the $\epsilon$-consistency of \prbal. We define
\begin{eqnarray*}
    \mu &=& \sum_{\bar \gamma^+ = N - C}^N \rp{1 - \rp{1 - \tilde{\mu}(\bar \gamma^+)}^C} \prob{\bar \gamma^+} \\
    \tilde{\mu}(\bar \gamma^+) &=& \sum_{\bar D = \hat D}^D \bin{D}{\frac{\bar \gamma^+}{N}}{\bar D}
\end{eqnarray*}

\begin{theorem}
\label{theorem:prbconsistency}
\prbal\ satisfies $\epsilon_c$-consistency, with
\begin{equation*}
\begin{split}
    \epsilon_c &\leq \epsilon^{\pcbin}_c + \rp{1 - \epsilon^{\pcbin}_c} \mu
\end{split}
\end{equation*}
if the underlying abstraction of \pcbin\ satisfies $\epsilon^{\pcbin}_c$-consistency.

\begin{proof}
We start by noting that $\tilde{\mu}(\bar \gamma^+)$ represents the probability that a specific correct process will eventually collect enough {\tt Ready}($m$) messages to deliver $m$, given the number $\bar \gamma^+$ of processes that eventually issue a {\tt Ready}($m$) message. 

Indeed, since every correct process picks its delivery sample independently, each of the $D$ elements of a correct process' delivery sample has a probability $\bar \gamma^+ / N$ of issuing a {\tt Ready}($m$) message.

We then note that $\mu$ represents the probability of any correct process eventually collecting enough {\tt Ready}($m$) messages to deliver $m$. $\mu$ is obtained by applying the law of total probability to $\mu(\bar \gamma^+)$.

Finally, $\epsilon_c$ is obtained by the assumption that, if the consistency of the underlying \pcbin\ instance is compromised, the totality of \prbal\ is compromised as well.
\end{proof}
\end{theorem}

\subsection{Totality}
\label{subsection:prbtotality}

In this section, we compute a bound on the $\epsilon$-totality of \prbal. As we discussed in \cref{subsubsection:prbadversarymodel}, here we bound the probability of compromising the totality of \prbal\ by assuming that, if the consistency of the \pcbin\ instance used in \prbal\ is compromised, the consistency of \prbal\ is compromised as well.

\subsubsection{Minimal operations}

Let $m^*$ be the only message that any correct process can \pcbin.Deliver. As we discussed in \cref{subsubsection:prbadversarymodel}, throughout an execution of \prbal, an adversary performs a sequence of minimal operations on the system, i.e., it either causes a correct process to \pcbin.Deliver $m^*$, or it causes a Byzantine process to send an arbitrary {\tt Ready}($m$) message to a correct process.

We further relax the bound by assuming that, if the adversary can cause any message $m \neq m^*$ to be delivered by at least one correct process, the totality of \prbal\ is compromised as well.

Under the assumption that no correct process can eventually collect enough {\tt Ready}($m$) messages to deliver any message $m$ different from $m^*$, causing a Byzantine process to send a {\tt Ready}($m$) message has no effect on the totality of the system. 

This reduces the set of adversarial operations that have a non-null effect on the totality of the system to:
\begin{itemize}
    \item Causing an arbitrary correct process to \pcbin.Deliver $m^*$.
    \item Causing a Byzantine adversary to send a {\tt Ready}($m^*$) message to a correct process.
\end{itemize}

We now prove a lemma to further reduce the set of minimal operations of an optimal adversary.

\begin{lemma}
\label{lemma:ratherpcbthanready}
Let $m^*$ be the only message that any correct process can potentially \pcbin.Deliver. Let $\pi$ be a correct process, let $\xi$ be a Byzantine process in $\pi$'s ready sample. An optimal adversary never causes $\xi$ to send a {\tt Ready}($m^*$) message to $\pi$.

\begin{proof}
As a result of receiving a {\tt Ready}($m^*$) message from $\xi$, $\pi$ can either:
\begin{itemize}
    \item Have collected less than $\hat R$ {\tt Ready}($m^*$) messages from its ready sample. The operation has no effect.
    \item Have collected exactly $\hat R$ {\tt Ready}($m^*$) messages from is ready sample. Then $\pi$ becomes ready for $m^*$. However, the same outcome could have been achieved deterministically by causing $\pi$ to \pcbin.Deliver $m^*$.
\end{itemize}

Since every outcome of $\xi$'s {\tt Ready}($m^*$) message to $\pi$ can be deterministically emulated by causing $\pi$ to \pcbin.Deliver (or not \pcbin.Deliver) $m^*$, the operation is useless to an optimal adversary.
\end{proof}
\end{lemma}

\subsubsection{Delivery probability}

Let $\gamma^-$ denote the random variable counting the number of correct processes that are eventually ready for $m^*$. In this section, we study the probability of totality being compromised, given the value of $\gamma^-$.

By definition, totality is compromised if at least one correct process delivers $m^*$ and one correct process does not deliver $m^*$.

Let $\pi$ be a correct process. We introduce the following events:
\begin{itemize}
    \item $A_\pi$: process $\pi$ delivers $m^*$.
    \item $A$: all correct processes deliver $m^*$.
    \item $\tilde A$: no correct process delivers $m^*$.
    \item $T$: the totality of the system is compromised.
\end{itemize}

Given $\bar \gamma^-$, the probability of $A_\pi$ is bound by
\begin{equation*}
    \alpha^-_\pi(\bar \gamma^-) \leq \prob{A_\pi \mid \bar \gamma^-} \leq \alpha^+_\pi(\bar \gamma^-)
\end{equation*}
with
\begin{eqnarray*}
    \alpha^-_\pi(\bar \gamma^-) &=& \sum_{\bar D = \hat D}^D \bin{D}{\frac{\bar \gamma^-}{N}}{\bar D} \\
    \alpha^+_\pi(\bar \gamma^-) &=& \sum_{\bar D = \hat D}^D \bin{D}{\frac{\bar \gamma^- + \rp{N - C}}{N}}{\bar D}
\end{eqnarray*}

The lower bound is attained when none of the Byzantine processes issue a {\tt Ready}($m^*$) message, and the upper bound is attained when all Byzantine processes issue a {\tt Ready}($m^*$) message.

Noting that each correct process independently picks its delivery sample, we can compute, given $\bar \gamma^-$, a lower bound for the probability of $A$:
\begin{equation*}
    \prob{A \mid \bar \gamma^-} \geq \rp{\alpha^-(\bar \gamma^-)}^C
\end{equation*}
and a lower bound for the probability of $\tilde A$:
\begin{equation*}
    \prob{\tilde A \mid \bar \gamma^-} \geq \rp{1 - \alpha^+(\bar \gamma^-)}^C
\end{equation*}

The above allow us to compute, given $\bar \gamma^-$, an upper bound for the probability of $T$:
\begin{eqnarray*}
    \prob{T \mid \bar \gamma^-} &=& \prob{\cancel{A}, \cancel{\tilde A} \mid \bar \gamma^-} \\
    &\leq& 1 - \prob{A \mid \bar \gamma^-} - \prob{\tilde A \mid \bar \gamma^-} \\
    &\leq& \alpha(\bar \gamma^-)
\end{eqnarray*}
with
\begin{equation*}
    \alpha(\bar \gamma^-) = 1 - \rp{\alpha^-(\bar \gamma^-)}^C - \rp{1 - \alpha^+(\bar \gamma^-)}^C
\end{equation*}

\subsubsection{C-step \contagion}
\label{subsubsection:cstepthresholdcontagion}

Due to \cref{lemma:ratherpcbthanready}, the minimal set of operations for an optimal adversary reduces to
\begin{itemize}
    \item Causing an arbitrary correct process to \pcbin.Deliver $m^*$.
    \item Causing a Byzantine process $\xi$ in the delivery sample of a correct process $\pi$ to send a {\tt Ready}($m^*$) message to $\pi$.
\end{itemize}

It is immediate to see that the latter operation has no effect over which correct processes eventually become {\tt Ready} for $m^*$. In the previous section, we computed an upper bound on the probability of compromising the totality of the system, given the number of correct processes that are eventually ready for $m^*$.

In this section, we prove a final constraint on the optimal adversarial strategy, and finally compute a bound on the $\epsilon$-totality of \prbal.

\begin{lemma}
\label{lemma:blackjack}
Let $m^*$ denote the only message that any correct process can \pcbin.Deliver. An optimal adversary executes in $C$ rounds. At every round, the adversary causes one correct process to \pcbin.Deliver $m^*$, then waits until all the resulting {\tt Ready} messages are delivered.

\begin{proof}
Due to \cref{lemma:unassistedequalscontagion}, the outcome of the execution is not affected by network scheduling: causing one correct process at a time to \pcbin.Deliver $m^*$ has the same effect, e.g., as causing any set of correct processes to simultaneously \pcbin.Deliver $m^*$.
\end{proof}
\end{lemma}

Following from \cref{lemma:blackjack}, we can intuitively see an adversarial execution whose goal is to compromise the totality of \prbal\ as a game similar to \emph{blackjack}. The game unfolds in $C$ rounds. At every round, the adversary causes one more correct process to \pcbin.Deliver $m^*$. With high probability, this will have two possible negative outcomes for the player:
\begin{itemize}
    \item Nothing happens: no correct process is able to deliver $m^*$, even if the Byzantine processes in its delivery sample issue a {\tt Ready}($m^*$) message. The only possible move is to play again.
    \item The execution is \emph{busted}: a feedback loop is generated that eventually causes, with high probability, every correct process to deliver $m^*$, even if no Byzantine process issues any {\tt Ready}($m^*$) message. The adversary fails in compromising the totality of the system.
\end{itemize}

If the adversary is lucky enough, however, one of the rounds will result in a configuration where no feedback loop occurred, but at least one correct process can deliver $m^*$. In that case, the adversary causes that process to deliver $m^*$, and stops: totality is compromised.

Following from \cref{lemma:unassistedequalsthresholdcontagion}, the probability distribution underlying the number of correct processes that are ready for $m^*$ at the end of the $n$-th step is
\begin{equation*}
    \prob{\bar \gamma^-_n} = \prob{\gamma(C, R, 1 - f, C, 1, \hat R)}
\end{equation*}

We can finally compute a bound on the $\epsilon$-totality of \prbal.

\begin{theorem}
\label{theorem:prbtotality}
\prbal\ satisfies $\epsilon_t$-totality, with
\begin{equation*}
\begin{split}
    \epsilon_t &\leq \epsilon_c^{\pcbin} + \mu + \epsilon_b \\ 
    \epsilon_b &= \sum_{n = 0}^C \sum_{\bar \gamma^-_n = 0}^C \prob{\gamma^-_n} \alpha(\bar \gamma^-_n)
\end{split}
\end{equation*}
if the underlying abstraction of \pcbin\ satisfies $\epsilon_c^{\pcbin}$-consistency.
\begin{proof}
Let $T_n$ denote the event of totality being compromised at the end of round $T_n$.

Under the assumption that the consistency of \pcbin\ is satisfied, and no message other than $m^*$ can be delivered by any correct process, the probability of $T_n$ with $n > 1$ is
\begin{equation*}
    \prob{T_n} = \sum_{\bar \gamma^-_n = 0}^C \prob{T_n \mid \bar \gamma^-_n} \prob{\bar \gamma^-_n \mid \cancel{T_{n - 1}}}
\end{equation*}

Indeed, the adversary will proceed to round $n$ only if round $n - 1$ was unsuccessful in compromising the totality of the system. We can use the law of total probability to get
\begin{eqnarray*}
    \prob{T_n} &\leq& \sum_{\bar \gamma^-_n = 0}^C \prob{T_n \mid \bar \gamma^-_n} \rp{\prob{\bar \gamma^-_n \mid \cancel{T_{n - 1}}} + \prob{\bar \gamma^-_n \mid T_{n - 1}}} \\
    &=& \sum_{\bar \gamma^-_n = 0}^C \prob{T_n \mid \bar \gamma^-_n} \prob{\bar \gamma^-_n}
\end{eqnarray*}

We can use Boole's inequality to get
\begin{equation*}
    \prob{T} \leq \sum_{n = 0}^C \prob{T_n}
\end{equation*}
and since
\begin{equation*}
    \prob{T_n \mid \bar \gamma^-_n} \leq \alpha(\gamma^-_n)
\end{equation*}
we have that $\epsilon_b$ bounds the probability of compromising totality, if the consistency of \pcbin\ is satisfied, and no message other than $m^*$ can be delivered by any correct process.

The value provided for $\epsilon_t$ follows from applying again Boole's inequality to include $\epsilon^{\pcbin}_c$ and $\mu$ (which, in \cref{subsection:prb-consistency}, we proved to bound the probability of any correct process delivering a message other than $m^*$).
\end{proof}
\end{theorem}

\clearpage
\section{Decorators}
\label{appendix:decorators}

In this appendix, we provide the proof that each of the sets of \cobin\ adversaries presented in \cref{subsection:twophaseadversaries} is optimal.

\subsection{Auto-echo adversary}
\label{subsection:autoechoadversaryproof}

\begin{algorithm}
\begin{algorithmic}[1]
\Implements
    \Instance{AutoEchoAdversary + \Cobin System}{aeadv}
\EndImplements

\Uses
    \InstanceSystem{\Cobin Adversary}{adv}{aeadv}
    \Instance{\Cobin System}{sys}
\EndUses

\Procedure{aeadv.Init}{{}}
    \State $queue = \emptyset$;
    \State
    \ForAll{\pi}{\Pi_C}
        \ForAll{m}{\mathcal{M}}
            \ForAll{\xi}{\Pi \setminus \Pi_C}
                \State $queue \leftarrow queue \cup \cp{(\pi, m, \xi)}$;
            \EndForAll
        \EndForAll
    \EndForAll
    \State
    \State $echoes = \cp{\bot}^{C \times C \times N}$; \Comment{$C \times C \times N$ table filled with $\bot$.}
    \State $executed = \false$;
    \State $adv.Init()$;
\EndProcedure

\algstore{autoechodecorator}
\end{algorithmic}
\caption{Auto-echo decorator}
\label{algorithm:autoechodecorator}
\end{algorithm}

\begin{algorithm}
\begin{algorithmic}[1]
\algrestore{autoechodecorator}

\Procedure{aeadv.Step}{{}}
    \If{$queue \neq \emptyset$}
        \State $(\pi, m, \xi) = queue[1]$;
        \State $sys.Echo(\pi, m, \xi, m)$;
        \State $queue \leftarrow queue \setminus \cp{(\xi, \pi, m)}$;
    \Else
        \State $executed \leftarrow \false$;
        \While{executed = \false}
            \State $adv.Step()$;
        \EndWhile
    \EndIf
\EndProcedure

\Procedure{aeadv.Byzantine}{process}
    \State \Return $sys.Byzantine(process)$;
\EndProcedure

\Procedure{aeadv.State}{{}}
    \State $state = \emptyset$;
    \State
    \ForAll{(\pi, m)}{sys.State()}
        \State $n = 0$;
        \State
        \ForAll{\rho}{sys.Sample(\pi, m)}
            \If{$echoes[\pi][m][\rho] = m$}
                \State $n \leftarrow n + 1$;
            \EndIf
        \EndForAll
        \State
        \If{$n \geq \hat E$}
            \State $state \leftarrow state \cup \cp{(\pi, m)}$;
        \EndIf
    \EndForAll
    \State
    \State \Return $state$;
\EndProcedure

\algstore{autoechodecorator}
\end{algorithmic}
\end{algorithm}

\begin{algorithm}
\begin{algorithmic}[1]
\algrestore{autoechodecorator}

\Procedure{aeadv.Sample}{process, message}
    \State $sample = \emptyset$;
    \State
    \ForAll{\rho}{sys.Sample(process, message)}
        \If{$echoes[process][message][\rho] \neq \bot$}
            \State $sample \leftarrow sample \cup \cp{\rho}$;
        \EndIf
    \EndForAll
    \State
    \State \Return $sample$;
\EndProcedure

\Procedure{aeadv.Deliver}{process, message}
    \State $executed = \true$;
    \State
    \ForAll{\pi}{\Pi_C}
        \ForAll{m}{\mathcal{M}}
            \State $echoes[\pi][m][process] = message$;
        \EndForAll
    \EndForAll
    \State
    \State $sys.Deliver(process, message, flag)$;
\EndProcedure

\Procedure{aeadv.Echo}{process, sample, source, message}
    \State $echoes[process][sample][source] = message$;
\EndProcedure

\Procedure{aeadv.End}{{}}
    \State $executed = \true$;
    \State $sys.End()$;
\EndProcedure
\end{algorithmic}
\end{algorithm}

\newpage
\begin{lemma}
The set of auto-echo adversaries $\mathcal{A}_{ae}$ is optimal.
\begin{proof}
We prove the result using a \textbf{decorator}, i.e., an algorithm that acts as an interface between an adversary and a system. An adversary coupled with a decorator effectively implements an adversary. Here we show that a decorator $\Delta_{ae}$ exists such that, for every $\alpha \in \mathcal{A}$, the adversary $\alpha' = \Delta_{ae}(\alpha)$ is an auto-echo adversary, and more powerful than $\alpha$. If this is true, then the lemma is proved: let $\alpha^*$ be an optimal adversary, then the auto-echo adversary $\alpha^+ = \Delta_{ae}(\alpha^*)$ is optimal as well.

\paragraph{Decorator}

\Cref{algorithm:autoechodecorator} implements \textbf{Auto-echo decorator}, a decorator that transforms an adversary into an auto-echo adversary. Provided with an adversary $adv$, Auto-echo decorator acts an interface between $adv$ and a system $sys$, effectively implementing an auto-echo adversary $aeadv$. Auto-echo decorator exposes both the adversary and the system interfaces: the underlying adversary $adv$ uses $aeadv$ as its system.

Auto-echo decorator works as follows:
\begin{itemize}
    \item Procedure $aeadv.Init()$ initializes the following variables:
    \begin{itemize}
        \item A $queue$ list that contains every combination of $(\pi, m, \xi)$, $\pi$ being a correct process, $m$ being a message and $\xi$ being a Byzantine process: $queue$ is used to initially cause every Byzantine process $\xi$ to send an {\tt Echo}($m$, $m$) message to every correct process $\pi$, for every message $m$.
        \item An $echoes$ table, initialized with $\bot$ values: $echoes$ is used to keep track of all the {\tt Echo} messages that would have been sent to each correct process in $sys$, if $adv$ was playing instead of $aeadv$.
    \end{itemize}
    \item Procedure $aeadv.Step()$ checks if $queue$ is not empty. If it is not empty, it pops (i.e., picks and removes) its first element $(\pi, m, \xi)$, with $\xi \in \Pi \setminus \Pi_C$, $\pi \in \Pi_C$ and $m \in \mathcal{M}$. It then causes $\xi$ to send $\pi$ an {\tt Echo}($m$, $m$) message.

    If $queue$ is empty instead, the procedure calls $adv.Step()$ until either $sys.Deliver(\ldots)$ or $sys.End()$ are called: this is achieved using the $executed$ flag.
    \item Procedure $aeadv.Byzantine(process)$ simply forwards the call to \\ $sys.Byzantine(process)$.
    \item Procedure $aeadv.State()$ returns a list of pairs $(\pi \in \Pi_C, m \in \mathcal{M})$ such that $\pi$ delivered $m$ in $sys$, and $\pi$ would have delivered $m$ in $sys$, if $adv$ was playing instead of $aeadv$. 
    
    This is achieved by querying $sys.State()$, then looping over each element $(\pi, m)$ of the response. For each $(\pi, m)$, the procedure loops over every element $\rho$ of $sys.Sample(\pi, m)$, and computes the number $n$ of {\tt Echo}($m$, $m$) messages that $\pi$ would have received from its echo sample for $m$ in $sys$, if $adv$ was playing instead of $aeadv$. This is achieved using the $echoes$ table. If $n$ is greater or equal to $\hat E$, $(\pi, m)$ is included in the list returned by the procedure.
    \item Procedure $aeadv.Sample(process, message)$ returns every process in \\ $sys.Sample(process, message)$ that would have sent an {\tt Echo}($message$, $message'$) message for some message $message'$ to $process$ in $sys$, if $adv$ was playing instead of $aeadv$. This is achieved using the $echoes$ table.
    \item Procedure $aeadv.Deliver(process, message)$ updates the $echo$ table to reflect all the {\tt Echo} messages that $process$ will send, as a result of having \pbin.Delivered $message$. It then forwards the call to \\ $sys.Deliver(process, message)$, causing $process$ to \pbin.Deliver $message$.
    \item Procedure $aeadv.Echo(process, sample, source, message)$ updates the $echo$ table to include the {\tt Echo}($sample$, $message$) message that $process$ would receive from $source$, if $adv$ was playing instead of $aeadv$.
    \item Procedure $aeadv.End()$ simply forwards the call to $sys.End()$.
\end{itemize}

\paragraph{Correctness}

We start by proving that no adversary, coupled with Auto-echo decorator, causes the execution to fail.

We start by establishing a preliminary result. Let $\pi \in \Pi_C$, let $m \in \mathcal{M}$. If $(\pi, m)$ is returned from $aeadv.State()$, then $\pi$ delivered $m$ in $sys$. Indeed, $(\pi, m)$ is returned from $aeadv.State()$ only if $(\pi, m)$ is returned from $sys.State()$.

Let $\pi \in \Pi_C$, let $m \in \mathcal{M}$. The following hold true:
\begin{itemize}
    \item An invocation to $aeadv.Step()$ results in one and only one call to \\ $sys.Deliver(\ldots)$, $sys.Echo(\ldots)$ or $sys.End()$. Indeed, if $queue$ is not empty, exactly one call to $sys.Echo(\ldots)$ is issued. Otherwise, $adv.Step()$ is called until $executed = \true$, and $executed$ is set to $\true$ only after an invocation to $sys.Deliver(\ldots)$ or $sys.End()$.
    \item Procedure $aeadv.State()$ never causes the execution to fail. Indeed, \\ $sys.Sample(\pi, m)$ is called only if $(\pi, m)$ is returned from $sys.State()$. This means that $sys.Sample(\pi, m)$ is called only if $\pi$ delivered $m$ in $sys$. Therefore, $sys.Sample(\pi, m)$ is never invoked from $aeadv.State()$ unless at least one correct process delivered $m$ in $sys$.
    \item No invocation of $aeadv.Sample(\ldots)$ causes the execution to fail.  Noting that $adv$ is correct, it will never invoke $aeadv.Sample(\pi, m)$ unless $(\pi', m)$ was returned from a previous invocation of $aeadv.State()$, for some $\pi' \in \Pi_C$. As we previously established, $(\pi', m)$ is returned from $aeadv.State()$ only if $\pi'$ delivered $m$ in $sys$. Therefore, $sys.Sample(\pi, \allowbreak m)$ is never invoked from $aeadv.Sample(\ldots)$ unless at least one correct process delivered $m$ in $sys$.
\end{itemize}

\paragraph{Auto-echo}

It is easy to prove that Auto-echo decorator always implements an auto-echo adversary. Indeed, every call to $aeadv.Step()$ results in a call to $sys.Echo(\pi, m, \xi, m)$, causing the Byzantine process $\xi$ to send an {\tt Echo}($m$, $m$) message to the correct process $\pi$, until $queue$ is exhausted. 

Therefore, only $sys.Echo(\ldots)$ is invoked until $\xi$ sent an {\tt Echo}($m$, $m$) message to $\pi$, for every $\pi \in \Pi_C$, every $m \in \mathcal{M}$, and every $\xi \in \Pi \setminus \Pi_C$.

\paragraph{Roadmap}

Let $\alpha \in \mathcal{A}$, let $\alpha' = \Delta_{ae}(\alpha)$. Let $\sigma$ be a system such that $\alpha$ compromises the consistency of $\sigma$. Let $\sigma'$ be an identical copy of $\sigma$. In order to prove that $\alpha'$ is more powerful than $\alpha$, we prove that $\alpha'$ compromises the consistency of $\sigma'$.

\paragraph{Trace}

We start by noting that, if we couple Auto-echo decorator with $\sigma'$, we effectively obtain a system instance $\delta$ with which $\alpha$ directly exchanges invocations and responses. Here we show that the trace $\tau(\alpha, \sigma)$ is identical to the trace $\tau(\alpha, \delta)$. Intuitively, this means that $\alpha$ has no way of \emph{distinguishing} whether it has been coupled directly with $\sigma$, or it has been coupled with $\sigma'$, with Auto-echo decorator acting as an interface. We prove this by induction.

Let us assume
\begin{eqnarray*}
    \tau(\alpha, \sigma) &=& ((i_1, r_1), \ldots) \\
    \tau(\alpha, \delta) &=& ((i'_1, r'_1), \ldots) \\
    i_j = i'_j, r_j = r'_j && \forall j \leq n
\end{eqnarray*}

We start by noting that, since $\alpha$ is a deterministic algorithm, we immediately have
\begin{equation*}
    i_{n + 1} = i'_{n + 1}
\end{equation*}
and we need to prove that $r_{n + 1} = r'_{n + 1}$. \\

Let us assume that $i_{n + 1} = ({\tt Byzantine}, \pi)$. Since $aeadv.Byzantine(\pi)$ simply forwards the call to $sys.Byzantine(\pi)$, and $\sigma'$ is an identical copy of $\sigma$, we immediately have $r_{n + 1} = r'_{n + 1}$. \\

Before considering the remaining possible values of $i_{n + 1}$, we prove some auxiliary results. Let $\pi$ be a correct process, let $\xi$ be a Byzantine process, let $\rho$ be a process, let $s, m$ be messages. For every $j \leq n + 1$, as we established, we have $i_j = i'_j$. Therefore, after the $(n + 1)$-th invocation, the following hold true:
\begin{itemize}
    \item $\rho$ sent an {\tt Echo}($s$, $m$) message to $\pi$ in $\sigma$ if and only if $echoes[\pi][s][\rho] = m$. Indeed, if $\rho$ is correct, $\rho$ sent an {\tt Echo}($s$, $m$) message to $\pi$ in $\sigma$ if and only if $aeadv.Deliver(\rho, m)$ was invoked. In turn, $echo[\pi'][s'][\rho]$ was set to $m$ for every $\pi' \in \Pi_C$, $s' \in \mathcal{M}$ if and only if $aeadv.Deliver(\rho, m)$ was invoked. If $\rho$ is Byzantine, $\rho$ sent an {\tt Echo}($s$, $m$) message to $\pi$ in $\sigma$ if and only if $aeadv.Echo(\pi, s, \rho, m)$ was invoked. In turn, $echo[\pi][s][\rho]$ was set to $m$ if and only if $aeadv.Echo(\pi, s, \rho, m)$ was invoked.
    \item If $\rho$ sent an {\tt Echo}($m$, $m'$) message to $\pi$ in $\sigma$ for some $m' \in \mathcal{M}$, then $\rho$ sent an {\tt Echo}($m$, $m''$) message to $\pi$ in $\sigma'$ as well, for some $m'' \in \mathcal{M}$. Indeed, if $\rho$ is correct, then $aeadv.Deliver(\rho, m')$ was invoked. As a result, $sys.Deliver(\rho, m')$ was called, and $\rho$ sent an {\tt Echo}($s'$, $m'$) message to $\pi'$ for every $\pi' \in \Pi_C$, $s' \in \mathcal{M}$. If $\rho$ is Byzantine, then it sent an {\tt Echo}($m'''$, $m'''$) message to $\pi'$, for every $\pi' \in \Pi_C$, $m''' \in \mathcal{M}$.
    \item If $\rho$ sent an {\tt Echo}($m$, $m$) message to $\pi$ in $\sigma$, then $\rho$ sent an {\tt Echo}($m$, $m$) message to $\pi$ in $\sigma'$ as well. Indeed, if $\rho$ is correct, then $aeadv.Deliver(\rho, \allowbreak m)$ was invoked. As a result, $sys.Deliver(\rho, m)$ was called, and $\rho$ sent an {\tt Echo}($s'$, $m$) message to $\pi'$ for every $\pi' \in \Pi_C$, $s' \in \mathcal{M}$. If $\rho$ is Byzantine, then it sent an {\tt Echo}($m'$, $m'$) message to $\pi'$, for every $\pi' \in \Pi_C$, $m' \in \mathcal{M}$.
    \item If $\pi$ delivered $m$ in $\sigma$, then $\pi$ delivered $m$ in $\sigma'$ as well. This follows from the above and the fact that $\sigma'$ is an identical copy of $\sigma$ (i.e., $\pi$'s echo sample for $m$ in $\sigma$ is identical to $\pi$'s echo sample in $\sigma'$.
\end{itemize}

Let us assume that $i_{n + 1} = ({\tt State})$. Let $\pi$ be a correct process, let $m$ be message. We start by noting that $aeadv.State()$ returns $(\pi, m)$ if and only if $\pi$ delivered $m$ in $\sigma'$, and $\pi$ delivered $m$ in $\sigma$. Indeed, $(\pi, m)$ is added to the return list of $aeadv.State()$ if and only if $(\pi, m)$ is returned from $sys.State()$, and at least $\hat E$ processes sent an {\tt Echo}($m$, $m$) message to $\pi$ in $\sigma$. If $(\pi, m) \in r_{n + 1}$, then $\pi$ delivered $m$ in $\sigma$, and $\pi$ delivered $m$ in $\sigma'$ as well. Therefore $(\pi, m) \in r'_{n + 1}$. If $(\pi, m) \in r'_{n + 1}$, then we immediately have that $\pi$ delivered $m$ in $\sigma$, and $(\pi, m) \in r_{n + 1}$.

Let us assume that $i_{n + 1} = ({\tt Sample}, \pi, m)$. Let $\rho$ be a process. We start by noting that $aeadv.Sample(\pi, m)$ returns $\rho$ if and only $\rho$ sent an {\tt Echo}($m$, $m''$) message to $\pi$ in $\sigma'$ for some $m'' \in \mathcal{M}$, and $echoes[\pi][m][\rho] \neq \bot$.  If $\rho \in r_{n + 1}$, then $\rho$ sent an {\tt Echo}($m$, $m'$) message to $\pi$ in $\sigma$, for some $m' \in \mathcal{M}$. Therefore, $\rho$ sent an {\tt Echo}($m$, $m''$) message to $\pi$ in $\sigma'$, for some $m'' \in \mathcal{M}$, and $echoes[\pi][m][\rho] = m' \neq \bot$. Consequently, $\rho \in r'_{n + 1}$. If $\rho \in r'_{n + 1}$, then $echoes[\pi][m][\rho] = m' \neq \bot$ for some $m' \in \mathcal{M}$. Therefore, $\rho$ sent an {\tt Echo}($m$, $m'$) message to $\pi$ in $\sigma$, and $\rho \in r_{n + 1}$. 

Noting that procedures $Deliver(\ldots)$ and $Echo(\ldots)$ never return a value, we trivially have that if $i_{n + 1} = ({\tt Deliver}, \pi, m)$ or $i_{n + 1} = ({\tt Echo}, \pi, s, \xi, m)$ then $r_{n + 1} = \bot = r'_{n + 1}$. By induction, we have $\tau(\alpha, \sigma) = \tau(\alpha, \delta)$.

\paragraph{Consistency of $\sigma'$}

We proved that $\tau(\alpha, \sigma) = \tau(\alpha, \delta)$. Moreover, we proved that if a correct process $\pi$ eventually delivers a message $m$ in $\sigma$, then $\pi$ also delivers $m$ in $\sigma'$.

Since $\alpha$ compromises the consistency of $\sigma$, two correct processes $\pi$, $\pi'$ and two distinct messages $m$, $m' \neq m$ exist such that, in $\sigma$, $\pi$ delivered $m$ and $\pi'$ delivered $m'$. Therefore, in $\sigma'$, $\pi$ delivered $m$ and $\pi'$ delivered $m'$. Therefore $\alpha'$ compromises the consistency of $\sigma'$.

Consequently, the adversarial power of $\alpha$ is smaller or equal to the adversarial power of $\alpha' = \Delta_{ae}(a)$, and the lemma is proved.
\end{proof}
\end{lemma}

\subsection{Process-sequential adversary}
\label{subsection:processsequentialadversaryproof}

\begin{algorithm}
\begin{algorithmic}[1]
\Implements
    \Instance{ProcessSequentialAdversary + \Cobin System}{psadv}
\EndImplements

\Uses
    \InstanceSystem{AutoEchoAdversary}{aeadv}{psadv}
    \Instance{\Cobin System}{sys}
\EndUses

\Procedure{psadv.Init}{{}}
    \State $perm = \cp{\bot}^C$; \tabto*{3.5cm} $cursor = 1$;
    \State $aeadv.Init()$;
\EndProcedure

\Procedure{psadv.Step}{{}}
    \State $aeadv.Step()$;
\EndProcedure

\Procedure{psadv.Byzantine}{process}
    \State \Return $sys.Byzantine(process)$;
\EndProcedure

\Procedure{psadv.State}{{}}
    \State \Return $sys.State()$;
\EndProcedure

\Procedure{psadv.Sample}{process, message}
    \State $sample = \emptyset$;
    \State
    \ForAll{\rho}{sys.Sample(process, message)}
        \If{$\rho \in \Pi_C$}
            \State $sample \leftarrow sample \cup \cp{\zeta(perm[\zeta^{-1}(\rho)])}$
        \Else
            \State $sample \leftarrow sample \cup \cp{\rho}$;
        \EndIf
    \EndForAll
    \State
    \State \Return $sample$;
\EndProcedure

\Procedure{psadv.Deliver}{process, message}
    \State $perm[cursor] = \zeta^{-1}(process)$;
    \State $sys.Deliver(\zeta(cursor), message)$;
    \State $cursor \leftarrow cursor + 1$;
\EndProcedure

\algstore{processsequentialodecorator}
\end{algorithmic}
\caption{Process-sequential decorator}
\label{algorithm:processsequentialdecorator}
\end{algorithm}

\begin{algorithm}
\begin{algorithmic}[1]
\algrestore{processsequentialodecorator}

\Procedure{psadv.Echo}{process, sample, source, message}
    \State $sys.Echo(process, sample, source, message)$;
\EndProcedure

\Procedure{psadv.End}{{}}
    \State $sys.End()$;
\EndProcedure

\end{algorithmic}
\end{algorithm}

\begin{lemma}
The set of process-sequential adversaries $\mathcal{A}_{ps}$ is optimal.
\begin{proof}
We again prove the result using a decorator, i.e., an algorithm that acts as an interface between an adversary and a system. An adversary coupled with a decorator effectively implements an adversary. Here we show that a decorator $\Delta_{ps}$ exists such that, for every $\alpha \in \mathcal{A}_{ae}$, the adversary $\alpha' = \Delta_{ps}(\alpha)$ is a process-sequential adversary, and as powerful as $\alpha$. If this is true, then the lemma is proved: let $\alpha^*$ be an optimal adversary, then the process-sequential $\alpha^+ = \Delta_{ps}(\alpha^*)$ is optimal as well.

\paragraph{Decorator}

\cref{algorithm:processsequentialdecorator} implements \textbf{Process-sequential decorator}, a decorator that transforms an auto-echo adversary into a process-sequential adversary. Provided with an auto-echo adversary $aeadv$, Process-sequential decorator acts as an interface between $aeadv$ and a system $sys$, effectively implementing a process-sequential adversary $psadv$. Process-sequential decorator exposes both the adversary and the system interfaces: the underlying adversary $aeadv$ uses $psadv$ as its system.

Process-sequential decorator works as follows:
\begin{itemize}
    \item Procedure $psadv.Init()$ initializes the following variables:
    \begin{itemize}
        \item A $perm$ array of $C$ elements: $perm$ is used to consistently translate process identifiers between $aeadv$ and $sys$.
        \item A $cursor$ variable, initially set to $1$: at any time, $cursor$ identifies the next process that will \pbin.Deliver a message in $sys$.
    \end{itemize}
    \item Procedure $psadv.Step()$ simply forwards the call to $aeadv.Step()$.
    \item Procedure $psadv.Byzantine(process)$ simply forwards the call to \\ $sys.Byzantine(process)$.
    \item Procedure $psadv.State()$ simply forwards the call to $sys.State()$.
    \item Procedure $psadv.Sample(process, message)$ returns the list of processes returned by $sys.Sample(process, \allowbreak message)$, translated through $perm$. More specifically, for every process $\rho$ in $sys.Sample(process, \allowbreak message)$: if $\rho$ is correct, it is translated to $\zeta(perm[\zeta^{-1}(\rho)])$; if $\rho$ is Byzantine, it is left unchanged.
    \item Procedure $psadv.Deliver(process, message)$ sets $perm[cursor]$ to $\zeta^{-1}(\allowbreak process)$, then forwards the call to $sys.Deliver(\zeta(cursor), message)$. Finally, it increments $cursor$. This serves the purpose to sequentially cause $\zeta(1)$, $\zeta(2)$, $\ldots$ to deliver a message, while storing the translation in $perm$ in order for $psadv.Sample(\ldots)$ to provide a response consistent with any previous invocation of $psadv.Deliver(\ldots)$.
    \item Procedure $psadv.Echo(process, sample, source, message)$ simply forwards the call to $sys.Echo(process, sample, source, message)$.
    \item Procedure $pasdv.End()$ simply forwards the call to $sys.End()$.
\end{itemize}

\begin{figure}
\centering
\includegraphics[scale=0.50]{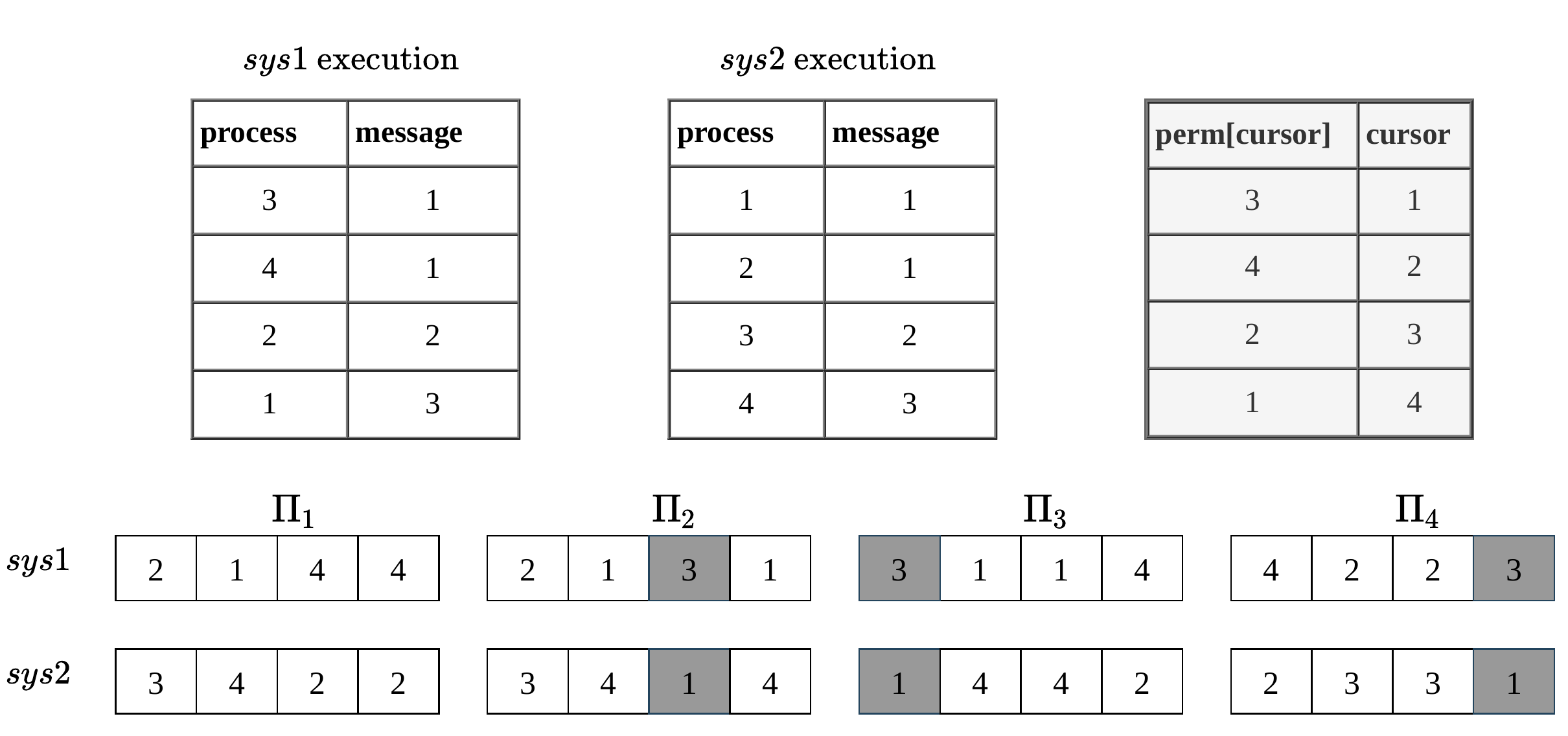}
\caption{Two systems with (one of) their respective echo samples. The table on the right shows the permutation from $sys1$ to $sys2$. Clearly both systems are equally likely. Moreover, the effect of process 3 delivering message 1 (grey) in $sys1$, is equal to process 1 delivering the same message in $sys2$. It can be seen that this holds for all further message deliveries. Intuitively this shows why we can restrict the adversary to always deliver in sequence.}
\label{figure:sieve:processsequential}
\end{figure}

\paragraph{Correctness}

We start by proving that no adversary, coupled with Process-sequential decorator, causes the execution to fail.

The following hold true:
\begin{itemize}
    \item No invocation of $psadv.Sample(\ldots)$ causes the execution to fail. Noting that $aeadv$ is correct, it will never invoke $psadv.Sample(\pi, m)$ unless $(\pi', m)$ was returned from a previous invocation of $psadv.State()$, for some $\pi' \in \Pi_C$. Moreover, since $psadv.State()$ simply forwards the call to $sys.State()$, if $(pi', m)$ was returned from $psadv.State()$, then $\pi'$ delivered $m$ in $sys$. Therefore, $sys.Sample(\pi, m)$ is never invoked from $psadv.Sample(\ldots)$ unless at least one correct process delivered $m$ in $sys$.
    \item Procedure $sys.Sample(\ldots)$ never calls $\zeta(\bot)$. We defer the proof of this result to a later section of this lemma.
    \item Procedure $sys.Deliver(\ldots)$ is never invoked twice on the same process. Indeed, by definition, $\zeta$ is a bijection between $1..C$ and $\Pi_C$, and $cursor$ is incremented every time $sys.Deliver(\ldots)$ is called.
\end{itemize}

\paragraph{Process-sequential}

It is easy to prove that Process-sequential decorator always implements a process-sequential adversary. Indeed, $sys.Deliver(\ldots)$ is invoked sequentially on $\zeta(1)$, $\zeta(2)$, $\ldots$ as $cursor$ is incremented, regardless of the process originally provided to $psadv.Deliver(\ldots)$.

\paragraph{System translation}

Let $\alpha$ be an adversary. We start by noting that, since $\alpha$ is correct, $\alpha$ always causes every correct process to \pbin.Deliver a message. We can therefore define a function
\begin{equation*}
    \mu: \mathcal{A} \times \mathcal{S} \times \Pi_C \rightarrow 1..C
\end{equation*}
such that $\mu(\alpha, \sigma, \pi) = d$ if and only if $\pi$ is the $d$-th process that $\alpha$ causes to \pbin.Deliver a message, when $\alpha$ is coupled with $\sigma$. We additionally define
\begin{equation*}
    \rp{\mu^{-1}(\alpha, \sigma, d) = \pi} \xLeftrightarrow[]{def} \rp{\mu(\alpha, \sigma, \pi) = d}
\end{equation*}

We then define a \textbf{system translation function} $\Psi[\alpha]: \mathcal{S} \rightarrow \mathcal{S}$ such that, for every system $\sigma$, every correct process $\pi$, every message $m$, and every $e \in 1..E$,
\begin{equation*}
    \Psi[\alpha](\sigma)[\pi][m][e] = 
    \begin{cases}
        \zeta(\mu(\alpha, \sigma, \sigma[\pi][m][e])) &\text{iff}\; \sigma[\pi][m][e] \in \Pi_C \\
        \sigma[\pi][m][e] &\text{otherwise}
    \end{cases}
\end{equation*}

Let $\sigma$ be a system, let $\sigma' = \Psi[\alpha](\sigma)$. Intuitively, $\sigma'$ is obtained from $\sigma$ simply by relabeling every correct process in every echo sample. Whenever a correct process $\pi$ appears in an echo sample in $\sigma$, it is replaced with $\zeta(d)$, $d$ being the position of $\pi$ in the ordered list of processes that $\alpha$ causes to \pbin.Deliver a message, when coupled with $\sigma$. Byzantine processes are left unchanged.

\paragraph{Roadmap}

Let $\alpha \in \mathcal{A}_{ae}$, let $\alpha' = \Delta_{ps}(\alpha)$. Let $\sigma \in \mathcal{S}$ such that $\alpha$ compromises the consistency of $\sigma$. In order to prove that $\alpha'$ is as powerful as $\alpha$, we prove that:
\begin{itemize}
    \item $\alpha'$ compromises the consistency of $\sigma' = \Psi[\alpha](\sigma)$.
    \item $\Psi[\alpha](\sigma)$ is a permutation on $\mathcal{S}$.
\end{itemize}

Indeed, if the above are true, then the probability of $\alpha'$ compromising the consistency of a random system $\sigma'$ is equal to the probability of $\alpha$ compromising the consistency of a random system $\sigma$, and the lemma is proved.

\paragraph{Trace}

We start by noting that, if we couple Process-sequential decorator with $\sigma'$, we effectively obtain a system interface $\delta$ with which $\alpha$ directly exchanges invocations and responses. Here we show that the trace $\tau(\alpha, \sigma)$ is identical to the trace $\tau(\alpha, \delta)$. Intuitively, this means that $\alpha$ has no way of \emph{distinguishing} whether it has been coupled directly with $\sigma$, or it has been coupled with $\sigma'$, with Process-sequential decorator acting as an interface. We prove this by induction.

Let us assume
\begin{eqnarray*}
    \tau(\alpha, \sigma) &=& ((i_1, r_1), \ldots) \\
    \tau(\alpha, \delta) &=& ((i'_1, r'_1), \ldots) \\
    i_j = i'_j, r_j = r'_j && \forall j \leq n
\end{eqnarray*}
with $n \geq 0$ (here $n = 0$ means that this is $\alpha$'s first invocation). We start by noting that, since $a$ is a deterministic algorithm, we immediately have
\begin{equation*}
    i_{n + 1} = i'_{n + 1}
\end{equation*}
and we need to prove that $r_{n + 1} = r'_{n + 1}$. \\

Let us assume that $i_{n + 1} = ({\tt Byzantine}, \pi)$. Let $\xi$ be a Byzantine process. If $\xi \in r_{n + 1}$ then, by definition, $\xi \in \sigma[\pi][1]$, i.e., for at least one $e \in 1..E$, $\sigma[\pi][1][e] = \xi$. Therefore, $\sigma'[\pi][1][e] = \xi$, and $\xi \in r'_{n + 1}$. If $\xi \notin r_{n + 1}$ then, for all $e \in 1..E$, $\sigma[\pi][1][e] \neq \xi$. If $\sigma[\pi][1][e] \in \Pi_C$, then $\sigma'[\pi][1][e] \in \Pi_C$ as well, so $\sigma'[\pi][1][e] \neq \xi$. If $\sigma[\pi][1][e] \in \Pi \setminus \Pi_C$, then $\sigma'[\pi][1][e] = \sigma[\pi][1][e] \neq \xi$. Therefore, $\xi \notin r'_{n + 1}$. \\

Before considering the remaining possible values of $i_{n + 1}$, we prove some auxiliary results. Let $\pi$ be a correct process, let $m$ be a message, let $d \in 1..C$, let $e \in 1..E$. For every $j \leq n + 1$, as we established, we have $i_j = i'_j$. Therefore, after the $(n + 1)$-th invocation, the following hold true:
\begin{itemize}
    \item $\pi$ \pbin.Delivered $m$ in $\sigma$ if and only if $\zeta(\mu(\alpha, \sigma, \pi))$ \pbin.Delivered $m$ in $\sigma'$. Indeed: 
    \begin{itemize}
        \item If $\pi$ \pbin.Delivered $m$ in $\sigma$, then $psadv.Deliver(\pi, m)$ was invoked. Moreover, by definition, $psadv.Deliver(\pi, m)$ was the $\mu(\alpha, \sigma, \pi)$-th invocation of $psadv.Deliver(\ldots)$. Noting that $cursor$ is incremented at each invocation of $psadv.Deliver(\ldots)$, when $psadv.\allowbreak Deliver(\pi, m)$ was invoked we had $cursor = \mu(\alpha, \sigma, \pi)$. Finally, $psadv.Deliver(\pi, m)$ forwards the call to $sys.Deliver(\zeta(cursor), \allowbreak m)$. Consequently, $\zeta(\mu(\alpha, \sigma, \pi))$ \pbin.Delivered $m$ in $\sigma'$.
        \item If $\zeta(\mu(\alpha, \sigma, \pi))$ \pbin.Delivered $m$ in $\sigma'$ then $sys.Deliver(\zeta(cursor), \allowbreak m)$ was invoked, with $cursor = \mu(\alpha, \sigma, \pi)$. Noting that $cursor$ is incremented after each invocation of $sys.Deliver(\ldots)$, we have that $psadv.Deliver(\ldots)$ was invoked at least $\mu(\alpha, \sigma, \pi)$ times. By definition, this means that $psadv.Deliver(\pi, \allowbreak m)$ was invoked. Consequently, $\pi$ \pbin.Delivered $m$ in $\sigma$.
    \end{itemize}
    \item If $\pi$ \pbin.Delivered a message in $\sigma'$, then $perm[\zeta^{-1}(\pi)] \neq \bot$. Indeed, noting that $cursor$ is incremented every time $psadv.Deliver(\ldots)$ is invoked, we have that $\rho$ \pbin.Delivered a message in $\sigma'$ as a result of the $\zeta^{-1}(\pi)$-th invocation of $psadv.Deliver(\ldots)$. As a result, $perm[\zeta^{-1}(\pi)]$ was set to a value other than $\bot$. From this follows that procedure $sys.Sample(\ldots)$ never calls $\zeta(\bot)$.
    \item If $perm[d] \neq \bot$, then $perm[d] = \zeta^{-1}(\mu^{-1}(\alpha, \sigma, d))$. Indeed, noting that $cursor$ is incremented every time $psadv.Deliver(\ldots)$ is invoked, $perm[d]$ was set to a value other than $\bot$ upon the $d$-th invocation of \\ $psadv.Deliver(\ldots)$. By the definition of $\mu$, the $d$-th invocation of $psadv.\allowbreak Deliver(\ldots)$ is $psadv.Deliver(\mu^{-1}(\alpha, \sigma, d), m')$, for some $m' \in \mathcal{M}$.

    \item $\sigma[\pi][m][e]$ sent an {\tt Echo}($m$, $m$) message to $\pi$ in $\sigma$ if and only if $\sigma'[\pi][m][e]$ sent an {\tt Echo}($m$, $m$) message to $\pi$ in $\sigma'$. Indeed, if $\sigma[\pi][m][e] \in \Pi_C$, then $\sigma'[\pi][m][e] = \zeta(\mu(\alpha, \sigma, \sigma[\pi][m][e]))$. Therefore, $\sigma[\pi][m][e]$ \pbin.Delivered $m$ in $\sigma$ if and only if $\sigma'[\pi][m][e]$ \pbin.Delivered $m$ in $\sigma'$. Noting that $\alpha$ is an auto-echo adversary, if $\sigma[\pi][m][e] \in \Pi \setminus \Pi_C$, then $\sigma'[\pi][m][e] = \sigma[\pi][m][e]$, and both sent an {\tt Echo}($m$, $m$) message to $\pi$ (in $\sigma$ and $\sigma'$, respectively).
    \item $\pi$ delivered $m$ in $\sigma$ if and only if $\pi$ delivered $m$ in $\sigma'$. This immediately follows from the above.
\end{itemize}

Let us assume $i_{n + 1} = ({\tt State})$. From the above immediately follows $r_{n + 1} = r'_{n + 1}$.

Let us assume $i_{n + 1} = ({\tt Sample}, \pi, m)$. Let $\rho$ be a process. The following hold true:
\begin{itemize}
    \item If $\rho \in r_{n + 1}$, then $\rho \in r'_{n + 1}$. Indeed, if $\rho \in \Pi_C$, then $\rho \in \sigma[\pi][m]$ and $\rho$ sent an {\tt Echo}($m$, $m'$) message to $\pi$ in $\sigma$, for some $m' \in \mathcal{M}$. Therefore, $\rho$ delivered $m'$ in $\sigma$. By definition, $\zeta(\mu(\alpha, \sigma, \rho)) \in \sigma'[\pi][m]$. Moreover, $\zeta(\mu(\alpha, \sigma, \rho))$ delivered $m'$ in $\sigma'$ and, as a result, it sent an {\tt Echo}($m$, $m'$) message to $\pi$ in $\sigma'$. Therefore, $\zeta(\mu(\alpha, \sigma, \rho)) \in sys.Sample(\pi, m)$. Finally, $perm[\mu(\alpha, \sigma, \rho)] = \zeta^{-1}(\rho)$. Consequently, $\rho \in r'_{n + 1}$. If $\rho \in \Pi \setminus \Pi_C$ then $\rho \in \sigma[\pi][m]$ and $\rho \in \sigma'[\pi][m]$. Moreover, $\rho$ sent an {\tt Echo}($m$, $m'$) message to $\pi$, for some $m' \in \mathcal{M}$, both in $\sigma$ and $\sigma'$. Consequently, $\rho \in r'_{n + 1}$.
    \item If $\rho \in r'_{n + 1}$, then $\rho \in r_{n + 1}$. Indeed, if $\rho \in \Pi_C$, then $\zeta(perm^{-1}[\zeta^{-1}(\rho)])$\footnote{Noting that $perm$ is injective, we define $perm^{-1}[b] = a \Longleftrightarrow perm[a] = b$.} was returned from $sys.Sample(\pi, m)$, in other words $\zeta(perm^{-1}[\zeta^{-1}(\rho)])$ \pbin.Delivered some message $m' \in \mathcal{M}$ in $\sigma'$. Moreover, using our auxiliary result on $perm$ we obtain
    \begin{equation*}
        \zeta(perm^{-1}[\zeta^{-1}(\rho)]) = \zeta(\mu(\alpha, \sigma, \rho))
    \end{equation*}
    therefore $\zeta(\mu(\alpha, \sigma, \rho))$ \pbin.Delivered $m'$ in $\sigma'$, and $\rho$ \pbin.Delivered $m'$ in $\sigma$. Finally, since $\zeta(\mu(\alpha, \sigma, \rho)) \in \sigma'[\pi][m]$, then by definition $\rho \in \sigma[\pi][m]$. Consequently, $\rho \in r_{n + 1}$.
\end{itemize}

Noting that procedures $Deliver(\ldots)$ and $Echo(\ldots)$ never return a value, we trivially have that if $i_{n + 1} = ({\tt Deliver}, \pi, m)$ or $i_{n + 1} = ({\tt Echo}, \pi, s, \xi, m)$ then $r_{n + 1} = \bot = r'_{n + 1}$. By induction, we have $\tau(\alpha, \sigma) = \tau(\alpha, \delta)$.

\paragraph{Consistency of $\sigma'$}

We proved that $\tau(\alpha, \sigma) = \tau(\alpha, \delta)$. Moreover, we proved that if a correct process $\pi$ eventually delivers a message $m$ in $\sigma$, then $\zeta(\mu(\alpha, \sigma, \pi))$ also delivers $m$ in $\sigma'$.

Since $\alpha$ compromises the consistency of $\sigma$, two correct processes $\pi$, $\pi'$ and two distinct messages $m$, $m' \neq m$ exist such that, in $\sigma$, $\pi$ delivered $m$ and $\pi'$ delivered $m'$. Therefore, in $\sigma'$, $\zeta(\mu(\alpha, \sigma, \pi))$ delivered $m$ and $\zeta(\mu(\alpha, \sigma, \pi'))$ delivered $m'$. Therefore $\alpha'$ compromises the consistency of $\sigma'$.

\paragraph{Translation permutation}

We now prove that, for any two $\sigma_a$, $\sigma_b \neq \sigma_a$, we have $\Psi[\alpha](\sigma_a) \neq \Psi[\alpha](\sigma_b)$. We prove this by contradiction. Suppose a system $\sigma'$ exists such that $\sigma' = \Psi[\alpha](\sigma_a) = \Psi[\alpha](\sigma_b)$. We want to prove that $\sigma_a = \sigma_b$.\\

We start by noting that, if $\tau(\alpha, \sigma_a) = \tau(\alpha, \sigma_b)$, then $\sigma_a = \sigma_b$. Indeed, if $\tau(\alpha, \sigma_a) = \tau(\alpha, \sigma_b)$, then for every $\pi \in \Pi_C$ and every $d \in 1..C$ we have
\begin{eqnarray*}
    \mu(\alpha, \sigma_a, \pi) &=& \mu(\alpha, \sigma_b, \pi) \\
    \mu^{-1}(\alpha, \sigma_a, d) &=& \mu^{-1}(\alpha, \sigma_b, d)
\end{eqnarray*}
and since, by definition, for every $\pi \in \Pi_C$, $m \in \mathcal{M}$ and $e \in 1..E$, we have
\begin{eqnarray*}
    \sigma_a[\pi][m][e] &=& 
    \begin{cases}
        \mu^{-1}(\alpha, \sigma_a, \zeta^{-1}(\sigma'[\pi][m][e])) &\text{iff}\; \sigma'[\pi][m][e] \in \Pi_C \\
        \sigma'[\pi][m][e] &\text{otherwise}
    \end{cases}
    \\
    \sigma_b[\pi][m][e] &=& 
    \begin{cases}
        \mu^{-1}(\alpha, \sigma_b, \zeta^{-1}(\sigma'[\pi][m][e])) &\text{iff}\; \sigma'[\pi][m][e] \in \Pi_C \\
        \sigma'[\pi][m][e] &\text{otherwise}
    \end{cases}
\end{eqnarray*}
we get
\begin{equation*}
    \sigma_a[\pi][m][e] = \sigma_b[\pi][m][e]
\end{equation*}
and $\sigma_a = \sigma_b$. \\

We prove that $\tau(\alpha, \sigma_a) = \tau(\alpha, \sigma_b)$ by induction. Let us assume
\begin{eqnarray*}
    \tau(\alpha, \sigma_a) &=& ((i_1, r_1), \ldots) \\
    \tau(\alpha, \sigma_b) &=& ((i'_1, r'_1), \ldots) \\
    i_j = i'_j, r_j = r'_j && \forall j \leq n
\end{eqnarray*}
with $n \geq 0$ (here $n = 0$ means that this is $\alpha$'s first invocation). We start by noting that, since $a$ is a deterministic algorithm, we immediately have
\begin{equation*}
    i_{n + 1} = i'_{n + 1}
\end{equation*}
and we need to prove that $r_{n + 1} = r'_{n + 1}$.\\ 

Let us assume that $i_{n + 1} = ({\tt Byzantine}, \pi)$. Let $\xi$ be a Byzantine process. if $\xi \in r_{n + 1}$, then for at least one $e \in 1..E$ we have $\sigma_a[\pi][m][e] = \xi$. Therefore, $\sigma'[\pi][m][e] = \xi$, and $\sigma_b[\pi][m][e] = \xi$. Consequently, $\xi \in r'_{n + 1}$. The argument can be reversed to prove $\xi \in r'_{n + 1} \implies \xi \in r_{n + 1}$. \\

Before considering the remaining possible values of $i_{n + 1}$, we prove some auxiliary result. Let $\pi$ be a correct process, let $m$ be a message, let $e \in 1..E$. For every $j \leq n + 1$, as we established, we have $i_j = i'_j$. Therefore, after the $(n + 1)$-th invocation, the following hold true:
\begin{itemize}
    \item $\pi$ \pbin.Delivered $m$ in $\sigma_a$ if and only if $\pi$ \pbin.Delivered $m$ in $\sigma_b$. Indeed, if $\pi$ \pbin.Delivered $m$ in $\sigma_a$, then some $j \leq (n + 1)$ exists such that $i_j = ({\tt Deliver}, \pi, m)$. Since $i'_j = i_j$, $\pi$ \pbin.Delivered $m$ in $\sigma_b$ as well. The argument can be reversed to prove that, if $\pi$ \pbin.Delivered $m$ in $\sigma_b$, then $\pi$ \pbin.Delivered $m$ in $\sigma_a$ as well.
    \item If $\pi$ \pbin.Delivered $m$ in $\sigma_a$ (or, equivalently, $\sigma_b$), then $\mu(\alpha, \sigma_a, \pi) = \mu(\alpha, \sigma_b, \pi)$. Indeed, some $j \leq (n + 1)$ exists such that $i_j = i'_j = ({\tt Deliver}, \pi, m)$. Since, for all $h < j$, we also have $i_h = i'_h$, then
    \begin{eqnarray*}
        \lhs \abs{\cp{h \in 1..(j - 1) \mid i_h = ({\tt Deliver}, \pi' \in \Pi_C, m' \in \mathcal{M})}} \\
        &=& \abs{\cp{h \in 1..(j - 1) \mid i'_h = ({\tt Deliver}, \pi' \in \Pi_C, m' \in \mathcal{M})}}
    \end{eqnarray*}
    \item $\sigma_a[\pi][m][e]$ sent an {\tt Echo}($m$, $m$) message to $\pi$ in $\sigma_a$ if and only if $\sigma_b[\pi][m][e]$ sent an {\tt Echo}($m$, $m$) message to $\pi$ in $\sigma_b$. We prove this by cases:
    \begin{itemize}
        \item Let us assume that $\sigma_a[\pi][m][e]$ is correct, and \pbin.Delivered $m$ in $\sigma_a$. By definition, we have
        \begin{eqnarray*}
            \sigma'[\pi][m][e] = \zeta(\mu(\alpha, \sigma_a, \sigma_a[\pi][m][e])) \\
            \sigma'[\pi][m][e] = \zeta(\mu(\alpha, \sigma_b, \sigma_b[\pi][m][e]))
        \end{eqnarray*}
        and from the above we have
        \begin{equation*}
            \zeta(\mu(\alpha, \sigma_a, \sigma_a[\pi][m][e])) = \zeta(\mu(\alpha, \sigma_b, \sigma_a[\pi][m][e]))
        \end{equation*}
        
        Equating the two above we get
        \begin{equation*}
            \zeta(\mu(\alpha, \sigma_b, \sigma_a[\pi][m][e])) = \zeta(\mu(\alpha, \sigma_b, \sigma_b[\pi][m][e]))
        \end{equation*}
        and noting that $\mu$ is always injective, we have $\sigma_a[\pi][m][e] = \sigma_b[\pi][m][e]$. Therefore $\sigma_b[\pi][m][e]$ \pbin.Delivered $m$ in $\sigma_b$. 
        
        The argument can be inverted to prove that, if $\sigma_b[\pi][m][e]$ is correct, and \pbin.Delivered $m$ in $\sigma_b$, then $\sigma_a[\pi][m][e]$ \pbin.Delivered $m$ in $\sigma_a$ as well.
        
        \item Let us assume that $\sigma_a[\pi][m][e]$ is correct, but did not \pbin.Deliver $m$. From the definition of $\Psi[\alpha]$, we know that $\sigma_b[\pi][m][e]$ is correct as well. By contradiction, following from the above, we have that if $\sigma_b[\pi][m][e]$ \pbin.Delivered $m$ in $\sigma_b$, $\sigma_a[\pi][m][e]$ would have \pbin.Delivered $m$ in $\sigma_a$ as well.
        
        The argument can be inverted to prove that, if $\sigma_b[\pi][m][e]$ is correct, but did not \pbin.Deliver $m$ in $\sigma_b$, then $\sigma_a[\pi][m][e]$ did not \pbin.Deliver $m$ in $\sigma_a$ either.
        
        \item Let us assume that $\sigma_a[\pi][m][e]$ is Byzantine. Then, from the definition of $\Psi[\alpha]$, we immediately have $\sigma_b[\pi][m][e] = \sigma_a[\pi][m][e]$ and, since $\alpha$ is an auto-echo adversary, both sent an {\tt Echo}($m$, $m$) message to $\pi$ (in their respective systems).
    \end{itemize}
    \item $\pi$ delivered $m$ in $\sigma_a$ if and only if $\pi$ delivered $m$ in $\sigma_b$ as well. This follows immediately from the above.
\end{itemize}

Let us assume $i_{n + 1} = ({\tt State})$. From the above immediately follows $r_{n + 1} = r'_{n + 1}$.

Let us assume $i_{n + 1} = ({\tt Sample}, \pi, m)$. Let $\rho$ be a process. If $\rho \in r_{n + 1}$, then for some $e \in 1..E$, $\sigma_a[\pi][m][e] = \rho$, and $\rho$ sent an {\tt Echo}($m$, $m'$) message to $\pi$ in $\sigma_a$, for some $m' \in \mathcal{M}$. Following from the above, we have $\sigma_b[\pi][m][e] = \rho$ as well, and $\rho$ sent an {\tt Echo}($m$, $m'$) message to $\pi$ in $\sigma_b$ as well. Therefore, $\rho \in r'_{n + 1}$. The argument can be inverted to prove that, if $\rho \in r'_{n + 1}$, then $\rho \in r_{n + 1}$ as well.

Noting that procedures $Deliver(\ldots)$ and $Echo(\ldots)$ never return a value, we trivially have that if $i_{n + 1} = ({\tt Deliver}, \pi, m)$ or $i_{n + 1} = ({\tt Echo}, \pi, s, \xi, m)$ then $r_{n + 1} = \bot = r'_{n + 1}$. By induction, we have $\tau(\alpha, \sigma_a) = \tau(\alpha, \sigma_b)$.

Therefore, $\sigma_a = \sigma_b$, which contradicts the hypothesis.
\end{proof}
\end{lemma}

\subsection{Sequential adversary}
\label{subsection:sequentialadversaryproof}

\begin{algorithm}
\begin{algorithmic}[1]
\Implements
    \Instance{SequentialAdversary + \Cobin System}{sqadv}
\EndImplements

\Uses
    \InstanceSystem{ProcessSequentialAdversary}{psadv}{sqadv}
    \Instance{\Cobin System}{sys}
\EndUses

\Procedure{sqadv.Init}{{}}
    \State $perm = \cp{\bot}^C$; \tabto*{3.5cm} $cursor = 1$; \tabto*{6cm} $step = 0$;
    \State
    \State $poisoned = \false$;
    \ForAll{\pi}{\Pi_C}
        \If{$\abs{sys.Byzantine(\pi)} \geq \hat E$}
            \State $poisoned \leftarrow \true$;
        \EndIf
    \EndForAll
    \State
    \State $psadv.Init()$;
\EndProcedure

\Procedure{sqadv.Step}{{}}
    \State $step \leftarrow step + 1$;
    \State
    \If{$poisoned = \false$ \textbf{or} $step \leq (N - C)C^2$}
        \State $psadv.Step()$;
    \ElsIf{$step \leq (N - C)C^2 + C$}
        \State $sys.Deliver(\zeta(step - (N - C)C^2), 1)$;
    \Else
        \State $sys.End()$;
    \EndIf
\EndProcedure

\Procedure{sqadv.Byzantine}{process}
    \State \Return $sys.Byzantine(process)$;
\EndProcedure

\algstore{sequentialdecorator}
\end{algorithmic}
\caption{Sequential decorator}
\label{algorithm:sequentialdecorator}
\end{algorithm}

\begin{algorithm}
\begin{algorithmic}[1]
\algrestore{sequentialdecorator}

\Procedure{sqadv.State}{{}}
    \State $state = \emptyset$;
    \State
    \ForAll{(\pi, m)}{sys.State()}
            \State $state \leftarrow state \cup \cp{(\pi, perm[m])}$;
    \EndForAll
    \State
    \State \Return $state$;
\EndProcedure

\Procedure{sqadv.Sample}{process, message}
    \State \Return $sys.Sample(process, perm^{-1}[message])$;
\EndProcedure

\Procedure{sqadv.Deliver}{process, message}
    \If{$message \in perm$}
        \State $sys.Deliver(process, perm^{-1}[message])$;
    \Else
        \State $perm[cursor] = message$;
        \State $sys.Deliver(process, cursor)$;
        \State $cursor \leftarrow cursor + 1$;
    \EndIf
\EndProcedure

\Procedure{sqadv.Echo}{process, sample, source, message}
    \State $sys.Echo(process, sample, source, message)$;
\EndProcedure

\Procedure{sqadv.End}{{}}
    \State $sys.End()$;
\EndProcedure
\end{algorithmic}
\end{algorithm}

\begin{lemma}
The set of sequential adversaries $\mathcal{A}_{sq}$ is optimal.

\begin{proof}
We again prove the result using a decorator. Here we show that a decorator $\Delta_{sq}$ exists such that, for every $\alpha \in \mathcal{A}_{ps}$, the adversary $\alpha' = \Delta_{sq}(\alpha)$ is a sequential adversary, and as powerful as $\alpha$. If this is true, then the lemma is proved: let $\alpha^*$ be an optimal adversary, then the sequential $\alpha^+ = \Delta_{sq}(\alpha^*)$ is optimal as well.

\paragraph{Decorator}

\cref{algorithm:sequentialdecorator} implements \textbf{Sequential decorator}, a decorator that transforms a process-sequential adversary into a sequential adversary. Provided with a process-sequential adversary $psadv$, Sequential decorator acts as an interface between $psadv$ and a system $sys$, effectively implementing a sequential adversary $sqadv$. Sequential decorator exposes both the adversary and the system interfaces: the underlying adversary $psadv$ uses $sqadv$ as its system.

Sequential decorator works as follows:
\begin{itemize}
    \item Procedure $sqadv.Init()$ initializes the following variables:
    \begin{itemize}
        \item A $perm$ array of $C$ elements: $perm$ is used to consistently translate messages between $psadv$ and $sys$.
        \item A $cursor$ variable, initially set to $1$: at any time, $cursor$ identifies the next message that will be \pbin.Delivered in $sys$, if $psadv$ will invoke the delivery of a process whose delivery $psadv$ never invoked before.
        \item A $poisoned$ variable: $poisoned$ is set to $\true$ if and only if at least one correct process in $sys$ is poisoned. This condition is verified by looping over $sys.Byzantine(\pi)$ for every correct process $\pi$.
        \item A $step$ variable, initially set to $0$: at any time, $step$ counts how many times $sqadv.Step()$ has been invoked.
    \end{itemize}
    \item Procedure $sqadv.Step()$ increments $step$, then implements two different behaviors depending on the value of $poisoned$:
    \begin{itemize}
        \item If $poisoned = \true$, it forwards the call to $psadv.Step()$ for the first $(N - C)C^2$ times. For the next $C$ steps, it sequentially invokes $sys.Deliver(\zeta(1), 1)$, $\ldots$, $sys.Deliver(\zeta(C), 1)$. Finally, it calls $sys.End()$.
        \item If $poisoned = \false$, it forwards the call to $psadv.Step()$.
    \end{itemize}
    \item Procedure $sqadv.Byzantine(process)$ simply forwards the call to \\ $sys.Byzantine(process)$.
    \item Procedure $sqadv.State()$ returns the list of process / message pairs returned by $sys.State()$, with each message translated through $perm$. More specifically, $sqadv.State()$ returns $(\pi, perm[m])$ for every $(\pi, m)$ in $sys.State()$.
    \item Procedure $sqadv.Sample(process, message)$ simply forwards the call to $sys.Sample(process, perm^{-1}[message])$.
    \item Procedure $sqadv.Deliver(process, message)$ checks if $psadv$ has already invoked the delivery of $message$ (this is achieved by checking if $message$ is in $perm$). If so, it forwards the call to $sys.Deliver(process, \allowbreak perm^{-1}[message])$. Otherwise, it sets $perm[cursor]$ to $message$, then forwards the call to $sys.Deliver(process, cursor)$. Finally, it increments $cursor$. This mechanism serves two purposes:
    \begin{itemize}
        \item To consistently translate a $sqadv.Deliver(\ldots)$ invocation to a \\ $sys.Deliver(\ldots)$ invocation. More specifically, the set of invocations $psadv.Deliver(\pi_1, m)$, $\ldots$, $psadv.Deliver(\pi_k, m)$ is always translated to $sys.Deliver(\pi_1, m')$, $\ldots$, $sys.Deliver(pi_k, m')$.
        \item To never cause the \pbin.Delivery of a message $b$ in $sys$ before every message $a < b$ has been \pbin.Delivered in $sys$ at least once.
    \end{itemize}
    \item Procedure $sqadv.Echo(process, sample, source, message)$ simply forwards the call to $sys.Echo(process, sample, source, message)$.
    \item Procedure $sqadv.End()$ simply forwards the call to $sys.End()$.
\end{itemize}

\paragraph{Correctness}

We start by proving that no adversary, coupled with Sequential decorator, causes the execution to fail. We distinguish two cases, based on the value of $poisoned$. 

Let us assume $poisoned = \true$. When $sqadv.Step()$ is invoked, the call is forwarded to $psadv.Step()$ only for the first $(N - C)C^2$ times. Noting that $psadv$ is an auto-echo adversary, every call to $psadv.Step()$ results in a call to $sqadv.Echo(\ldots)$. For the next $C$ steps, $sqadv.Step()$ sequentially causes $\zeta(1), \zeta(2), \ldots$ to \pbin.Deliver message $1$. Finally, $sqadv.Step()$ invokes $sys.End()$. Therefore, $sqadv$ never causes the execution to fail, and it implements a process-sequential adversary.

Let us assume $poisoned = \false$. Let $\pi$ be a correct process, let $m$ be a message. The following hold true:
\begin{itemize}
    \item Procedure $sqadv.State()$ never returns a $(\pi, \bot)$ pair. Indeed, if $(\pi, m) \in sys.State()$, then $\pi$ \pbin.Delivered $m$ in $sys$. Since $\pi$ is not poisoned, $\pi$ received at least one {\tt Echo}($m$, $m$) message from a correct process. Consequently, if $(\pi, m)$ is returned from $sys.State()$, then at least one correct process \pbin.Delivered $m$ in $sys$, i.e., $sys.Deliver(\pi', m)$ was invoked for some $\pi' \in \Pi_C$. The statement is proved by noting that, whenever $sys.Deliver(\pi', m)$ is invoked for some $\pi' \in \Pi_C$, we have $perm[m] \neq \bot$: indeed, either \\ $sys.Sample(process, perm^{-1}[message])$ is invoked, and $message \in perm$, or $sys.Sample(process, cursor)$ is invoked, and $perm[cursor] = message \neq \bot$.
    
    \item No invocation of $sqadv.Sample(\ldots)$ causes the execution to fail. Noting that $psadv$ is correct, it will never invoke $sqadv.Sample(\pi, m)$ unless $(\pi', m)$ was returned from a previous invocation of $sqadv.State()$, for some $\pi' \in \Pi_C$. Since $\pi'$ is not poisoned, $(\pi', perm^{-1}[m])$ was returned from $sys.State()$, therefore $\pi'$ delivered $perm^{-1}[m]$ in $sys$. Therefore $sys.Sample(\pi, m)$ is never invoked from $sqadv.Sample(\ldots)$ unless at least one correct process delivered $m$ in $sys$.
\end{itemize}

\paragraph{Sequential}

It is easy to prove that Sequential decorator always implements a sequential adversary. Indeed, if $poisoned = \true$, $sqadv$ simply causes every correct process to \pbin.Deliver message $1$ (which trivially implements a sequential adversary). If $poisoned = \false$, then whenever $sys.Deliver(\pi, m)$ is invoked, either of the following holds true:
\begin{itemize}
    \item $m = perm^{-1}[message]$ for some $message \in perm$. In this case $sys.Deliver(\ldots)$ was previously invoked on $m$ (i.e., some process $\pi'$ exists such that $sys.Deliver(\pi', m)$ was previously invoked).
    \item $m = cursor$. Then $sys.Deliver(\ldots)$ was never invoked on $m$. Noting that, whenever $sys.Deliver(\ldots)$ is invoked on a new message, $cursor$ is incremented, we have that every message $l < m$ was previously \pbin.Delivered by at least one correct process in $sys$.
\end{itemize}

\paragraph{System translation}

Let $\alpha$ be an adversary. We can define a function
\begin{equation*}
    \mu: \mathcal{A} \times \mathcal{S} \times \mathcal{M} \rightarrow 1..C \cup \cp{\bot}
\end{equation*}
such that:
\begin{itemize}
    \item $\mu(\alpha, \sigma, m) = (d \in 1..C)$ if and only if $m$ is the $d$-th distinct message that $\alpha$ causes at least one correct process to \pbin.Deliver, when $\alpha$ is coupled with $\sigma$.
    \item $\mu(\alpha, \sigma, m) = \bot$ if and only if $\alpha$ never causes any correct process to \pbin.Deliver $m$, when $\alpha$ is coupled with $\sigma$.
\end{itemize}

We additionally define $\nu: \mathcal{A} \times \mathcal{S} \rightarrow 1..C$ by
\begin{equation*}
    \nu(\alpha, \sigma) = \max_{m \in \mathcal{M}} \mu(\alpha, \sigma, m)
\end{equation*}
and
\begin{equation*}
    \rp{\mu^{-1}(\alpha, \sigma, d) = m} \xLeftrightarrow[]{def} \rp{\mu(\alpha, \sigma, m) = d}
\end{equation*}
for all $d \leq \nu(\alpha, \sigma)$. Here $\nu(\alpha, \sigma)$ counts the number of distinct messages that $\alpha$ causes at least one correct process to \pbin.Deliver, when coupled with $\sigma$. It is immediate to see that $\mu(\alpha, \sigma, d) = \bot$ for all $d > \nu(\alpha, \sigma)$.

We then define a \textbf{message permutation function} $\chi: \mathcal{A} \times \mathcal{S} \times \mathcal{M} \rightarrow \mathcal{M}$ as follows:
\begin{equation*}
    \chi(\alpha, \sigma, d) =
    \begin{cases}
        \mu^{-1}(\alpha, \sigma, d) &\text{iff}\; d \leq \nu(\alpha, \sigma) \\
        \max m \in \mathcal{M} \mid \\
        \quad \abs{\cp{l \leq m: \mu(\alpha, \sigma, l) = \bot}} = d - \nu(\alpha, \sigma) &\text{otherwise}
    \end{cases}
\end{equation*}

For a given $\alpha$ and $\sigma$, the permutation $\chi$ maps $d$ to the $d$-th distinct message that is \pbin.Delivered when $\alpha$ is coupled with $\sigma$, if such a message exists. If such a message does not exist, $\chi$ simply enumerates sequentially the messages that are never \pbin.Delivered when $\alpha$ is coupled with $\sigma$. \\

For example, let us consider the case where $C = 10$ and $\alpha$ coupled with $\sigma$ causes the \pbin.Delivery of messages $3, 7, 1, 4$ (in this order of first appearance). Then $\chi$ will assume the following values for $d \in 1..C$: $3, 7, 1, 4, 2, 5, 6, 8, 9, 10$.\\

Finally, we define a \textbf{system translation function} $\Psi[\alpha]: \mathcal{S} \rightarrow \mathcal{S}$ such that, for system $\sigma$, every correct process $\pi$ and every message $m$,
\begin{equation*}
    \Psi[\alpha](\sigma)[\pi][m] =
    \begin{cases}
        \sigma[\pi][m] &\text{iff}\; \exists \pi' \in \Pi_C \mid \pi' \;\text{is poisoned in}\;\sigma \\
        \sigma[\pi][\chi(\alpha, \sigma, m)] &\text{otherwise}
    \end{cases}
\end{equation*}

Let $\sigma$ be a system, let $\sigma' = \Psi[\alpha](\sigma)$. Intuitively, if at least one correct process is poisoned in $\sigma$, then $\sigma' = \sigma$. Otherwise, $\sigma'$ is obtained from $\sigma$ by permuting the echo samples of each correct process in $\sigma$ using $\chi$.

\paragraph{Roadmap}

Let $\alpha \in \mathcal{A}_{ps}$, let $\alpha' = \Delta_{sq}(\alpha)$.  Let $\sigma \in \mathcal{S}$ such that $\alpha$ compromises the consistency of $\sigma$. In order to prove that $\alpha'$ is as powerful as $\alpha$, we prove that:
\begin{itemize}
    \item $\alpha'$ compromises the consistency of $\sigma' = \Psi[\alpha](\sigma)$.
    \item $\Psi[\alpha](\sigma)$ is a permutation on $\mathcal{S}$.
\end{itemize}

Indeed, if the above are true, then the probability of $\alpha'$ compromising the consistency of a random system $\sigma'$ is equal to the probability of $\alpha$ compromising the consistency of a random system $\sigma$, and the lemma is proved.

\paragraph{Poisoned case}

We start by considering the case where $poisoned = \true$. Let $\pi$ be a correct process that is poisoned in $\sigma$. Noting that $psadv$ is an auto-echo adversary, $\pi$ eventually delivers every message. Indeed, every Byzantine process eventually sends to $\pi$ an {\tt Echo}($m$, $m$) message, for every $m \in \mathcal{M}$. Since all of $\pi$'s echo samples share the same set of at least $\hat E$ Byzantine processes, $\pi$ eventually delivers every message.

As a result, if at least one correct process in $\sigma$ is poisoned, the consistency of $\sigma$ is compromised by any auto-echo adversary. Noting that $\sigma' = \Psi[\alpha](\sigma) = \sigma$, and $\Delta_{sq}(\alpha)$ is an auto-echo adversary, we immediately have that $\alpha'$ compromises the consistency of $\sigma'$ as well.

In the next sections of this proof, we consider the case $poisoned = \false$.

\paragraph{Trace}

We start by noting that, if we couple Process-sequential decorator with $\sigma'$, we effectively obtain a system interface $\delta$ with which $\alpha$ directly exchanges invocations and responses. Here we show that, if $poisoned = \false$, the trace $\tau(\alpha, \sigma)$ is identical to the trace $\tau(\alpha, \delta)$. Intuitively, this means that, if $poisoned = \false$, $\alpha$ has no way of \emph{distinguishing} whether it has been coupled directly with $\sigma$, or it has been coupled with $\sigma'$, with Process-sequential decorator acting as an interface. We prove this by induction.

Let us assume $poisoned = \false$, and
\begin{eqnarray*}
    \tau(\alpha, \sigma) &=& ((i_1, r_1), \ldots) \\
    \tau(\alpha, \delta) &=& ((i'_1, r'_1), \ldots) \\
    i_j = i'_j, r_j = r'_j && \forall j \leq n
\end{eqnarray*}
with $n \geq 0$ (here $n = 0$ means that this is $\alpha$'s first invocation). We start by noting that, since $a$ is a deterministic algorithm, we immediately have
\begin{equation*}
    i_{n + 1} = i'_{n + 1}
\end{equation*}
and we need to prove that $r_{n + 1} = r'_{n + 1}$. \\

Let us assume that $i_{n + 1} = ({\tt Byzantine}, \pi)$. We can note that $sqadv.\allowbreak Byzantine(process)$ simply forwards the call to $sys.Byzantine(process)$, and $\chi$ defines a permutation over $1..C$. Therefore, a message $m$ exists such that $\pi$'s first echo sample for in $\sigma'$ is identical to $\pi$'s echo sample for $m$ in $\sigma$. Moreover, all of $\pi$'s echo samples in $\sigma$ share the same set of Byzantine processes. Consequently, $r_{n + 1} = r'_{n + 1}$. \\

Before considering the remaining possible values of $i_{n + 1}$, we prove some auxiliary results. We start by noting the following:
\begin{itemize}
    \item Let $d \in 1..C$. At any time, if $perm[d] \neq \bot$, then $perm[d] = \mu^{-1}(\alpha, \sigma, d)$. Indeed, at any time, a message $m$ is in $perm$ if and only if $sqadv.\allowbreak Deliver(\ldots)$ was previously invoked on $m$. Moreover, whenever $sqadv.\allowbreak Deliver(\ldots)$ is invoked on a message $m$ that is not in $perm$, $m$ is added to $perm$ and $cursor$ is incremented. Therefore $perm[cursor]$ is set to $m$ if and only if $sqadv.Deliver(\ldots)$ was never invoked on $m$, and $sqadv.Deliver(\ldots)$ was previously invoked on exactly $cursor - 1$ distinct messages. Moreover, by definition, when $sqadv.Deliver(\ldots)$ is invoked on $m$ for the first time, $sqadv.Deliver(\ldots)$ was previously invoked on exactly $\mu(\alpha, \sigma, m) - 1$ distinct messages. Consequently, $cursor = \mu(\alpha, \sigma, m)$, and $m = \mu^{-1}(\alpha, \sigma, cursor)$.
    \item No two values of $perm$ are equal to each other. Indeed, a message $m$ is added to $perm$ only if $m \notin perm$.
    \item Let $\pi \in \Pi_C$, let $m \in \mathcal{M}$. If $\pi$ delivered $m$, then at least one correct process \pbin.Delivered $m$. This separately holds true both in $\sigma$  and $\sigma'$. Indeed, if $\pi$ delivered $m$, then it received at least $\hat E$ {\tt Echo}($m$, $m$) messages from its echo sample for $m$ and, since no correct process is poisoned in neither $\sigma$ nor $\sigma'$, at least one of them must have come from a correct process.
\end{itemize}

Let $\pi$ be a correct process, let $\rho$ be a process, let $m$, $s$ be messages. For every $j \leq n + 1$, as we established, we have $i_j = i'_j$. By hypothesis, $\alpha$ is an auto-echo adversary, so $i_j = i'_j$, $r_j = r'_j = \bot$ for every $j \leq (N - C)C^2$. Let us consider the non-trivial case $n \geq (N - C)C^2$. After the $(n + 1)$-th invocation, the following hold true:
\begin{itemize}
    \item $\pi$ \pbin.Delivered $m$ in $\sigma$ if and only if $\pi$ \pbin.Delivered $\mu(\alpha, \sigma, m)$ in $\sigma'$. Indeed:
    \begin{itemize}
        \item If $\pi$ \pbin.Delivered $m$ in $\sigma$, then $sqadv.Deliver(\pi, m)$ was invoked.

        If $sqadv.Deliver(\pi, m)$ was the first invocation of $sqadv.Deliver(\allowbreak \ldots)$ on $m$, then $m$ was not in $perm$, $perm[cursor]$ was set to $m$, and $sys.Deliver(\pi, cursor)$ was invoked. As we previously proved, however, we have $perm[cursor] = \mu^{-1}(\alpha, \sigma, m)$, so $cursor \allowbreak = \mu(\alpha, \sigma, m)$. Consequently, $sys.Deliver(\pi, \mu(\alpha, \sigma, m))$ was invoked, and $\pi$ \\ \pbin.Delivered $\mu(\alpha, \sigma, m)$ in $\sigma'$. If $sqadv.Deliver(\pi, m)$ was not the first invocation of $sqadv.Deliver(\ldots)$ on $m$, then $m$ was in $perm$, and $sys.Deliver(\pi, perm^{-1}(m))$ was invoked. Due to the above, we have again $perm^{-1}[m] = \mu(\alpha, \sigma, m)$. Consequently, $sys.Deliver(\pi, \mu(\alpha, \sigma, m))$ was invoked, and $\pi$ \pbin.Delivered $\mu(\allowbreak \alpha, \allowbreak \sigma, \allowbreak m)$ in $\sigma'$.

        \item If $\pi$ \pbin.Delivered $\mu(\alpha, \sigma, m)$ in $\sigma'$, then $sys.Deliver(\pi, \mu(\alpha, \sigma, m))$ was invoked. If $sys.Deliver(\pi, cursor)$ was invoked, we have that $cursor = \mu(\alpha, \sigma, m)$, and $sqadv.Deliver(\pi, m')$ was invoked for some $m' \notin perm$. As a result, $perm[cursor]$ was set to $m'$. As we previously established, however,
        \begin{equation*}
            m' = \mu^{-1}(\alpha, \sigma, cursor) = \mu^{-1}(\alpha, \sigma, \mu(\alpha, \sigma, m)) = m
        \end{equation*}
        and $sqadv.Deliver(\pi, m)$ was invoked. As a result, $\pi$ \pbin.Delivered $m$ in $\sigma$. If $sys.Deliver(\pi, perm^{-1}(m'))$ was invoked for some $m' \in perm$, we have $perm^{-1}[m'] = \mu(\alpha, \sigma, m)$, and again $m' = m$. Consequently, $sqadv.Deliver(\pi, m)$ was invoked, and $\pi$ \pbin.Delivered $m$ in $\sigma$.
    \end{itemize}
    \item $\pi$ received an {\tt Echo}($m$, $m$) message from $\rho$ in $\sigma$ if and only if $\pi$ received an {\tt Echo}($\mu(\alpha, \sigma, m)$, $\mu(\alpha, \sigma, m)$) message from $\rho$ in $\sigma'$. Indeed:
    \begin{itemize}
        \item If $\rho$ is a correct process, from the above we have that $\pi$ \pbin.Delivered $m$ in $\sigma$ if and only if $\pi$ \pbin.Delivered $\mu(\alpha, \sigma, m)$ in $\sigma'$. Therefore, $\rho$ sent to $\pi$ an {\tt Echo}($m$, $m$) message if and only if $\rho$ sent to $\pi$ an {\tt Echo}($\mu(\alpha, \sigma, m)$, $\mu(\alpha, \sigma, m)$) message.
        \item If $\rho$ is a Byzantine process then, noting that $\alpha$ is an auto-echo adversary, $\rho$ sent to $\pi$ an {\tt Echo}($m$, $m$) both in $\sigma$ and $\sigma'$.
    \end{itemize}
    \item $\pi$ received an {\tt Echo}($s$, $m'$) message for some $m' \in \mathcal{M}$ from $\rho$ in $\sigma$ if and only if $\pi$ received an {\tt Echo}($\mu(\alpha, \sigma, s)$, $m''$) message for some $m'' \in \mathcal{M}$ from $\rho$ in $\sigma''$. Indeed:
    \begin{itemize}
        \item If $\rho$ is correct, it sent an {\tt Echo}($s'$, $m'$) message for every $s' \in \mathcal{S}$ and some $m' \in \mathcal{S}$ to $\pi$ in $\sigma$ if and only if $\rho$ \pbin.Delivered a message in $\sigma$. Moreover, $\rho$ \pbin.Delivered a message in $\sigma$ if and only if $\rho$ \pbin.Delivered a message in $\sigma'$. Finally, $\rho$ \pbin.Delivered a message in $\sigma'$ if and only if $\rho$ sent an {\tt Echo}($s''$, $m''$) message for every $s'' \in \mathcal{S}$ and some $m'' \in \mathcal{S}$ to $\pi$ in $\sigma'$.
        \item If $\rho$ is Byzantine, it sent an {\tt Echo}($m'$, $m'$) message for every $m' \in \mathcal{S}$ both in $\sigma$ and $\sigma'$.
    \end{itemize}
    \item $\pi$ delivered $m$ in $\sigma$ if and only if $\pi$ delivered $\mu(\alpha, \sigma, m)$ in $\sigma'$. Indeed, if $\pi$ delivered $m$ in $\sigma$, then at least one correct process \pbin.Delivered $m$ in $\sigma$, and at least one correct process \pbin.Delivered $\mu(\alpha, \sigma, m)$ in $\sigma'$; if $\pi$ delivered $\mu(\alpha, \sigma, m)$ in $\sigma'$, then at least one correct process \pbin.Delivered $\mu(\alpha, \sigma, m)$ in $\sigma'$, and at least one correct process \pbin.Delivered $m$ in $\sigma$. Following from the definition of $\chi$, $\pi$'s echo sample for $m$ in $\sigma$ is identical to $\pi$'s echo sample for $\mu(\alpha, \sigma, m)$ in $\sigma'$. Moreover, $\pi$ received an {\tt Echo}($m$, $m$) message from $\rho$ in $\sigma$ if and only if $\pi$ received an {\tt Echo}($\mu(\alpha, \sigma, m)$, $\mu(\alpha, \sigma, m)$) message from $\rho$ in $\sigma'$. Therefore $\pi$ delivered $m$ in $\sigma$ if and only if $\pi$ delivered $\mu(\alpha, \sigma, m)$ in $\sigma'$.
\end{itemize}

Let us assume $i_{n + 1} = ({\tt State})$. Let $\pi$ be a correct process, let $m$ be a message. The following hold true:
\begin{itemize}
    \item If $(\pi, m) \in r_{n + 1}$, then $(\pi, m) \in r'_{n + 1}$. Indeed, $\pi$ delivered $m$ in $\sigma$, therefore $\pi$ delivered $\mu(\alpha, \sigma, m)$ in $\sigma'$. Moreover, $sqadv.Deliver(\ldots)$ was invoked at least once on $m$, and $perm[\mu(\alpha, \sigma, m)] = m$. Finally, $perm[\mu(\alpha, \sigma, m)]$ was returned from $sqadv.State()$, i.e., $(\pi, m) \in r'_{n + 1}$.
    \item If $(\pi, m) \in r'_{n + 1}$, then $\pi$ delivered $perm^{-1}[m]$ in $\sigma'$. Since $perm[m] = \mu^{-1}(\alpha, \sigma, m)$, $\pi$ delivered $\mu(\alpha, \sigma, m)$ in $\sigma'$. Therefore $\pi$ delivered $m$ in $\sigma$, and $(\pi, m) \in r_{n + 1}$.
\end{itemize}

Let us assume $i_{n + 1} = ({\tt Sample}, \pi, m)$. At least one correct process delivered $m$ in $\sigma$. Since no correct process is poisoned, at least one correct process \pbin.Delivered $m$ in $\sigma$, and $sqadv.Deliver(\ldots)$ was invoked at least once on $m$. Therefore, $perm[m] = \mu^{-1}(\alpha, \sigma, m)$. Moreover, from the definition of $\chi$, we have that $\pi$'s echo sample for $m$ in $\sigma$ is identical to $\pi$'s echo sample for $\mu(\alpha, \sigma, m)$ in $\sigma'$. Finally, every process that sent an {\tt Echo}($s$, $m'$) for some $m' \in \mathcal{M}$ to $\pi$ in $\sigma$ sent an {\tt Echo}($\mu(\alpha, \sigma, s)$, $m''$) for some $m'' \in \mathcal{M}$ to $\pi$ in $\sigma'$. Since $sqadv.Sample(\pi, m)$ forwards the call to $sys.Sample(\pi, perm^{-1}(m))$, we again have $r_{n + 1} = r'_{n + 1}$.

Noting that procedures $Deliver(\ldots)$ and $Echo(\ldots)$ never return a value, we trivially have that if $i_{n + 1} = ({\tt Deliver}, \pi, m)$ or $i_{n + 1} = ({\tt Echo}, \pi, s, \xi, m)$ then $r_{n + 1} = \bot = r'_{n + 1}$. By induction, we have $\tau(\alpha, \sigma) = \tau(\alpha, \delta)$.

\paragraph{Consistency of $\sigma'$}

We proved that, if $poisoned = \false$, then $\tau(\alpha, \sigma) = \tau(\alpha, \delta)$. Moreover, we proved that if a correct process $\pi$ eventually delivers a message $m$ in $\sigma$, then $\pi$ delivers $\mu(\alpha, \sigma, m)$ in $\sigma'$.

Since $\alpha$ compromises the consistency of $\sigma$, two correct processes $\pi$, $\pi'$ and two distinct messages $m$, $m' \neq m$ exist such that, in $\sigma$, $\pi$ delivered $m$ and $\pi'$ delivered $m'$. Therefore, in $\sigma'$, $\pi$ delivered $\mu(\alpha, \sigma, m)$ and $\pi'$ delivered $\mu(\alpha, \sigma, m') \neq \mu(\alpha, \sigma, m)$ (since $\mu$ is a permutation). Therefore $\alpha'$ compromises the consistency of $\sigma'$.

\paragraph{Translation permutation}

We now prove that, for any two $\sigma_a$, $\sigma_b \neq \sigma_a$, we have $\Psi[\alpha](\sigma_a) \neq \Psi[\alpha](\sigma_b)$. We prove this by contradiction. Suppose a system $\sigma'$ exists such that $\sigma' = \Psi[\alpha](\sigma_a) = \Psi[\alpha](\sigma_b)$. We want to prove that $\sigma_a = \sigma_b$.\\

Following from the definition of $\Psi[\alpha]$, if at least one correct process in $\sigma'$ is poisoned, then we immediately have $\sigma_a = \sigma' = \sigma_b$. Consequently, no correct process in $\sigma'$ is poisoned.

We start by noting that, if $\tau(\alpha, \sigma_a) = \tau(\alpha, \sigma_b)$, then $\sigma_a = \sigma_b$. Indeed, if $\tau(\alpha, \sigma_a) = \tau(\alpha, \sigma_b)$, then for every $\pi \in \Pi_C$ and every $m \in \mathcal{M}$ we have
\begin{equation*}
    \mu(\alpha, \sigma_a, m) = \mu(\alpha, \sigma_b, m)
\end{equation*}
from which immediately follows
\begin{equation*}
    \chi(\alpha, \sigma_a, m) = \chi(\alpha, \sigma_b, m)
\end{equation*}
and, since no correct process is poisoned, for every $\pi \in \Pi_C$ and every $m \in \mathcal{M}$ we have
\begin{eqnarray*}
    \sigma_a[\pi][m] &=& \sigma'[\pi][\chi^{-1}(\alpha, \sigma_a, m)] \\
    &=& \sigma'[\pi][\chi^{-1}(\alpha, \sigma_b, m)] \\
    &=& \sigma_b[\pi][m]
\end{eqnarray*}
therefore $\sigma_a = \sigma_b$.

We prove that $\tau(\alpha, \sigma_a) = \tau(\alpha, \sigma_b)$ by induction. Let us assume
\begin{eqnarray*}
    \tau(\alpha, \sigma_a) &=& ((i_1, r_1), \ldots) \\
    \tau(\alpha, \sigma_b) &=& ((i'_1, r'_1), \ldots) \\
    i_j = i'_j, r_j = r'_j && \forall j \leq n
\end{eqnarray*}
with $n \geq 0$ (here $n = 0$ means that this is $\alpha$'s first invocation). We start by noting that, since $a$ is a deterministic algorithm, we immediately have
\begin{equation*}
    i_{n + 1} = i'_{n + 1}
\end{equation*}
and we need to prove that $r_{n + 1} = r'_{n + 1}$.\\

Let us assume that $i_{n + 1} = ({\tt Byzantine}, \pi)$. As we previously established, the Byzantine processes in $\pi$'s echo samples in $\sigma'$ are identical to the Byzantine processes in $\pi$'s echo samples in $\sigma_a$ and $\sigma_b$. Therefore $r_{n + 1} = r'_{n + 1}$.

Before considering the remaining possible values of $i_{n + 1}$, we prove some auxiliary result. Let $\pi$ be a correct process, let $\rho$ be a process, let $m$, $s$ be messages. For every $j \leq n + 1$, as we established, we have $i_j = i'_j$. By hypothesis, $\alpha$ is an auto-echo adversary, so $i_j = i'_j$, $r_j = r'_j = \bot$ for every $j \leq (N - C)C^2$. Let us consider the non-trivial case $n \geq (N - C)C^2$. After the $(n + 1)$-th invocation, the following hold true:
\begin{itemize}
    \item $\rho$ sent an {\tt Echo}($m$, $m$) message to $\pi$ in $\sigma_a$ if and only if $\rho$ sent an {\tt Echo}($m$, $m$) message to $\pi$ in $\sigma_b$. Indeed, if $\rho$ is a correct process, and $\rho$ \pbin.Delivered $m$ in $\sigma_a$, then some $j \leq (n + 1)$ exists such that $i_j = ({\tt Deliver}, \rho, m)$. Since $i'_j = i_j$, $\rho$ \pbin.Delivered $m$ in $\sigma_b$ as well. If $\rho$ is Byzantine, and $\rho$ sent an {\tt Echo}($m$, $m$) message to $\pi$ in $\sigma_a$, then some $j \leq (n + 1)$ exists such that $i_j = ({\tt Echo}, \pi, m, \rho, m)$. Since $i'_j = i_j$, $\rho$ sent an {\tt Echo}($m$, $m$) message to $\pi$ in $\sigma_b$ as well. Both arguments can be reversed to prove that, if $\rho$ sent an {\tt Echo}($m$, $m$) message to $\pi$ in $\sigma_b$, then $\rho$ sent an {\tt Echo}($m$, $m$) message to $\pi$ in $\sigma_a$ as well.
    \item $\rho$ sent an {\tt Echo}($s$, $m'$) for some $m' \in \mathcal{M}$ to $\pi$ in $\sigma_a$ if and only if $\rho$ sent an {\tt Echo}($s$, $m''$) message for some $m'' \in \mathcal{M}$ to $\pi$ in $\sigma_b$. Indeed:
    \begin{itemize}
        \item If $\rho$ is correct, and it sent an {\tt Echo}($s$, $m'$) message to $\pi$ in $\sigma_a$, then it \pbin.Delivered $m'$ in both $\sigma_a$ and $\sigma_b$. Consequently, $\rho$ sent an {\tt Echo}($s$, $m'$) message to $\pi$ in $\sigma_b$ as well. The argument can be inversed to prove that, if $\rho$ is correct and it sent an {\tt Echo}($s$, $m''$) message for some $m'' \in \mathcal{M}$ to $\pi$ in $\sigma_b$, then $\rho$ sent an {\tt Echo}($s$, $m'$) message for some $m' \in \mathcal{M}$ in $\sigma_a$.
        \item If $\rho$ is Byzantine, then it sent an {\tt Echo}($m'$, $m'$) message for every $m' \in \mathcal{M}$, both in $\sigma$ and $\sigma'$.
    \end{itemize} 
    \item If at least one correct process \pbin.Delivered $m$ in $\sigma_a$ (or, equivalently, $\sigma_b$), then $\mu(\alpha, \sigma_a, m) = \mu(\alpha, \sigma_b, m)$. Indeed, let $j$ be the minimum index such that $i_j = i'_j = ({\tt Deliver}, \pi', m)$ for some $\pi' \in \Pi_C$. By definition, we have
    \begin{eqnarray*}
        \mu(\alpha, \sigma_a, m) &=& \abs{\cp{m \in \mathcal{M} \mid \exists k \leq j, \pi' \in \Pi_C: i_k = ({\tt Deliver}, \pi', m)}} \\
        &=& \abs{\cp{m \in \mathcal{M} \mid \exists k \leq j, \pi' \in \Pi_C: i'_k = ({\tt Deliver}, \pi', m)}} \\
        &=& \mu(\alpha, \sigma_b, m)
    \end{eqnarray*}
    \item $\pi$ delivered $m$ in $\sigma_a$ if and only if $\pi$ delivered $m$ in $\sigma_b$. Indeed, if $\pi$ delivered $m$ in $\sigma_a$, then at least one correct process \pbin.Delivered $m$ both in $\sigma_a$ and $\sigma_b$, and $\mu(\alpha, \sigma_a, m) = \mu(\alpha, \sigma_b, m)$. From the definition of $\chi$, we immediately get $\chi(\alpha, \sigma_a, m) = \chi(\alpha, \sigma_b, m)$ and, as we previously established, $\pi$'s echo sample for $m$ in $\sigma_a$ is identical to $\pi$'s echo sample for $m$ in $\sigma_b$. Since $\pi$ received the same {\tt Echo}($m$, $m$) messages in $\sigma_a$ and $\sigma_b$, $\pi$ delivered $m$ in $\sigma_b$ as well. The argument can be reversed to prove that, if $\pi$ delivered $m$ in $\sigma_b$, then $\pi$ delivered $m$ in $\sigma_a$ as well.
\end{itemize}

Let us assume $i_{n + 1} = ({\tt State})$. From the above it immediately follows $r_{n + 1} = r'_{n + 1}$.

Let us assume $i_{n + 1} = ({\tt Sample}, \pi, m)$. As we established, $\pi$ receives an {\tt Echo}($m$, $m'$) message for some $m' \in \mathcal{M}$ from the same set of processes in $\sigma_a$ and $\sigma_b$. Moreover, since at least one correct process \pbin.Delivered $m$ in both $\sigma_a$ and $\sigma_b$, $\pi$'s echo sample for $m$ in $\sigma_a$ is identical to $\pi$'s echo sample for $m$ in $\sigma_b$. Therefore, $r_{n + 1} = r'_{n + 1}$.

Noting that procedures $Deliver(\ldots)$ and $Echo(\ldots)$ never return a value, we trivially have that if $i_{n + 1} = ({\tt Deliver}, \pi, m)$ or $i_{n + 1} = ({\tt Echo}, \pi, s, \xi, m)$ then $r_{n + 1} = \bot = r'_{n + 1}$. By induction, we have $\tau(\alpha, \sigma_a) = \tau(\alpha, \sigma_b)$.

Therefore, $\sigma_a = \sigma_b$, which contradicts the hypothesis.

\end{proof}
\end{lemma}

\subsection{Non-redundant adversary}
\label{subsection:nonredundantadversaryproof}

\begin{algorithm}
\begin{algorithmic}[1]
\Implements
    \Instance{NonRedundantAdversary + \Cobin System}{nradv}
\EndImplements

\Uses
    \InstanceSystem{SequentialAdversary}{sqadv}{nradv}
    \Instance{\Cobin System}{sys}
\EndUses

\Procedure{nradv.Init}{{}}
    \State $deliveries = \cp{\bot}^C$;
    \State $sqadv.Init()$;
\EndProcedure

\Procedure{nradv.Step}{{}}
    \State $sqadv.Step()$;
\EndProcedure

\Procedure{nradv.Byzantine}{process}
    \State \Return $sys.Byzantine(process)$;
\EndProcedure

\Procedure{nradv.State}{{}}
    \State $state = \emptyset$;
    \State
    \ForAll{(\cdot, m)}{sys.State()}
        \ForAll{\pi}{\Pi_C}
            \State $n = 0$;
            \State
            \ForAll{\rho}{sys.Sample(\pi, m)}
                \If{$\rho \in \Pi \setminus \Pi_C$ \textbf{or} $deliveries[\rho] = m$}
                    \State $n \leftarrow n + 1$;
                \EndIf
            \EndForAll
            \State
            \If{$n \geq \hat E$}
                \State $state \leftarrow state \cup \cp{(\pi, m)}$;
            \EndIf
        \EndForAll
    \EndForAll
    \State
    \State \Return $state$;
\EndProcedure

\algstore{nonredundantdecorator}
\end{algorithmic}
\caption{Non-redundant decorator}
\label{algorithm:nonredundantdecorator}
\end{algorithm}

\begin{algorithm}
\begin{algorithmic}[1]
\algrestore{nonredundantdecorator}
\Procedure{nradv.Sample}{process, message}
    \State \Return $sys.Sample(process, message)$;
\EndProcedure

\Procedure{nradv.Deliver}{process, message}
    \State $state = \emptyset$;
    \State
    \ForAll{(\cdot, m)}{sys.State()}
        \State $state \leftarrow state \cup \cp{m}$;
    \EndForAll
    \State
    \If{$state = \cp{message}$}
        \State $sys.Deliver(process, message + 1)$;
    \Else
        \State $sys.Deliver(process, message)$;
    \EndIf
    \State
    \State $deliveries[process] = message$;
\EndProcedure

\Procedure{nradv.Echo}{process, sample, source, message}
    \State $sys.Echo(process, sample, source, message)$;
\EndProcedure

\Procedure{nradv.End}{{}}
    \State $sys.End()$;
\EndProcedure
\end{algorithmic}
\end{algorithm}

\begin{lemma}
The set of non-redundant adversaries $\mathcal{A}_{nr}$ is optimal.

\begin{proof}

We again prove the result using a decorator. Here we show that a decorator $\Delta_{nr}$ exists such that, for every $\alpha \in \mathcal{A}_{sq}$, the adversary $\alpha' = \Delta_{sq}(\alpha)$ is a non-redundant adversary, and more powerful than $\alpha$. If this is true, then indeed the lemma is proved: let $\alpha^*$ be an optimal adversary, then the sequential $\alpha^+ = \Delta_{nr}(\alpha^*)$ is optimal as well.

\paragraph{Decorator}

\cref{algorithm:nonredundantdecorator} implements \textbf{Non-redundant decorator}, a decorator that transforms a sequential adversary into a non-redundant adversary. Provided with a sequential adversary $sqadv$, Non-redundant decorator acts as an interface between $sqadv$ and a system $sys$, effectively implementing a non-redundant adversary $nradv$. Non-redundant decorator exposes both the adversary and the system interfaces: the underlying adversary $sqadv$ uses $nradv$ as its system.

Non-redundant decorator works as follows:
\begin{itemize}
    \item Procedure $nradv.Init()$ initializes a $deliveries$ array that is used to keep track of the message each correct process would have delivered, if $sqadv$ was playing instead of $nradv$.
    \item Procedure $nradv.Step()$ simply forwards the call to $sqadv.Step()$;
    \item Procedure $nradv.Byzantine(process)$ simply forwards the call to \\ $sqadv.Byzantine(process)$.
    \item Procedure $nradv.State()$ returns a list of pairs $(\pi \in \Pi_C, m \in \mathcal{M})$ such at least one correct process delivered $m$ in $sys$, and $\pi$ would have delivered $m$ in $sys$, if $sqadv$ was playing instead of $nradv$.

    This is achieved by querying $sys.State()$, then looping over each message $m$ in the response. For every $\pi \in \Pi_C$, the procedure loops over every element $\rho$ of $sys.Sample(\pi, m)$, and computes the number $n$ of {\tt Echo}($m$, $m$) messages that $\pi$ would have received from its echo sample for $m$ in $sys$, if $sqadv$ was playing instead of $nradv$. This is achieved using the $deliveries$ table, and the hypothesis that $sqadv$ is an auto-echo adversary. If $n$ is greater or equal to $\hat E$, $(\pi, m)$ is included in the list returned by the procedure.

    \item Procedure $nradv.Sample(process, message)$ simply forwards the call to $sys.Sample(process, message)$.
    \item Procedure $nradv.Deliver(process, message)$ uses $sys.State()$ to determine which messages have been delivered by at least one correct process in $sys$. If $message$ is the only message that was delivered, the procedure forwards the call to $sys.Deliver(process, message + 1)$. Otherwise, it forwards the call to $sys.Deliver(process, message)$. Finally, it updates the $deliveries$ array to reflect the fact that $process$ would have \pbin.Delivered $message$ in $sys$, if $sqadv$ was playing instead of $nradv$.
    \item Procedure $nradv.Echo(process, sample, source, message)$ simply forwards the call to $sys.Echo(process, sample, source, message)$.
    \item Procedure $nradv.End()$ simply forwards the call to $sys.End()$.
\end{itemize}

\paragraph{Correctness}

We start by proving that no adversary, coupled with Non-redundant decorator, causes the execution to fail.

Let $\pi \in \Pi_C$, let $m \in \mathcal{M}$. The following hold true:
\begin{itemize}
    \item Procedure $nradv.State()$ never causes the execution to fail. Indeed, $sys.Sample(\pi, m)$ is called only if $(\pi', m)$ was returned from $sys.State()$, for some $\pi' \in \Pi_C$. This means that $sys.Sample(\pi, m)$ is called only if at least one correct process delivered $m$ in $sys$.
    \item No invocation of $nradv.Sample(\ldots)$ causes the execution to fail. Noting that $sqadv$ is correct, it will never invoke $nradv.Sample(\pi, m)$ unless $(\pi', m)$ was returned from a previous invocation of $nradv.State()$, for some $\pi' \in \Pi_C$. Moreover, $(\pi', m)$ is returned from $nradv.State()$ is and only if, for some $\pi'' \in \Pi_c$, $(\pi'', m)$ is returned from $sys.State()$. Therefore, $sys.Sample(\pi, m)$ is never invoked unless at least one correct process delivered $m$ in $sys$.
    \item Procedure $nradv.Deliver(\ldots)$ never calls $sys.Deliver(\ldots)$ on a message greater than $C$. Let $m \in \mathcal{M}$. If $m$ is the only message that was delivered in $sys$, then no correct process is poisoned in $sys$: indeed, as we proved, if a correct message was poisoned in $sys$, it would have delivered every message. Therefore, at least one correct process \pbin.Delivered $m$. Moreover, since $sqadv$ is a sequential adversary, it invokes $nradv.Deliver(\ldots)$ for the $n$-th time only on a message $m \leq n$. Since $nradv.Deliver(\ldots)$ is invoked at most $C$ times, we have $n \leq C$, and, since $m$ was delivered as a result of a previous invocation of $nradv.Deliver(\ldots)$, we have $m \leq n - 1$. Consequently, $m + 1 \leq C$.
\end{itemize}

We further prove that $nradv$ is a sequential adversary. Let $\pi \in \Pi_C$, let $m \in \mathcal{M}$. Since $sqadv$ is sequential, it invokes $nradv.Deliver(\pi, m)$ only if it previously invoked $nradv.Deliver(\ldots)$ on every message $l < m \in \mathcal{M}$. Therefore, $nradv$ can be a non-sequential adversary only as a result of a call to $sys.Deliver(\pi, m + 1)$. If $m$ is the only message that was delivered by at least one correct process in $sys$, then no correct process is poisoned in $sys$. Therefore, if $sys.Deliver(\pi, m + 1)$ is invoked, then, as we established, at least one correct process \pbin.Delivered $m$ in $sys$. Noting that the set of messages that are delivered by at least one correct process in $sys$ is non-decreasing, if no correct process is poisoned in $sys$ then $sqadv$ invoked $nradv.Deliver(\ldots)$ on $m$ at least once when no correct process had delivered $m$. Consequently, for every $l < m$, $nradv.Deliver(\ldots)$, and as a result $sys.Deliver(\ldots)$, was invoked on $l$.

\paragraph{Non-redundant}

It is easy to prove that Non-redundant decorator always implements a non-redundant adversary. Indeed, let $\pi \in \Pi_C$, let $m \in \mathcal{M}$, $sys.Deliver(\pi, m)$ is never invoked if $m$ is the only message that was delivered.

\paragraph{Roadmap}

Let $\alpha \in \mathcal{A}_{sq}$, let $\alpha' = \Delta_{nr}(\alpha)$. Let $\sigma$ be a system such that $\alpha$ compromises the consistency of $\sigma$. Let $\sigma'$ be an identical copy of $\sigma$. In order to prove that $\alpha'$ is more powerful than $\alpha$, we prove that $\alpha'$ compromises the consistency of $\sigma'$.

\paragraph{Trace}

We start by noting that, if we couple Non-redundant decorator with $\sigma'$, we effectively obtain a system instance $\delta$ with which $\alpha$ directly exchanges invocations and responses. Here we show that the trace $\tau(\alpha, \sigma)$ is identical to the trace $\tau(\alpha, \delta)$. Intuitively, this means that $\alpha$ has no way of \emph{distinguishing} whether it has been coupled directly with $\sigma$, or it has been coupled with $\sigma'$, with Non-redundant decorator acting as an interface. We prove this by induction.

Let us assume
\begin{eqnarray*}
    \tau(\alpha, \sigma) &=& ((i_1, r_1), \ldots) \\
    \tau(\alpha, \delta) &=& ((i'_1, r'_1), \ldots) \\
    i_j = i'_j, r_j = r'_j && \forall j \leq n
\end{eqnarray*}

We start by noting that, since $\alpha$ is a deterministic algorithm, we immediately have
\begin{equation*}
    i_{n + 1} = i'_{n + 1}
\end{equation*}
and we need to prove that $r_{n + 1} = r'_{n + 1}$. \\

Let us assume that $i_{n + 1} = ({\tt Byzantine}, \pi)$. Since $nradv.Byzantine(\pi)$ simply forwards the call to $sys.Byzantine(\pi)$, and $\sigma'$ is an identical copy of $\sigma$, we immediately have $r_{n + 1} = r'_{n + 1}$. \\

Before considering the remaining possible values of $i_{n + 1}$, we prove some auxiliary results. Let $\pi$ be a correct process, let $\rho$ be a process, let $m$ be a message . For every $j \leq n + 1$, as we established, we have $i_j = i'_j$. Therefore, after the $(n + 1)$-th invocation, the following hold true:
\begin{itemize}
    \item $\pi$ \pbin.Delivered $m$ in $\sigma$ if and only if $deliveries[\pi] = m$. This follows immediately from the fact that, whenever $nradv.Deliver(\pi, m)$ is invoked, $deliveries[\pi]$ is set to $m$.
    \item $\pi$ \pbin.Delivered a message in $\sigma$ if and only if $\pi$ \pbin.Delivered a message in $\sigma'$. This follows immediately from the fact that every $nradv.Deliver(\pi, \allowbreak m)$ is always either forwarded to $sys.Deliver(\pi, m)$ or $sys.Deliver(\pi, \allowbreak m + 1)$.
    \item If $m$ was delivered by at least one correct process in $\sigma$, then $m$ was delivered by at least one correct process in $\sigma'$ as well. Indeed:
    \begin{itemize}
        \item If at least one correct process is poisoned, then it delivered every message both in $\sigma$ and $\sigma'$.
        \item If no correct process is poisoned then, for some $j^* \leq n + 1$, after the $j$-th invocation, exactly one message $m^*$ was delivered by at least one correct process in $\sigma$. This follows from the fact that a non-poisoned process delivers $m$ only as a result of receiving an {\tt Echo}($m$, $m$) message, and no two messages {\tt Echo}($m$, $m$), {\tt Echo}($m' \neq m$, $m'$) are ever issued as a result of a single invocation.
        \item If no correct process is poisoned, and $m = m^*$, then some correct process $\pi^*$ delivered $m$ in $\sigma$ as a result of the $j^*$-th invocation. It is easy to see that, up to the $j^*$-th invocation, every call to $nradv.Deliver(\pi, m)$ was simply forwarded to $sys.Deliver(\pi, m)$. Therefore, noting that $\pi^*$'s echo sample for $m$ is identical in $\sigma$ and $\sigma'$, $\pi^*$ delivered $m$ in $sys$ as well.
        \item If no correct process is poisoned, and $m \neq m^*$, then no invocation of $nradv.Deliver(\ldots)$ sees $m$ as the only message delivered by at least one correct process in $sys$. Therefore, all calls to $nradv.Deliver(\pi, m)$ are simply forwarded to $sys.Deliver(\pi, m)$. Consequently, noting that $\pi$'s echo sample for $m$ is identical in $\sigma$ and $\sigma'$, if $\pi$ delivered $m$ in $\sigma$, then $\pi$ delivered $m$ in $\sigma'$ as well.
    \end{itemize}
\end{itemize}

Let us assume that $i_{n + 1} = ({\tt State})$. Let $\pi$ be a correct process, let $m$ be message. The following hold true:
\begin{itemize}
    \item If $(\pi, m) \in r_{n + 1}$, then $\pi$ delivered $m$ in $\sigma$. Therefore, at least one correct process delivered $m$ in $\sigma'$. Let $\pi'$ be a correct process in $\pi$'s echo sample for $m$ that \pbin.Delivered $m$ in $\sigma$: as we established, we have $deliveries[\pi'] = m$ and $\pi'$ \pbin.Delivered a message in $\sigma'$. Therefore, $\pi' \in sys.Sample(\pi, m)$. Since $nradv.State(\ldots)$ counts the processes in $sys.Sample(\pi, m)$ that are either Byzantine or have their $deliveries$ value set to $m$, we have $(\pi, m) \in r'_{n + 1}$.
    \item If $(\pi, m) \notin r_{n + 1}$, then less than $\hat E$ processes in $\pi$'s echo sample for $m$ are either Byzantine or have \pbin.Delivered $m$. Therefore, less than $\hat E$ processes in $\pi$'s echo sample are either Byzantine or have their $deliveries$ value set to $m$. Since $sys.Sample(\pi, m)$ is a subset of $\pi$'s echo sample for $m$, and since $nradv.State(\ldots)$ counts the processes in $sys.Sample(\pi, m)$ that are either Byzantine or have their $deliveries$ value set to $m$, $(\pi, m) \notin r'_{n + 1}$.
\end{itemize}

Let us assume that $i_{n + 1} = ({\tt Sample}, \pi, n)$. By hypothesis, $\pi$'s echo sample for $m$ is identical in $\sigma$ and $\sigma'$. Moreover, the set of processes that \pbin.Delivered a message is identical in $\sigma$ and $\sigma'$. Noting that $nradv.Sample(\pi, \allowbreak m)$ simply forwards the call to $sys.Sample(\pi, \allowbreak m)$, we immediately get $r_{n + 1} = r'_{n + 1}$.

Noting that procedures $Deliver(\ldots)$ and $Echo(\ldots)$ never return a value, we trivially have that if $i_{n + 1} = ({\tt Deliver}, \pi, m)$ or $i_{n + 1} = ({\tt Echo}, \pi, s, \xi, m)$ then $r_{n + 1} = \bot = r'_{n + 1}$. By induction, we have $\tau(\alpha, \sigma) = \tau(\alpha, \delta)$.

\paragraph{Consistency of $\sigma'$}

We proved that $\tau(\alpha, \sigma) = \tau(\alpha, \delta)$. Moreover, we proved that if a message $m$ is eventually delivered by at least a correct process in $\sigma$, then $m$ is eventually delivered by at least a correct process in $\sigma'$ as well.

Since $\alpha$ compromises the consistency of $\sigma$, two distinct messages $m$, $m' \neq m$ exist such that, in $\sigma$, both $m$ and $m'$ are delivered by at least one correct process. Therefore, in $\sigma'$, both $m$ and $m'$ are delivered by at least one correct process as well. Therefore, $\alpha'$ compromises the consistency of $\sigma'$.

Consequently, the adversarial power of $\alpha$ is smaller or equal to the adversarial power of $\alpha' = \Delta_{nr}(\alpha)$, and the lemma is proved.
\end{proof}
\end{lemma}

\subsection{Sample-blind adversary}
\label{subsection:sampleblindadversaryproof}

\begin{algorithm}
\begin{algorithmic}[1]
\Implements
    \Instance{SampleMaskedAdversary + \Cobin System}{smadv}
\EndImplements

\Uses
    \InstanceSystem{NonRedundantAdversary}{nradv}{smadv}
    \Instance{\Cobin System}{sys}
\EndUses

\Procedure{smadv.Init}{{}}
    \State $index = \bot$; \tabto*{3.5cm} $cache = \bot$;
    \State $trace = [\,]$;
    \State $deliveries = \cp{\bot}^C$;
    \State $nradv.Init()$;
\EndProcedure

\Procedure{optimize}{process, message}
    \State $smadv.Byzantine(process)$;
    \State $best.sample = \bot$; \tabto*{4cm} $best.probability = 0$;
    \State
    \ForAll{sample}{\Pi^E}
        \State $systems = 0$;
        \State $compromissions = 0$;
        \ForAll{\sigma}{\mathcal{S}}
            \If{$trace \sim \sigma$ \textbf{and} $\sigma[process][message] = sample$}
                \State $systems \leftarrow systems + 1$;
                \If{$NonRedundantAdversary \searrow \sigma$}
                    \State $compromissions \leftarrow compromissions + 1$;
                \EndIf
            \EndIf
        \EndForAll
        \State
        \If{$systems > 0$ \textbf{and} $compromissions / systems > best.probability$}
            \State $best.sample = sample$;
            \State $best.probability = compromissions / systems$;
        \EndIf
    \EndForAll
    \State
    \State \Return $best.sample$;
\EndProcedure

\algstore{samplemaskingdecorator}
\end{algorithmic}
\caption{Sample-masking decorator}
\label{algorithm:sampleblinddecorator}
\end{algorithm}

\begin{algorithm}
\begin{algorithmic}[1]
\algrestore{samplemaskingdecorator}

\Procedure{smadv.Step}{{}}
    \State $nradv.Step()$;
\EndProcedure

\Procedure{smadv.Byzantine}{process}
    \State $trace \leftarrow trace + \qp{(\texttt{Byzantine}, process, sys.Byzantine(process))}$;
    \State \Return $sys.Byzantine(process)$;
\EndProcedure

\Procedure{smadv.State}{{}}
    \State $trace \leftarrow trace + \qp{(\texttt{State}, sys.State())}$;
    \State $state = sys.State() \setminus \cp{index}$;
    \State
    \If{$index \neq \bot$}
        \State $(\pi, m) = index$;
        \State
        \State $n = 0$;
        \ForAll{\rho}{cache}
            \If{$\rho \in \Pi \setminus \Pi_C$ \textbf{or} $deliveries[\rho] = m$}
                \State $n \leftarrow n + 1$;
            \EndIf
        \EndForAll
        \State
        \If{$n \geq \hat E$}
            \State $state \leftarrow state \cup \cp{(\pi, m)}$;
        \EndIf
    \EndIf
    \State
    \State \Return $state$;
\EndProcedure

\algstore{samplemaskingdecorator}
\end{algorithmic}
\end{algorithm}

\begin{algorithm}
\begin{algorithmic}[1]
\algrestore{samplemaskingdecorator}

\Procedure{smadv.Sample}{process, message}
    \If{$index = \bot$}
        \State $index \leftarrow (process, message)$;
        \State $cache \leftarrow optimize(process, message)$;
    \EndIf
    \State
    \If{$(process, message) = index$}
        \State $sample = \cp{}$;
        \ForAll{\pi}{cache}
            \If{$deliveries[\pi] \neq \bot$}
                \State $sample \leftarrow sample \cup \cp{\pi}$;
            \EndIf
        \EndForAll
        \State \Return $sample$;
    \Else
        \State \Return $sys.Sample(process, message)$;
    \EndIf
\EndProcedure

\Procedure{smadv.Deliver}{process, message}
    \State $trace \leftarrow trace + \qp{(\texttt{Deliver}, (process, message))}$;
    \State $deliveries[process] = message$;
    \State $sys.Deliver(process, message)$;
\EndProcedure

\Procedure{smadv.Echo}{process, sample, source, message}
    \State $trace \leftarrow trace + \qp{(\texttt{Echo}, (process, sample, source, message))}$;
    \State $sys.Echo(process, sample, source, message)$;
\EndProcedure

\Procedure{smadv.End}{{}}
    \State $sys.End()$;
\EndProcedure
\end{algorithmic}
\end{algorithm}

\begin{lemma}
The set of sample-blind adversaries $\mathcal{A}_{sb}$ is optimal.

\begin{proof}
We again prove the result using a decorator. Here we show that a decorator $\Delta_{sm}$ exists such that, for every $\alpha \in \mathcal{A}_{nr}$, the adversary $\alpha' = \Delta^{C^2}_{sm}(\alpha)$ is a sample-blind adversary, and more powerful than $\alpha$. If this is true, then the lemma is proved: let $\alpha^*$ be an optimal adversary, then the sample-blind $\alpha^+ = \Delta^{C^2}_{sm}(\alpha^*)$ is optimal as well.

\paragraph{Decorator}

\cref{algorithm:sampleblinddecorator} implements \textbf{Sample-masking decorator}, a decorator that masks every invocation of $Sample(\pi, m)$ issued by a non-redundant adversary, if $Sample(\pi, m)$ is the first invocation of $Sample(\ldots)$ issued by that adversary. 

Provided with a non-redundant adversary $nradv$, Sample-masking decorator acts as an interface between $nradv$ and a system $sys$. Sample-masking decorator is only guaranteed to mask any invocation to $sys.Sample(\pi, m)$, for one process $\pi$ and one message $m$. Noting that $\abs{\Pi_C} = C$ and $\abs{\mathcal{M}} = C$, we have that, for every $\alpha \in \mathcal{A}_{nr}$, $\alpha' = \Delta^{C^2}_{sm}(\alpha)$ is a sample-blind adversary: indeed, all of $\alpha$'s $C^2$ possible calls to $Sample(\ldots)$ are necessarily masked.

Sample-masking decorator exposes both the adversary and the system interfaces: the underlying adversary $nradv$ uses $smadv$ as its system. Sample-masking decorator works as follows:
\begin{itemize}
    \item Procedure $smadv.Init()$ initializes the following variables:
    \begin{itemize}
        \item An $index$ and a $cache$ variable, both initially set to $\bot$: $index$ is used to store the pair $(\pi \in \Pi_C, m \in \mathcal{M})$ that was provided as argument to the first invocation of $smadv.Sample(\ldots)$; $cache$ is used to store the content of the echo sample $smadv$ generates for $(\pi, m)$ when $smadv.Sample(\ldots)$ is invoked for the first time. This guarantees that subsequent invocations of $smadv.Sample(\pi, m)$ are provided with consistent responses throughout the entire adversarial execution.
        \item A $trace$ array: $trace$ is used to store the sequence of invocations and responses exchanged between $nradv$ and $sys$.
        \item A $deliveries$ array of $C$ elements: $deliveries$ is used to track the message \pbin.Delivered in $sys$ by each correct process.
    \end{itemize}
    \item Procedure $optimize(process, message)$ returns the sample $sample$ for \\ $(process, message)$ that maximizes the probability of $nradv$ winning against a random system $\sigma$ that is compatible with $trace$, and satisfies \\ $\sigma[process][message] = sample$. This is achieved as follows:
    \begin{itemize}
        \item The procedure calls $smadv.Byzantine(process)$, causing an invocation to $sys.Byzantine(process)$ to be appended to $trace$ along with its response. This is necessary because, if $smadv.Byzantine(\allowbreak process)$ was never invoked before, the set of Byzantine processes in the generated $sample$ might differ from the Byzantine processes in $process$' echo sample for $message$ in $sys$. Noting that all of $process$' echo samples in $sys$ share the same set of Byzantine processes, a subsequent call to $smadv.Byzantine(process)$ could return a set of Byzantine processes that is inconsistent with the $sample$, causing undefined behavior on $nradv$.
        \item The procedure loops over every possible value of $sample$. For each value of $sample$, it counts the number $systems$ of systems $\sigma$ that are compatible with $trace$, and satisfy $\sigma[process][message] = sample$. Among the systems that satisfy those two constraints, the procedure counts the number $compromissions$ of systems whose consistency the adversary would compromise.
        \item The procedure returns the value of $sample$ that satisfies $systems > 0$, and maximizes $compromissions / systems$. In other words, the procedure returns a sample $sample$ that is compatible with at least with one of the systems that are compatible with $trace$, and maximizes the probability that the adversary would compromise the consistency of a randomly selected system compatible with $trace$, picked among those where $process$' echo sample for $message$ is $sample$.
    \end{itemize}
    \item Procedure $smadv.Byzantine(process)$ appends to $trace$ the invocation of $sys.Byzantine(process)$ along with its response. It then forwards the call to $sys.Byzantine(process)$.
    \item Procedure $smadv.State()$ appends to $trace$ the invocation of $sys.State()$ along with its response. It then returns the response of $sys.State()$, modified to be compatible with any previous masked invocation of $sys.Sample(\ldots)$. More specifically, if $index = (\pi, m) \neq \bot$ (i.e., $nradv$'s first invocation of $smadv.Sample(\ldots)$ was $smadv.Sample(\pi, \allowbreak m)$), then $(\pi, m)$ is included in the set of pairs returned by $smadv.\allowbreak State()$ only if $\pi$ would have delivered $m$ in $sys$, if $\pi$'s echo sample for $m$ was $cache$. This is achieved by looping over every process in $cache$, and counting the number $n$ of those processes that are either Byzantine, or \pbin.Delivered $m$ (this is achieved using the $deliveries$ array).
    \item Procedure $smadv.Sample(process, message)$ determines whether \\ $smadv.Sample(\ldots)$ has ever been invoked before by checking the value of $index$. If it has not, it sets $index$ to $(process, message)$, and generates a sample for $(process, message)$ by setting $cache$ to the value returned by $optimize(\allowbreak process, \allowbreak sample)$. 

    If $(process, message)$ is equal to $index$, the procedure returns the set of processes in $cache$ that \pbin.Delivered a message in $sys$. This is achieved by looping over every process $\rho$ in $cache$, and adding $\rho$ to the response if $\rho$ is either Byzantine, or satisfy $deliveries[\rho] \neq \bot$.
    
    If $(process, message)$ is not equal to $index$, the call is forwarded to \\ $sys.Sample(process, message)$.
    \item Procedure $smadv.Deliver(process, message)$ appends to $trace$ the invocation of $sys.Deliver(process, message)$. To reflect the fact that $process$ \pbin.Delivered $m$ in $sys$, it then updates the $deliveries$ array. Finally, it forwards the call to $sys.Deliver(process, message)$.

    \item Procedure $smadv.Echo(process, sample, source, message)$ appends to $trace$ the invocation of $sys.Echo(process, sample, source, message)$. It then forwards the call to $sys.Echo(process, sample, source, message)$.
    \item Procedure $smadv.End()$ simply forwards the call to $sys.End()$.
\end{itemize}

\paragraph{Correctness}
We start by proving that no adversary has undefined behavior when coupled with {\tt Sample-masked decorator}. An adversary has undefined behavior if, at any point, the sequence of invocations and responses it exchanges with $smadv$ is incompatible with every system. \\

Let $\pi \in \Pi_C$, let $m \in \mathcal{M}$, let us assume that the first invocation to $smadv.Sample(\ldots)$ is $smadv.Sample(\pi, m)$.
We start by noting that every invocation in $smadv$ is forwarded to the corresponding invocation in $sys$ except for $smadv.State()$ and $smadv.Sample(\pi, m)$. Moreover, before the first invocation of $smadv.Sample(\pi, m)$, $index$ is set to $\bot$ and, as a result, $smadv.State()$ effectively forwards to $sys.State()$. Therefore, the trace exchanged between $nradv$ and $smadv$ is trivially compatible with $sys$ before the first invocation of $smadv.Sample(\pi, m)$.

When $smadv.Sample(\pi, m)$ is invoked for the first time, $cache$ is set to $optimize(\pi, m)$. When $optimize(\pi, m)$ is called, it calls $smadv.Byzantine(\pi)$, which appends the invocation and the corresponding response to $trace$. After that, the set of systems that are compatible with $trace$ is non empty, as it trivially includes $sys$. The procedure $optimize(\pi, m)$ returns a sample $sample$ only if at least one system $\sigma$ is compatible with $trace$, and satisfies $\sigma[\pi][m] = sample$. Since $sys.Byzantine(\pi)$ is in $trace$, the Byzantine component of $sample$ is identical to $sys.Byzantine(\pi)$: indeed, any system $\sigma$ where the Byzantine component of $\sigma[\pi][m]$ is different from $sys.Byzantine(\pi)$ is incompatible with $\sigma$. 

Therefore, the system obtained by replacing $\pi$'s echo sample for $m$ in $sys$ with $cache$ is a valid system, and it is compatible with $trace$ up to the first invocation of $smadv.Sample(\pi, m)$. Moreover, $trace$ will always be compatible with such system. Indeed:
\begin{itemize}
    \item Every subsequent call to $smadv.Sample(\pi, m)$ uses the $deliveries$ table to determine which processes in $cache$ \pbin.Delivered a message in $sys$, thus returning a response that is consistent with $\pi$'s echo sample for $m$ being $cache$.
    \item Every subsequent call to $smadv.State()$ includes $(\pi, m)$ in its response only if at least $\hat E$ processes in $cache$ are either Byzantine or \pbin.Delivered $m$ in $sys$ (this is verified using the $deliveries$ table).
\end{itemize}
This proves that that no adversary, coupled with {\tt Sample-masked decorator}, has undefined behavior.
\paragraph{Sample-blind}

It is easy to see that Sample-masking decorator masks the first invocation to $Sample(\ldots)$ issued by the decorated adversary. Indeed, if $smadv.Sample(\pi, m)$ is the first invocation of $smadv.Sample(\ldots)$ issued by $nradv$, then $index$ is set to $(\pi, m)$, and $sys.Sample(\pi, m)$ is never be invoked.

Let $\alpha$ be a non-redundant adversary, we have that $\Delta_{sb}(\alpha)$ issues calls to $Sample(\ldots)$ for at most $C^2 - 1$ pairs $(\pi' \in \Pi_C, m' \in \mathcal{M})$. The same argument can be applied again to see that, by composing Sample-masking decorator with itself $C^2$ times, all possible calls to $Sample(\ldots)$ are masked. Therefore, $\alpha' = \Delta^{C^2}_{sb}(\alpha)$ is a sample-blind adversary.

\paragraph{Sample replacement}

Let $\alpha$ be an adversary, let $\sigma$ be a system. We define a function $\nu: \mathcal{A} \times \mathcal{S} \rightarrow \mathbb{N} \cup \cp{\bot}$ by
\begin{equation*}
    \nu(\alpha, \sigma) = \min \: n \mid \left( \tau(\alpha, \sigma)_n = (({\tt Sample}, \pi \in \Pi_C, m \in \mathcal{M}), \bot)\right)
\end{equation*} 
Intuitively, $\nu(\alpha, \sigma)$ returns the index of the first invocation of $Sample(\ldots)$ in $\tau(\alpha, \sigma)$ if such invocation exists, and $\bot$ otherwise. We additionally define $\pi(\alpha, \sigma)$ and $m(\alpha, \sigma)$ by
\begin{equation*}
    \tau(\alpha, \sigma)_{\nu(\alpha, \sigma)} = (({\tt Sample}, \pi(\alpha, \sigma), m(\alpha, \sigma)), \bot)
\end{equation*}
if $\nu(\alpha, \sigma) \neq \bot$, and by
\begin{equation*}
    \pi(\alpha, \sigma) = m(\alpha, \sigma) = \bot
\end{equation*}
if $\nu(\alpha, \sigma) = \bot$. Whenever at least an invocation to $Sample(\ldots)$ is issued when $\alpha$ is coupled with $\sigma$, $\pi(\alpha, \sigma)$ and $m(\alpha, \sigma)$ are the arguments to that invocation.

We then define $\sigma^-: \mathcal{A} \times \mathcal{S} \rightarrow \mathbb{N} \cup \cp{\bot}$, $\sigma^+: \mathcal{A} \times \mathcal{S} \rightarrow \mathbb{N} \cup \cp{\bot}$.
If $\nu(\alpha, \sigma) \neq \bot$
\begin{eqnarray*}
\sigma^-(\alpha, \sigma) &=&
    \max n < \nu(\alpha, \sigma) \mid \tau(\alpha, \sigma)_n = (({\tt State}), r_n), \cancel{\Psi(r_n, \alpha, \sigma)} \\
\sigma^+(\alpha, \sigma) &=&
    \min n < \nu(\alpha, \sigma) \mid \tau(\alpha, \sigma)_n = (({\tt State}), r_n),  \Psi(r_n, \alpha, \sigma)
\end{eqnarray*}
where $\Psi$ is a predicate defined as
\begin{eqnarray*}
\Psi(r_n, \alpha, \sigma) = (\pi(\alpha, \sigma), m(\alpha, \sigma)) \in r_n
\end{eqnarray*}
Otherwise, i.e. if $\nu(\alpha, \sigma) = \bot$
\begin{equation*}
    \sigma^-(\alpha, \sigma) = \sigma^+(\alpha, \sigma) = \bot
\end{equation*}
otherwise. Intuitively, when $\nu(\alpha, \sigma) \neq \bot$: $\sigma^-(\alpha, \sigma)$ returns the index of the last invocation of $State()$ prior to $\nu(\alpha, \sigma)$ that did not include $(\pi(\alpha, \sigma), \allowbreak m(\alpha, \sigma))$ in its response; $\sigma^+(\alpha, \sigma)$ returns the index of the first invocation of $State()$ prior to $\nu(\alpha, \sigma)$ that included $(\pi(\alpha, \sigma), m(\alpha, \sigma))$ in its response.

We additionally define $\delta: \mathcal{A} \times \mathcal{S} \times \mathcal{M} \times \mathbb{N} \rightarrow \powerset{\Pi_C}{}$ by
\begin{equation}
    \pi \in \delta(\alpha, \sigma, m, n) \xLeftrightarrow[]{def} \exists j < n \mid \tau(\alpha, \sigma)_j = (({\tt Deliver}, \pi, m), \bot)
\end{equation}
Intuitively, $\pi$ is in $\delta(\alpha, \sigma, m, n)$ if $\alpha$ invokes $Deliver(\pi, m)$ before the $n$-th invocation it issues, when coupled with $\sigma$. In other words, $\delta(\alpha, \sigma, m, n)$ represents the set of correct processes that \pbin.Deliver $m$ before the $n$-th invocation, when $\alpha$ is coupled with $\sigma$.

Finally, we define $\delta^-: \mathcal{A} \times \mathcal{S} \rightarrow \powerset{\Pi_C}{}$, $\delta^+: \mathcal{A} \times \mathcal{S} \rightarrow \powerset{\Pi_C}{}$ by
\begin{eqnarray*}
\delta^-(\alpha, \sigma) =
\begin{cases}
\delta(\alpha, \sigma, m(\alpha, \sigma), \sigma^-(\alpha, \sigma))) &\text{iff}\; \sigma^-(\alpha, \sigma) \neq \bot \\
\emptyset &\text{otherwise}
\end{cases} \\
\delta^+(\alpha, \sigma) =
\begin{cases}
\delta(\alpha, \sigma, m(\alpha, \sigma), \sigma^+(\alpha, \sigma))) &\text{iff}\; \sigma^+(\alpha, \sigma) \neq \bot \\
\Pi_C &\text{otherwise}
\end{cases}
\end{eqnarray*}
Intuitively: 
\begin{itemize}
    \item When $\sigma^-(\alpha, \sigma) \neq \bot$, $\delta^-(\alpha, \sigma)$ represents the set of processes that \pbin.Delivered $m(\alpha, \sigma)$ before $\sigma^-(\alpha, \sigma)$. Intuitively, $\delta^-$ is designed to guarantee that less than $\hat E$ elements of $\sigma[\pi(\alpha, \sigma)][m(\alpha, \sigma)]$ are either Byzantine or included in $\delta^-(\alpha, \sigma)$. If this was not the case, the $\sigma^-(\alpha, \sigma)$-th invocation of $State()$ would have included $(\pi(\alpha, \sigma), \allowbreak m(\alpha, \sigma))$ in its response.
    \item When $\sigma^+(\alpha, \sigma) \neq \bot$, $\delta^+(\alpha, \sigma)$ represents the set of processes that \pbin.Delivered $m(\alpha, \sigma)$ before $\sigma^+(\alpha, \sigma)$. Intuitively, $\delta^+$ is designed to guarantee that at least $\hat E$ elements of $\sigma[\pi(\alpha, \sigma)][m(\alpha, \sigma)]$ are either Byzantine or included in $\delta^+(\alpha, \sigma)$. If this was not then case, the $\sigma^+(\alpha, \sigma)$-th invocation of $State()$ would not have included $(\pi(\alpha, \sigma),\allowbreak m(\alpha, \sigma))$ in its response.
\end{itemize}

All the above definitions allow us to define a \textbf{sample replacement function} $\mathcal{E}[\alpha]: \mathcal{S} \rightarrow \powerset{\Pi^E}{}$ by
\begin{equation*}
    \mathcal{E}[\alpha](\sigma) = \emptyset
\end{equation*}
if $\nu(\alpha, \sigma) = \bot$ and
\begin{equation*}
    \bar E \in \mathcal{E}[\alpha](\sigma) \xLeftrightarrow[]{def}
    \begin{cases}
        \sigma[\pi(\alpha, \sigma)][m(\alpha, \sigma)][n] \in \Pi \setminus \Pi_C \implies \\
        \quad (\bar E[n] = \sigma[\pi(\alpha, \sigma)][m(\alpha, \sigma)][n]) \\
        \abs{\cp{n \in 1..E \mid \bar E[n] \in \delta^-(\alpha, \sigma) \cup (\Pi \setminus \Pi_C)}} < \hat E \\
        \abs{\cp{n \in 1..E \mid \bar E[n] \in \delta^+(\alpha, \sigma) \cup (\Pi \setminus \Pi_C)}} \geq \hat E
    \end{cases}
\end{equation*}
otherwise. Intuitively, $\mathcal{E}[\alpha]$ is designed so that, if $\alpha$ is non-redundant, when $\nu(\alpha, \sigma) \neq \bot$, a sample $E$ is in $\mathcal{E}[\alpha](\sigma)$ if, by replacing $\pi(\alpha, \sigma)$'s echo sample for $m(\alpha, \sigma)$ in $\sigma$ with $E$, we obtain a system $\sigma'$ that is \emph{interchangeable} with $\sigma$, i.e., a system that cannot be distinguished from $\sigma$ up to the $\nu(\alpha, \sigma)$-th invocation, and whose consistency is compromised by the same set of traces. We prove these two properties in the next section of this proof.

More specifically, a sample $\bar E$ is in $\mathcal{E}[\alpha](\sigma)$ if it satisfies the following conditions:
\begin{itemize}
    \item $\bar E$ shares the set of Byzantine processes in $\sigma[\pi(\alpha, \sigma)][m(\alpha, \sigma)]$.
    \item Less than $\hat E$ processes in $\bar E$ \pbin.Deliver $m(\alpha, \sigma)$ before the last invocation of $State(\ldots)$ in $\tau(\alpha, \sigma)$ (before $\nu(\alpha, \sigma)$) that does not include $(\pi(\alpha, \sigma), m(\alpha, \sigma))$ in its response.
    \item At least $\hat E$ processes in $\bar E$ \pbin.Deliver $m(\alpha, \sigma)$ before the first invocation of $State(\ldots)$ in $\tau(\alpha, \sigma)$ (before $\nu(\alpha, \sigma)$) that includes $(\pi(\alpha, \sigma), m(\alpha, \sigma))$ in its response.
\end{itemize}

\paragraph{Sample interchangeability}

Let $\alpha$ be a non-redundant adversary, let $\sigma$ be a system such that $\nu(\alpha, \sigma) \neq \bot$. Let $\pi^* = \pi(\alpha, \sigma)$, let $m^* = m(\alpha, \sigma)$. Let $\sigma'$ be a system such that, for every pair $(\pi, m) \neq (\pi^*, m^*)$ (i.e., $\pi \neq \pi^*$ or $m \neq m^*$), the two following statements hold:
\begin{eqnarray*}
    \sigma'[\pi^*][m^*] &\in& \mathcal{E}[\alpha](\sigma) \\
    \sigma'[\pi][m] &=& \sigma[\pi][m]
\end{eqnarray*}

In this section, we prove the following:
\begin{eqnarray*}
    \forall n < \nu(\alpha, \sigma), \tau(\alpha, \sigma)_n = \tau(\alpha, \sigma')_n \\
    (\alpha \searrow \sigma) \implies (\tau(\alpha, \sigma) \searrow \sigma')
\end{eqnarray*}

We establish the first result by induction. Let us assume
\begin{eqnarray*}
    \tau(\alpha, \sigma) &=& ((i_1, r_1), \ldots) \\
    \tau(\alpha, \sigma') &=& ((i'_1, r'_1), \ldots) \\
    i_j = i'_j, r_j = r'_j && \forall j \leq n
\end{eqnarray*}
with $n \geq 0$ (here $n = 0$ means that this is $\alpha$'s first invocation). We start by noting that, since $a$ is a deterministic algorithm, we immediately have
\begin{equation*}
    i_{n + 1} = i'_{n + 1}
\end{equation*}
and we need to prove that $r_{n + 1} = r'_{n + 1}$.\\

Let us consider the case $i_{n + 1} = ({\tt Byzantine}, \pi, m)$. Following from the definition of $\mathcal{E}[\alpha](\sigma)$, $\pi^*$'s echo sample for $m^*$ in $\sigma'$ includes the same set of Byzantine processes as $\pi^*$'s echo sample for $m^*$ in $\sigma$. Since all other echo samples are trivially identical in $\sigma$ and $\sigma'$, we have $r_{n + 1} = r'_{n + 1}$. \\

Let us consider the case $i_{n + 1} = ({\tt State})$. Let $\pi \in \Pi_C$, let $\rho \in \Pi$, let $m \in \mathcal{M}$. Noting that $i_j = i'_j \: \forall j \leq n + 1$, we trivially have that $\rho$ sent an {\tt Echo}($m$, $m$) message to $\pi$ in $\sigma$ if and only if $\rho$ sent an {\tt Echo}($m$, $m$) message to $\pi$ in $\sigma$. Noting that all echo samples but $\pi^*$'s echo sample for $m^*$ are identical in $\sigma$, we immediately get that the symmetric difference between $r_{n + 1}$ and $r'_{n + 1}$ can only include $(\pi^*, m^*)$. The following hold true:
\begin{itemize}
    \item If $(\pi^*, m^*) \in r_{n + 1}$, then $(\pi^*, m^*) \in r'_{n + 1}$. Indeed, if $(\pi^*, m^*) \in r_{n + 1}$, then by definition $\sigma^+(\alpha, \sigma) \leq n + 1$. Therefore, by definition, every correct process in $\delta^+(\alpha, \sigma)$ \pbin.Delivered $m^*$ (both in $\sigma$ and $\sigma'$). Noting that $\alpha$ is an auto-echo adversary, every process in $\delta^+(\alpha, \sigma) \cup (\Pi \setminus \Pi_C)$ sent an {\tt Echo}($m^*$, $m^*$) message to $\pi^*$, both in $\sigma$ and $\sigma'$. Finally, by definition, $\mathcal{E}[\alpha](\sigma)$ includes at least $\hat E$ processes in $\delta^+(\alpha, \sigma) \cup (\Pi \setminus \Pi_C)$. Therefore $\pi^*$ delivered $m^*$ in $\sigma'$, and $(\pi^*, m^*) \in r_{n + 1}$.
    \item If $(\pi^*, m^*) \notin r_{n + 1}$, then $(\pi^*, m^*) \notin r'_{n + 1}$. Indeed, if $(\pi^*, m^*) \in r_{n + 1}$, then by definition $\sigma^-(\alpha, \sigma) \geq n + 1$. Therefore, by definition, every correct process that \pbin.Delivered $m^*$ (both in $\sigma$ and $\sigma'$) is included in $\delta^-(\alpha, \sigma)$. Finally, by definition, $\mathcal{E}[\alpha](\sigma)$ includes less than $\hat E$ processes in $\delta^-(\alpha, \sigma) \cup (\Pi \setminus \Pi_C)$. Therefore $\pi^*$ did not deliver $m^*$ in $\sigma'$, and $(\pi^*, m^*) \notin r_{n + 1}$.
\end{itemize}
which proves $r_{n + 1} = r'_{n + 1}$.

Noting that, by definition, $n < \nu(\alpha, \sigma)$, $i_{n + 1}$ cannot be $({\tt Sample}, \pi, m)$.

Noting that procedures $Deliver(\ldots)$ and $Echo(\ldots)$ never return a value, we trivially have that if $i_{n + 1} = ({\tt Deliver}, \pi, m)$ or $i_{n + 1} = ({\tt Echo}, \pi, s, \xi, m)$ then $r_{n + 1} = \bot = r'_{n + 1}$. By induction, we have 
\begin{equation*}
    \forall n < \nu(\alpha, \sigma), \tau(\alpha, \sigma) = \tau(\alpha, \sigma')
\end{equation*}

Let us assume that $\alpha$ compromises the consistency of $\sigma$. We want to prove that $\tau(\alpha, \sigma)$ compromises the consistency of $\sigma'$. 

We start by noting that, since by definition $\alpha$'s $\nu(\alpha, \sigma)$-th invocation in $\tau(\alpha, \sigma)$ is $({\tt Sample}, \pi^*, m^*)$ then, since $\alpha$ is correct, for some $j < \nu(\alpha, \sigma)$, the $j$-th invocation in $\tau(\alpha, \sigma)$ is $({\tt State})$, and its response includes $(\pi, m^*)$ for some $\pi \in \Pi_C$. Therefore, before the $\nu(\alpha, \sigma)$-th invocation, at least one correct process in $\sigma$ delivered $m^*$. 

We previously proved, however, that since $j < \nu(\alpha, \sigma)$, we have $\tau(\alpha, \sigma)_j = \tau(\alpha, \sigma')_j$. Therefore, at least one correct process delivered $m^*$ in $\sigma'$ as well.

Since $\alpha$ compromises the consistency of $\sigma$, at least one correct process $\pi'$ eventually delivers a message $m' \neq m^*$ in $\sigma$. Noting that $\pi'$'s echo sample for $m'$ is identical in $\sigma$ and $\sigma'$, we immediately have that $\pi'$ delivers $m'$ in $\sigma'$ as well.

\paragraph{System optimization}

Let $\alpha$ be a non-redundant adversary, let $\sigma$ be a system. In the previous section of this proof, we proved that, if we replace $\pi(\alpha, \sigma)$'s echo sample for $m(\alpha, \sigma)$ in $\sigma$ with any sample in $\mathcal{E}[\alpha](\sigma)$, we obtain a system $\sigma'$ such that $\tau(\alpha, \sigma)_n = \tau(\alpha, \sigma')_n$ for all $n < \nu(\alpha, \sigma)$.

We start by defining a function $\mathcal{N}: \mathcal{A} \rightarrow \powerset{\mathcal{S}}{}$ by
\begin{equation*}
    \mathcal{N}(\alpha) = \cp{\sigma \in \mathcal{S} \mid \nu(\alpha, \sigma) \neq \bot}
\end{equation*}
Provided with an adversary $\alpha$, $\mathcal{N}$ returns the set of systems coupled with which $\alpha$ issues at least one invocation to $Sample(\ldots)$. 

We then define a function $\mathcal{S}[\alpha]: \mathcal{N}(\alpha) \rightarrow \powerset{\mathcal{N}(\alpha)}{}$ by
\begin{equation*}
    \mathcal{S}[\alpha](\sigma) = \cp{\sigma' \in \mathcal{S} \mid \tau(\alpha, \sigma)_{1..(\nu(\alpha, \sigma) - 1)} \sim \sigma'}
\end{equation*}
Intuitively, when $\nu(\alpha, \sigma) \neq \bot$, $\mathcal(S)[\alpha](\sigma)$ returns the set of systems that $\alpha$ cannot distinguish from $\sigma$, before the first invocation of $Sample(\ldots)$.

Let $\sigma$ be a system such that $\nu(\alpha, \sigma) \neq \bot$, let $\sigma' \in \mathcal{S}[\alpha](\sigma)$. Noting that $\alpha$ is a deterministic adversary, we immediately get
\begin{equation*}
    \tau(\alpha, \sigma')_n = \tau(\alpha, \sigma)_n \; \forall n < \nu(\alpha, \sigma)
\end{equation*}
and
\begin{equation*}
    \nu(\alpha, \sigma') = \nu(\alpha, \sigma)
\end{equation*}
from which immediately follows
\begin{equation*}
    \mathcal{S}[\alpha](\sigma') = \mathcal{S}[\alpha](\sigma)
\end{equation*}

Let $\alpha$ be a non-redundant adversary, let $\sigma$, $\sigma'$ be systems in $\mathcal{N}(\alpha)$. Let $\sigma \; \mathcal{S}[\alpha]\; \sigma'$ denote the relationship
\begin{equation*}
    \sigma' \in \mathcal{S}[\alpha](\sigma)
\end{equation*}
Since $\tau(\alpha, \sigma) \sim \sigma$, we immediately have that $\mathcal{S}[\alpha]$ is reflexive. Since we established $\mathcal{S}[\alpha](\sigma') = \mathcal{S}[\alpha](\sigma)$, $\mathcal{S}[\alpha]$ is also symmetric and transitive. Therefore, $\mathcal{S}[\alpha]$ is an equivalence relation on $\mathcal{N}(\alpha)$.

Let
\begin{equation*}
    \mathcal{S}[\alpha]_1, \ldots \mathcal{S}[\alpha]_h = \frac{\mathcal{N}(\alpha)}{\mathcal{S}[\alpha]}
\end{equation*}
intuitively, each $\mathcal{S}[\alpha]_i$ is a distinct set of systems that are indistinguishable to $\alpha$, before the first invocation of $Sample(\ldots)$. \\

Let $i \in 1..h$. Let $\sigma \in \mathcal{S}[\alpha]_i$, let $E \in \mathcal{E}[\alpha](\sigma)$, let $\sigma'$ be identical to $\sigma$, with the exception of $\pi(\alpha, \sigma)$'s echo sample for $m(\alpha, \sigma)$, which is replaced with $E$. As we previously proved, $\tau(\alpha, \sigma)_{1..(\nu(\alpha, \sigma) - 1)} \sim \sigma'$, therefore have $\sigma' \in \mathcal{S}[\alpha]_i$. Moreover, we proved that for every $\sigma$ in $\mathcal{S}[\alpha]_i$, $\mathcal{E}[\alpha](\sigma)$ yields the same set of samples.

Let $\sigma, \sigma'$ be systems in $\mathcal{S}[\alpha]_i$, let $\pi^* = \pi(\alpha, \sigma) = \pi(\alpha, \sigma')$, let $m^* = m(\alpha, \sigma) = m(\alpha, \sigma')$. Let $\sigma\;\mathcal{E}[\alpha]\;\sigma'$ denote the relationship
\begin{equation*}
    \sigma[\pi^*][m^*] = \sigma'[\pi^*][m^*] \in (\mathcal{E}[\alpha](\sigma) = \mathcal{E}[\alpha](\sigma'))
\end{equation*}
from its definition we can immediately see that $\mathcal{E}[\alpha]$ is an equivalence relation, and we can partition
\begin{equation*}
    \mathcal{E}[\alpha]^i_1, \ldots, \mathcal{E}[\alpha]^i_l = \frac{\mathcal{S}[\alpha]_i}{\mathcal{E}[\alpha]}
\end{equation*}
with
\begin{equation*}
    \abs{\mathcal{E}[\alpha]^i_1} = \ldots = \abs{\mathcal{E}[\alpha]^i_l}
\end{equation*}

Let $\mathcal{C}[\alpha]^i_1, \ldots, \mathcal{C}[\alpha]^i_l$ denote the probability of $\alpha$ compromising a random element of $\mathcal{E}[\alpha]^i_1, \ldots, \mathcal{E}[\alpha]^i_l$:
\begin{equation*}
    \mathcal{C}[\alpha]^i_j = \frac{\abs{\cp{\sigma \in \mathcal{E}[\alpha]^i_j \mid \alpha \searrow \sigma}}}{\abs{\mathcal{E}[\alpha]^i_j}}
\end{equation*}
we can determine the subset whose consistency $\alpha$ has the highest probability of compromising by
\begin{equation*}
    \mathcal{C}[\alpha]^i_* = \argmax_j \mathcal{C}[\alpha]^i_j
\end{equation*}

Finally, we define an \textbf{optimization function} $\mathcal{O}[\alpha]: \mathcal{N}(\alpha) \rightarrow \mathcal{N}(\alpha)$. Let $\sigma \in \mathcal{S}[\alpha]_i$, we define $\mathcal{O}[\alpha]$ by
\begin{equation*}
    \mathcal{O}[\alpha](\sigma)[\pi][m] =
    \begin{cases}
        \mathcal{E}[\alpha](\sigma)_{\mathcal{C}[\alpha]^i_*} &\text{iff}\; \pi = \pi(\alpha, \sigma), m = m(\alpha, \sigma) \\
        \sigma[\pi][m] &\text{otherwise}
    \end{cases}
\end{equation*}

As we previously proved, every $\mathcal{E}[\alpha]^i_j$ has the same number of elements. Moreover, $\mathcal{O}[\alpha]$ maps a system $\sigma$ in $\mathcal{E}[\alpha]^i_j$ to the corresponding system $\sigma'$ in $\mathcal{E}[\alpha]^i_{C[\alpha]^i_*}$ that is identical to $\sigma$, except for $\pi(\alpha, \sigma)$'s echo sample for $m(\alpha, \sigma)$, which is replaced with $\mathcal{E}[\alpha](\sigma)_{\mathcal{C}[\alpha]^i_*}$.

Therefore, for every $\sigma, \sigma' \in \mathcal{E}^i_{\mathcal{C}[\alpha]^i_*}$,
\begin{equation*}
    \abs{\mathcal{O}[\alpha]^{-1}(\sigma)} = \abs{\mathcal{O}[\alpha]^{-1}(\sigma')} = \frac{\abs{\mathcal{S}[\alpha]_i}}{\abs{\mathcal{E}[\alpha]^i_1}}
\end{equation*}

\paragraph{System masking}

Let $\alpha$ be a non-redundant adversary, let $\alpha' = \Delta_{sb}(\alpha)$, let $\sigma$ be a system. \\

We start by noting that, if $\nu(\alpha, \sigma) = \bot$, then $\tau(\alpha, \sigma) = \tau(\alpha', \sigma)$. Indeed, if $\alpha$ never invokes $Sample(\ldots)$ when coupled with $\sigma$, all calls to $smadv$ are simply forwarded to the corresponding calls in $sys$. Therefore, if $\alpha$ compromises the consistency of $\sigma$, then trivially $\alpha'$ compromises the consistency of $\sigma$ as well. \\

Let us assume that $\nu(\alpha, \sigma) \neq \bot$. Let $\sigma'$ be an identical copy of $\sigma$.  We start by noting that, if we couple Sample-masking decorator with $\sigma'$, we effectively obtain a system instance $\delta$ with which $\alpha$ directly exchanges invocations and responses. Here we show that the trace $\tau(\alpha, \mathcal{O}[\alpha](\sigma))$ is identical to the trace $\tau(\alpha, \delta)$. Intuitively, this means that $\alpha$ has no way of \emph{distinguishing} whether it has been coupled directly with $\mathcal{O}[\alpha](\sigma)$, or it has been coupled with $\sigma'$, with Non-redundant decorator acting as an interface.

We previously proved that the trace exchanged between $nradv$ and $smadv$ is identical to the trace that $nradv$ would exchange with $sys$, if $\pi(\alpha, \sigma)$'s echo sample for $m(\alpha, \sigma)$ in $sys$ was replaced with $cache$.

Let $i \in \mathbb{N}$ such that $\sigma \in \mathcal{S}[\alpha]_i$. Procedure $optimize$ explicitly loops over all possible values of $sample \in \Pi^E$. For every value of $sample$, if loops over all the systems $\bar \sigma$ that are compatible with $trace$, and satisfy $\bar \sigma[\pi(\alpha, \sigma)][m(\alpha, \sigma)] = sample$. If, at the end of the loop, $systems \neq 0$, then $compromissions$ effectively represents, for some $j$, the number of systems in $\mathcal{E}[\alpha]^i_j$ that $\alpha$ compromises. Since $optimize$ selects the value of $sample$ that maximizes $compromissions / system$, the value that is eventually assigned to $cache$ is effectively $\mathcal{E}[\alpha](\sigma)_{\mathcal{C}[\alpha]^i_*}$, which proves the statement.

We previously proved that, if $\alpha$ compromises the consistency of $\mathcal{O}[\alpha](\sigma)$, then $\tau(\alpha, \mathcal{O}[\alpha](\sigma))$ compromises the consistency of $\sigma$ as well. Noting that every invocation to $smadv.Deliver(\ldots)$ or $smadv.Echo(\ldots)$ is respectively forwarded to $sys.Deliver(\ldots)$ or $sys.Echo(\ldots)$, we finally obtain that if $\alpha$ compromises the consistency of $\mathcal{O}[\alpha](\sigma)$, then $\alpha'$ compromises the consistency of $\sigma$ as well.

\paragraph{Adversarial power}

We can finally show that the adversarial power of $\alpha'$ is greater than the adversarial power of $\alpha$. Let $\sigma$ be a system. \\

As we previously established, if $\sigma \notin \mathcal{N}(\alpha)$, then the probability of $\alpha$ compromising $\sigma$ is identical to the probability of $\alpha$ compromising $\sigma'$. \\

Let us assume that $\sigma \in \mathcal{N}(\alpha)$. Let $i, j \in \mathbb{N}$ such that $\sigma \in \mathcal{E}[\alpha]^i_j$. The probability of $\alpha$ compromising the consistency of $\sigma$ is
\begin{equation*}
    \prob{\alpha \searrow \sigma} = \mathcal{C}[\alpha]^i_j
\end{equation*}
and, since $\alpha'$ compromises the consistency of $\sigma$ if $\alpha$ compromises the consistency of $\mathcal{O}[\alpha](\sigma)$, the probability of $\alpha'$ compromising the consistency of $\sigma$ is
\begin{equation*}
    \prob{\alpha' \searrow \sigma} = \prob{\alpha \searrow \mathcal{O}[\alpha](\sigma)} = \mathcal{C}[\alpha]^i_{\mathcal{C}[\alpha]^i_*} \geq \mathcal{C}[\alpha]^i_j = \prob{\alpha \searrow \sigma}
\end{equation*}

Which proves that the adversarial power of $\alpha'$ is greater or equal to the adversarial power of $\alpha$. 
\end{proof}
\end{lemma}

\subsection{Byzantine-counting adversary}
\label{subsection:byzantinecountingadversaryproof}

\begin{algorithm}
\begin{algorithmic}[1]
\Implements
\Instance{ByzantineCountingAdversary + CobSystem}{bcadv}
\EndImplements

\Uses
\InstanceSystem{SampleBlindAdversary}{sbadv}{bcadv}
\Instance{\Cobin System}{sys}
\EndUses

\Procedure{bcadv.Init}{{}}
    \State $best.byzantine = \bot$; \tabto*{4.5cm} $best.compromissions = 0$;
    \State $space = \cp{\bot}^C$;
    \State
    \ForAll{\pi}{\Pi_C}
        \State $count = \abs{sys.Byzantine(\pi)}$;
        \State $space[\pi] = (\Pi \setminus \Pi_C)^{count}$;
    \EndForAll
    \State
    \ForAll{byzantine}{space[\pi_1] \times \ldots \times space[\pi_C]}
        \State $compromissions = 0$;
        \ForAll{\sigma}{\mathcal{S}}
            \State $match = \true$;
            \ForAll{\pi}{\Pi_C}
                \If{$\sigma.Byzantine(\pi) \neq byzantine[\pi]$}
                    \State $match \leftarrow \false$;
                \EndIf
            \EndForAll
            \State
            \If{$match$ \textbf{and} $SampleBlindAdversary \searrow \sigma$}
                \State $compromissions \leftarrow compromissions + 1$;
            \EndIf
        \EndForAll
        \State
        \If{$compromissions > best.compromissions$}
            \State $best.byzantine \leftarrow byzantine$;
            \State $best.compromissions = compromissions$;
        \EndIf
    \EndForAll
    \State $sbadv.Init()$;
\EndProcedure

\algstore{byzantinecountingdecorator}
\end{algorithmic}
\caption{Byzantine-counting decorator}
\label{algorithm:byzantinecountingdecorator}
\end{algorithm}

\begin{algorithm}
\begin{algorithmic}[1]
\algrestore{byzantinecountingdecorator}

\Procedure{bcadv.Step}{{}}
    \State $sbadv.Step()$;
\EndProcedure

\Procedure{bcadv.Byzantine}{process}
    \State \Return $best.byzantine[process]$;
\EndProcedure

\Procedure{bcadv.State}{{}}
    \State \Return $sys.State()$;
\EndProcedure

\Procedure{bcadv.Sample}{process, message}
    \State \textbf{raise error};
\EndProcedure

\Procedure{bcadv.Deliver}{process, message}
    \State $sys.Deliver(process, message)$;
\EndProcedure

\Procedure{bcadv.Echo}{process, sample, source, message}
    \State $sys.Echo(process, sample, source, message)$;
\EndProcedure

\Procedure{bcadv.End}{{}}
    \State $sys.End()$;
\EndProcedure
\end{algorithmic}
\end{algorithm}

\begin{lemma}
The set of Byzantine-counting adversaries $\mathcal{A}_{bc}$ is optimal.

\begin{proof}
We again prove the result using a decorator. Here we show that a decorator $\Delta_{bc}$ exists such that, for every $\alpha \in \mathcal{A}_{sb}$, the adversary $\alpha' = \Delta_{bc}(\alpha)$ is a Byzantine-counting adversary, and more powerful than $\alpha$. If this is true, then the lemma is proved: let $\alpha^*$ be an optimal adversary, then the Byzantine-counting $\alpha^+ = \Delta_{bc}(\alpha^*)$ is optimal as well.

\paragraph{Decorator}

\cref{algorithm:byzantinecountingdecorator} implements \textbf{Byzantine-counting decorator}, a decorator that transforms a sample-blind adversary into a Byzantine-counting adversary. Provided with a sample-blind adversary $sbadv$, Byzantine-counting decorator acts as an interface between $sbadv$ and a system $sys$, effectively implementing a Byzantine-counting adversary $bcadv$. Byzantine-counting decorator exposes both the adversary and the system interface: the underlying adversary $sbadv$ uses $bcadv$ as its system.

Byzantine-counting decorator works as follows:
\begin{itemize}
    \item Procedure $bcadv.Init()$ generates $best.byzantine$, an array of $C$ pre-computed responses that $bcadv$ will provide to any subsequent invocation of $bcadv.Byzantine(\ldots)$, optimized to maximize the probability of compromising $sys$. This is achieved as follows:
    \begin{itemize}
        \item The procedure loops over every correct process $\pi$, and queries \\ $\abs{sys.Byzantine(\pi)}$ to determine how many Byzantine processes there are in the first echo sample of $\pi$. For each $\pi$, the procedure sets variable $space[\pi]$ to the set of all possible responses to $bcadv. \allowbreak Byzantine(\pi)$ that satisfy the condition $\abs{bcadv.Byzantine(\pi)} = \abs{sys.Byzantine(\pi)}$.
        \item The procedure loops over every possible array $byzantine$ of $C$ responses that, for every $\pi \in \Pi_C$, satisfies $byzantine[\pi] \in space[\pi]$. It then counts the number of systems $\sigma$ that are compatible with $byzantine$ (i.e., that satisfy, for every $\pi \in \Pi_C$, $\sigma.Byzantine(\pi) = byzantine[\pi]$) and whose consistency is compromised by the underlying adversary $SampleBlindAdversary$.
        \item The procedure sets $best.byzantine$ to the array $byzantine$ that maximizes the number of systems compatible with $byzantine$ whose consistency is compromised by $SampleBlindAdversary$.
    \end{itemize}
    \item Procedure $bcadv.Byzantine(process)$ simply returns $best.byzantine[\allowbreak process]$.
    \item Procedure $bcadv.State()$ simply forwards the call to $sys.State()$.
    \item Procedure $bcadv.Sample(process, message)$ is never called. This is due to the fact that $sbadv$ is sample-blind.
    \item Procedure $bcadv.Deliver(process, message)$ simply forwards the call to $sys.Deliver(process, message)$.
    \item Procedure $bcadv.Echo(process, sample, source, message)$ simply forwards the call to $sys.Echo(process, sample, source, message)$.
    \item Procedure $bcadv.End()$ simply forwards the call to $sys.End()$.
\end{itemize}

\paragraph{Correctness}

We start by proving that no adversary has undefined behavior when coupled with {\tt Byzantine-counting decorator}. An adversary has undefined behavior if, at any point, the sequence of invocations and responses it exchanges with $bcadv$ is incompatible with every system.

Upon initialization, $bcadv$ generates an array $best.byzantine$ of $C$ responses, one for every call to $bcadv.Byzantine(\pi \in \Pi_C)$. For every correct process $\pi$, $best.byzantine[\pi]$ contains only Byzantine processes and satisfies $\abs{best.byzantine[\pi]} = \abs{sys.Byzantine(\pi)}$. Let $sys'$ be the system obtained by replacing the Byzantine component of each correct process $\pi$'s echo samples in $sys$ with $best.byzantine[\pi]$. The trace exchanged between $sbadv$ and $bcadv$ is always compatible with $sys'$. Indeed:
\begin{itemize}
    \item Every call to $bcadv.Byzantine(\pi)$ returns $best.byzantine[\pi]$ which is equal, by definition, to $sys.Byzantine(\pi)$.
    \item Every call to $bcadv.State()$ is simply forwarded to $sys.State()$. Let $\pi$ be a correct process, let $m$ be a message. Since that $bcadv$ is an auto-echo adversary, when $bcadv.State()$ is invoked, every Byzantine process in $\pi$'s echo sample for $m$ has sent an {\tt Echo}($m$, $m$) message both in $sys$ and $sys'$. Moreover, the number of Byzantine processes in $\pi$'s echo sample for $m$ is identical in $sys$ and $sys'$. Finally, set of correct processes in $\pi$'s echo sample for $m$ is identical in $sys$ and $sys'$. Consequently, $bcadv.State() = sys.State() = sys'.State()$.
\end{itemize}

\paragraph{Byzantine-counting}

It is immediate to see that Byzantine-counting decorator always implements a Byzantine-counting adversary. Indeed, for any $\pi \in \Pi_C$, $sys.Byzantine(\pi)$ is only invoked from $\abs{sys.Byzantine(\pi)}$.

\paragraph{Byzantine interchangeability}

Let $\alpha$ be a sample-blind system. Let $\sigma$ be a system, let $\sigma'$ be a system such that, for every correct process $\pi$, every message $m$, and every $n \in 1..E$,
\begin{eqnarray*}
    (\sigma[\pi][m][n] \in \Pi_C) &\implies& (\sigma'[\pi][m][n] = \sigma[\pi][m][n]) \\
    (\sigma[\pi][m][n] \notin \Pi_C) &\implies& (\sigma'[\pi][m][n] \notin \Pi_C)
\end{eqnarray*}
In other words, for every $\pi \in \Pi_C$ and every $m \in \mathcal{M}$, the set of correct processes in $\pi$'s echo sample for $m$ is identical in $\sigma$ and $\sigma'$.

Here we prove that, if $\alpha$ compromises $\sigma$, then $\tau(\alpha, \sigma)$ compromises $\sigma'$. In order to do this, we first establish some auxiliary results.

Let us consider the case where $\alpha$ is run against $\sigma$ and $\tau(\alpha, \sigma)$ is applied to $\sigma'$. Let $\pi$ be a correct process, let $\rho$ be a process, let $m$ be a message. At the end of both adversarial executions, the following hold true:
\begin{itemize}
    \item If $\pi$ \pbin.Delivered $m$ in $\sigma$, then $\pi$ \pbin.Delivered $m$ in $\sigma'$ as well. This follows immediately from the fact that $\tau(\alpha, \sigma)$ is applied to $\sigma'$, and $(({\tt Deliver}, \pi, m), \bot) \in \tau(\alpha, \sigma)$.
    \item If $\rho$ sent an {\tt Echo}($m$, $m$) message to $\pi$ in $\sigma$, then $\rho$ sent an {\tt Echo}($m$, $m$) message to $\pi$ in $\sigma'$. Indeed, if $\rho$ is a correct process, and it sent an {\tt Echo}($m$, $m$) message to $\pi$ in $\sigma$, then it \pbin.Delivered $m$ both in $\sigma$ and $\sigma'$. Therefore, it sent an {\tt Echo}($m$, $m$) message to $\pi$ in $\sigma'$ as well. If $\rho$ is a Byzantine process then, since $\alpha$ is an auto-echo adversary, $\rho$ sent an {\tt Echo}($m$, $m$) message to $\pi$ both in $\sigma$ and $\sigma'$.
    \item If $\pi$ delivered $m$ in $\sigma$, then $\pi$ also delivered $m$ in $\sigma'$. This follows from the above, and the fact that the correct processes in $\pi$'s echo sample for $m$ are identical in $\sigma$ and $\sigma'$.
\end{itemize}

If $\alpha$ compromises the consistency of $\sigma$, then two correct processes $\pi$, $\pi'$ and two distinct messages $m$, $m' \neq m$ exist such that $\pi$ delivered $m$, and $\pi'$ delivered $m'$ in $\sigma$. From the above, however, $\pi$ delivered $m$, and $\pi'$ delivered $m'$, in $\sigma'$ as well. Consequently, $\tau(\alpha, \sigma)$ compromises the consistency of $\sigma'$.

\paragraph{System optimization}

Let $\sigma$, $\sigma'$ be systems. We define the relationship $\stackrel{\abs{F}}{\sim}$ by
\begin{equation*}
    \rp{\sigma \stackrel{\abs{F}}{\sim} \sigma'} \xLeftrightarrow[]{def} \rp{\forall \pi \in \Pi_C, \forall n \in 1..E, \sigma[\pi][1][n] \in \Pi_C \Leftrightarrow \sigma'[\pi][1][n] \in \Pi_C}
\end{equation*}
In other words, $\sigma \stackrel{\abs{F}}{\sim} \sigma'$ if, for every $\pi$ and for every $n \in 1..E$, the $n$-th element of the first of $\pi$'s echo samples is either correct both in $\sigma$ and $\sigma'$, or Byzantine both in $\sigma$ and $\sigma'$.

It is immediate to see that $\stackrel{\abs{F}}{\sim}$ is an equivalence relation. We can therefore partition $\mathcal{S}$ with $\stackrel{\abs{F}}{\sim}$ to obtain
\begin{equation*}
    \mathcal{S}_1, \ldots, \mathcal{S}_h = \frac{\mathcal{S}}{\stackrel{\abs{F}}{\sim}}
\end{equation*}

Let $\sigma$, $\sigma'$ be systems. We define the relationship $\stackrel{F}{\sim}$ by
\begin{equation*}
    \rp{\sigma \stackrel{F}{\sim} \sigma'} \xLeftrightarrow[]{def} \rp{\forall \pi \in \Pi_C, \forall n \in 1..E, \sigma[\pi][1][n] \notin \Pi_C \Leftrightarrow \sigma'[\pi][1][n] = \sigma[\pi][1][n]}
\end{equation*}
Intuitively, $\sigma \stackrel{F}{\sim} \sigma'$ if the Byzantine processes in each echo sample are identical in $\sigma$ and $\sigma'$. Again, $\stackrel{F}{\sim}$ is an equivalence relation that we can use to partition $\mathcal{S}_i$:
\begin{equation*}
    \mathcal{S}^i_1, \ldots, \mathcal{S}^i_l = \frac{\mathcal{S}_i}{\stackrel{F}{\sim}}
\end{equation*}
and noting that, in \cobal, every correct process selects independently the correct processes in its echo samples, we have
\begin{equation*}
    \abs{\mathcal{S}^i_1} = \ldots = \abs{\mathcal{S}^i_l}
\end{equation*}

Let $\alpha$ be a sample-blind adversary. We define $\mathcal{C}[\alpha]^i_j$ as the fraction of systems in $\mathcal{S}^i_j$ whose consistency is compromised by $\alpha$:
\begin{equation*}
    \mathcal{C}[\alpha]^i_j = \frac{\abs{\cp{\sigma \in \mathcal{S}^i_j \mid \alpha \searrow \sigma}}}{\abs{\mathcal{S}^i_j}}
\end{equation*}

From $\mathcal{C}[\alpha]^i_j$ we can define
\begin{equation*}
    \mathcal{C}[\alpha]^i_* = \argmax_j \: \mathcal{C}[\alpha]^i_j
\end{equation*}
Intuitively, $\mathcal{C}[\alpha]^i_*$ identifies the partition of $\mathcal{S}_i$ that $\alpha$ has the highest probability of compromising consistency.

Finally, we define an \textbf{optimization function} $\mathcal{O}[\alpha]: \mathcal{S} \rightarrow \mathcal{S}$. Let $\sigma \in \mathcal{S}_i$, we define $\mathcal{O}[\alpha]$ by
\begin{eqnarray*}
    \mathcal{O}[\alpha](\sigma) &\in& \mathcal{S}^i_{\mathcal{C}[\alpha]^i_*} \\
    \sigma[\pi][m][n] \in \Pi_C &\implies& \mathcal{O}[\alpha](\sigma)[\pi][m][n] = \sigma[\pi][m][n]
\end{eqnarray*}

As we previously proved, every $\mathcal{S}^i_j$ has the same number of elements. Moreover, $\mathcal{O}[\alpha]$ maps a system $\sigma$ in $\mathcal{S}^i_j$ to the corresponding system $\sigma'$ in $\mathcal{S}^i_{\mathcal{C}[\alpha]^i_*}$ such that every correct process in an echo sample in $\sigma$ is identical to the corresponding process in $\sigma'$.

Therefore, for every $\sigma, \sigma' \in \mathcal{S}^i_{C[\alpha]^i_*}$,
\begin{equation*}
    \abs{\mathcal{O}[\alpha]^{-1}(\sigma)} = \abs{\mathcal{O}[\alpha]^{-1}(\sigma')} = \frac{\abs{\mathcal{S}_i}}{\abs{\mathcal{S}^i_1}}
\end{equation*}

\paragraph{System masking}

Let $\alpha$ be a sample-blind adversary, let $\alpha' = \Delta_{bc}(\alpha)$, let $\sigma$ be a system, let $\sigma'$ be an identical copy of $\sigma$.  We start by noting that, if we couple Byzantine-counting decorator with $\sigma'$, we effectively obtain a system instance $\delta$ with which $\alpha$ directly exchanges invocations and responses. Here we show that the trace $\tau(\alpha, \mathcal{O}[\alpha](\sigma))$ is identical to the trace $\tau(\alpha, \delta)$. Intuitively, this means that $\alpha$ has no way of \emph{distinguishing} whether it has been coupled directly with $\mathcal{O}[\alpha](\sigma)$, or it has been coupled with $\sigma'$, with Byzantine-counting decorator acting as an interface.

We previously proved that the trace exchanged between $sbadv$ and $bcadv$ is identical to the trace that $sbadv$ would exchange with the system $sys'$ that is obtained by replacing the Byzantine component of each correct process $\pi$'s echo samples in $sys$ with $best.byzantine[\pi]$.

Let $i \in \mathbb{N}$ such that $\sigma \in \mathcal{S}_i$. Procedure $bcadv.Init()$ explicitly loops over all the possible values of $byzantine$ that satisfy the condition $\abs{byzantine[\pi]} = \allowbreak \abs{sys.Byzantine(\pi)}$ for all $\pi \in \Pi_C$. It then loops over every system $\sigma$ that satisfies $\sigma.Byzantine(\pi) = byzantine[\pi]$, and counts the number of systems that $\alpha$ compromises. It finally selects the value of $byzantine$ that maximizes the number of compromissions. In doing so, $bcadv.Init()$ is effectively looping over every $\mathcal{S}^i_j$, and selecting the $j$ that maximizes the probability of $\alpha$ compromising a random element of $\mathcal{S}^i_j$. Since $bcadv.Init()$ is effectively masking $\sigma$ with the element of $\mathcal{S}^i_{\mathcal{C}[\alpha]^i_*}$ with which $\sigma$ shares the correct component of every sample, the trace $\tau(\alpha, \mathcal{O}[\alpha](\sigma))$ is identical to the trace $\tau(\alpha, \delta)$.

We previously proved that, if $\alpha$ compromises the consistency of $\mathcal{O}[\alpha](\sigma)$, then $\tau(\alpha, \mathcal{O}[\alpha](\sigma))$ compromises the consistency of $\sigma$ as well. Noting that every invocation to $bcadv.Deliver(\ldots)$ or $bcadv.Echo(\ldots)$ is respectively forwarded to $sys.Deliver(\ldots)$ or $sys.Echo(\ldots)$, we finally obtain that if $\alpha$ compromises the consistency of $\mathcal{O}[\alpha](\sigma)$, then $\alpha'$ compromises the consistency of $\sigma$ as well.

\paragraph{Adversarial power}

We can finally show that the adversarial power of $\alpha'$ is greater than the adversarial power of $\alpha$. Let $\sigma$ be a system. \\

Let $i, j \in \mathbb{N}$ such that $\sigma \in \mathcal{S}^i_j$. The probability of $\alpha$ compromising the consistency of $\sigma$ is
\begin{equation*}
    \prob{\alpha \searrow \sigma} = \mathcal{C}[\alpha]^i_j
\end{equation*}
and, since $\alpha'$ compromises the consistency of $\sigma$ if $\alpha$ compromises the consistency of $\mathcal{O}[\alpha](\sigma)$, the probability of $\alpha'$ compromising the consistency of $\sigma$ is
\begin{equation*}
    \prob{\alpha' \searrow \sigma} = \prob{\alpha \searrow \mathcal{O}[\alpha](\sigma)} =  \mathcal{C}[\alpha]^i_{\mathcal{C}[\alpha]^i_*} \geq 
    \mathcal{C}[\alpha]^i_j = \prob{\alpha \searrow \sigma}
\end{equation*}

Which proves that the adversarial power of $\alpha'$ is greater or equal to the adversarial power of $\alpha$. 
\end{proof}
\end{lemma}

\subsection{Single-response adversary}
\label{subsection:singleresponseadversaryproof}

\begin{algorithm}
\begin{algorithmic}[1]
\Implements
    \Instance{SingleResponseAdversary + CobSystem}{sradv}
\EndImplements

\Uses
    \InstanceSystem{ByzantineCountingAdversary}{bcadv}{sradv}
    \Instance{\Cobin System}{sys}
\EndUses

\Procedure{sradv.Init}{{}}
    \State $cache = \emptyset$; \tabto*{3.3cm} $poisoned = \false$; \tabto*{7cm} $step = 0$;
    \State
    \ForAll{\pi}{\Pi_C}
        \If{$\abs{sys.Byzantine(\pi)} \geq \hat E$}
            \State $poisoned \leftarrow \true$;
        \EndIf
    \EndForAll
    \State
    \State $bcadv.Init()$;
\EndProcedure

\Procedure{sradv.Step}{{}}
    \State $step \leftarrow step + 1$;
    \State
    \If{$poisoned = \false$ \textbf{or} $step \leq (N - C)C^2$}
        \State $bcadv.Step()$;
    \ElsIf{$step \leq (N - C)C^2 + C$}
        \State $sys.Deliver(\zeta(step - (N - C)C^2), 1)$;
    \Else
        \State $sys.End()$;
    \EndIf
\EndProcedure

\Procedure{sradv.Byzantine}{\pi}
    \State $count = \abs{sys.Byzantine(\pi)}$;
    \State \Return $\cp{\bot}^{count}$;
\EndProcedure

\Procedure{sradv.State}{{}}
    \State \Return $cache$;
\EndProcedure

\algstore{singleresponsedecorator}

\end{algorithmic}
\caption{Single-response decorator}
\label{algorithm:singleresponsedecorator}
\end{algorithm}

\begin{algorithm}
\begin{algorithmic}[1]
\algrestore{singleresponsedecorator}

\Procedure{sradv.Sample}{process, message}
    \State \textbf{raise error};
\EndProcedure

\Procedure{sradv.Deliver}{process, message}
    \State $sys.Deliver(process, message)$;
    \State
    \If{$cache = \emptyset$}
        \State $cache \leftarrow sys.State()$;
    \EndIf
\EndProcedure

\Procedure{sradv.Echo}{process, sample, source, message}
    \State $sys.Echo(process, sample, source, message)$;
\EndProcedure

\Procedure{sradv.End}{{}}
    \State $sys.End()$;
\EndProcedure
\end{algorithmic}
\end{algorithm}

\begin{lemma}
The set of single-response adversaries $\mathcal{A}_{sr}$ is optimal.

\begin{proof}
We again prove the result using a decorator. Here we show that a decorator $\Delta_{sr}$ exists such that, for every $\alpha \in \mathcal{A}_{bc}$, the adversary $\alpha' = \Delta_{sr}(\alpha)$ is a single-response adversary, and as powerful as $\alpha$. If this is true, then the lemma is proved: let $\alpha^*$ be an optimal adversary, then the sequential $\alpha^+ = \Delta_{sr}(\alpha^*)$ is optimal as well.

\paragraph{Decorator}

\cref{algorithm:singleresponsedecorator} implements \textbf{Single-response decorator}, a decorator that transforms a Byzantine-counting adversary into a single-response adversary. Provided with a Byzantine-counting adversary $bcadv$, Single-response decorator acts as an interface between $bcadv$ and a system $sys$, effectively implementing a single-response adversary $sradv$. Single-response decorator exposes both the adversary and the system interfaces: the underlying adversary $bcadv$ uses $sradv$ as its system.

Single-response decorator works as follows:
\begin{itemize}
    \item Procedure $sradv.Init()$ initializes the following variables:
    \begin{itemize}
        \item A $cache$ set, initially empty: $cache$ is used to store the first non-empty set returned from $sys.State()$.
        \item A $poisoned$ variable: $poisoned$ is set to $\true$ if and only if at least one correct process in $sys$ is poisoned. This condition is verified by looping over $sys.Byzantine(\pi)$ for every correct process $\pi$.
        \item A $step$ variable, initially set to $0$: at any time, $step$ counts how many times $sradv.Step()$ has been invoked.
    \end{itemize}
    \item Procedure $sradv.Step()$ increments $step$, then implements two different behaviors depending on the value of $poisoned$:
    \begin{itemize}
        \item If $poisoned = \true$, it forwards the call to $bcadv.Step()$ for the first $(N - C)C^2$ times. For the next $C$ steps, it sequentially invokes $sys.Deliver(\zeta(1), 1)$, $\ldots$, $sys.Deliver(\zeta(C), 1)$. Finally, it calls $sys.End()$.
        \item If $poisoned = \false$, it forwards the call to $bcadv.Step()$.
    \end{itemize}
    \item Procedure $sradv.Byzantine(process)$ returns an array of $count$ elements, $count$ being the number of elements returned by $sys.Byzantine(\allowbreak process)$. The array is filled with $\bot$ values: since $bcadv$ is Byzantine-counting, the content of the array is irrelevant.
    \item Procedure $sradv.State()$ simply returns $cache$.
    \item Procedure $sradv.Sample(process, message)$ is never called. This is due to the fact that $bcadv$ is sample-blind.
    \item Procedure $sradv.Deliver(process, message)$ forwards the call to \\ $sys.Deliver(process, message)$. Then, if $cache$ is empty, it updates $cache$ with $sys.State()$.
    \item Procedure $sradv.Echo(process, sample, source, message)$ simply forwards the call to $sys.Echo(process, sample, source, message)$.
    \item Procedure $sradv.End()$ simply forwards the call to $sys.End()$.
\end{itemize}

\paragraph{Correctness}

Here we prove that every adversary, when coupled with {\tt Single-response adversary}:
\begin{itemize}
    \item Has a well-defined behavior. An adversary has undefined behavior if, at any point, the sequence of invocations and responses it exchanges with $sradv$ is incompatible with every system.
    \item Is process-sequential, sequential, and Byzantine-counting.
\end{itemize} 

We start by noting that $poisoned = \true$ if and only if $sys$ is poisoned. Indeed, $sradv.Init()$ explicitly checks if any correct process has at least $\hat E$ Byzantine processes in its first echo sample. \\

We distinguish two cases, based on the value of $poisoned$. Let us assume $poisoned = \true$. When $sradv.Step()$ is invoked, the call is forwarded to $bcadv.Step()$ only for the first $(N - C)C^2$ times. Noting that $bcadv$ is an auto-echo adversary, every call to $bcadv.Step()$ results in a call to $sradv.Echo(\ldots)$. For the next $C$ steps, $sradv.Step()$ sequentially causes $\zeta(1), \zeta(2), \ldots$ to \pbin.Deliver message $1$. Finally, $sradv.Step()$ invokes $sys.End()$. Therefore, $sradv$ has a well defined behavior and implements a process-sequential adversary. Since it causes only message $1$ to be \pbin.Delivered, $sradv$ is also trivially sequential. \\

Let us assume $poisoned = \false$. As we proved in \cref{subsubsection:sequentialadversary}, since $sys$ is not poisoned, a correct process in $sys$ will only deliver a message $m$ as a result of an invocation to $sys.Deliver(\pi, m)$ for some $\pi \in \Pi_C$. Until $cache \neq \emptyset$, $cache$ is updated to $sys.State()$ after every call to $sys.Deliver(\ldots)$. Therefore, throughout the first phase, $sradv.State()$ is always identical to $sys.State()$. The trace exchanged between $bcadv$ and $sradv$ is, therefore, trivially compatible with $sys$. 

Throughout the second phase, we have $cache \neq \bot$. Since, throughout the first phase, $cache$ is updated after every call to $sys.Deliver(\ldots)$, only one message $m^*$ exists such that, for some $\pi^* \in \Pi_C$, $(\pi^*, m^*) \in cache$. Noting that $bcadv$ is a non-redundant adversary, it will never invoke $sradv.Deliver(\ldots)$ on $m^*$: indeed, the value returned from $sradv.State()$ never changes throughout the second phase. We define a system $sys'$ by
\begin{equation*}
    sys'[\pi][m][n] =
    \begin{cases}
        sys[\pi][m][n] &\text{iff}\; m = m^* \;\text{or}\; sys[\pi][m][n] \in \Pi \setminus \Pi_C\\
        \pi^* &\text{otherwise}
    \end{cases}
\end{equation*}

The trace exchanged between $bcadv$ and $sradv$ is compatible with $sys'$. Indeed, for every $\pi \in \Pi_C$, $\pi$'s sample for $m^*$ in $sys$ is identical to $\pi$'s echo sample for $m^*$ in $sys'$: at any moment, $\pi$ delivered $m^*$ in $sys$ if and only if $\pi$ delivered $m^*$ in $sys'$. For every $\pi \in \Pi_C$ and $m \neq m^* \in \mathcal{M}$, $\pi$'s every correct process in $\pi$'s sample for $m$ is $\pi^*$. However, $\pi^*$ \pbin.Delivered $m^* \neq m$. Therefore, since $sys'$ is not poisoned, no correct process in $sys'$ ever delivers a message other than $m^*$.

Every call to $sradv.Deliver(\ldots)$ and $sradv.Echo(\ldots)$ is respectively forwarded to $sys.Deliver(\ldots)$ and $sys.Echo(\ldots)$. Moreover, $bcadv$ is process-sequential and sequential. Therefore, if $poisoned = \true$, $sradv$ is also process-sequential and sequential.

It is immediate to see that Single-response decorator always implements a Byzantine-counting adversary. Indeed, for any $\pi \in \Pi_C$, $sys.Byzantine(\pi)$ is only invoked from $\abs{sys.Byzantine(\pi)}$.

\paragraph{Single-response}

It is immediate to see that Single-response decorator always implements a single-response adversary. Indeed, when $sys.State()$ returns a non-empty set for the first time, $cache$ is set to a non-empty set, and $sys.State()$ is never invoked again.

\paragraph{Roadmap}

Let $\alpha \in \mathcal{A}_{bc}$, let $\alpha' = \Delta_{sr}(\alpha)$. Let $\sigma$ be a system such that $\alpha$ compromises the consistency of $\sigma$. Let $\sigma'$ be an identical copy of $\sigma$. In order to prove that $\alpha'$ is as powerful as $\alpha$, we prove that $\alpha'$ compromises the consistency of $\sigma'$.

\paragraph{Poisoned case}

Noting that $\alpha'$ is an auto-echo adversary, if $\sigma$ is poisoned we immediately have that $\alpha'$ compromises the consistency of $\sigma'$.

\paragraph{Trace}

Let us assume that $\sigma$ is not poisoned. We start by noting that, if we couple Single-response decorator with $\sigma'$, we effectively obtain a system instance $\delta$ with which $\alpha$ directly exchanges invocations and responses.

We start by defining a boolean sequence $W$ by setting $W_n = \true$ if and only if, after the $n$-th invocation, two correct processes $\pi, \pi'$ and two distinct messages $m, m' \neq m$ exist such that $\pi$ delivered $m$ and $\pi'$ delivered $m'$ in $\sigma$. Since $\alpha$ compromises the consistency of $\sigma$, for some $n$ we have $W_n = \true$. Let
\begin{equation*}
    w = \min n \mid W_n = \true
\end{equation*}

Here we show that, for every $n \leq w$, the trace $\tau(\alpha, \sigma)_n$ is identical to the trace $\tau(\alpha, \delta)_n$. Intuitively, this means that, until the consistency of $\sigma$ is compromised, $\alpha$ has no way of \emph{distinguishing} whether it has been coupled directly with $\sigma$, or it has been coupled with $\sigma'$, with Single-response decorator acting as an interface. We prove this by induction.

Let us assume
\begin{eqnarray*}
    \tau(\alpha, \sigma) &=& ((i_1, r_1), \ldots) \\
    \tau(\alpha, \delta) &=& ((i'_1, r'_1), \ldots) \\
    i_j = i'_j, r_j = r'_j && \forall j \leq n
\end{eqnarray*}

We start by noting that, since $\alpha$ is a deterministic algorithm, we immediately have
\begin{equation*}
    i_{n + 1} = i'_{n + 1}
\end{equation*}
and we need to prove that $r_{n + 1} = r'_{n + 1}$. \\

Let us assume that $i_{n + 1} = ({\tt Byzantine}, \pi)$. Since procedure $sradv.\allowbreak Byzantine(\pi)$ forwards the call to $sys.Byzantine(\pi)$, $bcadv$ is a Byzantine-counting adversary, and $\sigma'$ is an identical copy of $\sigma$, with a minor abuse of notation we effectively have $r_{n + 1} = r'_{n + 1}$. \\

Let us assume that $i_{n + 1} = ({\tt State})$. We start by noting that, since all calls to $sradv.Deliver(\ldots)$ and $sradv.Echo(\ldots)$ are respectively forwarded to $sys.Deliver(\ldots)$ and $sys.Echo(\ldots)$, a correct process $\pi$ delivered $m^*$ in $\sigma$ if and only if it delivered $m^*$ in $\sigma'$ as well. As we proved, throughout the first phase, $sradv.State()$ always returns the same value as $sys.State()$. Let us assume $n > \abs{\eta(\alpha, \sigma)}$. Let $m^*$ be the only message that was delivered by at least one correct process in $\sigma$. Noting that a correct process delivers a message only as a result of a call to $sys.Deliver(\ldots)$, we have $n < w$. Therefore, by definition, no correct process in $\sigma$ delivered a message other than $m^*$. Since $\alpha$ is a non-redundant adversary, it never causes any correct process to \pbin.Deliver $m^*$ throughout the second phase. As a result, no correct process delivers $m^*$ in $\sigma$ throughout the second phase. Therefore, all the processes that delivered $m^*$ in $\sigma$ are represented in $cache$, and no other process delivered a message $m \neq m^*$. Consequently, $r_{n + 1} = r'_{n + 1}$.

Noting that procedures $Deliver(\ldots)$ and $Echo(\ldots)$ never return a value, we trivially have that if $i_{n + 1} = ({\tt Deliver}, \pi, m)$ or $i_{n + 1} = ({\tt Echo}, \pi, s, \xi, m)$ then $r_{n + 1} = \bot = r'_{n + 1}$. By induction, we have that, for every $n \leq w$, $\tau(\alpha, \sigma)_n = \tau(\alpha, \delta)_n$.

\paragraph{Consistency of $\sigma'$}

We proved that, for all $n \leq w$, $\tau(\alpha, \sigma)_n = \tau(\alpha, \delta)_n$. Moreover, we proved that if a correct process $\pi$ eventually delivers a message $m$ in $\sigma$ before the $w$-th invocation, then $\pi$ also delivers $m$ in $\sigma'$ before the $w$-th invocation.

Since $\alpha$ compromises the consistency of $\sigma$ after the $w$-th invocation, two correct processes $\pi$, $\pi'$ and two distinct messages $m$, $m' \neq m$ exist such that, in $\sigma$, $\pi$ delivered $m$ and $\pi'$ delivered $m'$ before the $w$-th invocation. Therefore, in $\sigma'$, $\pi$ delivered $m$ and $\pi'$ delivered $m'$ before the $w$-th invocation. Therefore $\alpha'$ compromises the consistency of $\sigma'$.

Consequently, the adversarial power of $\alpha$ is equal to the adversarial power of $\alpha' = \Delta_{sr}(a)$, and the lemma is proved.

\end{proof}
\end{lemma}

\subsection{Two-phase adversary}
\label{subsection:twophaseadversaryproof}

\begin{algorithm}
\begin{algorithmic}[1]
\Implements
    \Instance{TwoPhaseAdversary + \Cobin System}{tpadv}
\EndImplements

\Uses
    \InstanceSystem{StatePollingAdversary}{spadv}{tpadv}
    \Instance{\Cobin System}{sys}
\EndUses

\Procedure{tpadv.Init}{{}}
    \State $invocations = 0$;
    \State $spadv.Init()$;
\EndProcedure

\Procedure{compatible}{invocations}
    \State $systems = \emptyset$;
    \State
    \ForAll{\sigma}{\mathcal{S}}
        \State $match = \true$;
        \ForAll{\pi}{\Pi_C}
            \If{$\abs{\sigma.Byzantine(\pi)} \neq \abs{sys.Byzantine(\pi)}$}
                \State $match = \false$;
            \EndIf
        \EndForAll
        \State
        \If{$match = \true$ \textbf{and} $\abs{\eta(\alpha, \sigma)} = invocations$}
            \State $systems \leftarrow systems \cup \cp{\sigma}$;
        \EndIf
    \EndForAll
    \State
    \State \Return $systems$;
\EndProcedure

\Procedure{tpadv.Step}{{}}
    \State $spadv.Step()$;
\EndProcedure

\Procedure{tpadv.Byzantine}{process}
    \State $invocations \leftarrow invocations + 1$;
    \State $count = sys.Byzantine(process)$;
    \State \Return $\cp{\bot}^{count}$;
\EndProcedure

\algstore{twophasedecorator}
\end{algorithmic}
\caption{Two-phase decorator}
\label{algorithm:twophasedecorator}
\end{algorithm}

\begin{algorithm}
\begin{algorithmic}[1]
\algrestore{twophasedecorator}

\Procedure{tpadv.State}{{}}
    \State $invocations \leftarrow invocations + 1$;
    \State
    \If{$sys.State() \neq \emptyset$}
        \State $outcomes = \emptyset$;
        \ForAll{\sigma}{compatible(invocations)}
            \State $(invocation, response) = \tau(\alpha, \sigma)_{invocations}$;
            \State $outcomes \leftarrow outcomes \cup \cp{\rp{response, \tau(\alpha, \sigma)}}$;
        \EndForAll
        \State
        \State $best.response = \bot$; \tabto*{4.5cm} $best.compromissions = 0$;
        \State
        \ForAll{(response, fulltrace)}{outcomes}
            \State $compromissions = 0$;
            \State
            \ForAll{\sigma}{compatible(invocations)}
                \If{$fulltrace \searrow \sigma$}
                    \State $compromissions \leftarrow compromissions + 1$;
                \EndIf
            \EndForAll
            \State
            \If{$compromissions > best.compromissions$}
                \State $best.response \leftarrow response$;
                \State $best.compromissions = compromissions$;
            \EndIf
        \EndForAll
        \State
        \State \Return $best.response$;
    \Else
        \State \Return $\emptyset$;
    \EndIf
\EndProcedure

\Procedure{tpadv.Sample}{process, message}
    \State \textbf{raise error};
\EndProcedure

\Procedure{tpadv.Deliver}{process, message}
    \State $invocations \leftarrow invocations + 1$;
    \State $sys.Deliver(process, message)$;
\EndProcedure

\algstore{twophasedecorator}
\end{algorithmic}
\end{algorithm}

\begin{algorithm}
\begin{algorithmic}[1]
\algrestore{twophasedecorator}

\Procedure{tpadv.Echo}{process, sample, source, message}
    \State $invocations \leftarrow invocations + 1$;
    \State $sys.Echo(process, sample, source, message)$;
\EndProcedure

\Procedure{tpadv.End}{{}}
    \State $sys.End()$;
\EndProcedure
\end{algorithmic}
\end{algorithm}

\begin{lemma}
\label{lemma:twophaseadversaries}
The set of two-phase adversaries $\mathcal{A}_{tp}$ is optimal.

\begin{proof}
We again prove the result using a decorator. Here we show that a decorator $\Delta_{tp}$ exists such that, for every $\alpha \in \mathcal{A}_{sm}$, the adversary $\alpha' = \Delta_{tp}(\alpha)$ is a two-phase adversary, and more powerful than $\alpha$. If this is true, then the lemma is proved: let $\alpha^*$ be an optimal adversary, then the sequential $\alpha^+ = \Delta_{tp}(\alpha^*)$ is optimal as well.

\paragraph{Decorator}

\cref{algorithm:twophasedecorator} implements \textbf{Two-phase decorator}, a decorator that transforms a state-polling adversary into a two-phase adversary. Provided with a state-polling adversary $spadv$, Two-phase decorator acts as an interface between $spadv$ and a system $sys$, effectively implementing a single-response adversary $tpadv$. Two-phase decorator exposes both the adversary and the system interfaces: the underlying adversary $spadv$ uses $tpadv$ as its system.

Two-phase decorator works as follows:
\begin{itemize}
    \item Procedure $tpadv.Init()$ initializes a $invocations$ variable: at any time, $invocations$ counts the number of invocations issued by $spadv$.
    \item Procedure $compatible(invocations)$ returns a set of systems $\sigma$ that satisfy the following properties:
    \begin{itemize}
        \item For every correct process $\pi$, the number of Byzantine processes in $\pi$'s first echo sample is identical in $\sigma$ and $sys$.
        \item The length $\abs{\eta(\alpha, \sigma)}$ of the first phase when $\alpha$ is coupled with $\sigma$ is equal to $invocations$.
    \end{itemize}
    \item Procedure $tpadv.Step()$ simply forwards the call to $spadv.Step()$.
    \item Procedure $tpadv.Byzantine(process)$ increments $invocations$, then returns an array of $count$ elements, $count$ being the number of elements returned from $sys.Byzantine(process)$. The array is filled with $\bot$ values: since $spadv$ is Byzantine-counting, the content of the array is irrelevant.
    \item Procedure $tpadv.State()$ increments $invocations$. It then returns an empty set if $sys.State()$ is empty. If $sys.State()$ is not empty, the procedure returns, among all the possible responses that are compatible with the trace exchanged between $spadv$ and $tpadv$, the one that maximizes the probability of $spadv$ compromising the consistency of $sys$. This is achieved as follows:
    \begin{itemize}
        \item The procedure loops over every system $\sigma$ in the set $compatible(\allowbreak invocations)$. In doing so, the procedure loops over every system $\sigma$ such that: $\sigma$ has the same Byzantine count as $sys$; when $\alpha$ is coupled with $\sigma$, it concludes the first phase in exactly $invocations$ invocations.
        \item For every process $\sigma$ in $compatible(invocations)$, the procedure stores in a set $outcome$ a $(response, fulltrace)$ pair, $response$ being the $State()$ of $\sigma$ at the end of the first phase ($response$ is extracted from $\tau(\alpha, \sigma)_{invocations}$), and $fulltrace$ being $\tau(\alpha, \sigma)$, the full trace exchanged between $\alpha$ and $\sigma$.
        \item For every $(response, fulltrace)$ in $outcomes$, the procedure loops over every system $\sigma$ in $compatible(invocations)$, and counts the number of systems whose consistency is compromised by $fulltrace$. The procedure returns the value of $response$ that maximizes the number of systems in $compatible(invocations)$ whose consistency is compromised by $fulltrace$.
    \end{itemize}
    \item Procedure $tpadv.Sample(process, message)$ is never called. This is due to the fact that $spadv$ is sample-blind.
    \item Procedure $tpadv.Deliver(process, message)$ increments $invocations$, then forwards the call to $sys.Deliver(process, message)$.
    \item Procedure $tpadv.Echo(process, \allowbreak sample, \allowbreak source, \allowbreak message)$ increments $invocations$, then forwards the call to $sys.Echo(process, \allowbreak sample, \allowbreak source, \allowbreak message)$.
    \item Procedure $tpadv.End()$ simply forwards the call to $sys.End()$.
\end{itemize}

\paragraph{Correctness}

Here we prove that every adversary, coupled with Two-phase decorator:
\begin{itemize}
    \item Has a well-defined behavior. An adversary has undefined behavior if, at any point, the sequence of invocations and responses it exchanges with $tpadv$ is incompatible with every system. 
    \item Is Byzantine-counting and single-response.
\end{itemize}

We start by noting that, since $invocations$ is incremented every time $spadv$ issues an invocation, when $tpadv.State()$ is invoked and $sys.State() \neq \emptyset$ we have $invocations = \abs{\eta(\alpha, \sigma)}$. 

Every invocation of a procedure in $tpadv$ is always forwarded to the corresponding procedure in $sys$, except for $tpadv.State()$. Whenever $sys.State() = \emptyset$, $tpadv.State()$ returns $\emptyset$ as well. Therefore, up to the $(\abs{\eta(\alpha, sys)} - 1)$-th invocation, the trace exchanged between $spadv$ and $tpadv$ is trivially compatible with $sys$.

Procedure $compatible(invocations)$ returns all systems $\sigma$ such that the condition $\abs{\sigma.Byzantine(\pi)} = \abs{sys.Byzantine(\pi)}$ holds for all $\pi \in \Pi_C$, and $\abs{\eta(\alpha, \sigma)} = invocations = \abs{\eta(\sigma, sys)}$. It is immediate to see that $compatible(invocations)$ is non-empty, as it includes $sys$. Every system $\sigma \in compatible(invocations)$ is compatible with the first $n - 1$ elements of the trace exchanged between $spadv$ and $tpadv$. Procedure $tpadv.State()$ then returns a response $response$, such that
\begin{equation*}
    \tau(\alpha, \sigma)_{invocations} = (({\tt State}), response)
\end{equation*}
for some $\sigma \in compatible(invocations)$. Therefore, the first $n$ elements of the trace exchanged between $spadv$ and $tpadv$ is compatible with $\sigma$. Due to \cref{lemma:firstphasematch}, the entire trace exchanged between $spadv$ and $tpadv$ is compatible with $\sigma$.

It is easy to see that $tpadv$ always implements a Byzantine-counting and single-response adversary. Indeed: whenever $tpadv$ invokes $sys.Byzantine(\allowbreak \pi)$, it invokes $\abs{sys.Byzantine(\pi)}$; $tpadv.State()$ returns a non-empty set if and only if $sys.State()$ returns a non-empty set, and $spadv$ is a single-response adversary.

\paragraph{Two-phase}

It is immediate to see that Two-phase decorator always implements a two-phase adversary. Indeed, whenever $tpadv$ invokes $sys.State()$, it invokes $(sys.State() \neq \emptyset)$.

\paragraph{System partitioning}

Let $\alpha$ be a state-polling adversary, let $\sigma$ be a system. Let us denote with $\mathcal{S}^*$ the set of non-poisoned systems. We denote with $\stackrel{\alpha}{\sim}$ the two conditions $\forall \pi \in \Pi_C, \forall m \in \mathcal{M}$,
\begin{equation*}
    \abs{\cp{n \in 1..E \mid \sigma[\pi][m][n] \in \Pi_C}} = \abs{\cp{n \in 1..E \mid \sigma'[\pi][m][n] \in \Pi_C}}
\end{equation*}
and
\begin{equation*}
    \abs{\eta(\alpha, \sigma)} = \abs{\eta(\alpha, \sigma')}
\end{equation*}

It is immediate to see that $\stackrel{\alpha}{\sim}$ is an equivalence relation, and we can use $\stackrel{\alpha}{\sim}$ to partition $\mathcal{S}^*$:
\begin{equation*}
    \mathcal{S}[\alpha]_1, \ldots, \mathcal{S}[\alpha]_h = \frac{\mathcal{S}}{\stackrel{\alpha}{\sim}}
\end{equation*}

Let $i \in 1..h$. Due to \cref{lemma:firstphaselength}, we have
\begin{equation*}
    \forall \sigma, \sigma' \in \mathcal{S}[\alpha]_i, \forall n < \abs{\eta(\alpha, \sigma)},\; \tau(\alpha, \sigma)_n = \tau(\alpha, \sigma')_n
\end{equation*}

Moreover, since $\sigma$ is not poisoned, $\eta(\alpha, \sigma)$ includes at least one call to $Deliver(\ldots)$. Therefore, for every $i \in 1..h$, let $\sigma \in \mathcal{S}[\alpha]_i$, we can define a function $\delta[\alpha]_i: \mathcal{M} \times 1..(\abs{\eta(\alpha, \sigma})$ by
\begin{equation*}
    \pi \in \delta[\alpha]_i(m, n) \xLeftrightarrow[]{def} \exists j < n \mid \tau(\alpha, \sigma)_j = (({\tt Deliver}, \pi), \bot)
\end{equation*}
Intuitively, $\delta[\alpha]_i(m, n)$ represents the set of correct processes that $\alpha$ causes to \pbin.Deliver $m$ before the $n$-th invocation, when $\alpha$ is coupled with any $\sigma \in \mathcal{S}[\alpha]_i$.

We additionally define $\pi[\alpha]_i: \mathcal{M} \rightarrow \powerset{\Pi_C}{}$, $\pi^-[\alpha]_i: \mathcal{M} \rightarrow \powerset{\Pi_C}{}$ by, let $\sigma \in \mathcal{S}[\alpha]_i$,
\begin{eqnarray*}
    \pi[\alpha]_i(m) &=& \delta[\alpha]_i(m,  \abs{\eta(\alpha, \sigma)}) \\
    \pi^-[\alpha]_i(m) &=& \delta[\alpha]_i(m,  \abs{\eta(\alpha, \sigma)} - 1))
\end{eqnarray*}
Intuitively, $\pi[\alpha]_i(m)$ represents the set of correct processes that $\alpha$ causes to \pbin.Deliver $m$ throughout the first phase, when $\alpha$ is coupled with any $\sigma \in \mathcal{S}[\alpha]_i$. Noting that $\alpha$ is a state-polling adversary, and $\sigma$ is not poisoned, then $\pi^-[\alpha]_i(m)$ represents the set of correct processes that $\alpha$ causes to \pbin.Deliver $m$ throughout the first phase when $\alpha$ is coupled with any $\sigma \in \mathcal{S}[\alpha]_i$, excluding the last invocation to $Deliver(\ldots)$ in $\eta(\alpha, \sigma)$.

Finally, we define $m(\alpha)_i$ by, let $\sigma \in \mathcal{S}[\alpha]_i$
\begin{equation*}
    \tau(\alpha, \sigma)_{\abs{\eta(\alpha, \sigma)}} = (({\tt Deliver}, \pi \in \Pi_C, m), \bot)
\end{equation*}
Intuitively, $m(\alpha)_i$ is the last message that $\alpha$ causes a correct process to \pbin.Deliver throughout the first phase, when $\alpha$ is coupled with any $\sigma \in \mathcal{S}[\alpha]_i$. Noting that $\alpha$ is a state-polling adversary, and that $\sigma$ is not poisoned, $m$ is the only message delivered by at least one correct process at the end $\eta(\alpha, \sigma)$.

Let $\sigma, \sigma' \in \mathcal{S}[\alpha]_i$. We can prove that $\stackrel{\alpha}{\sim}$ can be equivalently restated as $\forall \pi \in \Pi_C, \forall m \in \mathcal{M}$
\begin{equation*}
\abs{\cp{n \in 1..E \mid \sigma[\pi][m][n] \in \Pi_C}} = \abs{\cp{n \in 1..E \mid \sigma'[\pi][m][n] \in \Pi_C}}
\end{equation*}
and
\begin{eqnarray*}
&\nexists& \pi \in \Pi_C \mid \\
&&\abs{\cp{n \in 1..E \mid \sigma[\pi][m(\alpha)_i][n] \in (\pi^-[\alpha]_i(m(\alpha)_i) \cup (\Pi \setminus \Pi_C)))}} \geq \hat E \\
&\exists& \pi \in \Pi_C \mid \\
&&\abs{\cp{n \in 1..E \mid \sigma[\pi][m(\alpha)_i][n] \in (\pi[\alpha]_i(m(\alpha)_i) \cup (\Pi \setminus \Pi_C)))}} \geq \hat E \\
&\nexists& m \neq m(\alpha)_i, \pi \in \Pi_C \mid \\
&& \abs{\cp{n \in 1..E \mid \sigma[\pi][m][n] \in (\pi[\alpha]_i(m) \cup (\Pi \setminus \Pi_C))}} \geq E
\end{eqnarray*}

Indeed, we are restating the condition $\abs{\eta(\alpha, \sigma)} = \abs{\eta(\alpha, \sigma')}$ with the following conditions:
\begin{itemize}
    \item No correct process has, in its echo sample for $m(\alpha)_i$, at least $\hat E$ processes that are either Byzantine, or \pbin.Deliver $m(\alpha)_i$ as a result of any invocation of $Deliver(\ldots)$ in $\eta(\alpha, \sigma)$ except the last. This encodes the condition that no correct process delivers $m(\alpha)_i$ before the last invocation of $Deliver(\ldots)$ in $\eta(\alpha, \sigma)$.
    \item At least one correct process has, in its echo sample for $m(\alpha)_i$, at least $\hat E$ processes that are either Byzantine, or \pbin.Deliver $m(\alpha)_i$ throughout the first phase when $\alpha$ is coupled with $\sigma$. This encodes the condition that at least one correct process delivers $m(\alpha)$ after the last invocation of $Deliver(\ldots)$ in $\eta(\alpha, \sigma)$.
    \item No correct process has, in its echo sample for $m \neq m(\alpha)_i$, at least $\hat E$ processes that are either Byzantine, or \pbin.Deliver $m$ throughout the first phase when $\alpha$ is coupled with $\sigma$. This encodes the condition that no message is delivered before $m(\alpha)_i$.
\end{itemize}

Let $\sigma, \sigma' \in \mathcal{S}[\alpha]_i$. We denote with $\stackrel{m}{\sim}$ the condition $\forall \pi \in \Pi_C$,
\begin{equation*}
    \sigma[\pi][m(\alpha)_i] = \sigma'[\pi][m(\alpha)_i]
\end{equation*}

Again, $\stackrel{m}{\sim}$ is an equivalence relation, and can be used to partition $\mathcal{S}[\alpha]_i$:
\begin{equation*}
    \mathcal{S}[\alpha]^i_1, \ldots \mathcal{S}[\alpha]^i_l = \frac{\mathcal{S}[\alpha]_i}{\stackrel{m}{\sim}}
\end{equation*}

Let $\bar \sigma \in \mathcal{S}[\alpha]_i$, let $\tau = \tau(\alpha, \bar \sigma)$. For any $\sigma \in \mathcal{S}[\alpha]_i$, $\tau$ compromises the consistency of $\sigma$ if $\tau$ causes at least one correct process to deliver a message $m' \neq m(\alpha)_i$ throughout the second phase. Since this condition is independent from the echo sample for $m(\alpha)_i$ of any correct process, we finally have that, for every $j \in 1..l$,
\begin{equation*}
    \prob{\tau \searrow \rp{\sigma \in \mathcal{S}[\alpha]^i_j}} = \prob{\tau \searrow \rp{\sigma \in \mathcal{S}[\alpha]_i}}
\end{equation*}

\paragraph{Adversarial power}

Here we prove that $\alpha' = \Delta_{tp}(\alpha)$ is more powerful than $\alpha$. Let $\bar \sigma$ denote a random system in $\mathcal{S}$.

Let us assume that $\bar \sigma$ is poisoned. Since both $\alpha$ and $\alpha'$ are auto-echo adversaries, both compromise $\bar \sigma$.

Let us assume that $\bar \sigma$ is not poisoned. For some $i, j$, we therefore have $\sigma \in \mathcal{S}[\alpha]^i_j$. When $tpadv.State()$ is invoked and $sys.State() \neq \emptyset$, the procedure returns a response $best.response$ such that the trace $\tau^*$ that $\alpha$ issues as a result of $best.response$ satisfies
\begin{equation*}
    \tau^* = \argmax_\tau \prob{\tau \searrow \rp{\sigma \in \mathcal{S}[\alpha]_i}}
\end{equation*}

As we proved in the previous section, we therefore have
\begin{equation*}
    \prob{\tau^* \searrow \rp{\sigma \in \mathcal{S}[\alpha]_i}} \geq \prob{\tau \searrow \rp{\sigma \in \mathcal{S}[\alpha]_i}} = \prob{\tau \searrow \rp{\sigma \in \mathcal{S}[\alpha]^i_j}}
\end{equation*}
which proves that, if $\sigma$ is not poisoned, then the probability of $\alpha'$ compromising $\sigma$ is greater or equal to the probability of $\alpha$ compromising $\sigma$.

The adversarial power of $\alpha'$ is therefore greater or equal to the adversarial power of $\alpha$, and the lemma is proved.
\end{proof}
\end{lemma}

\clearpage
\section{\contagion}
\label{appendix:contagion}

In this appendix we discuss \emph{epidemic processes}, mimicking the spread of a disease in a population, and the \contagion\ game, which gives a player the possibility to actively infect parts of a population.

As we discuss in \cref{subsection:prbalgorithm}, in \prbal, when a correct process receives enough {\tt Ready} messages from its \emph{ready sample} for the same message $m$,  it issues itself a {\tt Ready} message for $m$.
This produces a feedback mechanism that, in Appendix \ref{appendix:prbal}, we show to be isomorphic to an epidemic process as we define it below.

\contagion\ is a game where a player iteratively applies the epidemic process to chosen inputs. We use \contagion\ for modeling and analyzing our \prbal\ algorithm.

\subsection{Epidemic processes}

An epidemic process models the spreading of a disease in a population.

\subsubsection{Preliminary definitions}

\begin{definition}[Directed multigraph]
\label{definition:directedmultigraph}
A \textbf{directed multigraph} is a pair $g = (v, e)$, where $v$ is a set and $e: v^2 \rightarrow \mathbb{N}$ is a multiset whose elements are pairs of elements of $v$. We call the elements of $v$ the \textbf{vertices} (or \textbf{nodes}) of $g$. We call the elements of $e$ the \textbf{edges} of $g$.
\end{definition}

Following from \cref{definition:directedmultigraph}, a directed multigraph allows self-loops (let $a \in v$, $(a, a)$ can be an element of $e$) and multiple edges (let $a, b \in v$, the multiplicity of $(a, b)$ in $e$ can be greater than one).

\subsubsection{Contagion state}

The spreading of a disease is represented by a \textbf{contagion state}.

\begin{definition}[Contagion state]
\label{definition:contagionstate}
A \textbf{contagion state} is a pair $s = (g, w)$, where $g = (v, e)$ is a multigraph and $w \in \powerset{v}{}$. We call the elements of $w$ the \textbf{infected nodes} of $s$.
\end{definition}

Let $s = ((v, e), w)$ be a contagion state. In an epidemic process:
\begin{itemize}
\item Each node in $v$ corresponds to one individual member of the population.
\item A node is always in one of two possible states: \textbf{healthy} or \textbf{infected}. We do not consider any cure---once a node becomes infected, it stays infected forever. The set $w$ represents the nodes that are infected.
\item Edges model interactions between the members of the population. The multiset of edges $e$ represents the "can infect" relation. Note that this relation is not symmetric. A directed edge $(a \rightarrow b)$ between nodes $a$ and $b$ means that $a$ can infect $b$, but not that $b$ can infect $a$. 
\end{itemize}

\begin{definition}[Predecessors]
\label{definition:predecessors}
Let $g = (v, e)$ be a multigraph, let $a \in v$. Then the \textbf{predecessors} of $a$ in $g$ form the multiset $\mathfrak{p}\qp{a}: v \rightarrow \mathbb{N}$ defined by
\begin{equation*}
    \mathfrak{p}\qp{a}\rp{x \in v} = e(x, a)
\end{equation*}
\end{definition}

Following from \cref{definition:predecessors}, the predecessors of a node $a$ in a multigraph $g$ form the multiset of nodes that have an edge to $a$. If a node has multiple edges to $a$, then it has a multiplicity greater than one in $\mathfrak{p}\qp{a}$.

\subsubsection{Contagion rule}
\label{subsubsection:contagionrule}

In an epidemic process, the infection of healthy nodes follows a single rule.
\begin{itemize}
    \item \textbf{Contagion rule}: A healthy node becomes infected if the number of its infected predecessors reaches a critical threshold.
\end{itemize}

The input to an epidemic process is a contagion state $s$. The epidemic process repeatedly applies the contagion rule to $s$ until either all nodes are infected or no healthy node has enough infected predecessors to become infected itself. The epidemic process outputs the resulting contagion state.

\subsection{\contagion}
\label{subsection:pebbling}

\contagion\ is a game played on the nodes of a random directed multigraph $g$. \contagion\ consists of one or more \textbf{rounds}. Each round inputs a contagion state $s$ and outputs a contagion state $s'$. The input to the first round is the contagion state $(g, \emptyset)$, i.e., the contagion state with no infected nodes whose multigraph is $g$. The input to any other round is the output of the previous round.

A round with input $s$ is played as follows:
\begin{itemize}
    \item The \textbf{player} infects a fixed-size subset of the healthy nodes of the contagion state $s$. This results in a contagion state $s'$.
    \item The contagion state $s'$ is provided as input to an epidemic process. The output $s''$ of the epidemic process is returned.
\end{itemize} 

\subsection{Rules}
\label{subsection:rules}

In this section, we formally define the rules of \contagion\ and introduce its \textbf{parameters}. \\ 

\contagion\ is played on the nodes of a random, directed multigraph $g = (v, e)$. The in-degree of each node $n$ in $v$ is independently binomially distributed; each predecessor of $n$ is uniformly picked with replacement from $v$.

\subsubsection{Parameters}

A game of \contagion\ depends on the following numerical parameters:
\begin{itemize}
    \item \textbf{Node count} ($N \in \mathbb{N}$): Represents the number of nodes in the multigraph ($N = \abs{v}$).
    \item \textbf{Sample size} ($R \in \mathbb{N}$): Represents the maximum in-degree of a node in the multigraph.
    \item \textbf{Link probability} ($l \in [0, 1]$): Represents the probability of a predecessor link being successfully established. The in-degree of a node follows the distribution $\text{Bin}[R, l]$.
    \item \textbf{Round count} ($K \in \mathbb{N}_{> 0}$): Represents the number of rounds in the game.
    \item \textbf{Infection batch} ($(S < N) \in \mathbb{N}_{> 0}$): Represents the number of healthy nodes the player infects at the beginning of each round.
    \item \textbf{Contagion threshold} ($(\hat R \leq R) \in \mathbb{N}$): Represents the number of infected predecessors that will cause an healthy node to become infected (see Contagion rule).
\end{itemize}

\subsubsection{Game}
\label{subsubsection:game}

A game of \contagion\ is played as follows:

\begin{itemize}

\item A random, directed multigraph $g = (v, e)$ with $N$ nodes is built. For every node $n$ in $v$:
\begin{itemize}
    \item $R$ times:
    \begin{itemize}
        \item A Bernoulli random variable $\bar B \leftarrow \text{Bern}\qp{l}$ is sampled.
        \item If $\bar B = 1$, then a random node $m$ is selected with uniform probability from $v$, and the edge $m \rightarrow n$ is added to $e$ (i.e., $m$ is added to the predecessors of $n$).
    \end{itemize}
\end{itemize}
\item Let $s = (g, w = \emptyset)$ be a contagion state. For $K$ \textbf{rounds}:
\begin{itemize}
\item If at least $S$ nodes in $v$ are healthy (i.e., they are not in the set of infected nodes $w$), the player selects $S$ distinct nodes and infects them. The player \textbf{cannot} inform this choice with knowledge of the topology of $g$.
\item An epidemic process is run on $s$: until either every node in $v$ is infected (i.e., $v = w$), or no healthy node in $v$ has at least $\hat R$ infected predecessors, the following \textbf{contagion step} is iterated:
\begin{itemize}
\item Every node in $v$ with at least $\hat R$ infected predecessors is infected, i.e., it is added to $w$.
\end{itemize}
\end{itemize}
\end{itemize}

\begin{figure}

{
    \centering
    \includegraphics[scale=0.63]{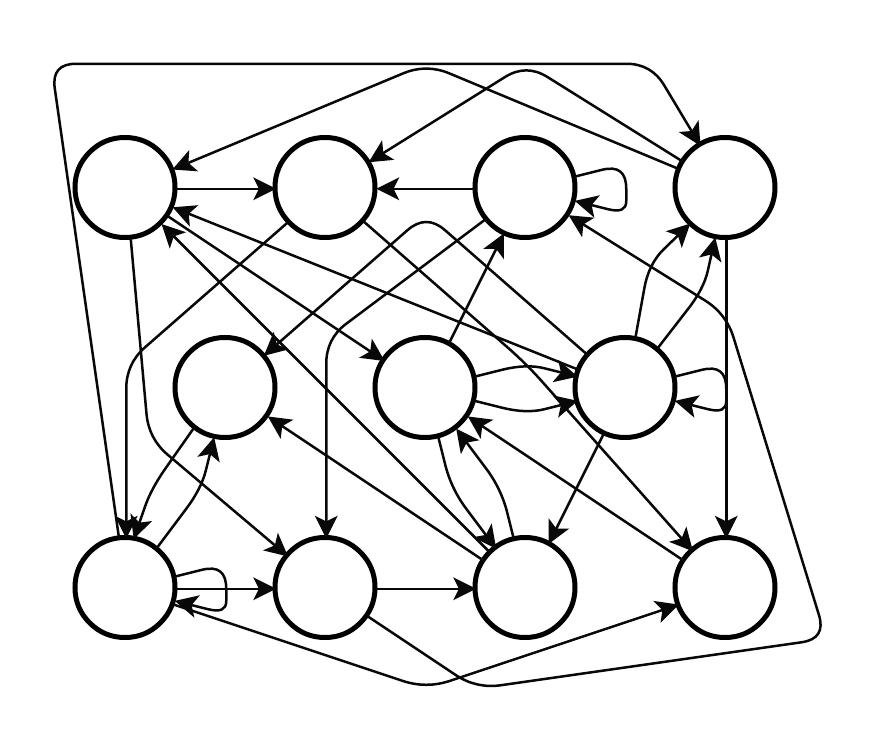}
    \includegraphics[scale=0.63]{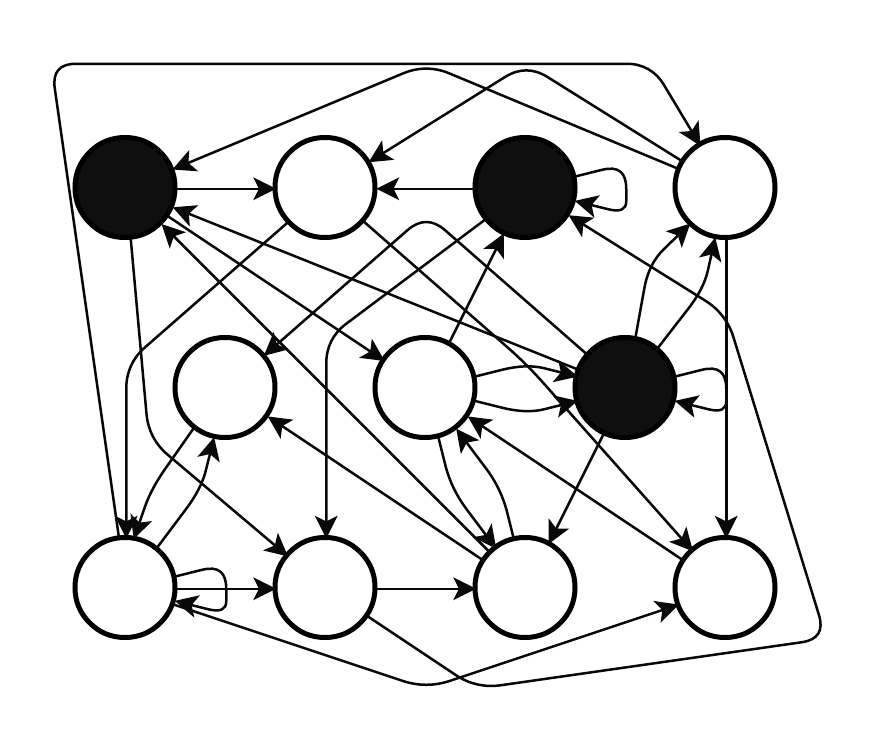}
}
\vspace{0.5mm}
\tabto*{0.20\textwidth}(1)\tabto*{0.65\textwidth}(2)
\vspace{2mm}

{
    \centering
    \includegraphics[scale=0.63]{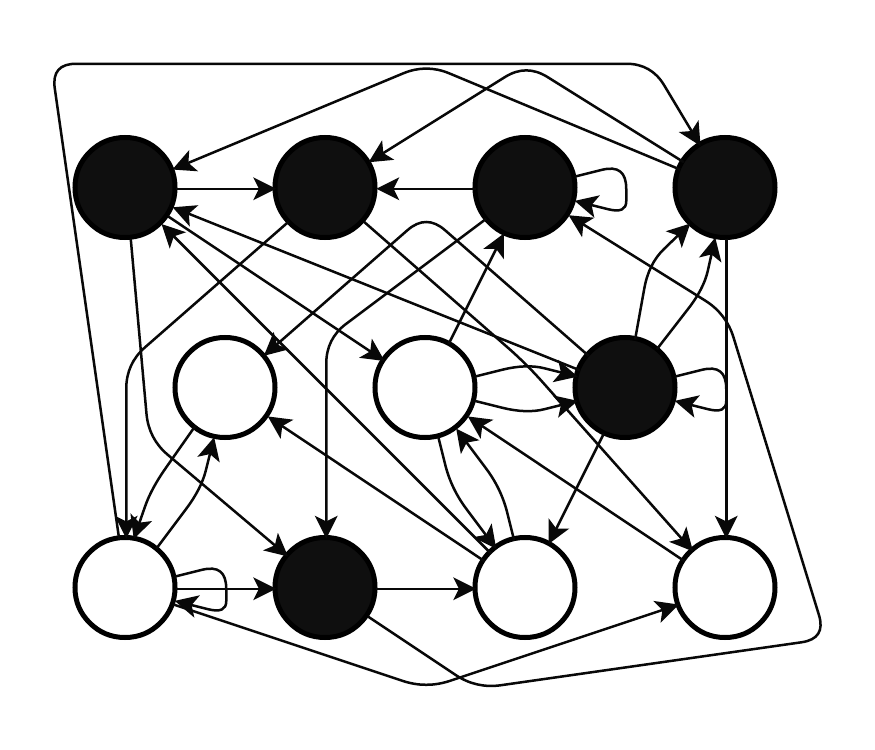}
    \includegraphics[scale=0.63]{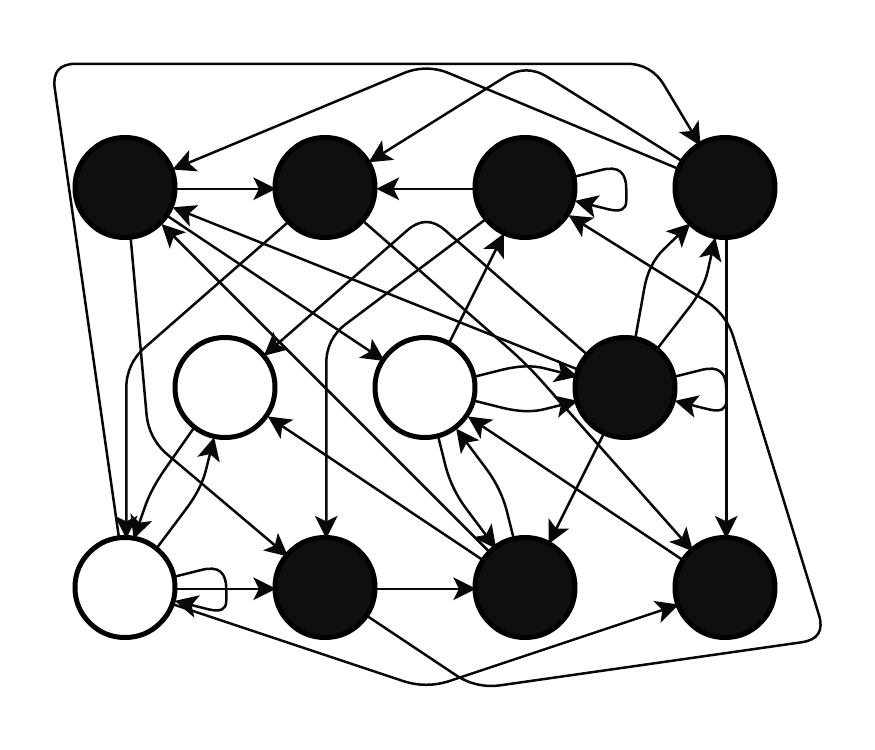}
}
\vspace{0.5mm}
\tabto*{0.20\textwidth}(3)\tabto*{0.65\textwidth}(4)
\vspace{2mm}

{
    \centering
    \includegraphics[scale=0.63]{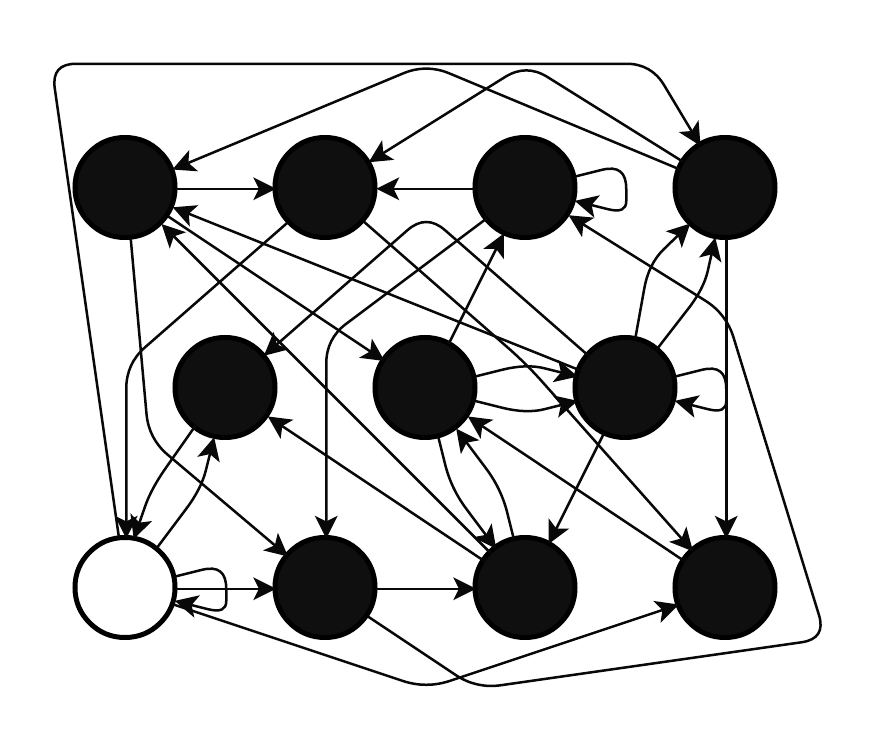}
    \includegraphics[scale=0.63]{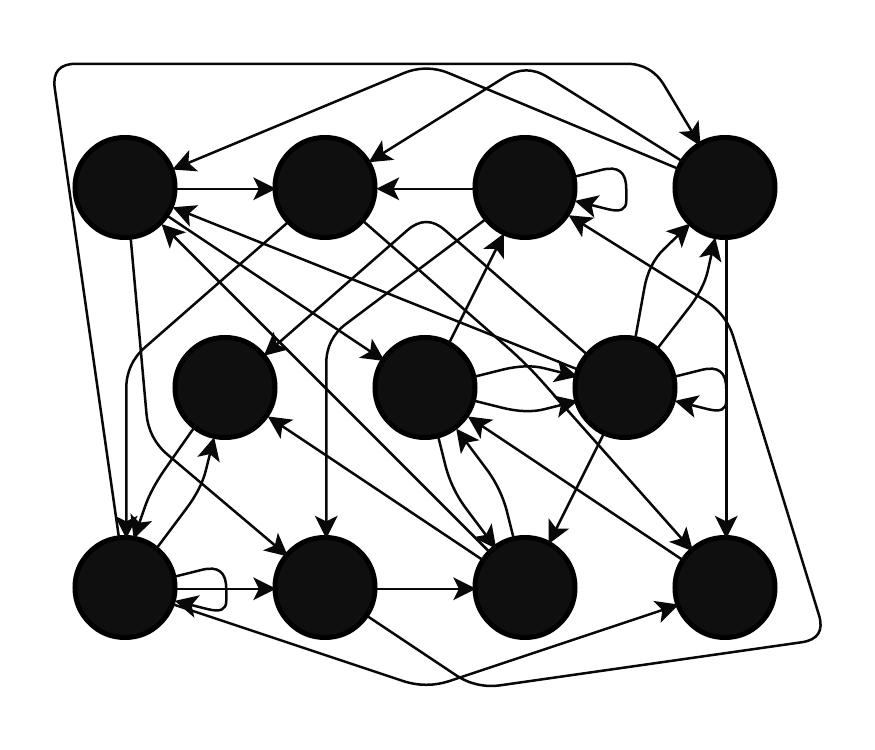}
}
\vspace{0.5mm}
\tabto*{0.2\textwidth}(5)\tabto*{0.65\textwidth}(6)

\caption{An example game of \contagion. Here $N = 11$, $l = 1$, $R = 3$, $\hat R = 2$, $K = 1$ and $S = 3$. Notice how nodes can be linked to themselves, form loops, or be linked more than once. An initial set of $S$ nodes (1) is infected by the player (2). The game then unfolds in contagion steps (3 to 6): whenever a node has at least $\hat R$ infected predecessors, it becomes infected. This example shows how easily a game of \contagion\ can converge to a fully-infected configuration.}
\label{figure:pebblinggame}
\end{figure}

\cref{figure:pebblinggame} shows an example game of \contagion\ with small parameters.

\subsection{Random variables}
\label{subsection:randomvariables}

We introduce the following random variables, which we discuss in more formal detail in the next sections:
\begin{itemize}
    \item \textbf{Infection size} $N^r_i$: represents the number of infected nodes at round $r$ and step $i$.
    \item \textbf{Frontier size} $U^r_i$: represents the number of nodes that are infected at round $r$ and step $i$, but are not infected at round $r$ and step $i - 1$.
    \item \textbf{Infection status} $W^r_i[j]$: represents whether or not the $j$-th node is infected at round $r$ and step $i$. We use $W^r_i[j]$ to signify that the node is infected, and $\cancel{W^r_i[j]}$ to signify that the node is not infected.
    \item \textbf{Infected predecessors count} $V^r_i[j]$: represents the number of predecessors of the $j$-th node that are infected at round $r$ and step $i$.
\end{itemize}

\textbf{Remark}: for the sake of readability, the round number and/or the node index (for $W$ and $V$) will be omitted whenever it can be unequivocally inferred from the context.

\subsection{Goal}
\label{subsection:goal}

The goal of this appendix is to compute the probability distribution underlying the number of infected nodes at the end of a game of \contagion.

\begin{lemma}
\label{lemma:pebblingconvergence}
For any $r$, the random variables $N^r_i$, $U^r_i$, $W^r_i[j]$, and $V^r_i$ converge in a finite number of steps.
\begin{proof}
We note the following:
\begin{itemize}
    \item $N^r_i$ is a non-decreasing function of $i$, and $N^r_i \leq N$.
    \item $U^r_{i > 0} = N^r_i - N^r_{i - 1}$.
    \item For any $j$, $W^r_i[j] \implies W^r_{i + 1}[j]$.
    \item $V^r_i$ is a non-decreasing function of $i$ and $V^r_i \leq R$.
\end{itemize}

From the above follows that all random variables converge for $i \rightarrow \infty$.

The codomains of $N$, $U$, $W$ and $V$ are all finite. Therefore, all random variables converge in a finite number of steps.
\end{proof}
\end{lemma}

\begin{corollary}
\label{corollary:finiterounds}
All rounds terminate in a finite number of contagion steps.
\end{corollary}

\begin{notation}[End of round]
We use $N^r_\infty$, $U^r_\infty$, $W^r_\infty[j]$, $V^r_\infty$ to denote the values of $N$, $U$, $W$, $V$ at the end of round $r$.
\end{notation}

The goal of this appendix is to compute the probability distribution underlying the random variable
\begin{equation}
\label{equation:pebblinggame}
\gamma(N, R, l, K, S, \hat R) = N^K_{\infty}
\end{equation}
i.e., the probability of a game of \contagion\ resulting in $\bar N^K_\infty$ nodes ultimately being infected. \Cref{lemma:pebblingconvergence} proves that $\Gamma$ is a well defined variable (i.e., the limit exists) and, since $K$ is finite, can be computed in a finite total number of steps.

\subsection{Sample space}
\label{subsection:samplespace}

In this section, we define a sample space for \contagion, i.e., the set of all possible outcomes of a \contagion\ game. As we described in \cref{subsection:pebbling}, the outcome of a game of \contagion\ is completely determined by two factors:
\begin{enumerate}
    \item The topology of the random multigraph $g$ on which \contagion\ is played. The probability distribution underlying $g$ is known, and we compute it in this section.
    \item The player's infection strategy, i.e., the nodes the player chooses to infect at the beginning of each round. The probability distribution underlying the player's choices is \textbf{unknown} and arbitrary. In this section, we only formalize their sample space.
\end{enumerate}

Thus, an element of the sample space is a pair consisting of a multigraph (1.) and an infection strategy (2.).

\subsubsection{Multigraph}
\label{subsubsection:multigraph}

As discussed in \cref{subsubsection:game}, a game of \contagion\ is played on the nodes of a multigraph $g = (v, e)$ allowing multi-edges and loops. Every node in $v$ has at most $R$ predecessors. Therefore, $g$ can be represented by a \textbf{predecessor matrix} as we define it below.

We start by explicitly labeling the elements of $v$.

\begin{notation}[Vertices]
Let $g = (v, e)$ be a multigraph, with $\abs{v} = N$. Without loss of generality, we label the elements of $v$ using natural numbers:
\begin{equation*}
    v = 1..N
\end{equation*}
\end{notation}

Since every node in $g$ has at most $R$ predecessors, for every $j \in v$ we can represent the elements of $ \mathfrak{p}\rp{j}$ as the components of a \emph{predecessor vector}.

\begin{definition}[Predecessor vector]
A \textbf{predecessor vector} is an element of the set
\begin{equation*}
    \mathcal{R} = \rp{\cp{\bot} \cup v}^R
\end{equation*}

In a multigraph $g = (v, e)$, whose in-degree is bound by $R$, we use a predecessor vector to represent the predecessors of a node.
Let $r \in \mathcal{R}$ be the predecessor vector of a node $j \in v$. If $r_k = \bot$, we say that the $k$-th predecessor of $j$ is \textbf{missing}.
\end{definition}

As discussed in \cref{subsubsection:game}, the predecessors of each node in $v$ are generated by independently sampling $R$ times a value $\bar B$ from a Bernoulli variable; whenever $\bar B = 1$, an additional predecessor is uniformly picked with replacement from the elements of $v$. 
We call a vector of predecessors selected this way a \emph{random predecessor vector}, as formally defined in \cref{definition:randompredecessorvector}.

\begin{definition}[Random predecessor vector]
\label{definition:randompredecessorvector}
A \textbf{random predecessor vector} is a predecessor vector generated by the procedure described in \cref{subsubsection:game}.

Let $r$ be a random predecessor vector. For every $k \in \cp{1, \ldots, R}$, $\bar B \leftarrow \text{Bern}\qp{l}$ is independently sampled; if $\bar B = 0$, $r_k$ is set to $\bot$, otherwise $r_k$ is set to an element of $v$, picked independently with uniform probability.
\end{definition}

\begin{lemma}
\label{lemma:randompredecessorvector}
Let $r$ be a random predecessor vector. Then
\begin{eqnarray*}
    \prob{\bar r} &=& \prod_{k = 1}^R \prob{\bar r_k} \\
    \prob{r_k = \bot} &=& \rp{1 - l} \\
    \prob{r_k = \rp{\bar r_k \in v}} &=& \frac{l}{N}
\end{eqnarray*}
\begin{proof}
Following from \cref{definition:randompredecessorvector}, each component of $r$ is independently sampled. Each component has a probability $(1 - l)$ of being missing. Each non-missing component of $r$ has an equal probability of being equal to any element of $v$.
\end{proof}
\end{lemma}

As we discussed in \cref{subsubsection:game}, the multigraph $g = (v, e)$ is constructed by independently generating the predecessors for each node in $v$. Therefore, the topology of $g$ is completely determined by $N$ predecessor vectors, that can be organized in a \emph{predecessor matrix}.

\begin{definition}[Predecessor matrix]
\label{definition:predecessormatrix}
A \textbf{predecessor matrix} is an element of the set
\begin{equation*}
    \mathcal{G} = \mathcal{R}^N
\end{equation*}
\end{definition}

\begin{notation}[Predecessor matrix]
Since a predecessor matrix uniquely identifies a multigraph, we interchangebly use $g$ to denote a predecessor matrix and its corresponding multigraph.
Let $g$ be a predecessor matrix defining a multigraph $(v, e)$, then $g_j$ is the predecessor vector of node $j \in v$.
\end{notation}

\begin{definition}[Random predecessor matrix]
\label{definition:randompredecessormatrix}
A \textbf{random predecessor matrix} is a predecessor matrix representing the outcome of the multigraph generation process described in \cref{subsubsection:game}. More formally, a random predecessor matrix consists of $N$ independent random predecessor vectors.

\end{definition}

\begin{lemma}
\label{lemma:randompredecessormatrix}
Let $g$ be a random predecessor matrix. Then
\begin{equation}
\label{equation:randompredecessormatrix}
    \prob{\bar g} = \prod_{j = 1}^N \prob{\bar g_j}
\end{equation}
\begin{proof}
It follows immediately from \cref{definition:randompredecessormatrix}.
\end{proof}
\end{lemma}

\subsubsection{Sub-threshold predecessor set}

As discussed in \cref{subsection:rules}, an epidemic process consists of a sequence of contagion steps. Let $s = ((v, e), w)$ be a contagion state. In a contagion step, a healthy node $j \in v$ ($j \in w$) becomes infected if at least $\hat R$ of its predecessors are infected, i.e., if 
\begin{equation*}
\label{equation:pebblingcondition}
    \abs{\mathfrak{p}\rp{j} \cap w} \geq \hat R
\end{equation*}

Given the set $w$, the set of predecessor vectors that do not satisfy the condition above is uniquely defined. We define \emph{sub-threshold predecessor sets} to capture this notion.

\begin{definition}[Sub-threshold predecessor set]
\label{definition:subthresholdpredecessorset}
Let $g$ be a predecessor matrix defining a multigraph $(v, e)$. Let $X \subseteq v$. The \textbf{sub-threshold predecessor set} of $X$ is the set
\begin{equation*}
    \tilde{\mathcal{R}}^X = \cp{r \in \mathcal{R} \mid \abs{\cp{k \in 1..R \mid r_k \in X}} < \hat R}
\end{equation*}

$\tilde{\mathcal{R}}^X$ contains all the predecessor vectors in $\mathcal{R}$ that have less than $\hat R$ components in $X$.
\end{definition}

\begin{figure}
\centering
\includegraphics[scale=0.65]{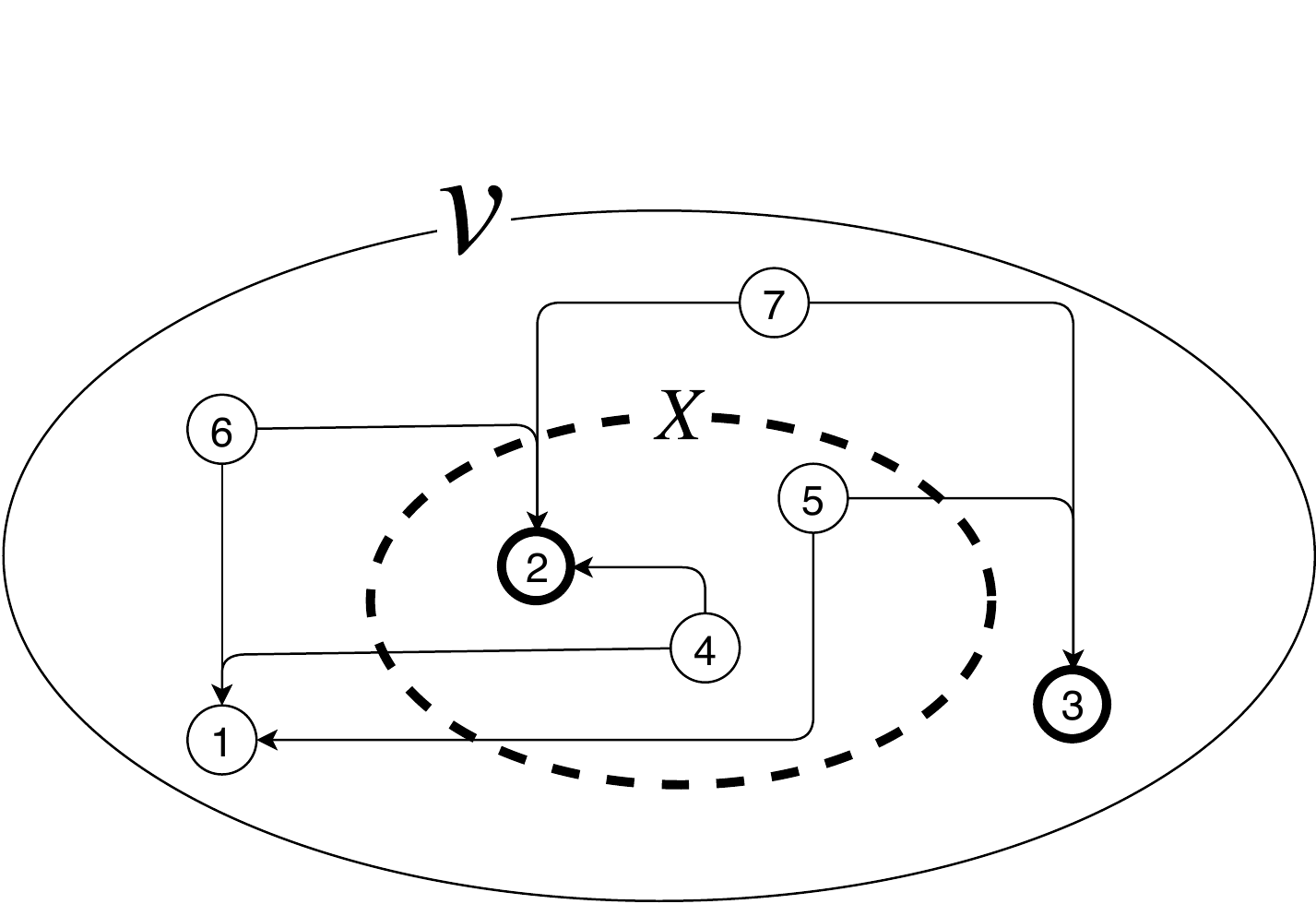}
\caption{An example multigraph $g = (v, e)$ with $7$ nodes. A subset $X \subseteq v$ is highlighted. Numbered dots represent the elements of $v$, and the edges to nodes $1$, $2$ and $3$ are displayed. With $R = 3$ and $\hat R = 2$, we have $g_1 \notin \tilde{\mathcal{R}}^X$, $g_2 \in \tilde{\mathcal{R}}^X$, and $g_3 \in \tilde{\mathcal{R}}^X$. Note how the predecessor vector of node $2$ is in the sub-threshold predecessor set of $X$ even if node $2$ is in $X$. Note how node $3$ has one missing predecessor (i.e., one of the elements in $g_3$ is $\bot$). The nodes whose predecessor vectors are in $\tilde{\mathcal{R}}^X$ are highlighted.}
\label{figure:subthreshold}
\end{figure}

\cref{figure:subthreshold} shows an example multigraph where the predecessors of three nodes are displayed, two of which are in the sub-threshold predecessor set of a given set $X$.

\subsubsection{Player's strategy}
\label{subsubsection:playerstrategy}

As discussed in \cref{subsection:rules}, at the beginning of each round of \contagion\ the player selects, if possible, $S$ distinct healthy nodes and infects them. These are the only $K$ choices the player makes throughout \contagion. Moreover, the player has no knowledge of the topology of the multigraph $g$ on which \contagion\ is played.

The player's choices can be expressed in an \emph{infection strategy}, as we formally define it in this section. Together with the topology of the multigraph on which the game is played, an infection strategy uniquely determines the outcome of an instance of \contagion.

Let $g = (v, e)$ be the multigraph on which \contagion\ is played. At the beginning of round $r$, the player knows the value of $W^{r'}_i[j]$ for every $r' < r$, every $i \in \mathbb{N}$ and every $j \in v$, which we encode in an \emph{infection history}. The player chooses a set of $S$ of the nodes that are healthy at the beginning of round $r$. We model this choice with an \emph{infection function}. We call \emph{infection strategy} the sequence of choices the player makes throughout the game.

\begin{definition}[Infection history]
An \textbf{infection history} for round $r > 0$ is an element of the set
\begin{equation*}
    \mathcal{H}_r = \rp{\rp{\cp{\bot, \top}^N}^\infty}^{r}
\end{equation*}

An infection history is a table with three indices. The first represents the round, the second represents the step, the third represents the node. Let $h \in \mathcal{H}_r$, then $h^{r'}_{i'}[j] = \top$ signifies that node $j$ is infected at round $r'$ and step $i'$.
\end{definition}

\begin{notation}[Round and step order]
Let $r, r'$ be round numbers, let $i, i'$ be step numbers. We say that $(r, i) < (r', i')$ if $(r, i)$ \emph{temporally precedes} $(r', i')$. More formally
\begin{equation*}
    (r, i) < (r', i') \Longleftrightarrow \rp{r' > r} \vee (r' = r \wedge i' > i)
\end{equation*}
\end{notation}

\begin{definition}[Valid infection history]
A \textbf{valid infection history} for round $r > 0$ is an element of the set
\begin{equation*}
    \mathcal{H}^*_r = \cp{h \in \mathcal{H}_r \mid h^{r'}_{i'}[j] = \top \implies
    h^{r''}_{i''}[j] = \top \; \forall\, (r'', i'') > (r', i')}
\end{equation*}

A valid infection history is an infection history where a node is never healed. If a node is infected at round $r'$ and step $i'$, then it also infected at any subsequent round $r''$ and step $i''$.
\end{definition}

\begin{definition}[Incomplete infection history] An \textbf{incomplete infection history} for round $r > 0$ is an element of the set
\begin{equation*}
    \mathcal{H}^+_r = \cp{
        h \in \mathcal{H}^*_r \mid 
        \abs{
            \cp{
                j \in 1..N \mid h^{r - 1}_\infty[j] = \bot
            }
        } \geq S
    }
\end{equation*}

An incomplete infection history is a valid infection history with at least $S$ healthy nodes at the end of round $r - 1$.
\end{definition}

\begin{definition}[Infection function]
An \textbf{infection function} for round $r$ is an element of the set
\begin{equation*}
    \mathcal{F}_r = \cp{f: \mathcal{H}^+_r \rightarrow \powerset{\cp{1..N}}{S} \mid \forall x \in f(h), h^{r - 1}_\infty[x] = \bot}
\end{equation*}

An infection function is a function that inputs an incomplete infection history and outputs a set of $S$ nodes, all of which are healthy at the end of round $r - 1$.
\end{definition}

\begin{definition}[Infection strategy]
\label{definition:infectionstrategy}
An \textbf{infection strategy} is an element of the set
\begin{equation*}
    \mathcal{F} = \powerset{S}{1..N} \times \prod_{r = 1}^{R-1} \mathcal{F}_r
\end{equation*}

The first element of an infection strategy is a set of $S$ nodes to infect at the beginning of round $0$. Let $r > 0$, the $r$-th element of an infection strategy is an infection function for round $r$.
\end{definition}

An infection strategy encodes all the choices a player makes during a game of \contagion:
\begin{itemize}
    \item At the beginning of round $0$, the player has no information available. All nodes are healthy, and its choice reduces to selecting $S$ of them to infect.
    \item At the beginning of round $r \geq 1$, the information available to the player is the propagation of the infection throughout all previous rounds. Such information is input to the $r$-th infection function, which returns a set of $S$ healthy nodes to infect.
\end{itemize}

\subsubsection{Sample space}

In \cref{subsection:samplespace}, we noticed how the outcome of a game of \contagion\ is completely determined once both the topology of the multigraph and the strategy of the player are known. 

In \cref{subsubsection:multigraph}, we showed how a multigraph can be expressed with a predecessor matrix, defined the space of predecessor matrices and derived the probability distribution underlying random predecessor matrices. 

In \cref{subsubsection:playerstrategy}, we showed how the choices that a player makes at the beginning of each round in response to the infection history can be encoded in infection strategies. We then defined the space of infection strategies. Unlike random multigraphs, infection strategies are under the control of the player. Therefore, a probability distribution over the space of infection strategies is not available.

As we discussed in \cref{subsection:samplespace}, an element of the sample space is a pair of a multigraph and an infection strategy.

\begin{definition}[Sample space]
The \textbf{sample space} for \contagion\ is the set $\Omega = \mathcal{G} \times \mathcal{F}$.
\end{definition}

\begin{lemma}
Let $\omega = (g, f)$ be a random element of $\Omega$. Then $\prob{\bar g, \bar f} = \prob{\bar g} \prob{\bar f}$, i.e., $g$ and $f$ are independent.
\begin{proof}
It immediately follows from the fact that the player has no knowledge of the topology of the multigraph $g$.
\end{proof}
\end{lemma}

\subsection{Random variables as sample functions}
\label{subsection:randomvariablesassamplefunctions}

In \cref{subsection:randomvariables} we intuitively defined a set of random variables to capture useful properties of a game of \contagion. In the next sections, we use those random variables to compute the probability distribution underlying the number of infected nodes at the end of a game.

In \cref{subsection:samplespace} we formally defined the sample space of a game of \contagion. We started by showing that an instance of the game is completely determined once the topology of the multigraph and the strategy of the player are known. We also computed the probability of any specific multigraph topology occurring.

In this section, we rigorously re-define the random variables we defined in \cref{subsection:randomvariables} by expressing them as functions on the sample space as defined in \cref{subsection:samplespace}.

\subsubsection{Infection history}

As discussed in \cref{subsubsection:playerstrategy}, an infection function for round $r$ inputs an incomplete infection history for round $r$ and outputs a set of $S$ nodes to infect out of those that are healthy at the end of round $r - 1$.

We introduce two useful functions to manipulate infection histories.

\begin{definition}[Sample history function]
\label{definition:samplehistoryfunction}
The \textbf{sample history function} for round $r$ is the function $\mathfrak{h}_r: \Omega \rightarrow \mathcal{H}^*_r$ defined by
\begin{equation*}
    \rp{\mathfrak{h}_r\rp{\omega}}^{r'}_i[j] = W^{r'}_i[j]\rp{\omega}
\end{equation*}

The sample history function for round $r$ inputs a sample $\omega$ and outputs the valid infection history for round $r$ produced by $\omega$.
\end{definition}

Note how the definition of sample history function relies on the definition of the infection status $W$. We introduced $W$ in \cref{subsection:randomvariables}, and we formally define it in the next section.

\begin{definition}[Sample completion function]
The \textbf{sample completion function} for round $r$ is the function $\mathfrak{c}_r: \Omega \rightarrow \cp{\top, \bot}$ defined by
\begin{equation*}
    \mathfrak{c}_r\rp{\omega} =
    \begin{cases}
        \bot &\text{iff}\; \mathfrak{h}_r\rp{\omega} \in \mathcal{H}^+_r \\
        \top &\text{otherwise}
    \end{cases}
\end{equation*}

The sample completion function for round $r$ inputs a sample and outputs $\top$ if the infection history of the sample is complete at round $r$, and $\bot$ otherwise.
\end{definition}

\subsubsection{Infection status}
\label{subsubsection:infectionstatus}

As stated in \cref{subsubsection:game}, the infection status is defined as follows:
\begin{itemize}
    \item At the beginning of the game, all the nodes are healthy.
    \item During the first step of each round, the player selects a set of $S$ healthy nodes and infects them.
    \item During every subsequent step, every healthy node that has at least $\hat R$ infected predecessors is infected.
    \item The infection state at the end of a round is carried without change to the beginning of the next round.
\end{itemize}

In order to formalize the above in the definition of infection status, we preliminarly define \emph{infection sets}.

\begin{definition}[Infection set]
\label{definition:infectionset}
The \textbf{infection set} at round $r$ and step $i$ is the random variable $\hat W^r_i: \Omega \rightarrow \powerset{1..N}{}$ defined by
\begin{equation*}
    \hat W^r_i(\omega) = \cp{j \in 1..N \mid W^r_i[j](\omega) = \top}
\end{equation*}

The infection set $\hat W^r_i(\omega)$ represents the set of nodes that are infected in $\omega$ at round $r$ and step $i$.
\end{definition}

Like the sample history function, the definition of infection set relies on the definition of infection status $W$, which we can now define by cases.

\begin{definition}[Infection status]
\label{definition:infectionstatus}
Let $\omega = \rp{g, f} \in \Omega$. The \textbf{infection status} for round $r$, step $i$ and node $j$ is the random variable $W^r_i[j]: \Omega \rightarrow \cp{\top, \bot}$ defined by
\begin{align}
\label{equation:W00}
W^0_0[j](\omega) &\;=\; \bot \\
\label{equation:Wr0}
W^{r > 0}_0[j](\omega) &\;=\; W^{r - 1}_\infty[j](\omega) \\
\label{equation:W01}
W^0_1[j](\omega) &\;=\;
\begin{cases}
\top &\text{iff}\; j \in f_0 \\
W^0_0[j](\omega) &\text{otherwise}
\end{cases}\\
\label{equation:Wr1}
W^{r > 0}_1[j](\omega) &\;=\; 
\begin{cases}
\top &\text{iff}\;\mathfrak{c}_r(\omega) = \bot \;\wedge\; j \in f_r\rp{\mathfrak{h}_r(\omega)} \\
W^r_0[j](\omega) &\text{otherwise}
\end{cases}\\
\label{equation:Wri}
W^r_{i > 1}[j](\omega) &\;=\;
\begin{cases}
W^r_{i - 1}[j](\omega) &\text{iff}\; g_j \in \tilde{\mathcal{R}}^{\hat W^r_{i - 1}(\omega)} \\
\top &\text{otherwise}
\end{cases}
\end{align}

The above equations encode the following properties:
\begin{itemize}
    \item At the beginning of the game (\cref{equation:W00}), all nodes are healthy.
    \item The infection status at the beginning of round $r > 0$ (\cref{equation:Wr0}) is equal to the infection status at the end of round $r - 1$.
    \item During step $1$ of round $0$ (\cref{equation:W01}), all the nodes in $f_0$ are infected. Intuitively, the player selects $S$ nodes and infects them. Note how this choice is not informed by any history (following from \cref{definition:infectionstrategy}, $f_0$ is a set and not a function).
    \item During step $1$ of round $r > 0$ (\cref{equation:Wr1}), if $\omega$ is not complete (i.e., there are at least $S$ healthy nodes at the beginning of round $r$), all the nodes in $f_r\rp{\mathfrak{h}_r(\omega)}$ are infected. Intuitively, the player selects $S$ healthy nodes and infects them. This choice is informed by the infection history for round $r$ (see \cref{definition:samplehistoryfunction}).
    \item During step $i > 0$ of any round $r$ (\cref{equation:Wri}), all the nodes whose predecessor vector is not in the sub-threshold predecessor set (see \cref{definition:subthresholdpredecessorset}) of the infection set at step $i - 1$ are infected. In other words, the contagion rule (see \cref{subsubsection:contagionrule}) is applied, and all the nodes that have at least $\hat R$ infected predecessors are infected.
\end{itemize}
\end{definition}

Following from \cref{definition:infectionstatus}, we prove that nodes are never healed in a game of \contagion.

\begin{lemma}
\label{lemma:nohealing}
Let $j \in 1..N$, $r, r' \in 1..K$, $i, i' \in \mathbb{N}$, let $\omega \in \Omega$. If $(r', i') \geq (r, i)$, then
\begin{equation*}
    W^r_i[j](\omega) = \top \implies W^{r'}_{i'}[j](\omega)
\end{equation*}
\begin{proof}
Let $r'' \in 1..K$, $i'' \in \mathbb{N}$. Following from \cref{equation:W00,equation:Wr0,equation:W01,equation:Wr1,equation:Wri}, we have
\begin{equation}
\label{equation:nohealing}
    W^{r''}_{i'' + 1}[j](\omega) \neq W^{r''}_{i''}[j](\omega) \implies W^{r''}_{i'' + 1}[j](\omega) = \top
\end{equation}

The lemma is proved by induction on \cref{equation:Wr0,equation:nohealing}.
\end{proof}
\end{lemma}

\begin{corollary}
The infection set $\hat W^r_i(\omega)$ is non-decreasing in $(r, i)$.
\end{corollary}

\subsubsection{Infection size, frontier size and infected predecessors count}

In \cref{subsubsection:infectionstatus}, we defined the infection status $W^r_i[j]$ as a function on the sample space (see \cref{definition:infectionstatus}). We also defined the infection set $\hat W^r_i$ as the set of nodes for which $W^r_i = \top$ (see \cref{definition:infectionset}).

As stated in \cref{subsection:randomvariables}, the infection size $N^r_i$ represents the number of infected nodes at round $r$ and step $i$, and the frontier size $U^r_{i > 0}$ represents the number of nodes that are infected at round $r$ and step $i$, but not at step $i - 1$. We can formalize the above in the following definitions.

\begin{definition}[Infection size]
The \textbf{infection size} for round $r$ and step $i$ is the random variable $N^r_i: \Omega \rightarrow 0..N$ defined by
\begin{equation*}
    N^r_i(\omega) = \abs{\hat W^r_i(\omega)}
\end{equation*}

The infection size counts the infected nodes at step $(r, i)$.
\end{definition}

\begin{definition}[Frontier size]
The \textbf{frontier size} for round $r$ and step $i$ is the random variable $U^r_i: \Omega \rightarrow 0..N$ defined by
\begin{equation*}
    U^r_{i > 0}(\omega) = N^r_i(\omega) - N^r_{i - 1}(\omega)
\end{equation*}

The infection size counts the nodes that are infected at step $(r, i)$, but not at step $(r, i - 1)$.
\end{definition}

As stated in \cref{subsection:randomvariables}, the infected predecessors count of node $j$ for round $r$ and step $i$ represents the number of predecessors of node $j$ that are infected at round $r$ and step $i$. We can formalize this definition in the following.

\begin{definition}[Infected predecessors count]
\label{definition:infectedpredecessorcount}
Let $\omega = (g, f) \in \Omega$. The \textbf{infected predecessors count} of node $j$ for round $r$ and step $i$ is the random variable $V^r_i[j]: \Omega \rightarrow 0..R$ defined by
\begin{equation*}
    V^r_i[j](\omega) = \abs{\cp{k \mid \rp{g_{j, k}}\in \hat W^r_i(\omega)}}
\end{equation*}

The infected predecessors count counts the number of predecessors of node $j$ that are infected at step $(r, i)$.
\end{definition}

\begin{lemma}
\label{lemma:subthresholdinfectedpredecessorcount}
Let $\omega = (g, f) \in \Omega$, let $j \in 1..N$, $r \in 1..K$, $i \in \mathbb{N}$. Then
\begin{equation*}
    g \in \tilde{\mathcal{R}}^{\hat W^r_i[j]} \Longleftrightarrow V^r_i[j](\omega) \leq \hat R
\end{equation*}
\begin{proof}
It follows immediately from \cref{definition:subthresholdpredecessorset,definition:infectedpredecessorcount}.
\end{proof}
\end{lemma}

\subsection{Contagion step}
\label{subsection:contagionstep}

In \cref{subsection:randomvariablesassamplefunctions}, we expressed the random variables we introduced in \cref{subsection:randomvariables} as functions over the elements of the sample space we defined in \cref{subsection:samplespace}. As we established in \cref{subsection:goal}, the goal of this appendix is to compute the distribution underlying $N^K_\infty$ (see \cref{equation:pebblinggame}).

Here we focus on the contagion steps of a round of \contagion. As per \cref{equation:Wri}, at every step $(r, i)$ such that $i > 1$, all the healthy nodes that have at least $\hat R$ infected predecessors become infected. 

In this section, we show that a contagion step defines a Markov chain with states $(\bar N^r_i, \bar U^r_i)$. More formally, we show that a transition matrix $\mathcal{M}$ exists such that, for every $(\bar N, \bar U)$, $(\bar N', \bar U')$ and for every $r \in 1..K$, $i \geq 1$,
\begin{equation}
    \label{equation:transitionmatrix}
    \mathcal{M}^{\bar N', \bar U'}_{\bar N, \bar U} = \prob{N^r_{i + 1} = \bar N', U^r_{i + 1} = \bar U' \mid N^r_i = \bar N, U^r_i = \bar U}
\end{equation}

Intuitively, this means that, once the infection size and the frontier size at step $(r, i)$ are determined, no other knowledge is needed to compute the probability distribution underlying the frontier size at step $(r, i + 1)$. This means, in particular, that the player's infection strategy does not affect the end result of the game. This result is somewhat unsurprising: since the player has no knowledge of the multigraph on which \contagion\ is played, the player has no way to meaningfully distinguish two nodes by the effect that their infection will have on the system. Since the number of infected nodes per round is determined, every choice of the player can be shown to be effectively equivalent to the infection of $S$ random healthy nodes.

\subsubsection{Roadmap}

\begin{notation}[Markov states]
We use $\ap{\bar N^r_i, \bar U^r_i}$ to denote the subset of the sample space $\Omega$ that satisfies $N^r_i(\omega \in \Omega) = \bar N^r_i, U^r_i(\omega \in \Omega) = \bar U^r_i$. 

Equivalently,
\begin{equation*}
    \ap{\bar N^r_i, \bar U^r_i} = \rp{N^r_i}^{-1}\rp{\bar N^r_i} \cap \rp{U^r_i}^{-1}\rp{\bar U^r_i}
\end{equation*}
\end{notation}

In order to show that a infection step defines a Markov chain with states $(\bar N^r_i, \bar U^r_i)$, we:
\begin{itemize}
    \item Define a set of \emph{partition functions} $\mathcal{S}^r_i: \Omega \rightarrow \powerset{\Omega}{}$ that map elements of $\Omega$ into well-structured subsets of $\Omega$. Intuitively, $\mathcal{S}^r_i$ maps a sample $\omega = (g, f)$ to a set of samples that are \emph{similar to it} (by a notion of similarity that we define later).
    \item Let $\omega' \in \mathcal{S}^r_i(\omega)$. We show that $\omega$ and $\omega'$ result in the same infection history up to step $(r, i)$.
    \item We show that $\mathcal{S}^r_i$ can be used to define an equivalence relation on the sample space $\Omega$.
    \item Let $\omega$ be equivalent to $\omega'$ through $\mathcal{S}^r_i$. We show that, since $N^r_i(\omega) = N^r_i(\omega')$ and $U^r_i(\omega) = U^r_i(\omega')$, then $\mathcal{S}^r_i$ can be used to quotient $\ap{\bar N^r_i, \bar U^r_i}$.
    \item Let $r \in 1..K$, $i > 1$. We use $\mathcal{S}^r_i$ to partition $\ap{\bar N^r_i, \bar U^r_i}$ in $s_1, \ldots, s_q$. We show that the probability of $\omega$ being in $\ap{\bar N^r_{i + 1}, \bar U^r_{i + 1}}$ given that $\omega$ is in $s_h$ is analitically computable and independent of $h$.
    \item We use the independence across partitions to compute the terms of $\mathcal{M}^{\bar N', \bar U'}_{\bar N, \bar U}$
\end{itemize}

\begin{figure}
\centering
\includegraphics[scale=0.50]{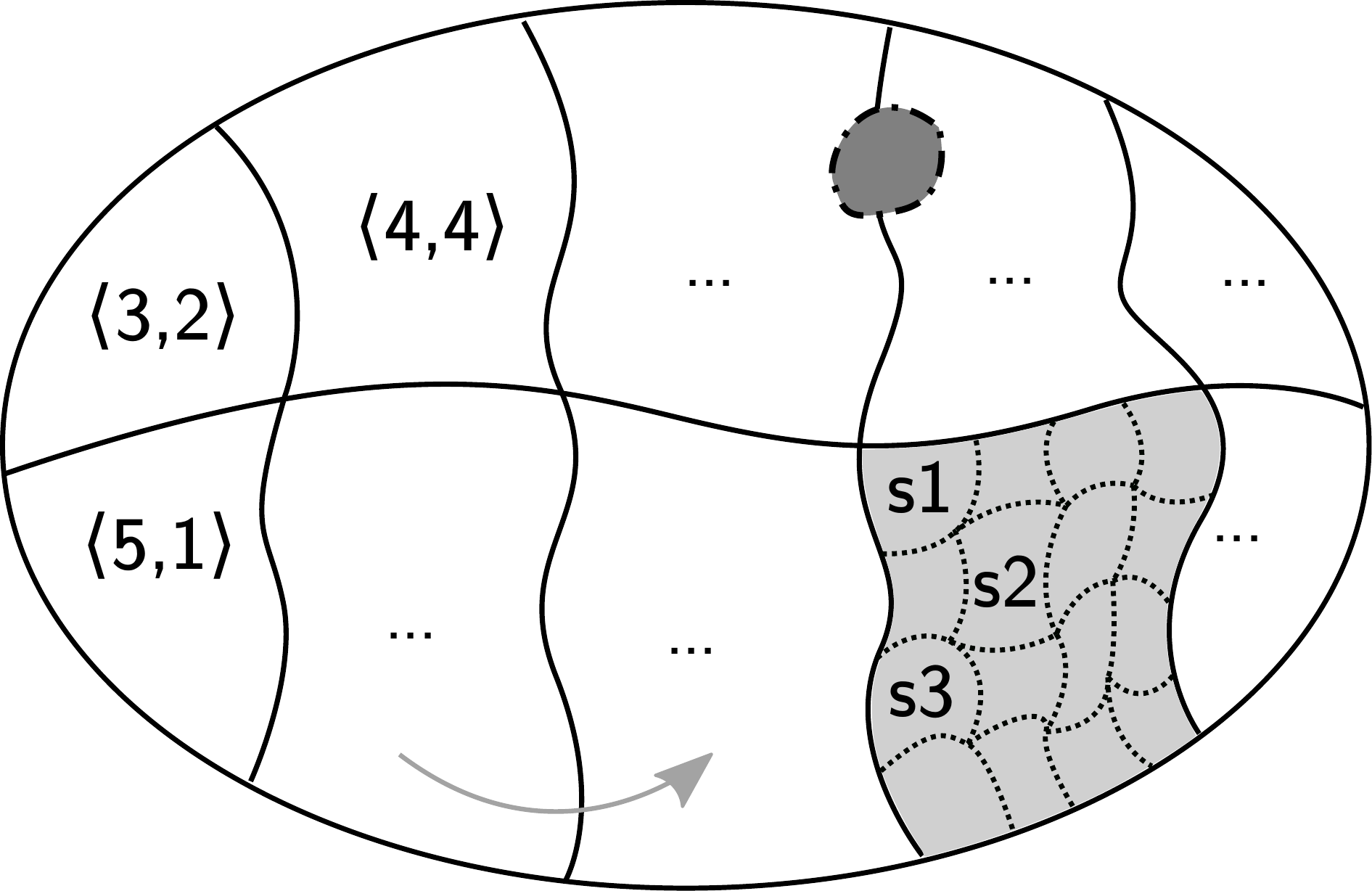}
\caption{An illustration of sample space and the steps needed to show that a contagion step defines a Markov chain. The grey arrow represents a transition from a state to another. One of the states is further partitioned by $\mathcal{S}^r_i$. The dark grey area represents a case that we prove won't happen.}
\label{figure:contagion:contagion-6}
\end{figure}

\subsubsection{Partition functions}
\label{subsubsection:partitionfunctions}

We start by defining a set of \emph{partition functions} $\mathcal{S}^r_i: \Omega \rightarrow \powerset{\Omega}{}$ that map elements of $\Omega$ into subsets of $\Omega$. Intuitively, a partition function maps a sample to a set of samples that are \emph{similar} to it. 

Let $\omega = (g, f) \in \Omega$, let $\omega' = (g', f') \in \mathcal{S}^r_i(\omega)$. We define $\mathcal{S}^r_i$ such that the following hold:
\begin{itemize}
    \item $f' = f$, i.e., the player's strategy is left unchanged by $\mathcal{S}^r_i$.
    \item Let $j \in 1..N$ be a node. If $j$ is infected in $\omega$ at step $(r, i)$, then $g'_j = g_j$. In other words, the predecessors of a node that is infected at step $(r, i)$ in $\omega$ are left unchanged by $\mathcal{S}^r_i$.
    \item Let $j \in 1..N$ be a node. If $j$ is not infected in $\omega$ at step $(r, i)$, then $g'_j$ is an element of the sub-threshold predecessor set of $\hat W^r_{i - 1}(\omega)$. In other words, the predecessors of a node that is not infected at step $(r, i)$ in $\omega$ can be changed by $\mathcal{S}^r_i$, as long as no more than $\hat R$ of them are infected in $\omega'$ at step $(r, i - 1)$. Intuitively, we allow the predecessors of $j$ to change in a way that does not make it infected in $\omega$ at step $(r, i)$.
\end{itemize}

We formalize the above in the following definition.
\begin{definition}[Partition function]
\label{definition:partitionfunction}
Let $r \in 1..K$, $i \geq 1$, let $\omega = (g, f) \in \Omega$. The \textbf{partition function} for round $r$ and step $i$ is the function $\mathcal{S}^r_i: \Omega \rightarrow \powerset{\Omega}{}$ defined by
\begin{eqnarray}
    \mathcal{S}^r_i(\omega) &=& \rp{\prod_{j = 1}^N \mathcal{S}^r_i[j](\omega)} \times \cp{f} \\
    \mathcal{S}^r_i[j](\omega) &=&
    \begin{cases}
        \cp{g_j} &\text{iff}\; W^r_i[j](\omega) = \top \\
        \tilde{R}^{\hat W^r_{i - 1}(\omega)} &\text{otherwise}
    \end{cases}
\end{eqnarray}
\end{definition}

\subsubsection{Infection history}
\label{subsubsection:infectionhistory}

In \cref{subsubsection:partitionfunctions} we defined a set of partitions functions that map a sample $\omega \in \Omega$ to a set of samples that are \emph{similar} to $\omega$. 

Let $\omega = (g, f) \in \Omega$. We designed $\mathcal{S}^r_i$ to leave unchanged the player's strategy and the predecessors of every node that is infected in $\omega$ at step $(r, i)$. The predecessors of the nodes that are not infected in $\omega$ at step $(r, i)$ can change, as long as less than $\hat R$ of them are among the nodes that are infected in $\omega$ at step $(r, i - 1)$.

Intuitively, $\mathcal{S}^r_i$ is designed to alter the topology of $g$ in a way that does not affect its infection history: since the predecessors of the nodes that are not infected in $\omega$ at step $(r, i)$ are not changed, they will still be infected in $\omega'$. Similarly, if a node is not infected in $\omega$ at step $(r, i)$, its predecessors are not changed in a way that makes it infected in $\omega'$ at step $(r, i)$.

In this section, we formally prove this intuitive result.
\begin{lemma}
\label{lemma:infectionhistory}
Let $j \in 1..N$, let $\omega, \omega' \in \Omega$. If $\omega' \in \mathcal{S}^r_i(\omega)$, then for every $(r', i') \leq (r, i)$
\begin{equation*}
    W^{r'}_{i'}[j](\omega') = W^{r'}_{i'}[j](\omega)
\end{equation*}
\begin{proof}
Let $\omega = (g, f)$ and $\omega' = (g', f')$. We prove the lemma by induction. We start by noting that, following from \cref{equation:W00},
\begin{equation*}
    W^0_0[j](\omega') = \bot = W^0_0[j](\omega)
\end{equation*}

Now, assume that $(r', i') < (r, i)$ and, for all $j \in 1..N$, $\hat W^{r'}_{i'}[j](\omega') = \hat W^{r'}_{i'}[j](\omega)$.\\

If $r' = 0$ and $i' = 0$, then from \cref{equation:W01} it follows that, if $j \in \rp{f_0 = f'_0}$,
\begin{equation*}
    W^0_1[j](\omega') = \top = W^0_1[j](\omega)
\end{equation*}
and, otherwise,
\begin{equation*}
    W^0_1[j](\omega') = W^0_0[j](\omega') = W^0_0[j](\omega) = W^0_1[j](\omega)
\end{equation*}

If $r' > 0$ and $i' = 0$, then $\mathfrak{h}_r(\omega') = \mathfrak{h}_r(\omega)$. Following from \cref{equation:Wr1}, if $\rp{\mathfrak{c}_r(\omega) = \mathfrak{c}_r(\omega')} = \bot$ and $j \in \rp{f_r(\mathfrak{h}_r(\omega)) = f'_r(\mathfrak{h}_r(\omega'))}$, then
\begin{equation*}
    W^{r'}_1[j](\omega') = \top = W^{r'}_1[j](\omega)
\end{equation*}
and otherwise
\begin{equation*}
    W^{r'}_1[j](\omega') = W^{r'}_0[j](\omega') = W^{r'}_0[j](\omega) = W^{r'}_1[j](\omega)
\end{equation*}

We now consider the case $i' \geq 1$. We start by noting that, since $\hat W^{r'}_{i'}(\omega') = \hat W^{r'}_{i'}(\omega)$, then $\tilde{\mathcal{R}}^{\hat W^{r'}_{i'}(\omega')} = \tilde{\mathcal{R}}^{\hat W^{r'}_{i'}(\omega)}$. \\

If $W^r_i[j](\omega) = \top$, then $g_j = g'_j$. Following from \cref{equation:Wri}, if $\rp{g'_j = g_j} \in \rp{\tilde{\mathcal{R}}^{\hat W^{r'}_{i'}(\omega')} = \tilde{\mathcal{R}}^{\hat W^{r'}_{i'}(\omega)}}$, then
\begin{equation*}
    W^{r'}_{i'+1}[j](\omega') = W^{r'}_{i'}[j](\omega') = W^{r'}_{i'}[j](\omega) = W^{r'}_{i' + 1}[j](\omega)
\end{equation*}
and otherwise
\begin{equation*}
    W^{r'}_{i' + 1}[j](\omega') = \top = W^{r'}_{i' + 1}[j](\omega)
\end{equation*}

If $W^r_i[j](\omega) = \bot$, then $g'_j \in \tilde{\mathcal{R}}^{\hat W^r_{i - 1}(\omega)}$. Noting that $(r', i') \leq (r, i - 1)$, from \cref{lemma:nohealing} it follows that $\hat W^{r'}_{i'}(\omega) \subseteq \hat W^r_{i - 1}(\omega)$, and consequently, from \cref{equation:Wri}, we have
\begin{equation*}
    g'_j \in \rp{\tilde{\mathcal{R}}^{\hat W^r_{i - 1}(\omega)} \subseteq \tilde{\mathcal{R}}^{\hat W^{r'}_{i'}(\omega)} = \tilde{\mathcal{R}}^{\hat W^{r'}_{i'}(\omega')}}
\end{equation*}
Moreover, since $W^r_i[j](\omega) = \bot$, from \cref{lemma:nohealing} it follows 
\begin{equation*}
    W^{r'}_{i' + 1}(\omega) = W^{r'}_{i'}(\omega) = W^{r'}_{i'}(\omega') = \bot
\end{equation*}
and therefore
\begin{equation*}
    W^{r'}_{i' + 1}(\omega') = W^{r'}_{i'}(\omega') = W^{r'}_{i' + 1}(\omega)
\end{equation*}

Finally, if $i = \infty$, then following from \cref{equation:Wr0} we have
\begin{equation*}
    W^{r' + 1}_0[j](\omega') = W^{r'}_\infty[j](\omega') = W^{r'}_\infty[j](\omega) = W^{r' + 1}_0[j](\omega)
\end{equation*}
\end{proof}
\end{lemma}

\begin{corollary}
\label{corollary:partitioninfectionfrontiersize}
Let $\omega, \omega' \in \Omega$. If $\omega' \in \mathcal{S}^r_i(\omega)$, then
\begin{eqnarray*}
    N^r_i(\omega') = N^r_i(\omega) \\
    U^r_i(\omega') = U^r_i(\omega)
\end{eqnarray*}
\end{corollary}

\subsubsection{Equivalence relation}
\label{subsubsection:equivalencerelation}

In \cref{subsubsection:partitionfunctions}, we introduced a set of functions $\mathcal{S}^r_i: \Omega \rightarrow \powerset{\Omega}{}$ that map a sample into a set of \emph{similar} samples. In \cref{subsubsection:infectionhistory}, we proved that, if $\omega \in \Omega$ and $\omega' \in \mathcal{S}^r_i(\omega)$, then $\omega$ and $\omega'$ produce the same infection history (i.e., the same values for $\hat W^r_i$) up to round $r$ and step $i$.

In this section, we show that $\mathcal{S}^r_i$ can be used to define a equivalence relation on $\Omega$.

\begin{lemma}
\label{lemma:sameoutputpartitionfunction}
Let $\omega, \omega' \in \Omega$. If $\omega' \in \mathcal{S}^r_i(\omega)$, then $\mathcal{S}^r_i(\omega') = \mathcal{S}^r_i(\omega)$.

\begin{proof}
Let $\omega = (g, f)$, $\omega' = (g', f')$. Following from \cref{lemma:infectionhistory}, for every $j$ we have
\begin{eqnarray*}
    W^r_i[j](\omega') = W^r_i[j](\omega) \\
    W^r_{i - 1}[j](\omega') = W^r_{i - 1}[j](\omega)
\end{eqnarray*}

Following from \cref{definition:partitionfunction}, if $W^r_i[j](\omega) = \top$, then $g'_j = g_j$. Consequently,
\begin{equation*}
    \mathcal{S}^r_i[j](\omega') = \cp{g'_j} = \cp{g_j} = \mathcal{S}^r_i[j](\omega)
\end{equation*}

If $W^r_i[j](\omega) = \bot$, then
\begin{equation*}
    \mathcal{S}^r_i[j](\omega') = \tilde{\mathcal{R}}^{\hat W^r_{i - 1}[j](\omega')} = \tilde{\mathcal{R}}^{\hat W^r_{i - 1}[j](\omega)} = \mathcal{S}^r_i[j](\omega)
\end{equation*}

Therefore,
\begin{equation*}
    \mathcal{S}^r_i(\omega') = \prod_{j = 1}^N \mathcal{S}^r_i[j](\omega') = \prod_{j = 1}^N \mathcal{S}^r_i[j](\omega) = \mathcal{S}^r_i(\omega)
\end{equation*}
\end{proof}
\end{lemma}

\begin{definition}[Partition relation]
Let $\omega, \omega' \in \Omega$. If $\omega' \in \mathcal{S}^r_i(\omega)$, then $\omega'$ has a \textbf{partition relation} with $\omega$ at round $r$ and step $i$:
\begin{equation*}
    \omega' \stackrel{(r, i)}{\sim} \omega
\end{equation*}
\end{definition}

\begin{lemma}
\label{lemma:equivalencerelation}
$\stackrel{(r, i)}{\sim}$ is an equivalence relation.
\begin{proof}
Let $j \in 1..N$, let $\omega \in \Omega$. Following from \cref{equation:Wri}, if $W^r_i[j](\omega) = \bot$, then $g_j \in \tilde{\mathcal{R}}^{\hat W^r_{i - 1}(\omega)}$. Consequently, following from \cref{definition:partitionfunction}, if $W^r_i[j](\omega) = \top$, then
\begin{equation*}
    g_j \in \cp{g_j} = \mathcal{S}^r_i[j](\omega)
\end{equation*}
and if $W^r_i[j](\omega) = \bot$, then
\begin{equation*}
    g_j \in \tilde{\mathcal{R}}^{\hat W^r_{i - 1}[j](\omega)} = \mathcal{S}^r_i[j](\omega)
\end{equation*}

Therefore, $\omega \in \mathcal{S}^r_i(\omega)$, and
\begin{equation*}
    \omega \stackrel{(r, i)}{\sim} \omega
\end{equation*}
therefore $\stackrel{(r, i)}{\sim}$ is reflexive.

Let $\omega' \in \mathcal{S}^r_i(\omega)$. By \cref{lemma:sameoutputpartitionfunction}, $\mathcal{S}^r_i(\omega') = \mathcal{S}^r_i(\omega)$. Consequently
\begin{equation*}
    \omega \in \rp{\mathcal{S}^r_i(\omega) = \mathcal{S}^r_i(\omega')}
\end{equation*}
and
\begin{equation*}
    \omega' \stackrel{(r, i)}{\sim} \omega \implies \omega \stackrel{(r, i)}{\sim} \omega'
\end{equation*}
therefore $\stackrel{(r, i)}{\sim}$ is symmetric.

Let $\omega'' \in \mathcal{S}^r_i(\omega')$. Again by \cref{lemma:sameoutputpartitionfunction},
\begin{equation*}
    \omega'' \in \rp{\mathcal{S}^r_i(\omega') = \mathcal{S}^r_i(\omega)}
\end{equation*}
and
\begin{equation*}
    \omega' \stackrel{(r, i)}{\sim} \omega, \omega'' \stackrel{(r, i)}{\sim} \omega' \implies \omega'' \stackrel{(r, i)}{\sim} \omega
\end{equation*}
therefore, $\stackrel{(r, i)}{\sim}$ is transitive.
\end{proof}
\end{lemma}

\subsubsection{Transition probabilities}

In \cref{subsubsection:equivalencerelation} we showed that the partition function we introduced in \cref{subsubsection:partitionfunctions} can be used to induce an equivalence relation on the sample space $\Omega$.

In this section, we use this result to show that a contagion step defines a Markov chain with states $(\bar N^r_i, \bar U^r_i)$, and compute the values of its associated transition matrix $\mathcal{M}$.

More formally, let $r \in 1..K$, $i \geq 1$. In this section, we compute
\begin{equation*}
    \prob{\bar N^r_{i + 1}, \bar U^r_{i + 1} \mid \bar N^r_i, \bar U^r_i}
\end{equation*}
and we show that its value is independent of the player's strategy.

As we established in \cref{lemma:equivalencerelation}, $\stackrel{(r, i)}{\sim}$ is an equivalence relation on $\Omega$. Moreover, let $\omega \in \Omega$, by \cref{corollary:partitioninfectionfrontiersize} we have $\mathcal{S}^r_i(\omega) \subseteq \ap{\bar N^r_i, \bar U^r_i}$.

We can therefore use $\stackrel{(r, i)}{\sim}$ to partition $\ap{\bar N^r_i, \bar U^r_i}$:
\begin{equation*}
    \cp{s_1, \ldots, s_q} = \frac{\ap{\bar N^r_i, \bar U^r_i}}{\stackrel{(r, i)}{\sim}}
\end{equation*}

By the law of total probability,
\begin{eqnarray*}
    \prob{\bar N^r_{i + 1}, \bar U^r_{i + 1} \mid \bar N^r_i, \bar U^r_i} &=& \prob{\bar N^r_{i + 1}, \bar U^r_{i + 1} \mid \ap{\bar N^r_i, \bar U^r_i}} \\
    &=& \sum_{l = 1}^q \prob{\bar N^r_{i + 1}, \bar U^r_{i + 1} \mid s_l} \prob{s_l \mid \ap{\bar N^r_i, \bar U^r_i}}
\end{eqnarray*}

Note how $\prob{s_l \mid \ap{\bar N^r_i, \bar U^r_i}}$ is unknown, as it depends on the probability distribution underlying the player's strategy. For a given $h$, we instead focus on computing $\prob{\bar N^r_{i + 1}, \bar U^r_{i + 1} \mid s_h}$. \\

\paragraph{Roadmap} In order to compute $\prob{\bar N^r_{i + 1}, \bar U^r_{i + 1} \mid s_h}$, we compute the probability for a node that is not infected in $s_h$ at step $(r, i)$ to become infected at time $(r, i + 1)$. Let $j$ be a node that is not infected in $s_h$ at step $(r, i)$. We compute the probability of it becoming infected at step $(r, i + 1)$ by first computing the probability distribution underlying $V^r_{i - 1}[j]$. Given $\bar V^r_{i - 1}[j]$, we then compute the probability distribution underlying $V^r_i[j]$, and threshold it with $\hat R$ to compute the probability of $j$ becoming infected at step $i + 1$. \\

\begin{notation}[Kronecker delta]
We use $\delta$ to denote the \textbf{Kronecker delta}. Let $i, j \in \mathbb{N}$, then
\begin{equation*}
    \delta_{i, j} = I(i = j)
\end{equation*}
\end{notation}

Let $\bar \omega = (\bar g, \bar f) \in s_h$ be an example of $s_h$. Let $W$, $\cancel W$ denote the set of nodes that are infected and not infected in $\bar \omega$ at step $(r, i)$, respectively:
\begin{eqnarray*}
    W = \cp{w_1, \ldots, w_n} &=& \hat W^i_r(\bar \omega) \\
    \cancel W = \cp{\cancel w_1, \ldots, \cancel w_m} &=& 1..N \setminus \hat W^i_r(\bar \omega)
\end{eqnarray*}
with $n = N^r_i(\bar \omega)$ and $m = N - n$. Let $\omega = (g, f)$, following from \cref{lemma:subthresholdinfectedpredecessorcount} we have
\begin{equation*}
    \rp{\omega \in s_h} \Longleftrightarrow \rp{g_{w_1} = \bar g_{w_1}, \ldots, g_{w_n} = \bar g_{w_n}, V^r_{i - 1}[\cancel w_1] \leq \hat R, \ldots, V^r_{i - 1}[\cancel w_m] \leq \hat R}
\end{equation*}

Let $j \in \cancel W$, i.e., $W^r_i[j] = \bot$. Using the independence of the distribution of each predecessor vector in $s_h$ (see \cref{equation:randompredecessormatrix,definition:partitionfunction}), we can compute the probability distribution underlying $V^r_{i - 1}[j]$ in $s_h$:
\begin{eqnarray*}
    \lhs \prob{\bar V^r_{i - 1}[j] \mid \cancel{W^r_i[j]}, s_h} \\
    &=& \prob{\bar V^r_{i - 1}[j] \mid \cancel{W^r_i[j]}, \bar g_{w_1}, \ldots, \bar g_{w_n}, r_{i - 1}[\cancel w_1] < \hat R, \ldots, V^r_{i - 1}[\cancel w_m] < \hat R} \\
    &=& \prob{\bar V^r_{i - 1}[j] \mid \cancel{W^r_i[j]}, V^r_{i - 1}[\cancel w_1] < \hat R, \ldots, V^r_{i - 1}[\cancel w_m] < \hat R} \\
    &=& \prob{\bar V^r_{i - 1}[j] \mid V^r_{i - 1}[j] < \hat R}
\end{eqnarray*}

Using Bayes' theorem we get
\begin{equation}
\label{equation:predecessorbayes}
    \prob{\bar V^r_{i - 1} \mid V^r_{i - 1} < \hat R} = \frac{\prob{V^r_{i - 1} < \hat R \mid \bar V^r_{i - 1}} \prob{\bar V^r_{i - 1}}}{\prob{V^r_{i - 1} < \hat R}}
\end{equation}

Following from \cref{lemma:randompredecessorvector}, each predecessor of $j$ is independently selected with uniform probability. Given $\bar N^r_{i - 1}$, each predecessor of $j$ has a probability $l \rp{\bar N^r_{i - 1} / N}$ of being in $\hat W^r_{i - 1}$. The unconditioned number of infected predecessors of $j$ is therefore binomially distributed:
\begin{equation}
\label{equation:unconditionedpredecessor}
    \prob{\bar V^r_{i - 1}} = \bin{E}{l \frac{\bar N^r_{i - 1}}{N}}{\bar V^r_{i - 1}}
\end{equation}

Plugging \cref{equation:unconditionedpredecessor} in \cref{equation:predecessorbayes} and noting that $N^r_{i - 1} = N^r_i - U^r_i$ we get
\begin{equation*}
    \prob{\bar V^r_{i - 1} \mid V^r_{i - 1} < \hat R} = \frac{I\rp{\bar V^r_{i - 1} < \hat R} \bin{R}{l \frac{\bar N^r_i - \bar U^r_i}{N}}{\bar V^r_{i - 1}}}{
    \sum_{\bar V = 0}^{\hat R - 1} \bin{R}{l \frac{\bar N^r_i - \bar U^r_i}{N}}{\bar V}}
\end{equation*}

We now compute the distribution underlying $V^r_i$, given $\bar V^r_{i - 1}$, $\cancel{W^r_i}$ and $s_h$. Given $\bar V^r_{i - 1}$, $\cancel{W^r_i}$ and $s_h$, $j$ has $E - \bar V^r_{i - 1}$ predecessors that are not in $\hat W^r_{i - 1}$. Let $g_{j, k}$ be a predecessor of $j$ that is not in $\hat W^r_{i - 1}$, we have
\begin{eqnarray*}
    \prob{g_{j, k} \in \hat W^r_i \mid g_{j, k} \notin \hat W^r_{i - 1}} &=& \frac{\prob{g_{j, k} \in \hat W^r_i, g_{j, k} \notin \hat W^r_{i - 1}}}{\prob{g_{j, k} \notin \hat W^r_{i - 1}}} \\
    &=& \frac{l \frac{\bar U_i}{N}}{1 - l \frac{\bar N^r_i - \bar U^r_i}{N}}
\end{eqnarray*}

Following from \cref{equation:randompredecessormatrix}, each predecessor of $j$ that is not in $\hat W^r_{i - 1}$ has an independent chance of being in $\hat W^r_i$. Therefore, the number of newly infected predecessors for $j$ at step $i$ is binomially distributed:
\begin{equation}
\label{equation:predecessordiff}
    \prob{\bar V^r_i \mid \bar V^r_{i - 1}, \cancel{W^r_i}, s_h} = \bin{R - \bar V^r_{i - 1}}{\frac{l \frac{\bar U_i}{N}}{1 - l \frac{\bar N^r_i - \bar U^r_i}{N}}}{\bar V^r_i - \bar V^r_{i - 1}}
\end{equation}

Using the law of total probability, we can now use \cref{equation:predecessorbayes,equation:predecessordiff} to compute the probability distribution underlying $V^r_i[j]$, given $\cancel{W^r_i}$ and $s_h$:
\begin{equation*}
    \prob{\bar V^r_i \mid \cancel{W^r_i}, s_h} = \sum_{\bar V^r_{i - 1} = 0}^{\hat R - 1} \prob{\bar V^r_i \mid \bar V^r_{i - 1}, \cancel{W^r_i}, s_h} \prob{\bar V^r_{i - 1} \mid \cancel{W^r_i}, s_h}
\end{equation*}

Finally, following from \cref{lemma:subthresholdinfectedpredecessorcount}, we get the probability of $W^r_i[j]$, given $\cancel{W^r_i}$ and $s_h$:
\begin{equation*}
    \prob{W^r_i \mid \cancel{W^r_i}, s_h} = \sum_{\bar V^r_i = \hat R}^R \prob{\bar V^r_i \mid \cancel{W^r_i}, s_h}
\end{equation*}

Since each of the $N - \bar N_i$ nodes in $\cancel W$ has an independent probability of becoming infected at round $i + 1$, the frontier size at step $i + 1$, given $s_h$ is binomially distributed:
\begin{equation*}
    \prob{\bar N^r_{i + 1}, \bar U^r_{i + 1} \mid s_h} = \bin{N - \bar N_i}{\prob{W^r_{i + 1} \mid \cancel{W^r_i}, s_h}}{\bar U^r_{i + 1}} \delta_{\bar N^r_{i + 1} - \bar N^r_i, \bar U^r_{i + 1}}
\end{equation*}

We can now note how, when computing $\prob{\bar N^r_{i + 1}, \bar U^r_{i + 1} \mid s_h}$, the condition on $s_h$ reduces only to a condition on the values of $\bar N^r_i$ and $\bar U^r_i$. Since $s_1, \ldots, s_q$ share the same values of $(\bar N^r_i, \bar U^r_i)$, the transition probability for the Markov chain underlying a contagion step reduces to
\begin{align}
    \label{equation:transitionprobability}
    \prob{\bar N^r_{i + 1}, \bar U^r_{i + 1} \mid \bar N^r_i, \bar U^r_i} &= \sum_{l = 1}^q \prob{\bar N^r_{i + 1}, \bar U^r_{i + 1} \mid s_l} \prob{s_l \mid \ap{\bar N^r_i, \bar U^r_i}} \\
    &= \prob{\bar N^r_{i + 1}, \bar U^r_{i + 1} \mid s_h} \sum_{l = 1}^q \prob{s_l \mid \ap{\bar N^r_i, \bar U^r_i}} \nonumber \\
    &= \prob{\bar N^r_{i + 1}, \bar U^r_{i + 1} \mid s_h} \nonumber
\end{align}

\subsection{Final infection size}

In \cref{subsection:contagionstep}, we showed that a contagion steps defines a Markov chain with states $(\bar N^r_i, \bar U^r_i)$, and we computed the values of its associated transition matrix $\mathcal{M}$. In this section, we use this result to achieve our goal to compute the probability distribution underlying the infection size at the end of a game of \contagion.

As we established in \cref{subsection:contagionstep}, provided with $\prob{\bar N^r_i, \bar U^r_i}$, we can compute $\prob{\bar N^r_{i + 1}, \bar U^r_{i + 1}}$. Moreover, following from \cref{corollary:finiterounds}, every configuration $\prob{\bar N^r_1, \bar U^r_1}$ converges in a finite number of steps $i^*$ to satisfy
\begin{eqnarray*}
    \prob{\bar N^r_i, \bar U^r_i} = \prob{\bar N^r_{i + 1}, \bar U^r_{i + 1}} &\;& \forall i \geq i^* \\
    \prob{U^r_i > 0} = 0 &\;& \forall i \geq i^*
\end{eqnarray*}

It is easy to see that the first step of each round (where the player selects $S$ healthy node and infects them) also defines a Markov chain that deterministically increases, if possible, the infection size by $S$.

Specifically, the transition probabilities from step $0$ to step $1$ in each round are defined by:
\begin{equation}
\label{equation:transitionbeginningofround}
\prob{\bar N^r_1, \bar U^r_1} =
\begin{cases}
\prob{N^r_0 = \bar N^r_1 - S} &\;\text{iff}\; \bar N^r_1 \geq S, \bar U^r_1 = S \\
\prob{N^r_0= \bar N^r_1} &\;\text{iff}\; \bar N^r_1 > (N - S), \bar U^r_1 = 0 \\
0 &\;\text{otherwise}
\end{cases}
\end{equation}

The distribution underlying the final infection size can be computed as follows:
\begin{itemize}
    \item The distribution underlying the first step of the game is known:
    \begin{equation*}
        \prob{\bar N^0_0, \bar U^0_0} = \delta_{\bar N^0_0, 0}\delta_{\bar U^0_0, 0}
    \end{equation*}
    \item For $K$ rounds:
    \begin{itemize}
        \item If $r > 0$, then $\prob{\bar N^r_0, \bar U^r_0} = \prob{\bar N^{r - 1}_\infty, \bar U^{r - 1}_\infty}$.
        \item Apply \cref{equation:transitionbeginningofround} to compute $\prob{\bar N^r_1, \bar U^r_1}$.
        \item Until convergence:
        \begin{itemize}
            \item Apply \cref{equation:transitionprobability} to compute $\prob{\bar N^r_i, \bar U^r_i}$.
        \end{itemize}
    \end{itemize}
\end{itemize}

\end{document}